\documentclass{optica-article}
\journal{opticajournal} 
\articletype{Review} 

\usepackage{lineno}
\usepackage{orcidlink}
\usepackage{float}
\usepackage{xcolor,tikz}
\usepackage{framed}
\newcommand{\mathsfit}[1]{\text{\sffamily\slshape #1}}
\newcommand{\slfrac}[2]{\ensuremath{{}^{\scriptscriptstyle #1}\!/\!_{\scriptscriptstyle #2}}}
\newcommand{\chapcite}[2]{(see~Chapter~#1~of~\cite{#2})}
\newcommand{\seccite}[2]{(see~Sec.~#1~of~\cite{#2})}


\begin{document}

\title{Photoemission and absorption under classical and entangled-photon-pair illumination}

\author{Malvin Carl Teich,\authormark{1,2,3,4,5,*}\orcidlink{0000-0001-8164-4622}
Mark C. Booth,\authormark{2,5} Francesco Lissandrin,\authormark{1,5,6}
and Bahaa E. A. Saleh\authormark{1,5,7}\orcidlink{0000-0002-3700-4713}}
\address{
\authormark{1}Department of Electrical \& Computer Engineering,
Boston University, Boston, Massachusetts 02215, USA\\
\authormark{2}Department of Biomedical Engineering,
Boston University, Boston, Massachusetts 02215, USA\\
\authormark{3}Department of Physics,
Boston University, Boston, Massachusetts 02215, USA\\
\authormark{4}Department of Electrical Engineering,
Columbia University, New York, New York 10027, USA\\
\authormark{5}Photonics Center,
Boston University, Boston, Massachusetts 02215, USA\\
\authormark{6}Dipartimento di Ingegneria dell'Informazione,
Universit{\`a} degli Studi di Padova, I-35131 Padova, Italy\\
\authormark{7}CREOL,
The College of Optics \& Photonics, University of Central Florida, Orlando, Florida 32816, USA\\
\authormark{*}teich@bu.edu}
\begin{abstract*}
Subthreshold photoemission and absorption under classical and
entangled-photon-pair illumination are reviewed, and the generation and properties
of entangled-photon pairs are surveyed. Three prominent forms of subthreshold
photoemission are examined: one-photon Fermi-tail photoemission (FTP), two-photon
photoemission (TPP), and entangled-two-photon photoemission (ETPP). Experimental
methods for measuring subthreshold photocurrents and photoelectron count rates
are discussed, along with strategies for enhancing selected contributions.
Experimental observations of FTP from a CsK$_2$Sb photocathode in a
photomultiplier tube (PMT), under both coherent and entangled-photon-pair
illumination, are reviewed, and the role of FTP as a noise source in two-photon
measurements is elucidated. TPP from Na and CsK$_2$Sb photocathodes in a PMT
under classical-light illumination is considered, as are TPP and ETPP from a CsK$_2$Sb
photocathode in a channel photomultiplier (CPM) under coherent and
entangled-photon-pair illumination. The observation of ETPP is facilitated by the use
of a CPM exhibiting negligible FTP, and by low-intensity illumination, which minimizes
TPP. Quantum models of TPP are in reasonable accord with experiment, and ETPP
theory and experiment can be reconciled for suitable values of the material-response
parameters. Entangled-two-photon absorption (ETPA) is analyzed, as are its
applications in entangled-two-photon fluorescence microscopy (ETPFM) and
entangled-two-photon spectroscopy (ETPS). The three principal forms of subthreshold
absorption parallel those of subthreshold photoemission: single-photon-induced
Boltzmann-tail absorption (BTA); cousin-/singleton-pair-induced two-photon
absorption (TPA); and twin-induced ETPA. Heuristic and fully quantum models of
these processes are compared, and experimental studies of ETPA and ETPFM,
together with methods for enhancing their observability, are summarized.
\end{abstract*}

\tableofcontents
\clearpage

\title{Photoemission and absorption under classical and entangled-photon-pair illumination}

\author{Malvin Carl Teich, Mark C. Booth, Francesco Lissandrin, and Bahaa E. A. Saleh}

\section{INTRODUCTION}\label{intro}

This review focuses on three principal forms of subthreshold photoemission:
1)~one-photon Fermi-tail photoemission (FTP); 2)~two-photon photoemission (TPP);
and 3)~entangled-two-photon photoemission (ETPP). The version that dominates can
often be established by judicious selection of the characteristics of the incident light,
the properties of the sample, and the specifications of the photodetection system. Of
particular interest is entangled-two-photon photoemission, the photoemissive
counterpart of entangled-two-photon absorption (ETPA), in which the absorption of
an entangled-photon pair results in the emission of an electron into vacuum. This
facilitates the observation of the phenomenon because the charge of electrons allows
them to be collected, multiplied, and detected with substantially higher efficiency
than fluorescence photons, which are emitted over large solid angles and undergo
losses associated with absorption, deflection, and scattering. Other, weaker forms of
subthreshold photoemission, e.g., higher-order multiphoton photoemission, are
beyond the scope of this review.

The notion that matter can undergo transitions via two-photon absorption was first
set forth by Maria G{\"o}ppert-Mayer in a theoretical paper published in
1931~\cite{GoeppertMayer31}. The invention of the laser by Maiman in
1960~\cite{maiman1960stimulated}, and the ensuing development of modern
nonlinear optics beginning in 1961, led to the observation of a number of two-photon
processes induced by classical light. Among these were the landmark experiments of
Franken~\emph{et al.}~\cite{franken1961} and Kaiser and
Garrett~\cite{kaisergarrett1961}  that reported the observation of optical
second-harmonic generation and fluorescence-mode two-photon absorption,
respectively. The first general theoretical treatment of light-wave mixing in a nonlinear
dielectric was formulated by Armstrong~\emph{et al.} in 1962~\cite{Armstrong62}.
This was followed in the same year by the analysis of Bloembergen and
Pershan~\cite{BloembergenPershan62}, which incorporated the boundary conditions
governing nonlinear reflection and refraction at interfaces. The theory of nonlinear
optics was presented in unified form in 1965 in a seminal monograph by
Bloembergen~\cite{bloembergen65}.

\begin{figure}[htb!]
\centering\includegraphics[width=3.in]{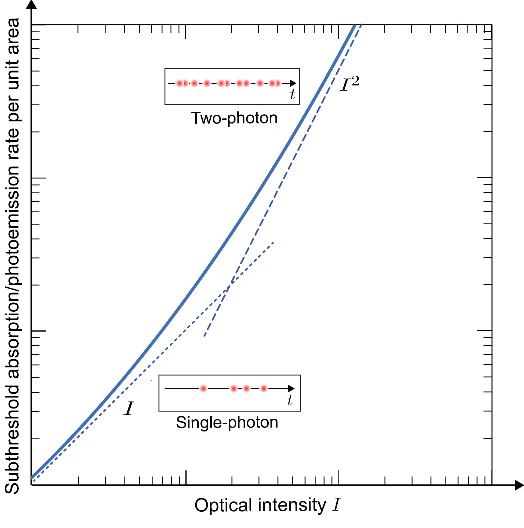}
\caption{The onset of two-photon absorption/photoemission for classical light is marked
by a shift in the rate (per unit area) from linear to quadratic scaling as the incident
optical intensity $I$ increases (blue curve). The linear region can result, for example, from
one-photon subthreshold effects such as Boltzmann-tail absorption or
Fermi-tail photoemission.
Since the probability for the absorption of a single photon in an incremental time
interval is proportional to $I$, the probability for the independent
absorption of two photons is proportional to $I^2$ for fixed photon statistics.
The insets portray idealized sample functions for both scenarios.}
\label{fig1}
\end{figure}

In 1964, two-photon photoemission was contemporaneously observed from Na metal
by Teich \emph{et al.}~\cite{Teich64,teich66PhD,Teich68} and from Cs$_3$Sb
semiconductor by Sonnenberg \emph{et al.}~\cite{Sonnenberg64}. The initial
observation from sodium metal in 1964~\cite{Teich64}  was subsequently expanded
into a formal theoretical and experimental treatment~\cite{teich66PhD,Teich68}. As
sketched in Fig.~\ref{fig1}, a hallmark of the onset of two-photon
absorption/photoemission is a shift in the rate from linear to quadratic scaling as the
optical intensity $I$ of the incident classical light increases~\chapcite{22}{saleh2019}.
Two-photon photoemission was later conjoined with optical heterodyning to form a
hybrid detection modality --- two-photon heterodyne photodetection --- in which dc
photomixing can occur~\cite{teich75JQE}.

Expanded opportunities for exploring the interaction between radiation and matter
were heralded by the development of sources of nonclassical
light~\cite{Teich88,teich1989squeezed}, including spontaneous optical parametric
downconversion (SPDC) in the late
1960s~\cite{harris67,magde67,klyshko67,Giallorenzi68,Klyshko69JETP,burnham70},
along with photon-number-squeezed
light~\cite{teich1983antibunching,teich1984JOSAB,teich1985subpoisson} and
quadrature-squeezed light~\cite{slusher85} in the mid-1980s. In particular, the
availability of time--frequency entangled-photon pairs created via spontaneous optical
parametric downconversion
(SPDC)~\cite{harris67,magde67,klyshko67,Giallorenzi68,Klyshko69JETP,burnham70}
fostered the exploration of a significant number of novel radiation--matter
interactions. SPDC, also referred to as optical parametric fluorescence~\cite{harris67},
spontaneous parametric scattering (SPS)~\cite{Giallorenzi68}, and parametric
luminescence~\cite{Klyshko69JETP}, is a nonlinear optical process whereby a small
fraction of the pump photons incident on a nonlinear optical material spontaneously
split into \textbf{twins} (also called \textbf{intrapair photons}) that are linked by the
conservation of energy and
momentum~\cite{Klyshko80,klyshko82,Teich90}~\chapcite{22}{saleh2019}.

\begin{figure}[htb!]
\centering\includegraphics[width=3.in]{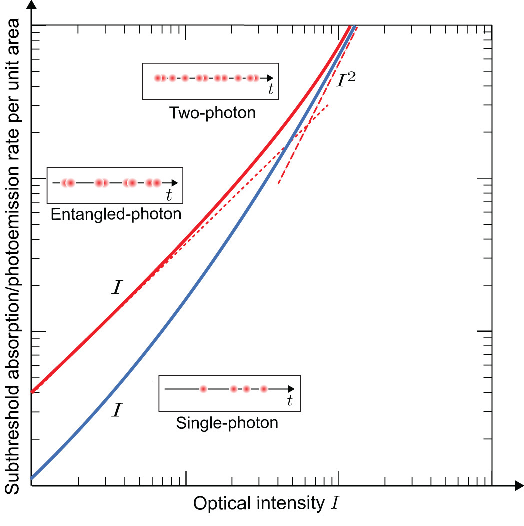}
\caption{Entangled-two-photon
absorption/photoemission in the isolated-photon-pair domain (red curve)
exhibits a rate (per unit area) that increases linearly with the incident
optical intensity $I$. In the ideal case,
when there is neither photon loss nor overcrowding, the rate scales linearly with $I$
since the presence of one twin signals the presence of the other and they are absorbed
as a pair. When the incident intensity exceeds a critical value, however,
twins overcrowd the allotted interaction volume, thereby increasing
the likelihood of contributions from interpair/accidental photon
pairs. The absorption of independent non-twin photon pairs
introduces a contribution proportional to $I^2$, just as for classical light
 (blue curve and
Fig.~\ref{fig1}). Insets are idealized sample functions for the various scenarios.
Refined versions of these sketches are portrayed in Figs.~\ref{fig12} and \ref{fig27}.
}\label{fig2}
\end{figure}

In the idealized case when loss and overcrowding are absent, the presence of one twin
signifies the presence of the other so the two-photon absorption rate scales linearly,
as sketched in Fig.~\ref{fig2}. Still, when the incident entangled-photon intensity
becomes sufficiently large, twins overcrowd the allotted spacetime interaction
volume. Two non-twin photons --- colloquially called \textbf{cousins} (or
\textbf{interpair photons})
--- can then be independently absorbed, giving rise to a quadratic contribution alongside
the linear contribution from twins (Fig.~\ref{fig2}).  In the presence of loss, each
surviving unpaired twin is referred to as a \textbf{singleton}. Singletons can induce
one-photon absorption (e.g., Fermi-tail photoemission or Boltzmann-tail absorption)
and can also team up to induce two-photon absorption.

Among the many forms of nonclassical light~\cite{sanchezsoto2026arxiv},
entangled-photon pairs are widely used for investigating light--matter interactions.
Their experimentally accessible and engineerable joint properties have motivated
many proposed and demonstrated
applications~\cite{leonmontiel2024QITreview,weiss25,patil2026}. While this review
focuses on entangled photons in the \textbf{isolated-photon-pair
domain}~\cite{Fei97,teich1997}, also called the two-photon limit or the
weak-squeezed-vacuum domain, we also consider the
\textbf{bright-squeezed-vacuum (BSV)}
domain~\cite{iskhakov2012,spasibko2017,sharapova2020}  in connection with
entangled-two-photon absorption.

\begin{framed}\noindent
The subsequent sections of this review explore various forms of subthreshold
photoemission, with particular attention devoted to their experimental signatures in
Na metal and CsK$_2$Sb semiconductor. The final section of the paper summarizes
the current status of subthreshold absorption and fluorescence microscopy under
entangled-photon-pair illumination.
\end{framed}

Section~\ref{sec:entphotmat} provides an overview of entangled-photon pairs and
their interactions with matter. Section~\ref{ssec:EPPs} contains a discussion of the
properties and generation of entangled-photon pairs. This is followed by brief
accounts of entangled-two-photon absorption (ETPA), entangled-two-photon
fluorescence microscopy (ETPFM), and entangled-two-photon spectroscopy (ETPS), in
Secs.~\ref{EPabsorp}, \ref{EPmicroscopy}, and \ref{EPspectroscopy}, respectively.
Entangled-two-photon photoemission (ETPP), which is of focal interest in this review,
is compared and contrasted with ETPA in Sec.~\ref{EPphoto}. Several alternative
paradigms and applications of entangled-photon pairs are catalogued in
Sec.~\ref{EPinteractions}.

Section~\ref{theory} provides an overview of various forms of photoemission. It
outlines suprathreshold (ordinary) photoemission (Sec.~\ref{singletheory}) and then
considers the origins of the three prominent forms of subthreshold photoemission:
one-photon Fermi-tail photoemission (Sec.~\ref{subthreshtheory}); two-photon
photoemission (Sec.~\ref{twophotontheory}); and entangled-two-photon
photoemission (Sec.~\ref{enttheory}). Optical power measures and the effects of
optical loss are considered in Sec.~\ref{optlosssys}.

Section~\ref{measiden} introduces a number of auxiliary techniques that facilitate
the measurement of subthreshold photocurrents and photoelectron count rates
(Sec.~\ref{measidenaux}). It also provides expressions for these quantities that
accommodate the distortions introduced by the various techniques for classical,
coherent, and entangled-photon-pair illumination (Sec.~\ref{measideniandmu}). This
approach enables CW-equivalent values of the responsivities and quantum
efficiencies to be extracted from the data. The linear-quadratic crossover intensities
and estimation protocols examined in Secs.~\ref{measidencross} and
\ref{paramestPE}, respectively, are useful for determining parameter values.
Methods for identifying various forms of subthreshold photoemission are outlined in
Sec.~\ref{measideniden}, and approaches for enhancing specific forms thereof are
suggested in Sec.~\ref{enhancingf}.

Section~\ref{subthreshold} is dedicated to an experimental examination of Fermi-tail
photoemission from CsK$_2$Sb and two other alkali-antimonide materials, under
coherent-light illumination. Fermi-tail photoemission acts as a source of noise when
measuring two-photon photoemission. The Fermi-tail photocurrent increases with
increasing sample temperature and decreasing illumination wavelength, as expected
from theory.

Section~\ref{twophoton} is devoted to the observation of two-photon photoemission,
a well-understood process that can provide useful information about photoemissive
electronic structure and transition mechanisms in matter. TPP from metallic sodium
has been examined in detail. Under classical-light illumination, the two-photon
quantum efficiency $\eta_{\scriptscriptstyle C}$ is roughly three orders of magnitude
greater for a thick Na sample than for a thin one at the same CW-equivalent intensity
--- the theoretical values predicted by volume and surface photoemission models are
in reasonable accord with the thick and thin measurements, respectively, although a
reduced-volume contribution to the thin-sample response cannot be excluded. The
two-photon quantum efficiency for CsK$_2$Sb has also been measured under
coherent-light illumination, and its temperature and wavelength dependencies
established --- it is about four orders of magnitude greater than that for the thick Na
sample. Analog and digital measurements yield consistent results within the
experimental uncertainty.

Section~\ref{entPMT} reports the results of Fermi-tail photoemission experiments
from CsK$_2$Sb under entangled-photon-pair illumination (in contrast to
Sec.~\ref{subthreshold}, where the illumination was coherent). The Fermi-tail
photoemission elicited by entangled-photon singletons is found to follow the same
one-photon behavior as that elicited by coherent photons. Under the
independent-loss model employed here, the fraction of photons passed by an optical
system of transmittance $\mathcal{T\/}$ remains $\mathcal{T\/}$, whether the
photons are coherent or singletons, whereas the fraction of entangled-photon pairs
that survive intact after transmission behaves as $\mathcal{T\/}^2$. Fermi-tail
photoemission, which can mask both two-photon photoemission and
entangled-two-photon photoemission, is difficult to avoid from photocathodes in
photomultiplier tubes.

Section~\ref{entchannel} analyzes experiments conducted on two-photon
photoemission from CsK$_2$Sb under entangled-photon-pair illumination (in contrast
to the classical illumination in Sec.~\ref{twophoton}). It turns out that
entangled-photon cousins, singleton pairs, and pairs of (independent) coherent
photons generate two-photon photoemission via the same independent-photon
mechanism. This section also details the observation of entangled-two-photon
photoemission from the CsK$_2$Sb photocathode of a channel photomultiplier (CPM)
module, and shows that quantum theory can reproduce the experimentally inferred
quantum efficiency for suitable values of the material-response parameters. Fermi-tail
photoemission from the CPM was below detectability, while low-intensity illumination
rendered conventional two-photon photoemission subdominant.

Section~\ref{entabsfluor} identifies the three prominent forms of subthreshold
absorption considered under entangled-photon-pair illumination:
single-photon-induced Boltzmann-tail absorption; cousin and singleton-pair induced
two-photon absorption; and twin-induced entangled-two-photon absorption. The
heuristic model used to describe these processes is elaborated in
Sec.~\ref{EPparticlemodel} and the results that it yields are compared with those
provided by its quantum counterparts in Sec.~\ref{ETPAquantummodel}.
Experimental studies related to subthreshold absorption and entangled-two-photon
fluorescence microscopy are discussed in Sec.~\ref{ssec:ETFAFM}, and several
methodologies for enhancing their observability in future studies are suggested in
Sec.~\ref{impabmic}.

\section{ENTANGLED-PHOTON PAIRS AND THEIR INTERACTIONS WITH MATTER}
\label{sec:entphotmat} The properties and generation of entangled-photon pairs are
summarized in Sec.~\ref{ssec:EPPs}. A brief overview of entangled-two-photon
absorption (ETPA) is provided in Sec.~\ref{EPabsorp}. This is followed, in
Secs.~\ref{EPmicroscopy} and \ref{EPspectroscopy}, respectively, with concise reviews
of entangled-two-photon fluorescence microscopy (ETPFM) and
entangled-two-photon spectroscopy (ETPS), two potentially powerful applications of
ETPA. While ETPA, ETPFM, and ETPS have attracted considerable attention in the
literature, a convincing, independently replicated demonstration of these effects in
organic molecules in solution has yet to be established. (As discussed in
Sec.~\ref{ssec:ETFAFM}, one-photon processes provide plausible explanations for
many of the reported observations.) Section~\ref{EPphoto} details the physical
mechanisms that facilitate the observation of entangled-two-photon photoemission
(ETPP) relative to ETPA. Finally, in Sec.~\ref{EPinteractions}, we catalog a number of
other entangled-photon-pair interactions and applications that have garnered
substantial interest but are not examined in this work.

\subsection{Properties and Generation of Entangled-Photon Pairs}\label{ssec:EPPs}

Entanglement in a composite quantum system is characterized by nonseparability of
its state. For pure states, as articulated by Schrödinger in 1935, this means that the
wavefunction cannot be factorized into independent wavefunctions for the
constituent subsystems~\cite{Schroedinger35}. Bohm subsequently formulated the
Einstein--Podolsky--Rosen argument for pairs of  spin-½ particles~\cite{bohm1951},
providing a framework that later underpinned Bell's theorem~\cite{Bell64,chsh1969}.
For a specified propagation direction, the two orthogonal polarization states of a
photon likewise span a two-dimensional Hilbert space and support entanglement
phenomena analogous to those of spin-½ particles. A coherently generated
two-photon state is commonly called a
\textbf{biphoton}~\cite{klyshko82}~\seccite{8.4}{saleh25}; when that state is
nonseparable with respect to the chosen photon subsystems, it represents an
\textbf{entangled-photon pair}.

The earliest general quantum treatment of vacuum-seeded parametric emission was
formulated by Louisell \emph{et al.} in 1961~\cite{Louisell61}. The implementation
of this effect in the optical region was contemporaneously effected in 1967 by Harris
\emph{et al.}, who reported tunable optical parametric fluorescence in lithium
niobate (LiNbO$_3$)~\cite{harris67}, and by Magde and Mahr, who carried out a
quantitative experimental study of the effect in ammonium dihydrogen phosphate
(NH$_4$H$_2$PO$_4$ or ADP)~\cite{magde67}. Submitting his theoretical paper
just a few days after Magde and Mahr, Klyshko also analyzed this effect, pointing to
the importance of measuring the mutual correlation of the emitted
fields~\cite{klyshko67}. The explicit recognition of the process as a source of
correlated photon pairs for quantum optics emerged in the subsequent work of
Zel'dovich and Klyshko in 1969~\cite{zeldovich69}, and was followed shortly
thereafter by the experimental observation of optical photon-pair simultaneity by
Burnham and Weinberg in 1970~\cite{burnham70}.

These early theoretical and experimental advances ultimately set the stage for the
direct observation of quantum entanglement in photonic systems. Landmark
experiments conducted by Alain Aspect and his collaborators in the early
1980s~\cite{aspect1981,aspect1982}, which made use of a double-transition
atomic-cascade source, provided more stringent tests of Bell's inequalities and
strengthened the experimental evidence for nonlocal quantum correlations
~\cite{PhysNobel2022}. Two-photon Einstein--Podolsky--Rosen (EPR)
states~\cite{Einstein35}, entangled in the parity of their one-dimensional transverse
spatial profiles, have  been shown to violate Bell's inequality in the spatial
domain~\cite{abouraddy2007violation,yarnall2007experimental}.

\subsubsection{Spontaneous Parametric Downconversion (SPDC)}\label{sssec:EPPs-prop}

\textbf{Spontaneous parametric downconversion} (SPDC) is a three-wave-mixing
process in an optically pumped second-order $\chi^{(2)}$ nonlinear medium that
generates photon pairs without an injected signal or idler field. In the low-gain
domain, its two-photon component can be entangled in one or more degrees of
freedom, depending on the pump, phase matching, and collection conditions.

At the level of the pump-quantum picture, each pair results from the conversion of
one pump quantum into signal and idler quanta. The \textbf{signal} and \textbf{idler}
photons, also denoted \textbf{1} and \textbf{2}, have central angular frequencies
$\omega_1^0$ and $\omega_2^0$, and central wavevectors $\mathbf{k}_1^0$ and
$\mathbf{k}_2^0$, respectively. For a pump centered at angular frequency
$\omega_p$, energy conservation at the central frequencies gives
\begin{equation}\label{encons}
  \omega_p = \omega_1^0 + \omega_2^0\,.
\end{equation}
For a homogeneous bulk medium, the corresponding central phase-matching
condition is
\begin{equation}\label{momcons}
 \mathbf{k}_p = \mathbf{k}_1^0 + \mathbf{k}_2^0\,.
\end{equation}
In quasi-phase-matched structures, Eq.~(\ref{momcons}) is modified by the
reciprocal-lattice vector associated with the nonlinear-domain pattern
(Sec.~\ref{sssec:EPPs-prop-QPM}).

For a source of finite bandwidth and interaction volume, the joint two-photon
amplitude is governed by the pump envelope and the phase-matching function, and is
further modified by propagation, filtering, and collection. With a sufficiently
narrowband pump, the sum-frequency distribution can be narrow even when the
individual photon spectra are broad, yielding strong frequency anticorrelations.
Similarly, a well-collimated pump can yield strong anticorrelations between the
transverse wavevectors of the two photons. In appropriate observation planes, the
photons can also exhibit tightly correlated arrival times and transverse positions,
characterized by the entanglement time $T_{\scriptscriptstyle E}$ and entanglement
area $A_{\scriptscriptstyle E}$ discussed in Sec.~\ref{sssec:TETA}. The strength and
widths of these correlations depend on the source and optical configuration; their
presence alone does not establish quantum entanglement.

\subsubsection{Birefringent Phase Matching (BPM)}
\label{sssec:EPPs-prop-birefring}

\begin{figure}[htb!]
\centering
\includegraphics[width=4.25in]{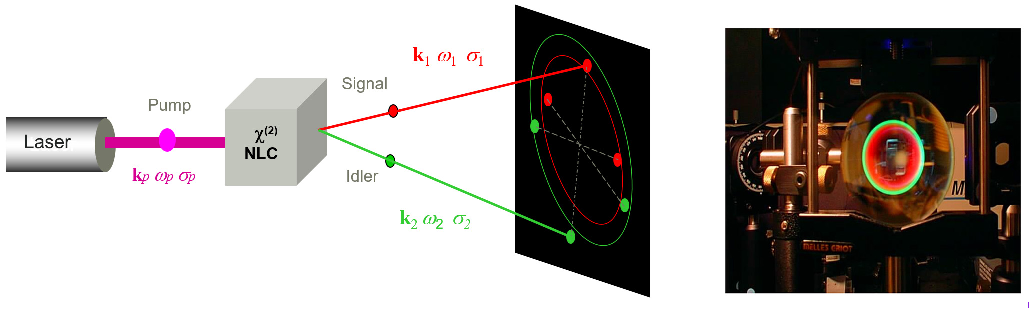}
\caption{
Generation of noncollinear SPDC in a nonlinear optical crystal
(NLC) via type-I birefringent phase matching.
 The red and green colors in the schematic distinguish signal and idler
photons. Their polarizations, labeled $\sigma_1$ and
$\sigma_2$, occupy the same polarization class. Emission occurs
over the angular ranges allowed by phase matching, producing
approximately conical emission patterns whose intersections
with an observation plane form rings. Different wavelengths
generally correspond to different cone angles. For a
well-collimated on-axis pump, the members of a pair have
approximately opposite transverse wavevectors; their polar
angles need not be equal for nondegenerate pairs. The photograph at right was taken
at Boston University in the course of an experiment conducted using LiIO$_3$.
}
\label{fig3}
\end{figure}

\textbf{Birefringent phase matching (BPM)} exploits the polarization- and
direction-dependent refractive indices of an anisotropic second-order nonlinear
optical crystal to compensate material dispersion and satisfy Eq.~(\ref{momcons}).
The crystal orientation is selected for the desired pump wavelength, signal and idler
wavelengths, and interaction geometry. The SPDC process may be \textbf{collinear},
in which case the signal and idler wavevectors are parallel and codirectional with that
of the pump, or \textbf{noncollinear}. Collinear downconversion can simplify optical
alignment and beam overlap and, with suitable mode matching, facilitate collection
into a limited aperture or a single spatial mode. A schematic noncollinear type-I
configuration and an example of its ring-shaped emission pattern are shown in
Fig.~\ref{fig3}.

The phase matching may be of type-I or type-II, depending on the polarization
relationships among the pump, signal, and idler waves. In common birefringently
phase-matched uniaxial-crystal configurations, \textbf{type-I phase matching}
produces signal and idler photons in the same polarization class, while the pump
occupies the other, conventionally written $e \rightarrow oo$ or $o \rightarrow ee$,
where $o$ and $e$ denote ordinary and extraordinary waves, respectively. For
noncollinear emission, a shared polarization class does not necessarily imply identical
polarization vectors for photons propagating in different directions. In geometries
exhibiting spatial--spectral coupling, the collected marginal spectra of the photons
and the joint spectrum of the pairs can be tailored by modifying the pump spatial
profile and the downconversion and collection
geometries~\cite{carrasco2004PRA,carrasco2006PRA,carrasco2006OL}.

\textbf{Type-II phase matching} produces signal and idler photons in different
polarization classes, one ordinary and the other extraordinary. This polarization
relationship does not, by itself, imply polarization entanglement. Polarization
entanglement can be obtained when alternative polarization assignments to the
collected photon modes are coherently superposed and distinguishing spatial,
spectral, and temporal information is removed or
compensated~\cite{kwiat1995-PRL}. The polarization configuration influences spatial
walk-off, phase-matching bandwidth, and the entanglement characteristics of the
collected state.

Early SPDC experiments generated photon pairs via three-wave mixing in bulk uniaxial
optical crystals with suitable nonlinear and birefringent properties. Birefringent phase
matching was implemented in materials such as lithium niobate
(LiNbO$_3$)~\cite{harris67,KlyshkoKrindach68}, ammonium dihydrogen phosphate
(NH$_4$H$_2$PO$_4$ or ADP)~\cite{magde67,burnham70}, potassium dihydrogen
phosphate (KH$_2$PO$_4$ or KDP)~\cite{Hong87}, deuterated potassium
dihydrogen phosphate (KD$_2$PO$_4$ or KD*P)~\cite{rarity1990-PRL,Larchuk93},
lithium iodate (LiIO$_3$)~\cite{Larchuk95}, and beta barium borate
($\beta$-BaB$_2$O$_4$ or BBO)~\cite{kwiat1995-PRL}. Lithium niobate is a widely
used nonlinear optical platform because of its availability, large second-order
nonlinearity, broad transparency range, and compatibility with waveguide fabrication
and ferroelectric-domain engineering~\cite{Fejer92}. The preferred material for a
particular SPDC source nevertheless depends on the operating wavelengths,
phase-matching geometry, and collection requirements.

\subsubsection{Quasi-Phase Matching (QPM), Periodic Poling (PP), and
Chirped Quasi-Phase Matching (CQPM)}\label{sssec:EPPs-prop-QPM}

The use of \textbf{quasi-phase matching (QPM)}, in which a position-dependent
nonlinearity is imposed on the nonlinear downconversion
structure~\cite{Armstrong62,Franken63,Giuseppe02,hum2007}, allows a chosen
phase mismatch to be compensated by an appropriate reciprocal-lattice vector
introduced through the poling pattern. Quasi-phase matching enables \textbf{type-0
phase matching}, in which the pump, signal, and idler share the same polarization. The
implementation of type-0 phase matching via birefringence, in contrast, is rarely
achievable because there is then no refractive-index difference to exploit. Moreover,
QPM enables access to the largest nonlinear tensor component of the material, which
can offer a substantial increase in the strength of the nonlinear interaction.

While quasi-phase matching compensates the phase mismatch at selected
frequencies, thereby enabling coherent longitudinal buildup of the SPDC amplitude, it
does not generally eliminate group-velocity mismatch (GVM). As a result, the GVM of
the interacting waves also contributes to the entanglement time $T_{\scriptscriptstyle
E}$ in quasi-phase-matched media; in many practically important type-0
implementations, in which all waves share the same polarization, the resulting value
of \(T_{\scriptscriptstyle E}\) depends on the signal–idler group-velocity mismatch,
higher-order dispersion, and the spectral modes admitted by the apparatus.

Quasi-phase matching can be achieved by utilizing \textbf{periodically poled (PP)}
ferroelectric crystals fabricated in such a way that the nonlinear coefficient of the
material alternates in sign, on a micrometer spatial scale, along the length of the
crystal~\cite{yariv1972-OC}. Bulk periodically poled crystals commonly used to
generate SPDC include lithium niobate
(PPLN)~\cite{myers1995,dechattelus04SPIE,dechattelus06OE,perina2009ppln},
stoichiometric lithium tantalate (PPSLT)~\cite{hum2007-SLT-JAP,lee2015-OE}, and
potassium titanyl phosphate (PPKTP)~\cite{kuklewicz2004-PRA}.

Time--frequency entangled-photon pairs with broad spectral widths can exhibit tightly
correlated arrival times when their spectral phase is suitably controlled, a property
that is desirable for observing quantum processes such as entangled-two-photon
absorption/photoemission~\cite{teich1989squeezed} and quantum-optical coherence
tomography (QOCT)~\cite{teich2012,Nasr03}. One way of fostering this property is to
simultaneously match the signal and idler group velocities and to operate at a point
where the total group-velocity dispersion for these waves is
minimal~\cite{nasr2005generation,javid2021PRL,silberhorn2024OE,silberhorn2024NJP}.

An approach for generating ultrabroadband entangled-photon pairs with greater
design flexibility is based on \textbf{chirped quasi-phase matching (CQPM)}, a
technique introduced by Carrasco \emph{et al.} in 2004~\cite{carrasco2004OL}. This
is implemented by using aperiodic (rather than periodic) poling that usually takes the
form of a linearly chirped poling period along the longitudinal dimension of the
nonlinear medium. Photon pairs phase-matched at different poling-period locations
comprise different combinations of wavenumber pairs. Many such signal and idler
pairs are therefore simultaneously phase-matched within the medium, resulting in
ultrabroadband SPDC. Chirped quasi-phase matching is readily implemented in bulk
materials such as CPPLN,
CPPSLT~\cite{nasr2008,nasr2008-SSPD,mohan2009,takeuchi2012OE},
CPPCLT~\cite{cushing2021JCP}, and CPPKTP (the initial ``C" signifies ``chirped'').
Alternatively, ultrabroadband entangled-photon pairs can be generated (at rates
comparable to those attained with CQPM) by using stochastic quasi-phase matching,
in which domains of different lengths are ordered randomly~\cite{perina2010}.

\subsubsection{Guided-Wave Phase Matching}\label{sssec:EPPs-prop-waveguide}

Spontaneous parametric downconversion can also be generated in nonlinear
waveguides fabricated on integrated-optical chips. Such structures provide spatial
confinement and can improve overlap among the interacting modes and the
collection efficiency. For copropagating guided-wave SPDC, the longitudinal
phase-matching condition is expressed in terms of the waveguide-mode propagation
constants, so that Eq.~(\ref{momcons}) becomes
\begin{equation}\label{momconswaveguide}
  \beta_{\!p} = \beta_1 + \beta_2.
\end{equation}
For quasi-phase-matched waveguides, the corresponding condition is
\(\beta_p=\beta_1+\beta_2+K\), where \(K\) is the signed reciprocal-lattice vector of
the poling pattern.

The propagation constants depend on the refractive indices of the waveguide material
and the polarizations of the modes, as well as on the waveguide geometry and
dimensions~\chapcite{7 and 22}{saleh2019}. The additional degrees of freedom offer
added flexibility in satisfying the phase-matching condition. SPDC from a LiNbO$_3$
waveguide structure was first reported by Hampel and Sohler in
1986~\cite{hampel1986}.

As with bulk birefringent crystals, key techniques that allow SPDC to be efficiently
generated in guided-wave structures include quasi-phase matching, periodic poling,
and chirped periodic
poling~\cite{saleh2009modal,saleh2010photonicCircuits,sipe2022AOP}. Integrated
nonlinear optical waveguides in
PPLN~\cite{Baldi93-EL,Baldi95,Tanzilli01-EL,Baldi2002-APL} are typically fabricated
by Ti-indiffusion or annealed proton exchange (APE). Stoichiometric lithium tantalate
(PPSLT) waveguides~\cite{Baldi95-OL}, often selected for their superior resistance to
photorefractive damage, are principally formed using APE.  Waveguides in
PPKTP~\cite{fiorentino2007-OE}, on the other hand, are generally realized using
rubidium ion exchange. In all of these platforms, the waveguide structure is
combined with electric-field poling to achieve quasi-phase matching. Alternatively,
ridge waveguides, which provide strong lateral confinement that facilitates the
generation of collinear SPDC, can be formed in any of these materials by precision
diamond dicing or dry etching techniques. Slab waveguides, which offer confinement
in only one dimension, support noncollinear SPDC generation.

Chirped versions of these poled guided-wave structures have been successfully
employed to generate broadband SPDC in CPPLN~\cite{fang2026} and
CPPSLT~\cite{takeuchi2021OE,takeuchi2023OE,takeuchi2024OPTICA}. Broad spectral
bandwidths can yield tightly time-correlated pairs when the spectral phase is suitably
controlled, making such pairs advantageous for entangled-two-photon interactions.
At the opposite end of the bandwidth scale, SPDC with reduced spectral width can be
generated by using counterpropagating photons in a waveguide comprising a
periodically poled second-order nonlinear medium such as
PPLN~\cite{Booth02,DeRossi02}.

\subsubsection{Photon-Pair Generation in Novel Media}
\label{sssec:EPPs-prop-novel}
Counterpropagating photon sources have been
demonstrated using a thin-film lithium-niobate (TFLN) [or lithium niobate on insulator
(LNOI)] platform~\cite{kellner2025OQ}. This serves to transform lithium niobate from
a low-contrast, weakly confining, diffusion-based waveguide platform into a
high-contrast, deeply etched, sub-micron integrated-photonics platform, thereby
enabling substantial improvements in confinement, nonlinear enhancement, and
integration density.

Photon pairs have also been generated via SPDC in atomic
monolayers~\cite{Lu2025NC}, nonlinear films~\cite{stich2026OE}, and
metasurfaces~\cite{2021chekhova-NL,SantiagoCruz2022Resonant,ma2024-AM,%
sukhorukov2025meta-SA,ndao2025APR,Zhong2025Enhanced}; via spontaneous
two-photon emission from quantum dots coupled to optical
microcavities~\cite{ota2011-2PQdot,LiuLiu2025Nature}; and via spontaneous
four-wave mixing in CMOS-compatible microring
resonators~\cite{takeuchi2025APLPhot}.  Chip-scale entangled-photon-pair sources
have also been developed~\cite{Kues2023}. Among these is an electrically pumped,
high-performance source of polarization-entangled photon pairs~\cite{pan2025}
formed from the hybrid integration of a TFLN chip that comprises a pair of PPLN
waveguides, a beamsplitter, and a polarization rotator/combiner, coupled with a DFB
laser chip --- precise wavelength tuning  is provided by thermal adjustment.

These platforms are promising for their potential compactness, integration, and novel
functionalities; their photon-pair generation performance varies substantially across
platforms and operating conditions.

\subsubsection{Coherence and Statistical Properties} \label{sssec:cohstat}

The coherence and statistical properties of SPDC in the isolated-photon-pair domain,
generated via birefringent phase matching, have been extensively
investigated~\cite{joobeur1994spatiotemporal,joobeur1996coherence,%
joobeur1998,saleh2005wolfEquations}. In the ideal low-gain limit, photons are
generated in pairs~\cite{Jost98,diGiuseppe2003direct}, so that the number
fluctuations in the signal and idler beams are strongly correlated. The coincidence
statistics therefore retain a clear imprint of the coherent pump, as manifested in
fourth-order interference and coincidence-counting experiments.

The marginal photon-number statistics in each arm depend on the number of
spatiotemporal modes collected and, for equally populated independent thermal
modes, follow a negative-binomial distribution, with the Bose--Einstein distribution as
its single-mode limit \cite{glaubmollow1967,yurke1987,joobeur1996coherence,
paleari2004,blausteiner2009}; when many weakly populated modes are collected, the
statistics approach Poisson form~\cite{Larchuk95}. The statistical properties of the
combined beam likewise depend on loss, modal structure, detector dead time, and
the counting window adopted in the
measurement~\cite{Larchuk95,cantor1975dead,teich1982effects,perina1983independent,saleh82,saleh1983PRA}.
The focusing and imaging properties of entangled-photon pairs have also been well
characterized~\cite{Klyshko88,Abouraddy02_JOSAB3,Nasr02}. The coherence
properties of downconverted beams at arbitrary levels of the photon flux have been
derived by Pe{\v r}ina~\cite{perina2014PRA,perina2015PRA,perina2016SR,
perina2016PRA}.

\subsubsection{Entanglement Time and Entanglement Area}
\label{sssec:TETA}

Entangled-photon pairs can be characterized by an \textbf{entanglement time}
$T_{\scriptscriptstyle E}$ and an \textbf{entanglement area} $A_{\scriptscriptstyle
E}$, representing characteristic widths of the pair contribution to the fourth-order
temporal and spatial coherence functions,
respectively~\cite{joobeur1996coherence,Fei97,Atature01-PRA}. The entanglement
time characterizes the spread of relative photon arrival times about a specified
interbeam delay, while the entanglement area characterizes the transverse region
over which the position of one photon predicts the position of its twin in a specified
observation plane. These quantities depend on the generated state and on
propagation, filtering, and collection; their numerical values also depend on the width
conventions adopted. \textbf{Partial entanglement} may be viewed as the dual of
partial coherence~\cite{Saleh00}.

The temporal correlations of SPDC are governed by the joint spectral amplitude,
including both its magnitude and phase. The pump spectrum and the phase-matching
function determine this amplitude at generation, while subsequent dispersion and
spectral filtering can modify the correlations delivered to a sample. The relative
signal--idler arrival time is to be distinguished from the arrival time of the pair relative
to the pump. Signal--idler group-velocity mismatch can broaden the
relative-arrival-time distribution because pairs generated at different positions in the
nonlinear medium accumulate different relative group delays. Pump--downconversion
group-velocity mismatch, in contrast, principally governs the timing of the pair relative
to the pump and is not directly identified with $T_{\scriptscriptstyle E}$. These
temporal degrees of freedom can nevertheless be coupled through the joint spectral
amplitude.

For collinear degenerate type-I or type-0 SPDC in the same spatial mode, the signal
and idler have equal group velocities at their central frequencies, so the first-order
signal--idler group-velocity mismatch vanishes. The finite relative-arrival-time
correlation width then depends on higher-order dispersion, interaction length, and
the spectral and spatial modes admitted by the apparatus. Quasi-phase matching
compensates a selected phase mismatch through the nonlinear-domain pattern, but
does not generally eliminate group-velocity mismatch or higher-order dispersion. The
phase-matching type alone therefore does not establish the magnitude of
$T_{\scriptscriptstyle E}$.

For absorption or photoemission proceeding through a real intermediate state with
characteristic lifetime $T_{\scriptscriptstyle A}$, a useful heuristic compares that
lifetime with the relative-arrival-time distribution at the sample. For a smooth
correlation distribution centered at an appropriate delay, when $T_{\scriptscriptstyle
E} \gg T_{\scriptscriptstyle A}$, the fraction of pairs arriving within the
intermediate-state lifetime is commonly of order $T_{\scriptscriptstyle
A}/T_{\scriptscriptstyle E}$. When $T_{\scriptscriptstyle E} \ll T_{\scriptscriptstyle A}$,
this particular temporal limitation is substantially reduced. Still, the actual transition
probability depends on the full two-photon amplitude and material response,
including interference, detuning, and relaxation; it is not generally determined by the
ratio of these times alone.

\subsubsection{Evolution and Design of Entangled-Photon-Pair Sources} \label{sssec:evo}

Bulk birefringent phase matching (Sec.~\ref{sssec:EPPs-prop-birefring})  provides a
relatively direct and flexible route to SPDC, but phase matching restricts the accessible
interaction geometries and polarizations. Spatial walk-off, group-velocity mismatch,
and emission over a range of angles can also complicate the collection of the
generated pairs. Collinear geometries simplify alignment and spatial-mode overlap,
while noncollinear geometries can provide useful access to spatial and polarization
correlations. Hence, choices that affect the generated state at the bulk crystal also
affect how much of that state can be collected and delivered to a sample.

Quasi-phase matching (Sec.~\ref{sssec:EPPs-prop-QPM}), commonly implemented via
periodic poling, addresses important limitations of birefringent phase matching. It
permits phase matching without relying on birefringence, thereby allowing access to
larger nonlinear-tensor components of the material. It also enables useful type-0
configurations and broader freedom in selecting operating wavelengths. These
advantages can increase pair-generation efficiency and simplify polarization
management. QPM does not, however, automatically eliminate GVM or guarantee a
desired temporal correlation, collection efficiency, or joint spectrum; those properties
still depend on the material, interaction length, pump, phase-matching design, and
collection optics. Chirped and other engineered poling patterns allow the
phase-matching response to be broadened and tailored, enabling ultrabroadband
pairs that can exhibit short temporal correlations when their spectral phase is suitably
controlled. Of course, the resulting bandwidth and spectral structure must be
appropriate to the transition being probed and to the filters and detectors in place.

Guided-wave sources (Sec.~\ref{sssec:EPPs-prop-waveguide}) address a different
practical problem: efficient interaction and collection into well-defined spatial modes.
Confinement in a waveguide improves pump/downconversion mode overlap and can
provide high-flux, collinear output that is readily coupled to a fiber or other
single-mode system. Periodically poled waveguides combine these benefits with QPM
and enable efficient sources in compact formats. Their advantages do not remove all
system-level constraints, however: propagation and coupling losses reduce the
photon flux delivered to the sample, and the spectral bandwidth and temporal
correlation continue to be determined by the waveguide dispersion, poling, and pump
conditions.

Integrated platforms (Sec.~\ref{sssec:EPPs-prop-novel}) extend the possibilities for
compactness and stability, and provide the ability to combine sources with other
optical components, but their practical performance must be assessed for the
particular device and operating regime, rather than inferred from integration alone.

The source properties discussed above are inextricably intertwined with the
interactions considered in Secs.~\ref{EPabsorp}--\ref{EPphoto} and the experiments
discussed in Secs.~\ref{entPMT}--\ref{entabsfluor}. From an experimental
perspective,  the relevant figure of merit is not simply the SPDC conversion efficiency
or the entangled-photon-pair rate at the nonlinear medium. Rather, it is the usable
photon flux delivered to the spatial, spectral, temporal, and polarization modes
admitted by the apparatus, and it must also accommodate the effects of optical loss.
Similarly, the entanglement time $T_{\scriptscriptstyle E}$ and entanglement area
$A_{\scriptscriptstyle E}$ are not independent settings that can be separately
optimized. Instead, the temporal and transverse photon correlations are
co-determined by the transmission, collection, and detection features of the
experiment, as discussed in Sec.~\ref{sssec:TETA}.

Photon flux must also be considered in relation to the regime in which the experiment
is conducted. At low values of the photon flux, increasing the rate of usable
entangled-photon pairs can improve observability while maintaining operation in the
isolated-photon-pair domain. At higher values of the flux, however, independent
cousins and loss-generated singletons can contribute to conventional two-photon
processes, producing a quadratic background that competes with the
twin-photon-induced linear signal. Pulsed pumping can increase the peak photon flux,
but this enhances independent-photon two-photon processes, which is
disadvantageous when the aim is to distinguish an entangled-photon-pair-induced
contribution. Moreover, the statistics and light--matter interactions in the high-gain
regime differ from those in the isolated-photon-pair domain.

It is critically important to minimize optical loss in experiments conducted with
entangled-photon pairs because the loss of either member destroys the intact-pair
contribution. Under the independent-loss model, when each photon experiences an
optical transmittance $\mathcal{T\/}$, the fraction of pairs transmitted intact is
$\mathcal{T\/}^2$. These requirements motivate source and system designs that
combine appropriate correlations with a low-loss path to the sample and efficient
collection --- the maximization of bandwidth or photon flux in isolation is not
sufficient. In fluorescence-mode ETPA, a useful source must deliver pairs with suitable
correlations and adequate photon flux, while the weak and diffuse fluorescence signal
must be collected against residual background. In ETPP, the emitted electrons are
collected and detected locally with high efficiency, but the measurement still requires
a source and optical path that provide sufficient entangled-photon-pair flux and the
suppression of competing one-photon and independent-photon contributions.

The evolution from bulk crystals through poled and guided-wave devices to integrated
sources has greatly expanded the available control over efficiency, modes, bandwidth,
and system coupling. However, it has not removed the need to match the source to
the sample, interaction timescale, and measurement protocol. The design of suitable
forms of source--interaction matching links Sec.~\ref{ssec:EPPs} with the absorption,
microscopy, spectroscopy, and photoemission experimental configurations
considered in subsequent sections.

\subsection{Entangled-Two-Photon Absorption (ETPA)}\label{EPabsorp}

Conventional two-photon absorption (TPA) is a well-established nonlinear optical
process whose rate under classical illumination is proportional to the square of the
incident optical intensity. Attaining appreciable TPA often requires high values of the
incident intensity, which can be undesirable for samples susceptible to photodamage.
In contrast, as illustrated in Fig.~\ref{fig2}, the twin-induced contribution to
absorption in the isolated-photon-pair domain scales linearly with incident
photon-pair flux, potentially offering an advantage at low intensities, although the
resulting absorption rate may remain below experimental detectability.

\begin{figure}[htb!]
\centering\includegraphics[width=2.25in]{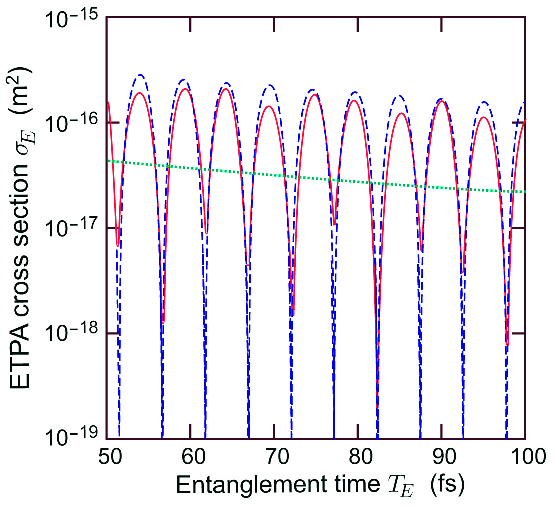}
\caption{Degenerate ($\omega_1^0 = \omega_2^0 = \tfrac12 \omega_p $)
entangled-two-photon
absorption cross section calculated for the 1S$\to$2S transition
in atomic hydrogen as a function of the entanglement time (solid red curve).
Also displayed are approximations that rely on a  single
intermediate state (dashed blue curve) and on a heuristic particle-like model
(dotted green curve).
(Adapted with permission from Fig.~1 of~\cite{Fei97}, \textcopyright 1997
American Physical Society.)
}\label{fig4}
\end{figure}

The concept of \textbf{entangled-two-photon absorption (ETPA)} was introduced in
the early 1980s~\cite{klyshko82}. Related two-photon processes were subsequently
investigated both  theoretically~\cite{geabanacloche1989,Javanainen90} and
experimentally~\cite{Friberg85_OC,georgiades1995PRL,dayan2005PRL}. In 1997, it
was shown that the ETPA cross section  $\sigma_{\!\scriptscriptstyle E}$ for simple
atoms could, in principle, be modified relative to the standard two-photon cross
section by varying the source entanglement time $T_{\!\scriptscriptstyle
E}$~\cite{Fei97,Perina98}. Unusual behavior could emerge, as illustrated by the
example provided in Fig.~\ref{fig4}: the 1S$\to$2S transition cross section in atomic
hydrogen oscillates with increasing entanglement time, exhibiting alternating regions
of cross-section enhancement and transparency~\cite{Fei97}. Still, it was understood
that demonstrating ETPA in the isolated-photon-pair domain would be experimentally
challenging~\cite{saleh1996ETPA,booth1998ETPA}.

Two standard approaches are commonly used for implementing
entangled-two-photon absorption. In \textbf{transmission-mode ETPA}, the
absorption is monitored by directly measuring the optical transmittance of the
system. In \textbf{fluorescence-mode ETPA}, the absorption leads to an excited state
whose subsequent decay via fluorescence is monitored. The latter mode can offer
greater sensitivity because fluorescence detection rejects transmitted excitation light,
although background fluorescence and low ETPA-induced fluorescence rates can limit
its effectiveness.  In fact, the first observation of standard two-photon absorption was
conducted in fluorescence-mode: in 1961, Kaiser and
Garrett~\cite{kaisergarrett1961} illuminated a crystal of CaF$_2$:Eu$^{2+}$ with
694.3-nm light from a ruby laser and observed blue fluorescence at 425.0~nm.

An important aspect of using time--frequency entangled-photon pairs for ETPA
stems from the fact that the sum-frequency spectrum can be narrow even though
the marginal photon spectra are broad. This is a consequence of the frequency
anticorrelations of the constituent photons, enabling joint spectrotemporal
behavior that is not available for a single classical field envelope subject to the usual
Fourier-pair
tradeoff~\cite{jost1997ETPS,saleh1998entangledPhotonSpectroscopy,dayan2004PRL,dayan2007PRA}.
These unusual properties have led to an array of potential scientific and
technological applications for such sources.

Entangled-two-photon absorption is examined in detail in Sec.~\ref{entabsfluor},
which considers both the heuristic framework generally used to describe these
processes and fully quantum formulations. Numerous reports detailing the
observation of ETPA have appeared in the literature, but independent replication has
not been achieved despite extensive efforts to do so (Sec.~\ref{ssec:ETFAFM}).

\subsection{Entangled-Two-Photon Fluorescence Microscopy (ETPFM)}
\label{EPmicroscopy}

\begin{figure}[htb!]
\centering
\includegraphics[width=4.50in]{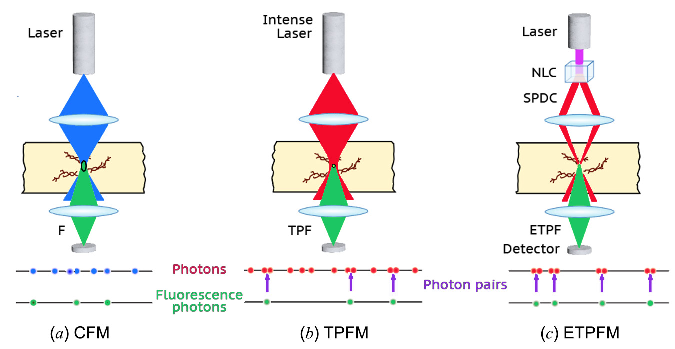}
\caption{Comparison of fluorescence microscopy using single photons,
independent photons, and entangled-photon pairs.
($a$)~Confocal fluorescence microscopy (CFM): shorter-wavelength
illumination (blue) excites single-photon fluorescence (F, green);
a confocal detection pinhole preferentially selects
fluorescence from the focal region.
($b$)~Two-photon fluorescence microscopy (TPFM): intense,
longer-wavelength illumination (red) produces two-photon fluorescence
(TPF, green), with excitation localized predominantly near the focus
by its quadratic dependence on intensity.
($c$)~Entangled-two-photon fluorescence microscopy (ETPFM):
a pump laser drives spontaneous parametric downconversion (SPDC)
in a nonlinear crystal (NLC), producing correlated photon pairs
(red) intended to excite entangled-two-photon fluorescence
(ETPF, green) at reduced incident intensity.
The lower diagrams schematically illustrate excitation-photon and
fluorescence-photon arrival times. Purple arrows mark accidental
close arrivals of independent photons in ($b$) and correlated
pair arrivals in ($c$). The diagrams illustrate the excitation
mechanisms, not relative fluorescence yields or demonstrated
imaging performance; absorption and fluorescence are probabilistic,
and only a tiny fraction of the incident pairs produces a detected fluorescence photon.}
\label{fig5}
\end{figure}

Fluorescence-mode entangled-two-photon absorption can be used, in principle, to
implement a quantum version of two-photon fluorescence microscopy known as
\textbf{entangled-two-photon fluorescence microscopy
(ETPFM)}~\cite{teich1997,teich1997ETPFM,us5796477,wo2003060610,thew2023ContempPhys}.
The principal characteristics of two established forms of fluorescence microscopy and
the proposed ETPFM approach are depicted in Fig.~\ref{fig5}:
 ($a$)~confocal fluorescence microscopy
(CFM),\, ($b$)~two-photon fluorescence microscopy (TPFM),\, and
($c$)~entangled-two-photon fluorescence microscopy (ETPFM). In CFM
(Fig.~\ref{fig5}($a$)), single-photon absorption excites fluorescence both within and
outside the focal region, but a detection pinhole preferentially transmits fluorescence
originating near the focal
plane~\cite{sheppard1977confocal,minsky1988confocal,pawley2006,paddock2014}.
This spatial filtering provides optical sectioning and suppresses out-of-focus
background, although it does not prevent excitation and associated photobleaching
outside the focal region. The shorter excitation wavelengths generally used for
single-photon fluorescence also tend to undergo greater scattering in biological tissue
than the longer wavelengths used for two-photon excitation.

In conventional TPFM (Fig.~\ref{fig5}($b$)), two photons jointly supply the energy
required to excite the
fluorophore~\cite{Denk90,masters2008nonlinear,konig2018multiphoton}. For
coherent illumination, their arrival times are uncorrelated, apart from the modulation
imposed by the laser pulse envelope, and the two-photon absorption rate depends
quadratically on the local intensity. A large value of the intensity is therefore required
to obtain appreciable fluorescence; this is commonly achieved by using tightly focused
ultrashort laser pulses. The quadratic dependence confines excitation predominantly
to the focal region and provides intrinsic optical sectioning without a detection
pinhole, thereby reducing out-of-focus fluorescence and photobleaching. Moreover,
the longer excitation wavelengths can provide greater penetration into biological
tissue. These advantages are accompanied by the cost and complexity of an ultrafast
laser system and by the possibility of photodamage at the high intensities required
near the focus.

ETPFM seeks to retain the benefits of long-wavelength two-photon excitation while
reducing the intensity level required to drive the absorption process. As illustrated in
Fig.~\ref{fig5}($c$), a nonlinear optical crystal (NLC) generates spontaneous
parametric downconversion (SPDC) that comprises photon pairs with correlated
arrival times. This avoids having to rely on accidental temporal coincidences between
independently arriving photons, as with TPFM. In the low-flux regime, absorption
involving the two members of the same pair can scale linearly, rather than
quadratically, with the incident intensity. Furthermore, the sum-frequency spectrum
can be narrow even when the individual photons have broad spectra, allowing
temporal correlation to coexist with selective excitation of a narrow two-photon
resonance (Sec.~\ref{EPabsorp}); the pump may be continuous-wave (CW) or pulsed.
These features motivate the pursuit of low-light-level, high-resolution fluorescence
imaging with reduced specimen damage. However, they do not, in and of themselves,
establish comparable image brightness, improved resolution, or the same intrinsic
optical sectioning as conventional TPFM: the spatial excitation profile depends on the
joint two-photon field at the specimen. Practical limitations include the available
photon-pair flux, optical losses, and uncertainties in the entangled-two-photon
absorption cross section and effective entanglement area. Reports of ETPFM have
appeared, but independent replication remains lacking (Sec.~\ref{ssec:ETFAFM}).

In closing, it is perhaps worthy of mention that while most implementations of
two-photon/nonlinear microscopy rely on photon emission from the sample, e.g.,
fluorescence or harmonic generation, some variants rely instead on non-emissive
processes, whereby the two-photon interaction alters a probe beam or produces
secondary non-photonic effects. Examples of such effects are nonlinear absorption,
refractive-index modulation, and photoacoustic signal generation.

\subsection{Entangled-Two-Photon Spectroscopy (ETPS)}\label{EPspectroscopy}

Nonlinear spectroscopy, which has been used extensively with coherent light in a
broad variety of configurations, can provide information about the energy levels of
a sample that cannot be accessed using conventional
spectroscopy~\cite{Mukamel1995PNOS,Cundiff2023OMCS}.

In particular, as mentioned in Sec.~\ref{EPabsorp}, time--frequency
entangled-photon-pair excitation has the merit that the biphoton state can
simultaneously exhibit strong temporal correlations and narrow sum-frequency
structure, thereby enabling forms of joint temporal--spectral interrogation that are
not achievable with a single classical pulse envelope. The ability to simultaneously
provide strong correlations in time and frequency can be particularly advantageous
in
spectroscopy~\cite{jost1997ETPS,saleh1998entangledPhotonSpectroscopy,kojima2004,
dayan2004PRL,dayan2007PRA,svozilík2018CP,svozilik2018JOSAB,
schlawin2016,schlawin2017JPB,schlawin2018,stefanov2021JCP,
mukamel2022JPC,chekhova2022OL,AlvarezMendoza2025,Fujihashi2026,Fan2026}.
A quantum-electrodynamic formulation of multidimensional spectroscopy has been
developed by Schlawin~\cite{schlawin2025JCP}.

\begin{figure}[htb!]
\centering\includegraphics[width=3.in]{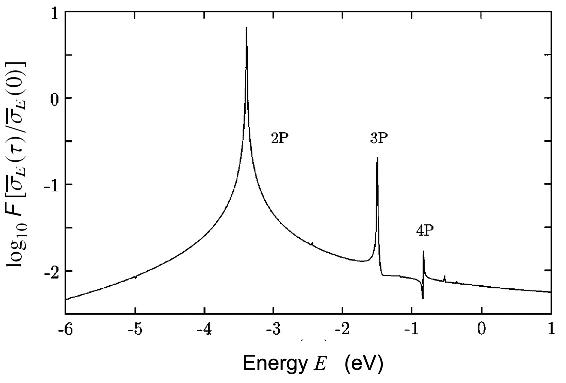}
\caption{Fourier transform of the normalized weighted-and-averaged atomic hydrogen
1S$\to$2S cross section,
$\mathsfit{F}\left[\overline{\sigma}_{\!\scriptscriptstyle E}(\tau)
/\overline{\sigma}_{\!\scriptscriptstyle E}(0)\right]$,
as a function of energy $E$ using an
entanglement-time range $1001 < T_{\scriptscriptstyle E} \leqslant 2001$~fs and an interbeam
delay-time range $0 \leqslant \tau \leqslant 1000$~fs, with increments of
$\delta T_{\scriptscriptstyle E}
= \delta \tau = 0.5$~fs. (Adapted with permission from Fig.~3
of~\cite{saleh1998entangledPhotonSpectroscopy},
\textcopyright 1998 American Physical Society.)
}\label{fig6}
\end{figure}

The initial prescription for carrying out \textbf{entangled-two-photon spectroscopy
(ETPS)}~\cite{saleh1998entangledPhotonSpectroscopy,Perina98} demonstrated that
the energy-level structure of a sample could be retrieved by carrying out a Fourier
analysis of sets of entangled-two-photon absorption probabilities for different
entanglement times or different time delays between photon arrivals. A numerical
illustration of this approach for the exactly calculable case of the 1S$\to$2S
two-photon transition in atomic hydrogen (which takes place via the intermediate  P
states 1S$\to$2P$\to$2S, 1S$\to$3P$\to$2S, etc.) is illustrated in Fig.~\ref{fig6}.
Both the virtual-state energy levels and the transition matrix elements (which
correspond to the heights of the peaks) clearly emerge. The energy levels are
determined from the locations of the peaks at $-3.40$, $-1.50$, and $-0.85$ eV,
which are identified with the principal quantum numbers $n = 2$, 3, and 4 and
correspond to the 2P, 3P, and 4P states, respectively. The values of the transition
matrix elements calculated from the heights of the peaks correspond to their actual
values.

Subsequent developments have shown  that some of the complexities associated with
implementing this approach can be mitigated by making use of auxiliary techniques,
such as modifying the temperature of the SPDC nonlinear optical
crystal~\cite{leonmontiel2019temp} or carrying out measurements at different pump
wavelengths~\cite{mertenskotter2021-JOSAB}. Eliminating the necessity of having to
control the correlations of the entangled-photon pairs is particularly valuable when
there are numerous intermediate states.

\subsection{Entangled-Two-Photon Photoemission (ETPP)}\label{EPphoto}

External photoemission from semiconductors and metals is governed by the nature
of the photoemissive material, principally its bandgap energy $E_{\!g}$, electron
affinity $\chi$, effective emission threshold, and quantum efficiency $\eta$; and by
the characteristics of the incident light, principally its photon energy $h\nu =
\hslash\omega$, angle of incidence, polarization, and statistical properties. For the
idealized intrinsic-semiconductor model adopted in this review, the emission
threshold is written as $\texttt{W} \approx E_{\!g} + \chi$, whereas for metals it is
identified with the work function ${\scriptstyle \mathcal{W}}$.

\begin{figure}[htb!]
\centering\includegraphics[width=4.25in]{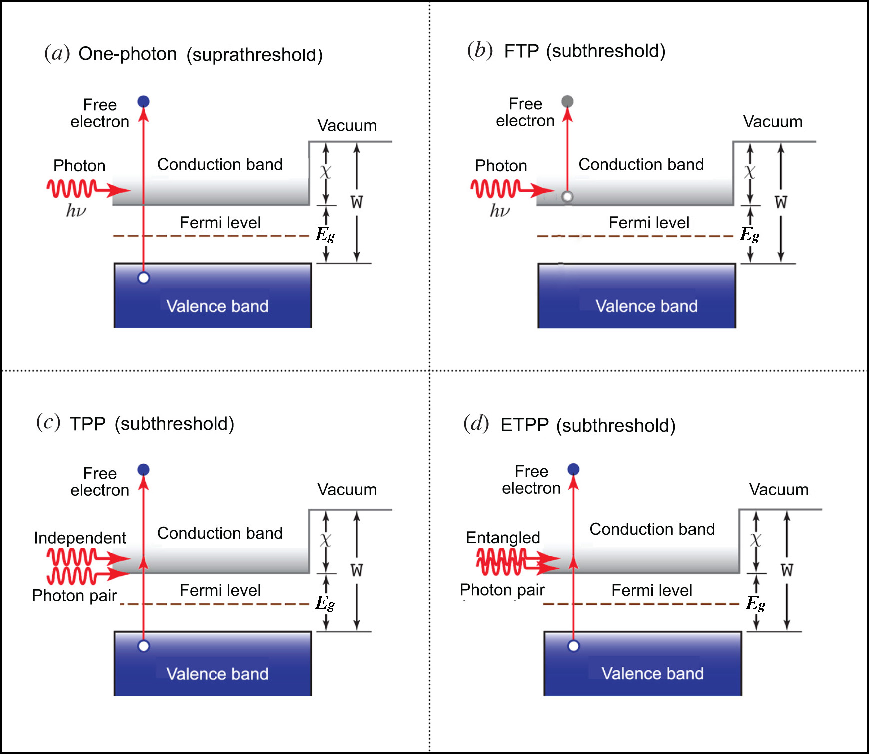}
\caption{Idealized heuristic energy-band diagrams for various forms of photoemission
from a semiconductor material at $\mathsfit{T\/} > 0$~K. The bandgap energy,
electron affinity, and ionization energy are denoted \,$E_{\!g}$, \,$\chi$, and \,\texttt{W},
respectively. \,($a$)~\emph{One-photon ordinary (suprathreshold) photoemission.}
When the photon energy $h\nu$ exceeds the ionization energy of the material,
it can impart sufficient energy to an electron in the valence band to escape to the
vacuum: $h\nu >  (E_{\!g} + \chi) = \texttt{W}$.\, ($b$)~\emph{One-photon
Fermi-tail (subthreshold) photoemission (FTP).} A photon of energy \,$h\nu < \texttt{W}$\,
can impart sufficient energy to an electron in the tail of the Fermi function to raise it
above the vacuum level.
\,($c$)~\emph{Two-photon (subthreshold) photoemission (TPP).}
When the energy $h\nu$ of each of two photons exceeds the bandgap energy,
but lies below the ionization energy of the material, two independent photons can
impart sufficient energy to an electron in the valence band to raise it to the
vacuum level: $E_{\!g} <  h\nu <  (E_{\!g} + \chi) < 2h\nu.$\, For simplicity,
energy-degenerate photons are displayed and the conduction-band intermediate state is taken to be
real.
\,($d$)~\emph{Entangled-two-photon (subthreshold) photoemission (ETPP).}
The energy conditions are the same as those for two-photon photoemission:
$E_{\!g} <  h\nu <  (E_{\!g} + \chi) < 2h\nu$.\, Again,  degenerate photon pairs are
displayed for simplicity and the conduction-band intermediate state is taken to be
real. Understanding the distinctions
among these four forms of photoemission is essential for being able to
discriminate among them.}
\label{fig7}
\end{figure}

In suprathreshold photoemission, the incident photon energy lies above the relevant
one-photon emission threshold, whereas in subthreshold photoemission it lies below
that threshold. Idealized band diagrams for suprathreshold photoemission and the
three most prominent forms of subthreshold photoemission from semiconductors are
portrayed in Fig.~\ref{fig7}, and elucidated in the caption thereof.

The study of the subthreshold process of \textbf{entangled-two-photon
photoemission (ETPP)}, illustrated in Fig.~\ref{fig7}($d$), was initiated at Boston
University in
1998~\cite{booth1998ETPA,Lissandrin99,teich00capri,Teich00MIT,Booth01,Booth04,lissandrin2004,booth2006}
and was taken up shortly thereafter by groups in Pasadena~\cite{strekalov02},
Moscow~\cite{chernov03}, and Kyoto~\cite{kobayashi2006,kobayashi2007}.

Particularly useful for observing ETPP are waveguide-based sources that employ
degenerate type-I or type-0 quasi-phase matching in periodically poled lithium niobate
(PPLN) or stoichiometric lithium tantalate (PPSLT). These sources generate collinear
SPDC with high efficiency, and they can incorporate chirped poling (CPPLN/CPPSLT) to
produce broad joint spectral bandwidths that, with suitable spectral-phase control,
can yield tightly time-correlated photon pairs, advantages for fostering efficient
entangled-photon interactions.

Entangled-two-photon photoemission is a useful analog of entangled-two-photon
absorption for a number of reasons, the overriding one being that it relies on the
emission of electrons rather than photons. The following advantages emerge when
comparing ETPP with fluorescence-mode ETPA (which can suppress transmitted
excitation light but remains susceptible to fluorescence backgrounds):
\begin{itemize}
    \item The interaction and detection processes in ETPP are co-located at a
        common locus, thereby avoiding the involvement of an additional component
        with its attendant sub-unity quantum efficiency.

    \item Photoemissive materials and excitation wavelengths can be selected such
        that the transition takes place via a real, rather than virtual, intermediate state;
        under suitable circumstances, this can enhance the transition probability and
        quantum efficiency.

    \item The collection efficiency for photoelectrons is significantly greater than that
        for fluorescence photons. Charged photoelectrons can be collected, multiplied,
        and detected with high efficiency, whereas fluorescence photons are emitted
        over a large solid angle and are readily absorbed, deflected, and scattered.
        Fluorescence collection efficiency depends strongly on the numerical aperture
        and collection geometry, and can be increased by using high-NA optics, multiple
        collection objectives, or an integrating sphere.

    \item Photoelectrons are not subject to loss through self-quenching and
        reabsorption, unlike fluorescence photons.

    \item Experiments can be conducted within the confines of a channel
        photomultiplier (CPM) module or a photomultiplier tube (PMT). The electron
        multipliers built into these devices provide large internal gain with little noise,
        thereby facilitating the detection of low-flux entangled-photon pairs.

    \item Fermi-tail photoemission can lie below detectability in CPM modules with
        CsK$_2$Sb photocathodes, whereas it acts as an insidious source of noise in
        PMTs with CsK$_2$Sb photocathodes.

    \item Entangled-two-photon photoemission in a semiconductor leverages the
        high absorber density inherent in condensed matter. This in turn enhances the
        overall interaction, effectively overcoming the detection limits that typically
        hinder observations in dilute atomic and molecular samples.
\end{itemize}

The observation of entangled-two-photon photoemission from the CsK$_2$Sb
photocathode of a channel photomultiplier (CPM) module is reported in
Sec.~\ref{KobalissSec}. As shown in Sec.~\ref{sssec:Kobalissandrin}, quantum theory
can reproduce the experimentally inferred entangled-two-photon quantum efficiency
for a suitable value of the material response ratio $\mathcal{Q}$.

From the point of view of applications, entangled-two-photon photoelectron
spectroscopy (ETPPS), proposed by Gu \emph{et al.}~\cite{gumukamel2023}, can
serve as a quantum-optical extension of conventional two-photon photoelectron
spectroscopy~\cite{Imamura68,Weinelt2002}, in which entangled-photon pairs,
rather than classical light pulses, excite and emit electrons from molecules, surfaces,
nanostructures, and solids. By virtue of the temporal correlations and spectral
anticorrelations inherent in time--frequency entangled-photon pairs, ETPPS can offer
enhanced spectral selectivity, sublifetime temporal resolution, and access to
coherence-driven phenomena that are not observable with classical light, thereby
enabling a new regime of photoemissive probing that is sensitive to quantum
correlations in both light and matter.

Although photoelectrons are usually treated classically, under special circumstances
their intrinsic quantum states can play a significant role~\cite{lhuillier25}. Indeed,
electron--ion entanglement in atomic and molecular photoemission can reveal the
dynamics of decoherence and entanglement at the attosecond
timescale~\cite{berkane25}. Photoemission elicited by bright squeezed vacuum
exhibits novel and interesting features~\cite{chekhova24,chekhova25}.

\subsection{Other Paradigms and Applications of
Entangled-Photon Pairs}\label{EPinteractions} Aside from absorption, fluorescence
microscopy, spectroscopy, and photoemission, numerous alternative paradigms and
applications that engage entangled-photon pairs have been proposed and
investigated. Some of these, catalogued below with brief descriptions, have been
implemented with varying degrees of success. Our emphasis in this section is on
effects that specifically involve entangled-photon pairs, rather than on the broader
scope of quantum interactions that include, for example, quadrature-squeezed
light~\cite{mckenzie04,takeno07,vahlbruch08,vahlbruch16} and
photon-number-squeezed
light~\cite{perina2013OE,perina2017PRA,perina2024,Thapliyal2026}. While the
applications we discuss are peripheral to subthreshold absorption, spectroscopy, and
photoemission, they nevertheless serve to indicate how nonclassical two-photon
correlations can be harnessed to realize capabilities beyond conventional
intensity-based methods.

\subsubsection{Entangled-Two-Photon Imaging}\label{EP-Imaging}

Entangled-photon pairs support a broad class of imaging modalities in which image
formation is governed by fourth-order field correlations (typically accessed via
coincidence detection), rather than by the second-order field correlations underlying
conventional intensity imaging. Early experiments established that an image can be
reconstructed nonlocally by correlating detections in the two arms of an SPDC
source~\cite{Pittman95,Lugiato02}. Subsequent theory placed these effects on an
\textbf{entangled-photon Fourier-optics} foundation, showing that the propagated
two-photon probability amplitude determines the imaging transfer characteristics,
and that the attainable resolution, coherence behavior, and system response depend
on the structure of the entangled-photon state and the optical propagation
geometry~\cite{Abouraddy02_JOSAB3}.

In this context, useful physical intuition is provided by Klyshko's advanced-wave
(unfolded) picture~\cite{Klyshko80}, in which the coincidence-imaging system is
reinterpreted as an equivalent one-photon imaging system with a reflected wave
propagating from one detector, through the optical system to the source, and thence
to the other arm, thereby clarifying the origin of image inversion, magnification, and
effective point-spread behavior. Also important is the distinction between generic
correlation and genuine entanglement: while entanglement enables certain
distributed quantum-imaging capabilities that cannot be reproduced with classically
correlated fields~\cite{abouraddy2001role}, ghost imaging itself can also be realized
using classical correlations~\cite{ErkmenShapiro2010AOP}.
 Analyses of
photon-correlation imaging also clarify the associated noise mechanisms and
signal-to-noise ratio (SNR) tradeoffs, and provide a framework for comparing
quantum and classical correlation-imaging protocols at fixed flux, bandwidth, and
detection conditions~\cite{saleh2008}.

Over the past two decades, imaging that makes use of entangled photons has evolved
to encompass \textbf{entangled-two-photon imaging}, ghost imaging, heralded
imaging, and related correlation-based schemes that exploit spatial, temporal,
spectral, and polarization correlations, including entanglement where
relevant~\seccite{9.3}{saleh25}. Recent reviews emphasize theoretical and
experimental advances, including those in single-photon-sensitive detector arrays and
cameras, improved coincidence detection, and expanding applications at low-light
levels and in
bioimaging~\cite{ErkmenShapiro2010AOP,zhang2024SA,yue2025-SupResImage-NP,
XinyiLi2025,lotfipour2025,boyd2025,simon2025,
padgett2026,qophthal2026,zhaomedimaging2026,Stewart2026,xraySPDCimaging2026}.
Performance measures are usually expressed in terms of spatial resolution, contrast,
acquisition time, and noise. Contemporary developments in quantum photonic
platforms and structured light introduce further quantum-imaging
modalities~\cite{ndao2025APR}.

\subsubsection{Entangled-Two-Photon Metrology}
\label{EP-Metrology}

Metrology is the science of measurement, including the definition of measurement
standards, the calibration of instruments, and the estimation of physical parameters
with quantified accuracy and precision. \textbf{Quantum metrology} employs
quantum states and measurement strategies involving light or matter to characterize
and potentially improve measurement precision beyond specified classical
benchmarks, and emphasizes the systematic study of quantum-enhanced parameter
estimation (often formalized through Fisher information and quantum
Cram{\'e}r--Rao bounds), sensitivity limits, and quantum advantages. It finds use over
a broad range of applications that stretches from atomic clocks and
gravitational-wave detectors to imaging resolution and
spectroscopy~\cite{Taylor2016PR,
tan2019,polino2020,schlawin26chapter}~\seccite{9.4}{saleh25}.

\textbf{Entangled-two-photon metrology} is most naturally viewed as an outgrowth of
two-photon interference, because the relevant observables are typically joint
detection statistics (especially coincidence rates and coincidence interferograms)
whose dependence on an unknown parameter is governed by biphoton amplitudes
and fourth-order interference
effects~\cite{simon2017,simon2025,zhangsussman2026}. A simple but important
early example is absolute detector calibration using correlated photon pairs from
SPDC (the Klyshko method), in which detector quantum efficiency is inferred from
singles and coincidence counts without the need for a separately calibrated optical
source~\cite{Klyshko77,Klyshko80-calib}. This calibration relies on photon-pair
number correlations rather than entanglement itself, and the inferred detection
efficiency includes intervening optical losses unless they are separately characterized.

More generally, entangled-photon metrology uses coincidence-domain interference
observables to estimate delays, phases, dispersive effects, or
polarization-dependent sample parameters, while heralding and coincidence gating
can suppress uncorrelated background counts in low-light operation. The
metrological significance of entangled-photon pairs lies not only in possible
precision enhancement, but also in the availability of self-referenced and
noise-rejecting measurement observables derived directly from two-photon
correlation
structure~\cite{MigdallCastellettoDegiovanniRastello2002,Giovannetti2004Science,
PolyakovMigdall2007,GiovannettiLloydMaccone2011}. Hyperentangled states can
provide a natural platform for multiparameter quantum metrology since distinct
physical parameters may be encoded in different entangled degrees of
freedom~\cite{Atature01-PRA,sergienko2003quantumMetrology,Szczykulskaetal2016}.

\subsubsection{Quantum Ellipsometry}\label{EP-Ellipsometry}

Ellipsometry is a polarization-based optical metrology technique in which material
properties (most commonly thin-film thickness and optical constants) are inferred
from the relative amplitude and phase changes imparted to the orthogonal
polarization components of an incident source of light, upon reflection from (or
transmission through) a sample~\cite{Azzam77}. Classical ellipsometers employ a
variety of polarization-analysis configurations, and accurate measurements generally
require calibration or characterization of the polarization optics and instrumental
response~\cite{Tompkins99,Fujiwara2007}.

\textbf{Quantum ellipsometry} (also termed twin-photon ellipsometry or
entangled-two-photon ellipsometry) is distinct in that the measurement is
performed with SPDC-generated photon pairs and coincidence detection, so that
the relevant observable is a fourth-order correlation rather than a single-beam
intensity. In the polarization-entangled realization, the entangled-photon state and
coincidence analyzer settings encode the sample-induced polarization amplitude
ratio and phase shift directly into the coincidence fringe, enabling extraction of the
ellipsometric parameters without the need for external source/detector calibration
and, in the original formulation, without a reference sample.

When the concept was introduced in
2001~\cite{Abouraddy2001OL,abouraddy2002JOSAB,us6822739}, it was
emphasized that polarization entanglement could effectively furnish the
interferometric functionality needed for \emph{ideal ellipsometry}, while reducing
the optical complexity of a conventional interferometer. Correlated-photon
ellipsometry experiments were subsequently carried out using semiconductor
samples such as Si and GaAs. The results of these calibration-free experiments
demonstrated that the ellipsometric parameters recovered were in good accord with
theory.

Signal-to-noise performance was also analyzed relative to that of conventional
schemes~\cite{toussaint2004}. This work also clarified that useful
quantum-ellipsometric advantages can arise from photon-number correlations
measured via coincidence counting, even when polarization entanglement itself is
absent or not explicitly exploited, thereby broadening the operational
interpretation of quantum ellipsometry. Summaries of the method and its
advantages have been provided by Simon \emph{et al.}~\cite{simon2017}, with an
emphasis on the reduction of relative parameter-estimation error and the
elimination of external-reference calibration.

\subsubsection{Entangled-Two-Photon Sensing}\label{EP-Sensing}

Sensing is the inference of a physical quantity or system state (e.g., displacement,
vibration, phase, absorption, or target presence) from measured observables ---
\textbf{entangled-two-photon sensing} encompasses protocols in which signal
extraction is governed by two-photon coincidence statistics and entangled-photon
correlations, rather than by single-photon intensity
measurements~\cite{Degen2017RMP,Pirandola2018QST}. This is to be
distinguished from entangled-two-photon metrology (Secs.~\ref{EP-Metrology} and
\ref{EP-Ellipsometry}), which emphasizes precision limits, estimator variance, and
quantum-enhanced parameter estimation. Entangled-two-photon sensing is
broader, encompassing detection and transduction tasks in which
entangled-photon pairs provide improvements in selectivity, robustness,
background rejection, timing discrimination, or contrast. In short, sensing exploits
the structure of entangled-photon correlations, whereas metrology focuses
specifically on attainable estimation precision under specified
statistical-performance criteria~\cite{simon2017,simon2025}.

From an experimental perspective, entangled-photon sensing accommodates
architectures in which the sensed quantity is encoded in the coincidence rate,
coincidence timing, joint spectral/temporal correlations, or two-photon interference
visibility. The nanovibrometry example investigated by Lualdi \emph{et
al.}~\cite{kwiat2025vibrometry} is representative: it demonstrates
entanglement-enhanced sensing of nanometer-scale vibrations using two-photon
interference with highly nondegenerate time--frequency entanglement, together with
a timestamp-based flux-probing spectral-analysis protocol that recovers vibrational
signals at frequencies up to 21~kHz. In separate experiments, the authors
demonstrate resilience to imbalanced optical loss and background, and report a
quantum advantage in amplitude estimation relative to classical single-photon
interference under these degraded conditions. More generally, this example
illustrates how entangled-two-photon sensing can offer robust signal recovery in
low-light, lossy, or background-contaminated environments. This emphasis on
performance in adverse backgrounds is also consonant with the broader
quantum-sensing/quantum-illumination
literature~\cite{Pirandola2018QST,Lopaeva2013PRL}.

\subsubsection{Quantum-Optical Coherence Tomography}\label{EP-QOCT}

\textbf{Quantum-optical coherence tomography (QOCT)}, the
entangled-two-photon counterpart of conventional optical coherence tomography
(OCT)~\cite{Huang1991OCT,Fercher2003RPP}, is a quantum-interferometric
modality that combines elements of both sensing and metrology. In standard
time-domain OCT, a classical broadband source is used in conjunction with a
Michelson interferometer to achieve axial sectioning via low-coherence
interferometry. In this approach, the axial point-spread function is broadened by
sample dispersion unless compensation is introduced. In QOCT, in contrast,
frequency-anticorrelated photon pairs generated by spontaneous parametric
down-conversion (SPDC) are caused to interfere in a Hong--Ou--Mandel (HOM)
configuration~\cite{Hong87,prasad1987,fearn1987,rarity2024} and the
tomographic signal is obtained from fourth-order (coincidence)
interference~\seccite{9.1}{saleh25}.

In its standard idealized form, with strongly frequency-anticorrelated photon pairs
generated using a narrowband pump, QOCT has two principal advantages relative to
OCT for comparable source bandwidth: 1)~approximately a twofold improvement in
axial resolution; and 2)~intrinsic cancellation of even-order sample dispersion
(including group-velocity dispersion)~\cite{abouraddy2002QOCT,
Nasr03,us6882431,teich2012}. Under these conditions, QOCT enables axial sectioning
that is immune to even-order dispersion and, in multilayer samples, access to
interstitial dispersion information through the structure of the coincidence
interferogram~\cite{nasr2004}. This framework initiated a broad line of investigations
that spanned theory and proof-of-principle experiments, and led to demonstrations of
resolution enhancement via source engineering (e.g., chirped quasi-phase matching),
and 3D imaging of biological specimens~\cite{carrasco2004OL,nasr2009}.

Subsequent developments substantially expanded the QOCT landscape.
Polarization-sensitive quantum-optical coherence tomography (PS-QOCT) generalized
the method to birefringent and polarization-structured samples while retaining the
quantum-interferometric dispersion advantages, and later experiments demonstrated
biological-sample polarization imaging and
characterization~\cite{booth2004,booth2011,sukharenko2024PSQOCT}. Parallel
efforts improved source and system performance, including ultrabroadband/tailored
sources and high-depth-resolution demonstrations in dispersive
samples~\cite{okano2015QOCT,hayama2022QOCT}. More recent work has addressed
practical obstacles that have hindered the deployment of QOCT. Limitations such as
low count rates and multilayer artifacts/cross-interference features have spurred the
development of alternative correlation-based implementations,
source/interferometer redesigns, and phase-engineered strategies for artifact
suppression and improved information extraction in complex
samples~\cite{kulik2025,katamadze2025,urenbarzanjeh25,
WeiLi2025PRL,kwiat2025,uren2026scireports,takeuchi2026qoctHiVol,Kolenderska2026Fast}.
Although QOCT is unlikely to supplant OCT as a general-purpose clinical tool, it serves
as a model for the development of new quantum architectures for precision probing
and imaging.

\subsubsection{Quantum Holography}\label{EP-Holography}

Holography, in the classical sense introduced by Gabor in
1948~\cite{Gabor1948,Gabor1949}, records an interference pattern between an
object field and a mutually coherent reference field, which encodes both amplitude
and phase information for subsequent wavefront reconstruction. Following Gabor's
invention, major practical and conceptual advances included the off-axis laser
holography approach of Leith and Upatnieks~\cite{Leith1962} and Denisyuk's
reflection-holography geometry~\cite{Denisyuk1962}.

\textbf{Quantum holography} retains the objective of phase-sensitive field
reconstruction. In entangled-two-photon implementations, holographic information is
retrieved from fourth-order field correlations measured through photon-coincidence
detection. Other implementations, notably holography with undetected light, exploit
induced first-order coherence in the detected field. In early entanglement-based
two-photon formulations, Belinskii and Klyshko~\cite{belinskii1994} developed a
broader two-photon optics framework that includes holography, emphasizing the
flexibility of Klyshko's advanced-wave (unfolded) picture~\cite{Klyshko80}. Building on
this approach, it was shown that entanglement can enable holographic information
retrieval in a configuration that is not possible classically: one photon interrogates a
remote object inside a chamber whose walls act as a spatially nonresolving bucket
detector (integrating sphere), while its entangled partner is detected with spatial
resolution in a separate arm. The chamber-integrated (i.e., marginal) coincidence-rate
distribution over the spatially resolved detector serves as the
hologram~\cite{abouraddy2001quantumHolography,2003qcmcholography}.

This form of quantum holography thus yields a holographically reconstructible record
even when the individual beams are not themselves suitable for ordinary first-order
interference, highlighting the central role of entanglement (and more generally
two-photon coherence) in the imaging
process~\cite{abouraddy2001quantumHolography,2003qcmcholography}. Quantum
holography is most naturally viewed as a phase-sensitive member of the broader
entangled-two-photon imaging family (Sec.~\ref{EP-Imaging}), but it also shares
methodological connections with the metrology and sensing modalities outlined in
Secs.~\ref{EP-Metrology} and \ref{EP-Sensing}, respectively.

Other approaches have been introduced more
recently~\cite{devaux2019,defienne2021} and quantum holography has broadened
from coincidence-scanning architectures to induced-coherence and
undetected-light implementations~\cite{Wang91JOSAB,zou1991,Wang91PRA}. It is
useful to distinguish the earlier entangled-two-photon holography (where the
holographic signal is reconstructed from coincidence measurements between
detected photon pairs) from the more recent quantum holography with undetected
light (where the object is probed at one wavelength while the hologram is inferred
from interference measured at a different detected wavelength via induced
coherence)~\cite{graefe2022SA,graefe2024OE}.

In the phase-shifting approach demonstrated by T{\"o}pfer \emph{et al.},
amplitude and phase of the object transmission/reflection are recovered from
multiple signal-photon images acquired at controlled phase steps in an SU(1,1)-type
nonlinear interferometer, with the object-field information encoded in interference
between alternative two-photon probability amplitudes, rather than in the direct
interference of the detected object and reference beams~\cite{graefe2022SA}.
Off-axis holographic imaging with undetected light has extended this to single-shot,
wide-field recovery of complex object information via Fourier off-axis processing,
thereby reducing the need for multi-frame phase stepping while improving
robustness for dynamic samples~\cite{graefe2024OE}.

As summarized in the treatment provided by Simon \emph{et
al.}~\cite{simon2017}, quantum holography is naturally viewed as part of the
broader family of correlation-based quantum imaging methods, but is distinguished
by its explicit wavefront-reconstruction objective and its use of nonclassical
correlations to encode phase information under conditions that are inaccessible or
impractical for standard holography (e.g., nonlocal detection, spectral separation,
or detection constraints). Ghost imaging, in contrast, is primarily associated with
correlation-based intensity imaging~\cite{ErkmenShapiro2010AOP,forbes2025}.

\section{SUPRATHRESHOLD AND SUBTHRESHOLD PHOTOEMISSION} \label{theory}

The idealized heuristic band diagrams associated with suprathreshold photoemission
and three forms of subthreshold photoemission from semiconductors were displayed
in Fig.~\ref{fig7}:
\begin{itemize}
  \item Fig.~\ref{fig7}($a$). Suprathreshold (one-photon) photoemission. By definition, this
      process is absent when \,$h\nu < \texttt{W}$.
  \item Fig.~\ref{fig7}($b$). Subthreshold (one-photon) Fermi-tail photoemission. FTP can
      be initiated by classical light or by entangled-photon pairs.
  \item Fig.~\ref{fig7}($c$). Subthreshold two-photon photoemission. TPP can be induced
      by classical light or by entangled-photon pairs.
  \item Fig.~\ref{fig7}($d$). Subthreshold entangled-two-photon photoemission. By
      definition, ETPP can be induced only by entangled-photon pairs.
  \end{itemize}

We present brief descriptions of these four processes in Secs.~\ref{singletheory},
\ref{subthreshtheory}, \ref{twophotontheory}, and \ref{enttheory}, in turn. A
background current $i_{\scriptscriptstyle D}$, arising from detector dark current and
electronic offsets, may also be present; its significance depends on the detector and
operating conditions.
 We
close in Sec.~\ref{optlosssys} by discussing measures of optical power and the
presence of optical loss.

\subsection{Suprathreshold (Ordinary) Photoemission} \label{singletheory}

The ordinary photoelectric effect (photoeffect) was discovered by Hertz in
1887~\cite{Hertz87} and the underlying physics  was ultimately explained by Einstein
in 1905~\cite{Einstein05PE}. If the energy of a photon illuminating a material is
sufficiently large, it can impart enough kinetic energy to an electron to allow it to
surmount the potential barrier at the material's surface and escape into the vacuum
as a free particle~\cite{Ganciu2026}. This process, called photoelectric emission or
photoemission~\chapcite{19}{saleh2019}, can involve transport, screening, and
scattering effects that influence the timing of electron emission.
 \textbf{Suprathreshold photoemission} is
the most common, and by far the strongest, form of photoemission. It is portrayed in
the band diagram of Fig.~\ref{fig7}($a$) for a generic intrinsic semiconductor.

\subsubsection{Einstein Photoemission Equation}\label{1PEinst}
Energy conservation requires that an electron emitted from the valence band,
where electrons are plentiful, have a maximum kinetic energy
\begin{equation}\label{eq17-0-2}
E_{\max} = h\nu - \texttt{W} = h\nu - (E_{\!g} + \chi)
\end{equation}
upon liberation. Here, $h\nu$, $E_{\!g}$, $\chi$, and \texttt{W} represent the photon
energy, bandgap energy, electron affinity, and ionization energy of the material,
respectively. Within the idealized intrinsic-semiconductor picture of
Fig.~\ref{fig7}($a$), the photoelectron attains the maximum value of the kinetic
energy specified in Eq.~(\ref{eq17-0-2}) only when it initially lies at the very top of the
valence band; a deeper lying electron must expend energy to reach the top of the
valence band, thereby reducing its kinetic energy on liberation.
Equation~(\ref{eq17-0-2}) is known as the \textbf{Einstein photoemission equation}.

\subsubsection{Volume vs. Surface Effects} \label{volvssurf}
Early theories of ordinary photoemission from metals were derived by Fan in
1945~\cite{Fan45}, and from semiconductors by Kane in 1962~\cite{kane62}. The
theories have matured significantly over the years. The currently accepted
foundational model for metals consists of a three-step process codified by Spicer
and Berglund for Cu and Ag~\cite{Spicer1964theory,Spicer1964exp,Sommer68}:
1)~optical excitation of an electron within the volume of the material; 2)~transport
of the excited carrier to the surface; and 3)~escape through the surface potential
barrier. Each stage is treated independently, with corresponding probabilities
whose product yields the overall quantum efficiency.

The excitation probability is governed by the joint density of states and the
occupation probabilities are dictated by the Fermi--Dirac distribution. The transport
stage considers both elastic and inelastic scattering processes; the mean free path
depends on electron energy and the material, and does not vary monotonically with
energy over all energy ranges.  The escape probability includes classical, over-barrier
emission but can also involve quantum tunneling. The escape depth $d$ is an effective
attenuation length that characterizes the depth below the surface from which a
photoexcited electron can still reach the surface, surmount the potential barrier, and
be emitted into vacuum. An exact solution to the time-dependent Schr{\"o}dinger
equation is  available for strong-field, ultrafast photoemission from biased-metal
surfaces~\cite{zhang16}.

Suprathreshold photoemission from alkali metals is generally also considered to be
predominantly a volume effect, particularly for photon energies well above
threshold and for sufficiently thick
samples~\cite{thomas56,thomas57a,thomas57b,meessen61}. The three-step
process for semiconductors often incorporates additional features such as carrier
diffusion, minority-carrier dynamics, and surface recombination
velocities~\chapcite{5}{Seznec2022}. Excited carriers in semiconductors can
thermalize via electron--phonon or electron--hole scattering during transport to the
surface, and surface band bending can enhance photoemission. Other refinements
include finite-temperature Fermi--Dirac distributions and Schottky-barrier lowering.

\subsubsection{Photocurrent}\label{1Pcurrent}

The suprathreshold photocurrent $i_{\scriptscriptstyle S}$ can be written as
\begin{equation}\label{iSphiA}
  i_{\scriptscriptstyle S} \propto \phi A \propto \Phi \propto P,
\end{equation}
indicating that it is proportional to the incident photon-flux density $\phi$
(photons/m$^2$-s), illumination area $A$ (m$^2$), incident photon flux $\Phi$
(photons/s), and optical power $P$ (W). Under comparable illumination conditions,
suprathreshold photoemission generally yields substantially larger photocurrents than
the subthreshold processes portrayed in Figs.~\ref{fig7}($b$)--($d$).

\subsubsection{Responsivity and Quantum Efficiency}\label{1PRQE}

The responsivity $\mathcal{R}_{\scriptscriptstyle S}$ (A/W) is the proportionality
constant that relates the photocurrent $i_{\scriptscriptstyle S}$ to the incident
optical power $P$ in Eq.~(\ref{iSphiA}), i.e.,
\begin{equation}\label{iSETAresp}
i_{\scriptscriptstyle S} = \mathcal{R}_{\scriptscriptstyle S} P.
\end{equation}
A closely related measure, the quantum efficiency $\eta_{\scriptscriptstyle S}$, is
linked to $\mathcal{R}_{\scriptscriptstyle S}$ via the electron and photon fluxes,
\,$\mu_{\scriptscriptstyle S} = i_{\scriptscriptstyle S}/e$\, and \,$\Phi = P/h\nu$,\,
respectively. For radiation of wavelength $\lambda = c/\nu$, the relationship may
be expressed as
\begin{equation}\label{iSRESPetaSuper}
\eta_{\scriptscriptstyle S}~\text{[electrons/photon]} \equiv \frac{\mu_{\scriptscriptstyle S}}
{\Phi} =    \frac{h\nu}{e}\,\mathcal{R}_{\scriptscriptstyle S} =
\frac{hc}{e\lambda}\cdot \mathcal{R}_{\scriptscriptstyle S}~~\text{[A/W]}.
\end{equation}
Values for $\mathcal{R}_{\scriptscriptstyle S}$ and  $\eta_{\scriptscriptstyle S}$ are
typically 2--4 orders of magnitude larger for semiconductors than for metals.

A generic version of Eq.~(\ref{iSRESPetaSuper}) is applicable for the various forms of
subthreshold photoemission illustrated in Figs.~\ref{fig7}($b$)--($d$):
\begin{equation}\label{iSRESPeta}
\eta_{\scriptscriptstyle [\cdot]}~\text{[electrons/photon]} =
\frac{hc}{e\lambda}\cdot \mathcal{R}_{\scriptscriptstyle [\cdot]}~~\text{[A/W]}.
\end{equation}
The quantum efficiency $\eta$ is dimensionless since the proportionality constant
$hc/e\lambda$ has units of J/C (or W/A), which cancels the  A/W units of the
responsivity.

In the classical photoemission literature of the 1930s--1960s, the quantum efficiency
\,$\eta$\, was commonly referred to as the photoelectric yield \,$Y$. The responsivity
\,$ \mathcal{R}$ \, has its origin in the analog photocurrent and optical-power
measurements of the day. With the advent of digital instrumentation and
photon/electron counting techniques in the late 1960s and 1970s, the use of
quantum efficiency --- the number of emitted electrons per incident photon ---
became dominant.

\subsubsection{Photoemission from CsK$_2$Sb} \label{singletheoryCsK2Sb}

CsK$_2$Sb is the sample of choice for most of the experiments reported in this work.
Discovered by Sommer in 1963~\cite{Sommer1963}, this material is often modeled
for phenomenological purposes as an intrinsic, direct-bandgap, bialkali-antimonide
semiconductor with a Fermi level lying near the center of the bandgap, so that its
energy-band diagram is qualitatively captured by the sketch displayed in
Fig.~\ref{fig7}. The optical absorption is volumetric, but the emission is limited by
carrier transport and escape from a near-surface region, consistent with the
three-step description set forth in Sec.~\ref{volvssurf}.

Representative experimental studies report that CsK$_2$Sb has a bandgap energy
$E_{\!g} \approx 1.0$~eV, an electron affinity $\chi \approx 1.1$~eV, and a
valence-band ionization energy $\texttt{W} \approx E_{\!g} + \chi \approx
2.1$~eV~\cite{Sommer68,Nathan70,Varma78,Ghosh80,Lissandrin99}; these values
are adopted in the present work. While the precise parameter values depend on
preparation and surface condition, a quantum efficiency   $\eta_{\scriptscriptstyle
S} \approx 0.25$ is attainable and corresponds to a responsivity
$\mathcal{R}_{\scriptscriptstyle S} = (e \lambda/hc) \,\eta_{\scriptscriptstyle S}
\approx 0.085$~A/W at $\lambda = 420$~nm, in accordance with
Eq.~(\ref{iSRESPetaSuper}). CsK$_2$Sb is widely used as a photon-to-electron
converter in many applications.

From a theoretical perspective, first-principles calculations yield a range of values
for the bandgap of CsK$_2$Sb. Within Kohn--Sham density functional theory
(DFT)~\cite{KohnSham1965} using a generalized gradient approximation (GGA), the
calculated bandgap turns out to be $E_{\!g}=0.92$~eV. Although this lies
reasonably close to the experimental estimate, it should be noted that semi-local
DFT eigenvalue bandgaps are not quasiparticle excitation energies and typically
underestimate the fundamental bandgap. Using a quasiparticle methodology,
Cocchi \textit{et al.}~\cite{cocchi2018CsK2Sb,cocchi2019CsK2Sb} obtained
$E_{\!g}=1.62$~eV, whereas Sharma \textit{et al.}~\cite{sharma2023CsK2Sb}
reported $E_{\!g}=1.44$~eV from a hybrid-functional treatment that includes
spin--orbit coupling. Differences among these theoretical values reflect, in part, the
distinct physical quantities targeted and methodological choices.

\subsubsection{Single-Photon Counting with a PMT} \label{singletheoryPMT}

\begin{figure}[htb!]
\centering\includegraphics[width=3.25in]{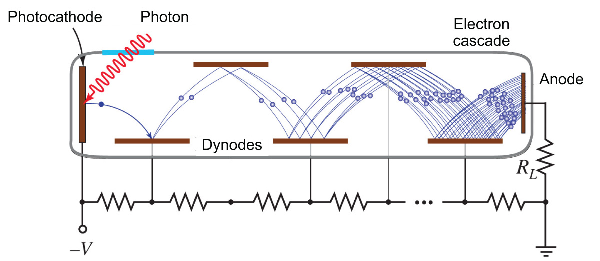}
\caption{Sketch of a photomultiplier tube (PMT), highlighting the
photocathode, dynodes, anode, and internal electron-multiplication arrangement.
PMTs convert individual photons into current pulses.
Electron gain generally increases with the number
of dynode stages, whereas transit time and timing
jitter also depend on the electrode geometry and applied voltages.
 PMTs can contain as many as
14 dynodes.} \label{fig8}
\end{figure}

For materials that can be prepared in the form of a photocathode, photoemission may
be conveniently observed by using a photodiode or a photomultiplier tube (PMT),
such as the one schematically portrayed in Fig.~\ref{fig8}~\chapcite{19}{saleh2019}. A
photon generates a photoelectron with probability $\eta_{\scriptscriptstyle S}$ in an
opaque or semitransparent photocathode. The photoelectron then enters the
electron-multiplier structure within the PMT and accelerates toward specially placed
cesiated-oxide or semiconductor surfaces called dynodes, which are maintained at
successively higher potentials~\cite{iams1935,zworykin1936,wright2017}. The
ensuing stochastic secondary emission from successive dynodes results in a cascade of
electrons that ultimately reaches the anode, which is maintained at the highest
electrical potential~\cite{teich1986}. The net result is an electron flux in the circuit,
called the photoelectric current (or photocurrent), that in linear analog operation is
proportional to the incident photon flux~\cite{shockley1938}; in photon-counting
operation, the detected count rate is proportional to the incident photon flux within
the usual limits imposed by dead time, thresholding, and saturation.

Visible and near-infrared PMTs can exhibit the following characteristics:
1)~suprathreshold spectral sensitivities that extend over the wavelength range
$150 \leqslant \lambda \leqslant 1000$~nm; 2)~quantum efficiencies as high as
$\eta_{\scriptscriptstyle S} \approx 0.40$; 3)~dark-count rates as low as 100
counts/s; 4)~low-noise electron amplification with gains as high as $\approx 10^8$;
5)~timing jitter of 300~ps; 6)~maximum count rates of 10~MHz; 7)~diameters
ranging from mm to ½~m; 8)~room-temperature operation; and 9)~limited
photon-number-resolving capability.

\subsubsection{Photoelectron Statistics} \label{1Qphotstats}

The statistical properties of the photons incident on the photocathode of a
photon-counting PMT are reflected in the statistical properties of the photoelectrons,
which in turn govern the statistical properties of the current pulses at the anode. For
classical light, ideal photoelectron detections follow a doubly stochastic Poisson point
process (DSPP) whose conditional rate is proportional to the instantaneous optical
intensity, while the conditional mean photoelectron count over a given observation
interval is proportional to the integrated intensity (energy) received during that
interval~\cite{saleh78}~\chapcite{13}{saleh2019}.

Notably, a single-photon avalanche diode (SPAD) can serve as a convenient solid-state
alternative to a PMT for photon-counting applications~\chapcite{19}{saleh2019}. The
SPAD is an avalanche photodiode (APD) biased slightly above its avalanche breakdown
voltage so that a single electron--hole pair generated by the absorption of a photon is
sufficient to precipitate avalanche breakdown, creating a large current pulse signifying
the arrival of the photon. The detector response is binarized in this mode of
operation, which serves to mitigate gain and circuit noise.

A superconducting nanowire single-photon detector (SNSPD) provides another
alternative for photon-counting applications~\chapcite{19}{saleh2019}. The nanowire
is maintained at a cryogenic temperature below its superconducting transition
temperature and is biased with a current below its critical current. The absorption of a
photon locally suppresses superconductivity, triggering a transient resistive region
that produces a voltage pulse that signifies a photon-detection event. Unlike a PMT or
SPAD, the SNSPD relies neither on photoelectron emission nor avalanche
multiplication. Relative to a SPAD, an SNSPD can offer lower dark-count rates,
reduced timing jitter, and higher detection efficiency, particularly at near-infrared
wavelengths, without avalanche-associated afterpulsing. Although a conventional
single-element SNSPD, like a SPAD, generally provides a binary response, specially
designed nanowire configurations and arrays can provide photon-number resolution.
These advantages come at the cost of cryogenic operation and the associated
complexity of the system. The statistics of the detection pulses can be used to infer
the incident photon statistics, when account is taken of the detection efficiency, dark
counts, and detector dead time.

\subsection{Subthreshold Fermi-Tail Photoemission (FTP)} \label{subthreshtheory}

As discussed in Sec.~\ref{singletheory}, suprathreshold (ordinary) photoemission
operates in the regime $h\nu > \texttt{W}$. However, the photocurrent does not
precipitously fall to zero when this condition is violated. That's because a subthreshold
photon of energy \,$h\nu < \texttt{W}$\, can impart sufficient energy to an electron
in the tail of the Fermi function located in the conduction band, and raise it above the
vacuum level as portrayed in Fig.~\ref{fig7}($b$). The observation of \textbf{Fermi-tail
photoemission} under coherent and entangled-photon-pair illumination is considered
in Secs.~\ref{subthreshold} and \ref{entPMT}, respectively.

\subsubsection{Fermi-Tail Photocurrent}\label{subFTcurrent}
As with suprathreshold photoemission, the Fermi-tail photocurrent
\,$i_{\scriptscriptstyle F}$\, follows
\begin{equation}\label{iFphiA}
  i_{\scriptscriptstyle F} \propto \phi A \propto \Phi \propto P,
\end{equation}
indicating that it too is proportional to $\phi$, $A$, $\Phi$, and $P$. Under
comparable illumination conditions, Fermi-tail photoemission generally produces
substantially smaller photocurrents than suprathreshold photoemission.

As will become apparent in the sequel, the Fermi-tail photocurrent
$i_{\scriptscriptstyle F}$ in Eq.~(\ref{iFphiA}) and the entangled-two-photon
photocurrent $i_{\scriptscriptstyle E}$ in Eq.~(\ref{iephiA}) both scale linearly with
incident optical power in the isolated-photon-pair domain, making Fermi-tail
photoemission particularly troublesome in experiments designed to detect ETPP. In
carrying out experiments designed to observe ETPP, it is therefore crucial to minimize
Fermi-tail photoemission to the greatest extent possible. This can be achieved
because the Fermi-tail photocurrent depends strongly on the temperature of the
sample, the wavelength of the illumination, and the detector architecture, while the
entangled-two-photon photocurrent is distinguished chiefly by its dependence on pair
correlations, loss, and the relevant area--time interaction scales.

\subsubsection{Responsivity and Quantum Efficiency}\label{FTRQE}

Following Eqs.~(\ref{iSRESPeta}) and (\ref{iFphiA}), the Fermi-tail responsivity
$\mathcal{R}_{\scriptscriptstyle F}$ \,(A/W)\, is defined as the proportionality
constant between $i_{\scriptscriptstyle F}$ and $P$,
\begin{equation} \label{iFETAresp}
i_{\scriptscriptstyle F} =  \mathcal{R}_{\scriptscriptstyle F}   P \,,
\end{equation}
and the Fermi-tail quantum efficiency $\eta_{\scriptscriptstyle F}$
(electrons/photon) is given by
\begin{equation}\label{iFRESPeta}
  \eta_{\scriptscriptstyle F} = \frac{h c}{e\lambda}\,\mathcal{R}_{\scriptscriptstyle F}.
\end{equation}

\subsubsection{Fowler Theory for Metals}\label{subsubfowler}
For metals, Fermi-tail photoemission can be phenomenologically described by the
venerable Fowler theory of photoemission~\cite{Fowler1931}, which provides an
analytical approximation for the quantum efficiency (and responsivity) near threshold.
In the limit ${\scriptstyle \mathcal{W}} - h\nu  \gg k_{\!\scriptscriptstyle
B}\!\mathsfit{T\/}$, where $k_{\!\scriptscriptstyle B}$ is Boltzmann's constant and
$\mathsfit{T\/}$ is the absolute temperature, the Fowler function $F(x)$ can be
approximated by a simple exponential that can be used to extract a useful
model-based estimate of the work function from a single responsivity measurement at
a subthreshold photon energy.~\cite{Teich1967}.

This approach leverages the logarithmic dependence of the inferred work function on
the measured quantum efficiency $\eta_{\scriptscriptstyle F}$ and on an empirical
constant $a$, whose exact value affects the result only weakly as a result of its
logarithmic placement. Expressing the yield in Eq.~(4) of~\cite{Teich1967} as the
quantum efficiency $\eta_{\scriptscriptstyle F}$, the work function \,${\scriptstyle
\mathcal{W}}$\, can be estimated as
\begin{equation}\label{WFest}
  {\scriptstyle \mathcal{W}} \approx h\nu + k_{\!\scriptscriptstyle B}\!\mathsfit{T\/}
  \ln (a \mathcal{A} \mathsfit{T\/}^2/\eta_{\scriptscriptstyle F}),
  \quad\qquad h\nu < {\scriptstyle \mathcal{W}}
\end{equation}
where $a \approx 4 \times 10^{-36}$~m$^2$-s/quantum, and $\mathcal{A} = 4\pi m
k_{\!\scriptscriptstyle B}^2/h^3$. This method has been successfully used to
estimate the work function of a vapor-deposited sodium-metal photocathode in a
photomultiplier tube~\cite{Teich1967}.

Although it is agnostic as to whether the photoemission arises from a volume or a
surface effect, the metallic Fowler theory is not generally applicable to
semiconductors without modification for a number of reasons: 1)~the initial states
depend on semiconductor band structure and carrier occupation, and can include
thermally populated conduction-band states; 2)~the relevant emission threshold is
governed by semiconductor band structure, state occupancy, and surface
electrostatics, rather than by the metallic Fermi-edge framework that underlies
Fowler theory; 3)~photoemission is strongly influenced by band-structure effects,
including the joint density of states and surface band bending; and 4)~in
nondegenerate semiconductors, carrier distributions near the band edges are
approximately Boltzmann distributed, rather than behaving as a degenerate Fermi
gas. Subthreshold photoemission from some semiconductor materials can also involve
mid-gap states introduced by traps, impurities, or disorder.

\subsection{Subthreshold Two-Photon Photoemission (TPP)}\label{twophotontheory}

\textbf{Two-photon photoemission (TPP)} was initially observed in 1964 from the
metal sodium (Na) by Teich \emph{et al.}~\cite{Teich64,teich66PhD,Teich68} and
from the semiconductor cesium antimonide (Cs$_3$Sb) by Sonnenberg \emph{et
al.}~\cite{Sonnenberg64}. In 1965, Pope \emph{et al.} demonstrated that TPP could
also be elicited from organic crystals such as anthracene, tetracene, and
perylene~\cite{pope65}. In the comparisons reported in~\cite{Teich68}, two-photon
responsivities for organic crystals were roughly 3 orders of magnitude greater than
those for semiconductors, which in turn were 2--5 orders of magnitude greater than
those for metals. Because two-photon responsivity depends on incident intensity,
these ratios are not universal material constants.  By way of comparison, one-photon
suprathreshold responsivities for semiconductors are typically 2--4 orders of
magnitude larger than those for metals (Sec.~\ref{1PRQE}).

An early and comprehensive review of multiphoton photoemission from metals,
semiconductors, and dielectrics was provided by Peter Barashev in
1972~\cite{barashev1972I}. Also of interest are recollections about multiphoton
processes offered by Aleksandr Prokhorov at the First International Conference on
Multiphoton Processes held at the University of Rochester (New York) in
1977~\cite{prokhorov78}, and by Nicolaas Bloembergen at the Sixth International
Conference on Multiphoton Processes held at Universit{\'e} Laval (Qu{\'e}bec City) in
1993~\cite{bloembergen94}. The observation of two-photon photoemission under
classical and entangled-photon-pair illumination is considered in detail in
Secs.~\ref{twophoton} and \ref{entchannel}, respectively. Multiphoton
photoemission, although beyond the scope of this review, is a fertile area of study
(see, e.g., \cite{farkas79,petek2020,petek2026}).

\subsubsection{Einstein Two-Photon Photoemission Equation}\label{2PEinst}
As illustrated in Fig.~\ref{fig7}($c$), two-photon photoemission (TPP) can take place
when an electron within a material is liberated into vacuum by the absorption of two
photons. Conservation of energy provides that the maximum kinetic energy that can
be imparted to the electron by the absorption of a pair of photons, each of energy
$h\nu$, is
\begin{equation}\label{2PKinenergy}
E_{\max} = 2h\nu - \texttt{W}= 2h\nu - (E_{\!g} + \chi) .
\end{equation}
Equation~(\ref{2PKinenergy}) is the \textbf{Einstein two-photon photoemission
equation}, which is analogous to Eq.~(\ref{eq17-0-2}) for the standard Einstein
photoemission equation, with $2h\nu$ replacing $h\nu$.

In a sequential pathway, the first photon promotes the electron to a real intermediate
state, while the second allows it to escape into vacuum. More generally, the
two-photon transition can proceed through virtual intermediate states without
sequential population of an intermediate level. Real intermediate states can include
conduction-band states or mid-gap states such as traps. The condition  $ E_{\!g} <
h\nu$, illustrated in Fig.~\ref{fig7}($c$) and specified in its caption, provides that the
two-photon transition take place via a real conduction-band state, which can, under
suitable spectral and dynamical conditions, provide a larger transition probability than
a purely virtual state~\cite{Hattori00}. Our principal interest lies not in the kinetic
energy of the emitted electrons, however, but rather in the magnitude of the
generated two-photon photocurrent, which we consider next.

\subsubsection{Two-Photon Photocurrent} \label{2pvpt}

Two-photon photoemission can, in principle, be manifested as a volume effect or a
surface effect. From a theoretical perspective, the two effects are distinguished by the
form of the potential $V(\mathbf{r})$ in the unperturbed Hamiltonian for the
Schr{\"o}dinger equation. The leading two-photon transition amplitude arises from
second-order perturbation theory involving the $\mathbf{A}\cdot\mathbf{p}$ term in
the interaction Hamiltonian; depending on the model, first-order contributions
involving the $\mathbf{A}^{\!2}$ term must also be considered ~\cite{teich66PhD},
where $\mathbf{A}$ is the vector potential of the field and $\mathbf{p}$ is the
electron momentum.

A CW optical source generates a steady (dc) two-photon photocurrent
$i_{\scriptscriptstyle C}$. The process is inherently nonlinear --- in the perturbative
regime, for fixed photon statistics and spatially uniform illumination over area $A$,
both volume and surface contributions to the photocurrent are proportional to the
square of the incident optical power $P=h\nu\Phi=h\nu\phi A$ and inversely
proportional to $A$. For classical light, the photocurrent follows the form
\begin{equation}\label{i2phiA}
  i_{\scriptscriptstyle C} \propto \phi^2 A \propto \Phi^2/A \propto P^2/A \propto IP,
\end{equation}
where $\Phi$ and $\phi$ are the incident photon flux (photons/s) and photon-flux
density (photons/m$^2$-s), respectively, and $I$ is the incident intensity
(W/m$^2$). While the photocurrent $\,i_{\scriptscriptstyle C}\,$ also depends on
the polarization, incidence angle, and pulse duration of the incident
light~\chapcite{5}{Seznec2022}, we do not discuss these dependencies here since
they are not crucial in the current context. In the presence of extremely strong
fields, effects such as surface heating and tunneling also give rise to electron
emission, which result in a violation of Eq.~(\ref{i2phiA})~\cite{keldysh65,farkas79}.

As with suprathreshold photoemission, two-photon photoemission is often
modeled as a volume effect for both metals and
semiconductors~\chapcite{5}{Seznec2022}, although the relative importance of
volume and surface contributions depends on the material, geometry, and
excitation conditions. Indeed, sodium metal appears to exhibit both effects. As
discussed in Sec.~\ref{subsub:VolSurf}, two-photon photoemission from a thick
sodium sample is associated with a volume effect, while two-photon photoemission
from a thin sodium sample appears to arise from a weak surface effect, and is more
difficult to observe. We provide a brief description of the underpinnings of
two-photon surface photoemission and then provide a more extensive discussion of
two-photon volume photoemission.

The theory of \textbf{two-photon surface photoemission} from a metal was devised
by Smith in 1962~\cite{Smith62}; the results he derived for the photocurrent were
subsequently corrected by Bowers~\cite{bowers64} and
Marinchuk~\cite{marinchuk66,marinchuk71}. Smith's model assumed that
monochromatic light was incident on a metal described by the Sommerfeld model,
which presupposes constant potentials inside and outside the surface, with a step
discontinuity at the surface. Smith made use of second-order, time-independent
perturbation theory to calculate the probability flux in an outward direction normal to
the surface. His Hamiltonian did not include the $\,\mathbf{A}^2\,$ term, but it was
later explicitly shown by Bowers that the $\,\mathbf{A}^2\,$ term vanishes within this
Smith-type surface-photoemission model~\cite{bowers64}. The theoretical expression
for the two-photon surface responsivity at $\mathsfit{T\/} = 0$~K, as derived by Smith
and corrected by Bowers and Marinchuk, was conveniently included as Eq.~(2.5) in
Barashev's review paper~\cite{barashev1972I}.

A more general surface formulation was subsequently advanced by
Adawi~\cite{Adawi64}, who used stationary scattering theory, which is equivalent
to perturbation theory. In the limit of a square-well potential, Adawi's photocurrent
reduces to Smith's corrected result. Formulations adopting a Sommerfeld-type
ansatz (plane-wave electron states) inherently exclude direct volume photoemission
since a free electron cannot absorb a photon while simultaneously conserving
energy and momentum.

The theory of \textbf{two-photon volume photoemission} from a metal was initially
devised by Bloch in 1964~\cite{Bloch64}. A theoretical expression for the two-photon
volume photocurrent $\,i_{\scriptscriptstyle C}$, based on Bloch's results but
corrected in several respects, was set forth by Teich in
1966~\cite{teich66PhD,Teich68,Lissandrin99,lissandrin2004}. The underlying model
assumes direct interband transitions, considers the electrons in the material to be
nearly free and to have Bloch-like wavefunctions for both the initial and final states,
and assumes a spherical Fermi surface. The final state is assumed to be the vacuum.
Time-dependent perturbation theory remains applicable for sufficiently weak optical
fields, including short pulses, whereas nonperturbative methods, such as
time-dependent density functional theory, may be required when the driving field
becomes sufficiently strong.

As discussed in detail in~\cite{teich66PhD}, Bloch's 1964 model has been modified
to incorporate: 1)~an electron escape depth that depends on the photoelectron
energy; 2)~optical reflection at the sample surface; and 3)~an interaction
{H}amiltonian that accommodates second-order transitions via the dominant
$\,\mathbf{A} \cdot \mathbf{p}\,$ term along with first-order transitions via the
$\,\mathbf{A}^{\!2}\,$ term. Bloch's formula has also been multiplied by the factor
$1/\sqrt{2\pi\,}$ to correct a numerical error. For convenience, the corrected TPP
photocurrent, as provided in Eqs.~(3.36) and (3.32) of~\cite{teich66PhD} (or
equivalently, Eq.~(1) of~\cite{Teich68}), is reproduced when the incident light is a
CW coherent beam in a single temporal and spatial mode,
\begin{eqnarray} \label{i2pcurrent}
i_{\scriptscriptstyle C} &=& \phi^2 A \cdot e\,\beta^2 d \,
\frac{ \hslash r_0^2 mc^2}{4 (\hslash \omega)^2}\, [M]
\, \frac{4\pi k_{\scriptscriptstyle F}}{3}\, \frac{E_{\scriptscriptstyle F}}{2\,\hslash\omega}\,
\left(1+ \frac{{\scriptstyle \mathcal{W}} -
2\,\hslash\omega }{E_{\scriptscriptstyle F}}\right)^{3/2} \nonumber \\[1mm]
 &=&\frac{P^2}{A} \cdot e\,\beta^2 d \,
 \frac{ \hslash r_0^2 mc^2}{4 (\hslash \omega)^4}\, [M]
\, \frac{4\pi k_{\scriptscriptstyle F}}{3}\, \frac{E_{\scriptscriptstyle F}}{2\,\hslash\omega}\,
\left(1+ \frac{{\scriptstyle \mathcal{W}} - 2\,\hslash\omega }
{E_{\scriptscriptstyle F}}\right)^{3/2} ,
\end{eqnarray}
where the two-photon oscillator strength $[M]$ is given by
\begin{equation}\label{matrixelement}
  [M] \equiv \:\left| \, \frac{2}{m} \, \frac{\langle f |
  \widehat{p}_x | j \rangle \langle j | \widehat{p}_x | i \rangle} {E_j - E_i -
  \hslash\omega} \, \right|^{\:2}.
\end{equation}
The expression for $i_{\scriptscriptstyle C} $ in Eq.~(\ref{i2pcurrent}) is suitable for
estimating the TPP photocurrent under the stated excitation conditions; it does not
describe the detailed behavior of the photocurrent arbitrarily close to the two-photon
threshold. The expression for $[M]$ in Eq.~(\ref{matrixelement}) replaces the
incorrect expression in square brackets in Eq.~(7) of~\cite{Bloch64}.

The parameters in Eqs.~(\ref{i2pcurrent}) and (\ref{matrixelement}) are defined as
follows: \,$i_{\scriptscriptstyle C}$ is the two-photon photocurrent (A); \,$\phi$ is the
incident CW photon-flux density (photons/m$^2$-s); \,$A$ is the area of illumination
(m$^2$); \,$e$ is the electronic charge (A-s); \,$R$ is the probability of reflection of a
photon from the surface of the sample; \,$\beta = 1-R$ is the probability of photon
passage to the interior of the sample; \,$d$ is the electron escape depth in the
material (m); \,$\hslash = h/2\pi$ is the reduced Planck constant (kg-m$^2/$s); \,$r_0
= e^2/4\pi\epsilon_0mc^2$ (m) is the classical electron radius; \,$m$ is the electron
mass (kg); \,$c$ is the speed of light in vacuum (m/s); \,$\omega = 2\pi \nu$ is the
angular frequency of the incident photon (rad/s); \,$\hslash\omega$ is the energy of
the incident photon (kg-m$^2$/s$^2$); \,$k_{\scriptscriptstyle F}$ is the electron
wavenumber associated with the Fermi energy (m$^{-1}$); \,$E_{\scriptscriptstyle F}$
is the Fermi energy (kg-m$^2$/s$^2$); \,${\scriptstyle \mathcal{W}}$ is the work
function of the material (kg-m$^2$/s$^2$); \,$[M]$ is the two-photon oscillator
strength; \,$i$, $j$, and $f$ designate the initial, intermediate, and final states,
respectively; \,$\widehat{p}_x$ is the momentum operator; \,$E_i$ is the energy of
the initial state $i$ (kg-m$^2$/s$^2$); and \,$E_j$ is the energy of the intermediate
state $j$ (kg-m$^2$/s$^2$). The quadratic dependence of $i_{\scriptscriptstyle C}$
on $\beta$ in Eq.~(\ref{i2pcurrent}) is a consequence of the quadratic dependence of
$i_{\scriptscriptstyle C}$ on $\phi$.

Although Bloch's model was designed for metals, where the electrons are initially in
the conduction band, with some reinterpretation and a measure of caution, it can
be used for semiconductors, where the electrons are initially in the valence
band~\cite{Ghosh80,Ghosh82,lissandrin2004}. This involves substituting the
electron energy and wavenumber at the top of the valence band,
$E_{i\:\text{max}}$ and $k_{i\:\text{max}}$, for the Fermi energy
$E_{\scriptscriptstyle F}$ and wavenumber $k_{\scriptscriptstyle F}$, respectively.
However, these parameters can be difficult to estimate accurately so it is preferable
to instead determine the photocurrent by numerically evaluating Eq.~(A15)
of~\cite{lissandrin2004}, which is suitable for both Na and CsK$_2$Sb.

The two-photon oscillator strength specified in Eq.~(\ref{matrixelement}) has been
estimated for Na metal in Eq.~(3.35) of~\cite{teich66PhD}; and for CsK$_2$Sb by
making use of Eqs.~(10c) and (28) of~\cite{lissandrin2004}:
\begin{align} \label{eq:oscstrength}
[M] & \approx \: \phantom{444} 8   \qquad \mbox{\,Na metal} \\
[M] & \approx \:  4448  \;  \qquad   \text{CsK}_2\text{Sb}\,.
\end{align}
Still, Eqs.~(\ref{i2pcurrent}) and (\ref{matrixelement}) should be used with caution,
in part because of uncertainties associated with calculating the matrix elements
inherent in $[M]$, and also because nondirect transitions have been excluded from
the model.

\subsubsection{Responsivity and Quantum Efficiency}\label{2PRQE}

The two-photon responsivity $\mathcal{R}_{\scriptscriptstyle C}$ \,(A/W)\, for
volume photoemission is defined by writing Eq.~(\ref{i2pcurrent}) in the form
\begin{equation} \label{i2pcurrentETA}
i_{\scriptscriptstyle C} = \mathcal{L}_{\scriptscriptstyle C} \frac{P^2}{A} =
\mathcal{L}_{\scriptscriptstyle C} I P  = \mathcal{R}_{\scriptscriptstyle C}  P \,,
\end{equation}
where $A$ is the illumination area. Here $P$ and $I$ represent the power and
intensity incident on the sample, respectively, and the two-photon responsivity
coefficient $\mathcal{L}_{\scriptscriptstyle C}$ has units of A\,m$^2$/W$^2$. For CW
operation, the intensity-dependent responsivity $\mathcal{R}_{\scriptscriptstyle C}$
(A/W) is defined as the ratio $i_{\scriptscriptstyle C}/P$:
\begin{equation}\label{eqRLI}
  \mathcal{R}_{\scriptscriptstyle C} = \mathcal{L}_{\scriptscriptstyle C}  I.
\end{equation}
Since two-photon photoemission is nonlinear, $\mathcal{R}_{\scriptscriptstyle C}$
is itself proportional to $I$, the intensity of the light impinging on the sample.

Following Eq.~(\ref{iSRESPeta}), the two-photon quantum efficiency
$\eta_{\scriptscriptstyle C}$ (electrons/photon) is obtained from the two-photon
responsivity  $\mathcal{R}_{\scriptscriptstyle C}$ for light of wavelength $\lambda$
via
\begin{equation}\label{iCRESPeta}
  \eta_{\scriptscriptstyle C} = \frac{h c}{e\lambda}\,\mathcal{R}_{\scriptscriptstyle C}\,.
\end{equation}
In the nascent two-photon photoemission literature of the 1960s, the two-photon
responsivity $\mathcal{R}_{\scriptscriptstyle C}$ was referred to as the
\emph{two-quantum yield} \,$\Lambda$\, (A/W), which meshed with the analog
measurements of the day~\cite{teich66PhD,Teich68}. With the development of
photon and electron counting techniques in the late 1960s and 1970s, the closely
related two-photon efficiency $\eta_{\scriptscriptstyle C}$ (electrons/photon)
gained currency.

\subsubsection{Illumination with Fluctuating Light}\label{optpowvar}

The two-photon photocurrent $i_{\scriptscriptstyle C}$ specified in
Eq.~(\ref{i2pcurrent}) is enhanced when the incident light power fluctuates because
peaks in the power (or intensity) are weighted more heavily by the nonlinearity.
Throughout this subsection, the quantity $g_2$ defined below characterizes the
intensity fluctuations of an unpulsed optical source. Enhancement arising from
independently imposed temporal modulation is considered separately in
Sec.~\ref{2Qphotstatspulse}.

As an example, we consider the power fluctuations associated with an optical beam
containing $\mathcal{M\/}$ temporal frequency components of equal power
$P_j=\overline{P}/\mathcal{M\/}$, whose individual phases $\phi_j$ are random and
statistically independent. The instantaneous power $P(t)$ may then be written as
\begin{equation}\label{instpwrmodes}
  P(t)=\sum_{j=1}^\mathcal{M\/} P_j
\;+\; 2\!\!\!\!\! \sum_{1\le j<k\le \mathcal{M\/}} \!\!\!\!\sqrt{P_jP_k}\,
\cos\!\big(\Delta\omega_{jk}\,t+\Delta\phi_{jk}\big),
\end{equation}
and $\Delta\phi_{jk}=\phi_j-\phi_k$ are the relative phases between components.
Averaging over a time interval larger than the beat periods, and over the random
phases, yields mean and mean-square powers given by
\begin{equation}\label{avgdmodes}
  \overline{P}=\sum_{j=1}^\mathcal{M\/} P_j,\qquad
\overline{P^2}=\Big(\sum_j P_j\Big)^2
+ 2\sum_{j<k} P_jP_k,
\end{equation}
where we have used the fact that averaging over $\cos(\cdot) \to 0$, averaging
over $\cos^2(\cdot) \to $ ½\,, and the cross-pair averages vanish.

Given that
\begin{equation}\label{avgdmodescross}
  \sum_{j<k} P_jP_k=\binom{\mathcal{M\/}}{2}\Big(\frac{\overline{P}}{\mathcal{M\/}}\Big)^2
=\frac{\mathcal{M\/}-1}{2\mathcal{M\/}}\,\,\overline{P}^{\,2},
\end{equation}
Eq.~(\ref{avgdmodes}) becomes
\begin{equation}\label{avgdmodes2}
  \overline{P^2}=\overline{P}^{\,2}\,\frac{2\mathcal{M\/}-1}{\mathcal{M\/}},
\end{equation}
which leads to
\begin{equation}\label{avgdmodes3}
  \frac{\overline{P^2}}{\overline{P}^{\,2}}=\frac{2\mathcal{M\/}-1}{\mathcal{M\/}}.
\end{equation}

To forge a connection with the terminology of optical coherence
theory~\cite{mandel65,saleh78,Teich88,perina1985,Perina91,mandel1995,born2002,wolf2007,goodman2015},
also referred to as statistical optics~\chapcite{12}{saleh2019}, we introduce the
definition
\begin{equation}\label{g2def}
  g_2 \equiv g^{(2)}(0) = \frac{\overline{P^2}}{\overline{P}^{\,2}}.
\end{equation}
For conciseness, we have abbreviated $g^{(2)}(\tau = 0)$, the normalized
second-order intensity correlation function at zero delay time (also called the
bunching parameter), as $g_2$. For the sum of $\mathcal{M\/}$ equal-amplitude,
random-phase temporal components coupled into a single detected mode, we thus
deduce that the two-photon photocurrent $i_{\scriptscriptstyle C}$ in
Eq.~(\ref{i2pcurrent}) will be enhanced by the factor
\begin{equation}\label{g2=2}
 g_2 = \frac{2\mathcal{M\/} - 1}{\mathcal{M\/}} \to 2 \;\;\;
 \text{as}\;\;\; \mathcal{M\/} \to \infty.
\end{equation}

This result applies to interfering random-phase components within the same detected
spatial mode. (It differs from the result $g_2=1+1/\mathcal{M\/}$ obtained when the
intensities of $\mathcal{M\/}$ statistically independent thermal modes are summed.)
Indeed, the expression provided in Eq.~(8) of the paper by Lissandrin \emph{et
al.}~\cite{lissandrin2004} is a factor of two larger than Eq.~(\ref{i2pcurrent}) precisely
because it assumes $\mathcal{M\/} \to \infty$ and \,$g_2 \to 2$. In the special case
of a single coherent mode ($\mathcal{M\/} = 1$), we have \,$g_2 = 1$. In general, for
illumination with light that fluctuates in an arbitrary manner, the factor $g_2$ allows
the CW-equivalent two-photon responsivity and quantum efficiency to be extracted
from the data, as exemplified in Sec.~\ref{sub:TPRespon}. Two-photon photoelectron
counting distributions and photocurrent spectra have also been calculated for
classical light with various statistical
properties~\cite{teichdiament69,diamentteich69,jaiswal69,barashev1972II}.

The emergence of quantum coherence theory in the early
1960s~\cite{glauber1963a,glauber1963b} provided an impetus for the further
examination of the role played by the statistical properties of classical and
nonclassical light in nonlinear interactions such as two-photon photoemission.
Shortly thereafter, it was established by Teich and Wolga~\cite{teich1966PRL},
Lambropoulos \emph{et al.}~\cite{lambropoulos1966PR}, and
Mollow~\cite{mollow1968PR} that the rate of multiphoton absorption is linked to
Glauber's higher-order, normally ordered quantum-field correlation
functions~\cite{titulaer1965}.

Excitation with a single detected chaotic (thermal) mode, which exhibits photon
bunching~\cite{Teich88} and for which $g_2=2$, thus results in a factor-of-two
enhancement of the photocurrent relative to coherent illumination at the same mean
intensity. This result is the same as that obtained for the random-phasor
multicomponent construction discussed above: as the number of statistically
independent phasor components increases, the resultant complex field approaches a
zero-mean Gaussian random variable, and the detected intensity corresponds to that
of a single effective thermal mode. This limiting behavior is codified by the Siegert
relation~\cite{siegert43,mandel1995,ferreira20siegert}, which expresses the
second-order intensity correlation of a Gaussian field in terms of the corresponding
first-order field correlation. As another example, illumination with single-mode
degenerate bright squeezed vacuum, considered in
Secs.~\ref{ETPAquantummodelBSV}--\ref{subsubsec:BSV}, exhibits $g_2\approx3$ in
the large per-mode-occupancy limit. This can provide an approximately threefold
enhancement for an instantaneous two-photon response at fixed mean intensity,
although the enhancement of an actual TPA transition also depends on the spectral
correlations of the light and the response of the absorbing
material~\cite{iskhakov2012, spasibko2017,sharapova2020}.

\subsubsection{Illumination with Pulsed Light} \label{2Qphotstatspulse}
The photocurrent $i_{\scriptscriptstyle C}$ in Eq.~(\ref{i2pcurrent}) is similarly
enhanced when the illumination arises from a pulsed source, such as a mode-locked
or pulsed laser. Here we consider temporal modulation imposed independently of the
source fluctuations characterized by $g_2$ in Sec.~\ref{optpowvar}. The power
waveform $P(t)$ used below describes the modulation-induced pulse envelope alone.
Because the two-photon photocurrent scales quadratically with optical power
[Eq.~(\ref{i2phiA})], a train of brief, high-intensity pulses is more effective in producing
two-photon photoemission than a CW beam of the same mean power. This
enhancement assumes a perturbative two-photon response that is effectively
instantaneous relative to the pulse envelope; population dynamics of long-lived
intermediate states can modify this behavior.

Succinctly stated, two-photon photoemission is initiated by the mean-square of the
light-power waveform rather than by the square-mean, which then leads to an
enhancement factor $\Gamma \;(\geqslant 1)$ of the form
\begin{equation}\label{eq:intenfactor}
    \Gamma \equiv
    \frac{\overline{P^2}}{\overline{P}^2}
    = \frac{\dfrac{1}{\tau_1}\int_0^{\tau_1} P^2(t)\,dt}
           {\left(\dfrac{1}{\tau_1}\int_0^{\tau_1} P(t)\,dt\right)^2}.
\end{equation}
The averaging is conveniently carried out over a single period $\tau_1$ of the periodic
optical waveform. In the absence of imposed temporal modulation, $\Gamma=1$,
irrespective of the intrinsic fluctuations of the unpulsed source.

It is convenient to express $\Gamma$ in terms of the duty cycle $\,\Delta\,$ of the
pulsed waveform,
\begin{equation}\label{eq:deltadef}
    \Delta \equiv \tau_0/\tau_1,
\end{equation}
where $\tau_0$ is the duration of an individual optical pulse. Assuming rectangular
optical pulses for simplicity, the peak power $\widehat{P}$ of a pulse with the same
energy per cycle as a CW beam of mean power $\overline{P}$ satisfies
\begin{equation}\label{eq:equivfactor1}
    \widehat{P}\,\tau_0 = \overline{P}\,\tau_1,
\end{equation}
so that
\begin{equation}\label{eq:equivfactor2}
    \widehat{P} = \frac{\overline{P}}{\Delta}.
\end{equation}
The mean-square power is then
\begin{equation}\label{eq:equivfactor3}
    \overline{P^2} = \frac{1}{\tau_1}\int_0^{\tau_0}
    \left(\frac{\overline{P}}{\Delta}\right)^2 dt
    = \frac{\overline{P}^2}{\Delta},
\end{equation}
while the square-mean power for the CW beam is simply $\overline{P}^2$.

For a rectangular optical pulse train, we conclude that the enhancement factor
$\Gamma$ specified in Eq.~(\ref{eq:intenfactor}) may be expressed as
\begin{equation}\label{eq:equivfactor5}
    \Gamma = \frac{\overline{P}^2/\Delta}{\overline{P}^2}
    = \frac{1}{\Delta}\,,
\end{equation}
so that $\Gamma$ varies inversely with the duty cycle $\Delta$. Decreasing $\Delta$,
which reflects a growing concentration of optical power into decreasing time
intervals, provides a proportional increase in the nonlinear enhancement factor
$\Gamma$. For illumination with pulsed light of arbitrary waveform, the factor
$\Gamma$ allows the CW-equivalent two-photon responsivity and quantum
efficiency to be extracted from the data. If a fluctuating light source is pulsed by an
independent process, the mean-square to square-mean ratio becomes
$\smash{\overline{P^2}/\overline{P}^2 = g_2 \, \Gamma}$, as exemplified in
Sec.~\ref{sub:TPRespon}.

\subsection{Subthreshold Entangled-Two-Photon Photoemission (ETPP)}\label{enttheory}

\textbf{Entangled-two-photon photoemission (ETPP)} takes place when a material
absorbs a pair of twin photons and liberates an electron into vacuum, as portrayed in
Fig.~\ref{fig7}($d$). The process is governed in part by the characteristics of the
incident entangled-photon pairs, e.g., their entanglement time $T_{\scriptscriptstyle
E}$ and entanglement area $A_{\scriptscriptstyle E}$. Despite the fact that ETPP
involves the absorption of pairs of photons, in the isolated-photon-pair domain the
entangled-two-photon photocurrent $i_{\scriptscriptstyle E}$ is directly proportional
to the incident photon-flux density $\phi$, i.e.,
\begin{equation}\label{iephiA}
  i_{\scriptscriptstyle E} \propto \phi A \propto \Phi \propto P.
\end{equation}
This linear scaling arises because the two photons belong to a preexisting correlated
pair, so that, for a fixed per-pair absorption probability, the joint absorption rate is
proportional to the incident pair flux.
 The linear scaling of $i_{\scriptscriptstyle
E}$ with $\phi$ is in contrast to the quadratic scaling of the two-photon
photocurrent with the photon-flux density, $i_{\scriptscriptstyle C} \propto
\phi^2$, expressed in Eq.~(\ref{i2phiA}). This latter behavior arises because the
arrival of one coherent photon signals nothing about the arrival of another. The
Einstein two-photon photoemission equation provided in Eq.~(\ref{2PKinenergy})
remains valid. The observation of ETPP is described in Sec.~\ref{entchannel}.

As pointed out in Sec.~\ref{subFTcurrent}, Fermi-tail photoemission (FTP) produces a
particularly deleterious background photocurrent that hinders the observation of
entangled-two-photon photoemission.
 This is because the form of the Fermi-tail
photocurrent in Eq.~(\ref{iFphiA}) is identical to that of the entangled-two-photon
photocurrent in Eq.~(\ref{iephiA}) and therefore easily masks it. In experiments
designed to observe ETPP, it is therefore crucial to reduce FTP to the greatest
extent possible, and preferentially to eliminate it.

\subsubsection{Entangled-Two-Photon Photocurrent}\label{entcurrent}

A quantum theory of entangled-two-photon photoemission was formulated by
Lissandrin in 1999 and expressed in final form in
2004~\cite{Lissandrin99,lissandrin2004}. The model assumes volume
photoemission, direct interband transitions, Bloch-like wavefunctions for the initial
and final states, and spherical Fermi surfaces. Although the model was principally
designed for describing ETPP in metals, a redefinition of some of its parameters
allowed it to provide a framework for describing ETPP in semiconductors.

The incident light was assumed to be entangled-photon pairs generated via type-I
collinear SPDC, with entanglement time $T_{\scriptscriptstyle E}$ and entanglement
area $A_{\scriptscriptstyle E}$. The nonlinear optical crystal was of length $L$ and
the pump was assumed to be a monochromatic laser with wave number $k_p$.
While a treatment of this kind is generally referred to as volume photoemission, it is
more aptly described as volume-initiated, surface-limited photoemission since the
photon absorption and excitation occur in the volume of the material but the
emission is constrained by the surface barrier.

The theoretical expression for the entangled-two-photon photocurrent
\,$i_{\scriptscriptstyle E}$,\, which is presented in Eqs.~(25)--(27) of Lissandrin
\emph{et al.}~\cite{lissandrin2004}, is reproduced here for convenience:
    \begin{align} \label{iepcurrent}
    i_{\scriptscriptstyle E} &= \phi A \cdot e\,\beta^2 d \,
    \frac{ r_0^2 m^4 c^2}{4 \pi^3 \hslash^5} \,
    \frac{\omega_p^2}{4 \omega_1^0 \omega_2^0} \;\xi \;
    \frac{1}{A_{\scriptscriptstyle E} T_{\scriptscriptstyle E}}
    \,\int_{k_{\text{min}}}^{k_{\scriptscriptstyle F}} \!
    \frac{F(k, T_{\scriptscriptstyle E})}{\sqrt{k^2 +
    (2m / \hslash) \omega_p} - k}  \, k \, dk,  \\[2.0mm]
    k &= \sqrt{2mE_i/\hslash^2} \,, \nonumber  \\[3.0mm] \label{iepcurrentF}
    F(k, T_{\scriptscriptstyle E}) &= \left| \, \int_{E_{j\:\text{min}}}^{E_{j\:\text{max}}}
    \left(
        \frac{1 - e^{-i (T_{\scriptscriptstyle E} / \hslash) (E_j - \hslash^2 k^2 / 2m -
        \hslash \omega_1^0) - T_{\scriptscriptstyle E} \kappa_j / 2}}{E_j -
        \hslash^2 k^2 / 2m - \hslash \omega_1^0 - i \hslash \kappa_j / 2}
        \right. \right. \nonumber \\
    &\quad + \left. \frac{1 - e^{-i (T_{\scriptscriptstyle E} / \hslash) (E_j -
    \hslash^2 k^2 / 2m - \hslash \omega_2^0) - T_{\scriptscriptstyle E} \kappa_j / 2}}{E_j -
    \hslash^2 k^2 / 2m - \hslash \omega_2^0 - i \hslash \kappa_j / 2}
    \right)  \sqrt{E_j - E_c} \: dE_j\: \Bigg|^{\:2}, \\[2.0mm] \label{iepcurrentmu}
     \xi &= \left\langle \:\left|\, \frac{4}{m} \, \frac{\langle f | \widehat{p}_x \,
     | j \rangle \langle j | \widehat{p}_x | i \rangle} {\hslash\omega_p} \, \right|^{\:2}
     \right\rangle \left/ N_j^2 \right..   \end{align}
Here
\,$i_{\scriptscriptstyle E}$ is the entangled-two-photon photocurrent (A);
\,$\phi$ is the incident photon-flux density (photons/m$^2$-s);
\,$A$ is the area of illumination (m$^2$);
\,$e$ is the electronic charge (A-s);
\,$\beta^2 = (1-R)^2$ is the probability of passage of an entangled-photon
pair to the sample interior;
\,$d$ is the electron escape depth in the material (m);
\,$r_0 = e^2/4\pi\epsilon_0mc^2$ (m) is the classical electron radius;
\,$m$ is the electron mass (kg);
\,$c$ is the speed of light in vacuum (m/s);
\,$\hslash$ is the reduced Planck constant (kg-m$^2/$s);
\,$\omega_p$ is the angular frequency of the SPDC pump (rad/s);
\,$\omega_1^0$ and $\omega_2^0$ are the central angular frequencies of the
SPDC signal and idler waves, respectively (rad/s);
\,$A_{\scriptscriptstyle E}$ is the entanglement area (m$^2$);
\,$T_{\scriptscriptstyle E}$ is the entanglement time (s);
\,$i$, $j$, and $f$ designate the initial, intermediate, and final states, respectively;
\,$k$ is the initial electron wavenumber (m$^{-1}$);
\,$k_\mathrm{min}$ is the electron wavenumber required to overcome
the material's ionization energy (m$^{-1}$);
\,$k_{\scriptscriptstyle F}$ is the electron wavenumber associated with the
Fermi energy (m$^{-1}$), replaced by the wavenumber $k_{i\:\text{max}}$
at the top of the valence band for semiconductors;
\,$F(k, T_{\scriptscriptstyle E})$ is the entanglement-electron energy overlap
function (kg-m$^2$/s$^2$);
\,$E_i$ is the energy of the initial state $i$ (kg-m$^2$/s$^2$);
\,$E_j$ is the energy of the intermediate state $j$ (kg-m$^2$/s$^2$);
\,$E_c$ is the conduction band edge for the intermediate-state
continuum (kg-m$^2$/s$^2$);
\,$\kappa_j$ is the linewidth of the intermediate state $j$ (1/s);
\,$\widehat{p}_x$ is the momentum operator; and \,$\xi$ is the mean-square
normalized two-photon-transition matrix element per
squared  intermediate-state number density $N_j$ (m$^6$).

The dependence of the entangled-two-photon photocurrent \,$i_{\scriptscriptstyle
E}$\, on \,$\beta^2$\, in Eq.~(\ref{iepcurrent}) is a consequence of the fact that the
loss of either of the twins via reflection at the sample surface results in the
destruction of the pair --- and each twin is independently subject to such loss.
For nondegenerate photon pairs with different surface reflectances
$R_1$ and $R_2$, the factor $\beta^2$ is replaced by $(1-R_1)(1-R_2)$.
Analogously, intrinsic optical loss between the source and sample introduces a
multiplicative factor $\mathcal{T\/}_{\!\!0}^2$, where $\mathcal{T\/}_{\!\!0}$ is the
intrinsic optical-system transmittance (Eq.~(\ref{iepcurrent}), as it stands, assumes
that $\mathcal{T\/}_{\!\!0} = 1$).
Also, by virtue of its proportionality to \,$\phi A =
\Phi$ (photons/s), $i_{\scriptscriptstyle E}$ is independent of the illumination area
(provided that the entanglement area
$A_{\scriptscriptstyle E}$ remains unchanged
as the illumination area is varied), a feature that
distinguishes $i_{\scriptscriptstyle E}$ from $i_{\scriptscriptstyle C}$. For the same
reason, $i_{\scriptscriptstyle E}$ is independent of the enhancement factors for
$i_{\scriptscriptstyle C}$ that derive from intensity fluctuations and pulsed-light
illumination, as considered in Secs.~\ref{optpowvar} and \ref{2Qphotstatspulse},
respectively. A more complete description of the range of validity of
Eqs.~(\ref{iepcurrent})--(\ref{iepcurrentmu}), along with a discussion of the
subtleties
 associated with their use, is provided in~\cite{lissandrin2004}.

\subsubsection{Responsivity and Quantum Efficiency}\label{entrespqe}
The entangled-two-photon responsivity $\mathcal{R}_{\scriptscriptstyle E}$
\,(A/W)\, is the proportionality constant that emerges when Eq.~(\ref{iepcurrent}) is
cast in the form of Eq.~(\ref{iephiA}), i.e.,
\begin{equation} \label{iepcurrentETA}
i_{\scriptscriptstyle E} =  \mathcal{R}_{\scriptscriptstyle E} P \,.
\end{equation}
The entangled-two-photon quantum efficiency $\eta_{\scriptscriptstyle E}$ is
defined as the ratio of the photoelectron flux $\mu_{\scriptscriptstyle E}$ to the
photon flux $\Phi$:
 \begin{equation} \label{efluxtopflux}
    \eta_{\scriptscriptstyle E} \equiv \frac{\mu_{\scriptscriptstyle E}}{\Phi} =
    \frac{i_{\scriptscriptstyle E}/e}{\phi A}.
 \end{equation}
 Here, $\Phi$ counts the individual photons belonging to
intact entangled-photon pairs incident on the sample,
and not the pairs themselves.
The value of  $\eta_{\scriptscriptstyle E}$ may therefore be calculated directly from
Eq.~(\ref{iepcurrent}). Additionally, for degenerate photon pairs of
wavelength $\lambda$, $\eta_{\scriptscriptstyle E}$
is related to the entangled-two-photon responsivity
$\mathcal{R}_{\scriptscriptstyle E}$ via
Eq.~(\ref{iSRESPeta}):

\begin{equation}\label{ieRESPeta}
  \eta_{\scriptscriptstyle E} = \frac{h c}{e\lambda}\,\mathcal{R}_{\scriptscriptstyle E}\,.
\end{equation}

While the photocurrent $i_{\scriptscriptstyle E}$ expressed in Eq.~(\ref{iepcurrent})
contains a factor explicitly proportional to
$1/A_{\scriptscriptstyle E} T_{\scriptscriptstyle E}$,
an additional dependence on $T_{\scriptscriptstyle E}$ is inherent in the function
$F(k, T_{\scriptscriptstyle E})$ provided in Eq.~(\ref{iepcurrentF}).
Since both $\mathcal{R}_{\scriptscriptstyle E}$ and
$\eta_{\scriptscriptstyle E}$ are proportional to $i_{\scriptscriptstyle E}$,
these quantities all share the same dependencies on $T_{\scriptscriptstyle E}$.
However, when using the
heuristic framework for entangled-two-photon photoemission and absorption
detailed in Sec.~\ref{EPparticlemodel}, or when the dependence
of $F(k, T_{\scriptscriptstyle E})$ on $T_{\scriptscriptstyle E}$ is negligible,
at fixed incident photon flux we can write
\begin{equation}\label{Retaentagl}
  i_{\scriptscriptstyle E} \propto \mathcal{R}_{\scriptscriptstyle E} \propto \eta_{\scriptscriptstyle E} \propto
  \frac{1}{A_{\scriptscriptstyle E} T_{\scriptscriptstyle E}}.
\end{equation}
Reported values of $\mathcal{R}_{\scriptscriptstyle E}$
and $\eta_{\scriptscriptstyle E}$ should then be accompanied by the associated value of
$A_{\scriptscriptstyle E} T_{\scriptscriptstyle E}$ or, alternatively, the quantities
$\mathcal{R}_{\scriptscriptstyle E}A_{\scriptscriptstyle E} T_{\scriptscriptstyle E}$
and $\eta_{\scriptscriptstyle E}A_{\scriptscriptstyle E} T_{\scriptscriptstyle E}$ should
be
reported in place of $\mathcal{R}_{\scriptscriptstyle E}$ and $\eta_{\scriptscriptstyle
E}$, respectively. This issue is discussed further in connection with the
entangled-two-photon cross section per primitive cell discussed in
Sec.~\ref{entxsect}.

\subsubsection{Cross Section per Primitive Cell}\label{entxsect}

From Eq.~(\ref{efluxtopflux}), the dependence of the photocurrent
$i_{\scriptscriptstyle E}$ on the entangled-two-photon quantum efficiency
$\eta_{\scriptscriptstyle E}$ takes the form
\begin{equation}\label{iEsigma}
  i_{\scriptscriptstyle E}  =  \phi A \,e \,\eta_{\scriptscriptstyle E} \,.
\end{equation}
The current
\,$i_{\scriptscriptstyle E}$ can also be written in terms of the
entangled-two-photon cross section per primitive cell of the photoemitter
$\sigma_{\!\scriptscriptstyle E}$ from a heuristic perspective.

The effective number of primitive cells sampled in
the escape layer is $NAd$, where $N$ is the number of primitive cells per unit
volume, $A$ is the illumination area, and $d$ is the escape depth. The emission
probability per incident photon (in the thin-target/low-probability limit) is thus
$\beta^2 N \sigma_{\!\scriptscriptstyle E} d$, where $\beta^2$ is the probability of
pair passage to the interior of the sample, yielding
\begin{equation}\label{iEsigma-new}
  i_{\scriptscriptstyle E}  =  \phi A\, e \, \beta^2  \sigma_{\!\scriptscriptstyle E} N d .
\end{equation}
Here $\sigma_{\!\scriptscriptstyle E}$ is defined relative to the incident photon
flux $\phi$, rather than the photon-pair flux $\phi/2$.
Combining Eqs.~(\ref{iEsigma}) and (\ref{iEsigma-new}) yields an expression that
relates $\sigma_{\!\scriptscriptstyle E}$ (a measure typically used in absorption
studies) to $\eta_{\scriptscriptstyle E}$ (a measure
typically used in photoemission studies),
\begin{equation}\label{iEsigma2}
  \sigma_{\!\scriptscriptstyle E}  = \frac{\eta_{\scriptscriptstyle E}} {\beta^2 N d},
\end{equation}
with
\begin{equation}\label{Retaentaglsig}
  \sigma_{\!\scriptscriptstyle E} \propto \frac{1}{A_{\scriptscriptstyle E}
  T_{\scriptscriptstyle E}}\,.
\end{equation}

For completeness, therefore, the value of $\sigma_{\!\scriptscriptstyle E}$ should be
accompanied by the value of $A_{\scriptscriptstyle E} T_{\scriptscriptstyle E}$. A
compact way of providing the combined information is to report the
normalized entangled-two-photon cross section per primitive cell
(or per molecule for a molecular absorber),
\begin{equation}\label{sigAETE}
  \delta_{\scriptscriptstyle E} \equiv \sigma_{\!\scriptscriptstyle E}
  A_{\scriptscriptstyle E} T_{\scriptscriptstyle E}.
\end{equation}
Equation~(\ref{sigAETE}) is algebraically analogous to Eq.~(\ref{sigdelAT}) for the
conventional two-photon absorption cross section $\sigma^{(2)}$. However,
Eq.~(\ref{sigAETE}) is designed simply to provide a complete description of
$\sigma_{\!\scriptscriptstyle E}$, and uses the symbol $\delta_{\scriptscriptstyle E}$
(m$^4$\,s) to do so, while  Eq.~(\ref{sigdelAT}) is designed to enable
$\sigma_{\!\scriptscriptstyle E}$ to be determined from an established value of
$\sigma^{(2)}$.

\subsubsection{Photoelectron Count Rates from CsK$_2$Sb and Na}\label{entcurrentCs}

\begin{figure}[htb!]
\centering\includegraphics[width=4.25in]{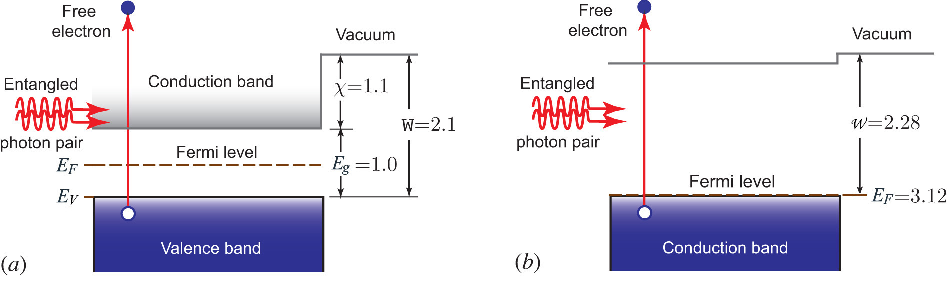}
\caption{($a$)~Simplified energy-band diagram for entangled-two-photon
photoemission from CsK$_2$Sb. Values for the bandgap energy $E_{\!g}$,
electron affinity $\chi$, and ionization energy \texttt{W} are specified
in eV.  \,($b$)~Simplified energy-band diagram for entangled-two-photon
photoemission from Na metal. Values for the Fermi level $E_{\scriptscriptstyle F}$
and work function ${\scriptstyle \mathcal{W}}$ are specified in eV.
} \label{fig9}
\end{figure}

We now revisit Eqs.~(\ref{iepcurrent})--(\ref{iepcurrentmu}), drawn from the model
developed by Lissandrin \emph{et al.}~\cite{Lissandrin99,lissandrin2004}, to
estimate the theoretical photoelectron count rates $\mu_{\scriptscriptstyle E}$ for
CsK$_2$Sb and Na metal. The calculations rely on: \,1)~the simplified energy-band
diagrams portrayed in Fig.~\ref{fig9}; \,2)~the SPDC pump and nonlinear-crystal
parameters displayed in Table~\ref{tab:spdcparams}; and \,3)~the
photoemissive-material parameters provided in
Table~\ref{tab:photoemissiveparams}. Note that the work function of a material
${\scriptstyle \mathcal{W}}$ is defined as the energy difference between the vacuum
level and the Fermi level $E_{\scriptscriptstyle F}$, as illustrated for the metal band
diagram sketched in Fig.~\ref{fig9}($b$). For the intrinsic-semiconductor band
diagram displayed in Fig.~\ref{fig9}($a$), on the other hand, the Fermi level lies near
midgap, i.e.\ $E_{\scriptscriptstyle F}-E_{\scriptscriptstyle V} \approx \slfrac{1}{2}
E_{\!g}$, so that the work function is expressed as ${\scriptstyle \mathcal{W}}
=\chi+\slfrac{1}{2}E_{\!g}$, whereas the ionization energy is given by
$\texttt{W}=\chi+E_{\!g}$.

\begin{table}[htb!]
\centering
\begin{small}
\caption{SPDC pump and nonlinear-crystal parameters used for calculating
the entangled-two-photon photocurrent with the model developed by
Lissandrin~\emph{et al.}~\cite{lissandrin2004}.
The crossover photon-flux density and corresponding intensity
are based on the original heuristic relation
$\phi_{\scriptscriptstyle EC}=1/(2A_{\scriptscriptstyle E}
T_{\scriptscriptstyle E})$ [Eq.~(16a) of~\cite{lissandrin2004}],
and are therefore a factor of two smaller than the values
that would be obtained using the modified heuristic model
presented in Sec.~\ref{EPparticlemodel}.}
\renewcommand{\arraystretch}{1.1}
\begin{tabular}{lc}
\hline
\noalign{\vskip 0.7mm}
Parameter (units) & Value \\[0.7mm]
\hline
\noalign{\vskip 0.7mm}
Pump wavelength $\lambda_p$ (nm) & 406 \\
Pump photon energy $\hbar \omega_p$ (eV) & 3.054 \\
Nondegeneracy ratios $\omega_1^0 / \omega_p$
& \slfrac{1}{2}\,, \slfrac{1}{3}\,, \slfrac{1}{8}  \\
Entanglement time $T_{\scriptscriptstyle E}$ (fs),  reference value & 10 \\
Entanglement area $A_{\scriptscriptstyle E}$ (m$^2$) & \,\,$1.0 \times 10^{-10}$ \\
Illumination area $A$ (m$^2$) & \,\,$1.0 \times 10^{-10}$ \\
Photon-flux density $\phi$ (photons/m$^2$·s) & $5.0 \times 10^{23}$ \\
Crossover photon-flux density $\phi_{\scriptscriptstyle EC}$ (photons/m$^2$·s) &
$5.0 \times 10^{23}$ \\
Crossover intensity (at 812~nm) $I_{\scriptscriptstyle EC}$ (W/m$^2$) &
$\! 1.2 \times 10^{5}$ \\[0.7mm]
\hline
\end{tabular}
\label{tab:spdcparams}
\end{small}
\end{table}

\begin{table}[htb!]
\centering
\begin{small}
\caption{Photoemissive-material parameters ($\mathsfit{T\/} = 300$~K) used for
calculating the entangled-two-photon photocurrent with the model developed
by  Lissandrin \emph{et al.}~\cite{lissandrin2004}. The data for CsK$_2$Sb are
drawn from Lissandrin \emph{et al.}~\cite{lissandrin2004}. The data for Na metal
are drawn from Sec.~III.D-5 of Teich~\cite{teich66PhD}.}
\renewcommand{\arraystretch}{1.1}
\begin{tabular}{lcc}
\hline
\noalign{\vskip 0.7mm}
Parameter (units) & CsK$_2$Sb & Na \\[0.7mm]
\hline
\noalign{\vskip 0.7mm}
Probability of photon passage to sample interior $\beta$ & 0.7 & 0.05 \\
Ionization energy \texttt{W}$/$Work function ${\scriptstyle \mathcal{W}}$  (eV)
& 2.1 & 2.28 \\
Primitive-cell density/Atomic density $N$ (m$^{-3}$) & $3.1 \times 10^{27}$
& $2.7 \times 10^{28}$ \\
Electron affinity $\chi$ (eV) & 1.1 & --- \\
Bandgap energy $E_{\!g}$ (eV) & 1.0 & --- \\
Fermi energy $E_{\scriptscriptstyle F}$ (eV) & --- & 3.12 \\
Fermi wavevector $k_{\scriptscriptstyle F}$ (m$^{-1}$) & --- & $9.0 \times 10^{9}$ \\
Intermediate-state lifetime $\tau_j$ (fs) & 270 & 10 \\
Intermediate-state linewidth $\kappa_j$ (s$^{-1}$) & $3.7 \times 10^{12}$
& $1.0 \times 10^{14}$ \\
Minimum intermediate-state energy $E_{j\,\mathrm{min}}$ (eV) & 2.5 & 5.18 \\
Maximum intermediate-state energy $E_{j\,\mathrm{max}}$ (eV) & 4.5 & 8.0 \\
Escape depth $d$ (nm) & 40 & 40 \\
Two-photon oscillator strength $[M]$ & 4448 & 8 \\
Average matrix element $\xi$ (m$^{6}$) & $2.7 \times 10^{-53}$ & $1.7 \times 10^{-56}$ \\
Quantum efficiency $\eta_{\scriptscriptstyle E}^\text{\sc lis}$ (electrons/photon)
& $1.6 \times 10^{-9}$ & $1.6 \times 10^{-15}$ \\
Responsivity $\mathcal{R}_{\scriptscriptstyle E}^\text{\sc lis}$ (A/W)
& $1.0 \times 10^{-9}$ & $1.0 \times 10^{-15}$ \\[0.7mm]
\hline
\end{tabular}
\label{tab:photoemissiveparams}
\end{small}
\end{table}
\begin{figure}[htb!]
\centering\includegraphics[width=3.25in]{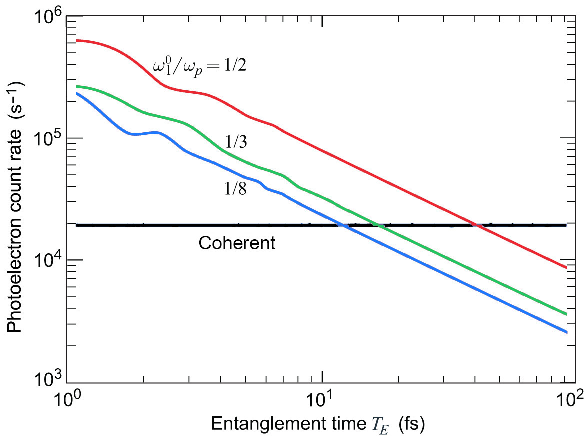}
\caption{Entangled-two-photon photoelectron count rate $\mu_{\scriptscriptstyle E} =
i_{\scriptscriptstyle E}/e$ for the semiconductor CsK$_2$Sb as a function of the
entanglement time $T_{\scriptscriptstyle E}$, for twins with nondegeneracy
ratios $\omega_1^0/\omega_p = 1 - \omega_2^0/\omega_p =
$ \slfrac{1}{2} (red, degenerate), \slfrac{1}{3} (green), and \slfrac{1}{8} (blue).
The parameter values
used in carrying out the calculations are provided in Tables~\ref{tab:spdcparams}
and \ref{tab:photoemissiveparams}.
The two-photon photoelectron count rate
$\mu_{\scriptscriptstyle C}=i_{\scriptscriptstyle C}/e$
for a coherent source is shown for comparison.
The calculated value of the count rate is approximately six orders of magnitude
greater than that for Na metal.
(Adapted with permission from Fig.~3 of~\cite{lissandrin2004},
\textcopyright 2004 American Physical Society.)} \label{fig10}
\end{figure}

\begin{figure}[htb!]
\centering\includegraphics[width=3.25in]{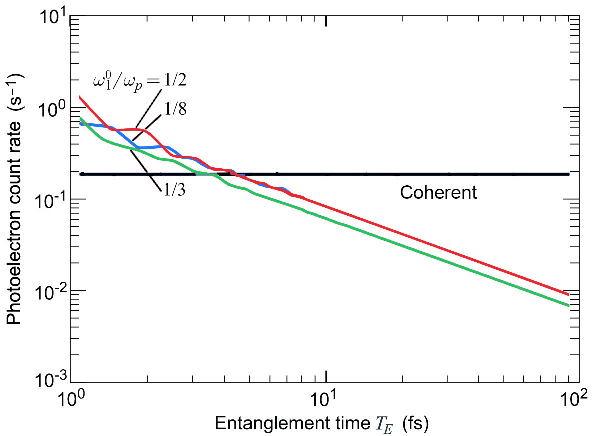}
\caption{Entangled-two-photon photoelectron count rate $\mu_{\scriptscriptstyle E} =
i_{\scriptscriptstyle E}/e$ for Na metal as a function of the entanglement time
$T_{\scriptscriptstyle E}$, for twins with nondegeneracy ratios
$\omega_1^0/\omega_p = 1 - \omega_2^0/\omega_p = $ \slfrac{1}{2} (red, degenerate),
\slfrac{1}{3} (green), and \slfrac{1}{8} (blue). The parameter values used in carrying out the
calculations are provided in Tables~\ref{tab:spdcparams} and
\ref{tab:photoemissiveparams}.
The two-photon photoelectron count rate
$\mu_{\scriptscriptstyle C}=i_{\scriptscriptstyle C}/e$
for a coherent source is shown for comparison.
(Adapted with permission from Fig.~2 of~\cite{lissandrin2004},
\textcopyright 2004 American Physical Society.)} \label{fig11}
\end{figure}

The resulting theoretical photoelectron count rates $\mu_{\scriptscriptstyle E} =
i_{\scriptscriptstyle E}/e$ \,for CsK$_2$Sb and Na are plotted vs. the entanglement
time $T_{\scriptscriptstyle E}$ in Figs.~\ref{fig10} and \ref{fig11}, respectively, for
several values of the nondegeneracy ratio:  $\omega_1^0/\omega_p =
\slfrac{1}{2}\,,\:\slfrac{1}{3}\,,\: \text{and}\: \slfrac{1}{8}$ [$\omega_p =
\omega_1^0 + \omega_2^0$, in accordance with Eq.~(\ref{encons})]. The curves for
the two materials are similar in character, but the magnitude of the count rate for Na
is smaller than that from  CsK$_2$Sb by a factor of $\approx 10^{6}$. It is apparent
that the  count rates depend only weakly on the energy nondegeneracy of the
entangled-photon pair, especially for Na. It has also been shown (in Figs.~4 and 5
of~\cite{lissandrin2004}) that, for energy-degenerate incident photons, varying the
intermediate-state lifetime over the range $10^{-15} <\tau_j < 10^{-13}$~s has little
effect on the curves displayed in Figs.~\ref{fig10} and \ref{fig11}.

The curves displayed in Figs.~\ref{fig10} and \ref{fig11} are roughly inversely
proportional to the entanglement time $T_{\scriptscriptstyle E}$, as specified in
Eq.~(\ref{iepcurrent}). The deviations from inverse proportionality (observed at
smaller values of $T_{\scriptscriptstyle E}$) are emblematic of the harmonic
interference terms inherent in Eq.~(\ref{iepcurrentF}), and they clearly reveal that
the entanglement characteristics of the source are intertwined with the parameters
of the sample in a generally nonseparable way. Curves based on the (nonentangled)
two-photon theory of Bloch~\cite{Bloch64} and Teich~\cite{teich66PhD}, labeled
``Coherent,'' are plotted for comparison. These curves were computed using
numerical integration to ensure consistency with the values computed for
$\mu_{\scriptscriptstyle E}$.

The entangled-two-photon quantum efficiency defined in Eq.~(\ref{efluxtopflux}),
\,$\eta_{\scriptscriptstyle E} = \mu_{\scriptscriptstyle E}/\Phi$,\, is estimated as
follows. Referring to Fig.~\ref{fig10}, the computed photoelectron count rate for
CsK$_2$Sb at $T_{\scriptscriptstyle E} = 10$~fs is seen to be
\,$\mu_{\scriptscriptstyle E}  \approx 8.0 \times 10^4$~s$^{-1}$. The photon flux
$\Phi = \phi A$ used in calculating these curves is specified by the values of the
photon-flux density $\phi$ and area $A$ reported in Table~\ref{tab:spdcparams}.
Carrying out the calculation leads to $\eta_{\scriptscriptstyle E}^\text{\sc lis} \approx
8.0 \times 10^4/[(5.0 \times 10^{23}) \times (1.0 \times 10^{-10})] = 1.6 \times
10^{-9}$~electrons/photon, as reported in Table~\ref{tab:photoemissiveparams}.
Notably, the heuristic model elaborated in Sec.~\ref{EPparticlemodel} leads to a
similar result, as explained by Lissandrin \emph{et al.}~\cite{lissandrin2004}.

An experiment demonstrating entangled-two-photon photoemission from a
CsK$_2$Sb photocathode in a channel photomultiplier module is reported in
Sec.~\ref{ssec:expCPM}. The theoretical results developed in this section are rescaled
in Sec.~\ref{sssec:Kobalissandrin}, where it is shown that agreement
with experiment can be achieved for a suitable value of the material
response ratio $\mathcal{Q}$.

\subsection{Measures of Optical Power and the Effects of Optical Loss}\label{optlosssys}

The results provided to this point rely on the tacit assumption that the transmittance
of the optical system is unity ($\mathcal{T\/} = 1$), signifying that all optical
components situated between the light source  and the photoemissive sample are
lossless.  Table~\ref{tab:optlosssys} lists various commonly used measures of power
in optical systems and displays the effects of optical loss ($\mathcal{T\/} < 1$) on
each. Loss resulting from the Bernoulli random deletion of photons is always
detrimental but not equally so for all sources of
light~\cite{teich1982effects,perina1983independent}. Optical loss is particularly
deleterious for entangled-two-photon photoemission and absorption because the
loss of either member of the entangled-photon pair eradicates their pairing. This
eliminates the intact pair contribution to twin-induced processes such as
ETPP and ETPA;
however, the surviving photon remains available as a singleton that can still
participate in one- or two-photon processes.
For the independent loss of each twin, the contribution
from intact photon pairs therefore scales as
$\mathcal{T\/}^2$, whereas the total transmitted
optical power, including surviving singletons,
scales as $\mathcal{T\/}$~\cite{Klyshko80-calib,Booth01,Booth04,lissandrin2004}.

\begin{table}[htb!]
\centering
\begin{small}
\caption{Various measures of power commonly used in optical systems and the
effect that optical loss ($\mathcal{T\/} < 1$) has on them.
The total transmitted optical power scales as $\mathcal{T}$\!,
whereas the power carried by surviving intact entangled-photon
pairs scales as $\mathcal{T}^2$ under independent photon loss.
The units associated with the various measures are indicated. Intensity, as well as
photon-flux measures, behave in an analogous manner.}
\label{tab:optlosssys}
\renewcommand{\arraystretch}{1.2}
\begin{tabular}{lll}
\hline
Optical-power measure & Symbol & Units \\[0.5mm]
\hline
\noalign{\vskip 0.5mm}
Power of a CW optical pump & $P_{\!\scriptscriptstyle P}$ & W \\
Optical power of a CW source of entangled-photon pairs &
$P_{\!\scriptscriptstyle E}$ & W \\
Instantaneous optical power of an arbitrary source of radiation & $P_0(t)$ & W \\
Optical power of a CW source of radiation & $P_0$ & W \\
Peak optical power of a pulsed source of radiation & $\widehat{P}_0$ & W \\
Mean-square optical power of a source of radiation & $\overline{P_0^2}$ & W$^2$ \\
Instantaneous incident optical power & $P(t) = \mathcal{T\/} P_0(t)$ & W \\
CW incident optical power & $P = \mathcal{T\/} P_0$ & W \\
Peak incident optical power of a pulsed source & $\widehat{P} =
\mathcal{T\/} \widehat{P}_0$ & W \\
Mean-square incident optical power & $\overline{P^2} =
\mathcal{T\/}^2\overline{P_0^2}$ & W$^2$ \\
CW incident optical power carried by intact pairs and singletons &
$P = \mathcal{T\/} P_{\!\scriptscriptstyle E}$ & W \\
CW incident optical power carried by intact entangled pairs &
$P_\mathrm{intact} = \mathcal{T\/}^2 P_{\!\scriptscriptstyle E}$ & W \\
CW incident optical power carried by singletons &
$P_\mathrm{singletons} = \mathcal{T\/}(1-\mathcal{T\/}) P_{\!\scriptscriptstyle E}$ & W \\ [0.5mm]
\hline
\end{tabular}
\end{small}
\end{table}

\section{MEASUREMENT AND IDENTIFICATION OF SUBTHRESHOLD\\ PHOTOEMISSION}
\label{measiden}

As detailed in Sec.~\ref{theory} (and portrayed in Fig.~\ref{fig7}), the photocurrents
observed in the course of carrying out subthreshold photoemission measurements in
semiconductors can arise from several sources:
\begin{enumerate}
  \item The one-photon Fermi-tail linear photocurrent, $\,i_{\scriptscriptstyle F} =
  \mathcal{R}_{\scriptscriptstyle F} P$,\, results from thermally excited Fermi-tail
  electrons in the conduction band [Eq.~(\ref{iFETAresp})].

 \item The two-photon quadratic photocurrent,
 $\,i_{\scriptscriptstyle C} =
 \mathcal{L}_{\scriptscriptstyle C} P^2/A$,\, results
 from the absorption of two photons, with the
 transition proceeding through real or virtual
 intermediate states
 [Eq.~(\ref{i2pcurrentETA})].

  \item The entangled-two-photon linear photocurrent, $\,i_{\scriptscriptstyle E} =
  \mathcal{R}_{\scriptscriptstyle E} P$,\, arises from the simultaneous absorption of
  entangled-photon pairs by valence-band electrons [Eq.~(\ref{iepcurrentETA})].

  \item The effective background current
  $i_{\scriptscriptstyle D}$ represents contributions
  from dark counts and electronic baseline offsets.
  Its magnitude depends on the detector and
  measurement conditions and is sometimes negligible, though it need not be.
\end{enumerate}

Section~\ref{measidenaux} introduces a number of auxiliary techniques that are
commonly used to facilitate the measurement of subthreshold photocurrents.
Section~\ref{measideniandmu} provides expressions for the measured photocurrents
and photoelectron count rates that accommodate the distortions introduced by
these techniques, under classical, coherent, and entangled-photon-pair illumination.
Section~\ref{measidencross} derives the crossover intensities at which the linear and
quadratic contributions to the  photocurrents are equal for each of the illumination
regimes. Section~\ref{paramestPE} discusses procedures for estimating the values of
subthreshold parameters of interest. Sections~\ref{measideniden} and
\ref{enhancingf} suggest approaches that can assist in identifying various forms of
subthreshold photoemission, and in enhancing them, respectively.

\subsection{Auxiliary Methodologies for Extracting Subthreshold Photocurrents}
\label{measidenaux}

A number of auxiliary measurement techniques are commonly employed to facilitate
the extraction of subthreshold photocurrents, which are always small:
\begin{enumerate}
  \item Lock-in detection employs signal modulation (e.g., imposed by a light chopper)
  and phase-sensitive demodulation, followed by low-pass filtering, to isolate weak signals
  buried in noise.
However, lock-in detection extracts a selected
Fourier component of the modulated photocurrent
(usually the first harmonic), from which the
underlying photocurrent can be inferred using
an appropriate calibration. A mechanical
chopper may also reduce the time-averaged
optical throughput.
  The lock-in amplitude-normalization factor depends on the waveform, reference
  power convention, and peak-versus-RMS calibration; it need not be less than
  unity. Linear and quadratic contributions may require different factors.

  \item The use of fluctuating light, such as a multimode laser or thermal source,
  affects linear and nonlinear photoemission differently. As outlined in
  Sec.~\ref{optpowvar}, because of its square-law power dependence the
  two-photon photocurrent $i_{\scriptscriptstyle C}$ is enhanced by the
  factor \,$g_2 \; (\geqslant 1$ for classical light) relative to its value for
  coherent light ($g_2 = 1$).

\item The use of a pulsed optical source, such as a pulsed or mode-locked
laser, also affects linear and nonlinear photoemission differently.
As explained in Sec.~\ref{2Qphotstatspulse}, by virtue of its square-law
power dependence the two-photon photocurrent $i_{\scriptscriptstyle C}$
is enhanced by the factor $\Gamma \; (\geqslant 1)$ relative to its value
in the absence of imposed temporal modulation, with the same mean
optical power and underlying photon statistics. In the absence of
imposed temporal modulation, $\Gamma = 1$.

  \item Reducing the area of illumination $A$ enhances $i_{\scriptscriptstyle C}$ relative
to the other forms of subthreshold photoemission because, for fixed incident optical power
at the sample, $i_{\scriptscriptstyle C} \propto 1/A$, whereas the linear contributions
are independent of $A$, provided that the entanglement area $A_{\scriptscriptstyle E}$
remains unchanged.

\end{enumerate}

\subsection{Subthreshold Photocurrents and Photoelectron Count Rates}
\label{measideniandmu}

The formulas for the subthreshold photocurrents
provided in Eqs.~(\ref{iFETAresp}),
(\ref{i2pcurrentETA}), and (\ref{iepcurrentETA})
describe idealized responses in terms of the
optical power incident on the sample, without
explicitly accounting for source-to-sample
optical losses or auxiliary measurement factors.
Here, we provide expressions for the measured photocurrents
and photoelectron count rates that include modifications  to accommodate the
distortions introduced by the auxiliary methodologies discussed in
Sec.~\ref{measidenaux}. Modifications are also included to accommodate
optical-system loss, which reduces the optical power or intensity at the source
($P_0$ or $I_0$) to that at the sample ($P$ or $I$) --- various measures of system
optical power, and the manner in which they are affected by the optical-system
transmittance $\mathcal{T\/} \; (\leqslant 1)$, were discussed in
Sec.~\ref{optlosssys} and summarized in Table~\ref{tab:optlosssys}. The expressions
provided in this section encompass classical, coherent, and entangled-photon-pair
illumination, and enable CW-equivalent values of the responsivities and quantum
efficiencies to be extracted from the measurements.
All three forms of light are used in the experiments reported in the sequel.

\subsubsection{Classical-Light Illumination}\label{fosteringcoh}

For illumination with classical light, the total measured mean photocurrent
$\,\overline{\imath}_{\scriptscriptstyle TC}\,$ generated at the sample includes an
effective dark/circuit current $i_{\scriptscriptstyle D}$ and the individual
contributions set forth in Eqs.~(\ref{iFETAresp}) and (\ref{i2pcurrentETA}) --- the
terms involving the responsivity $\mathcal{R}_{\scriptscriptstyle F}$ and the
responsivity coefficient $\mathcal{L}_{\scriptscriptstyle C}$ represent the Fermi-tail
and two-photon photocurrents, respectively. The dependencies on the wavelength
of the incident radiation  $\lambda$ and temperature of the sample $\mathsfit{T\/}$
are brought into relief by writing $\,\overline{\imath}_{\scriptscriptstyle TC}\,$ in the
form
\begin{equation}\label{totalcurrentfullC}
 \overline{\imath}_{\scriptscriptstyle TC}(\lambda,\mathsfit{T\/})  =
 i_{\scriptscriptstyle D} + \mathcal{F} \left[
\mathcal{R}_{\scriptscriptstyle F}(\lambda,\mathsfit{T\/})\;(\mathcal{T\/}
\overline{P_0})
+ \mathcal{L}_{\scriptscriptstyle C}(\lambda,\mathsfit{T\/})\;(\mathcal{T\/}
\overline{P_0})^2\,g_2\, \Gamma/A \right].
\end{equation}

The lock-in-detection factor \,$\mathcal{F}$ corresponds to item \#1 in
Sec.~\ref{measidenaux}.
The use of a common factor $\mathcal{F}$ for
the linear and quadratic contributions assumes
identical lock-in normalization factors, as
occurs for ideal binary on/off modulation.
For more general modulation waveforms,
separate factors may be required.

The combination \,$g_2 \Gamma/A$, where $g_2$ and
$\Gamma$ are defined in Eqs.~(\ref{g2def}) and (\ref{eq:intenfactor}), respectively,
serves as a convenient composite enhancement factor for the two-photon
contribution, although its constituent terms (which correspond to items \#2--4 in
Sec.~\ref{measidenaux}, respectively) have distinct physical origins: photon statistics,
temporal concentration, and transverse geometric concentration, respectively.
The factorization $g_2\Gamma$ assumes that
the intrinsic fluctuations of the unpulsed
source and the imposed temporal modulation
are statistically independent.

The
mean optical power $\overline{P}$ incident on the sample is explicitly written as
$\mathcal{T\/} \overline{P_0}$ to highlight the fact that the individual contributions
within $\overline{\imath}_{\scriptscriptstyle TC}$ depend differently on
$\mathcal{T\/}$. The dark/circuit background current $i_{\scriptscriptstyle D}$ is
included to phenomenologically represent the measurement baseline under the
bandwidth and detection conditions of the experiment.

\subsubsection{Coherent-Light Illumination}\label{fosteringcoherent}

Coherent light is a special case of classical
light; its intrinsic photon statistics are
Poisson and $g_2=1$, while independently
imposed temporal modulation is represented by
$\Gamma$. Equation~(\ref{totalcurrentfullC})
therefore reduces to
\begin{equation}\label{totalcurrentfullCoherent}
\overline{\imath}_{\scriptscriptstyle TC}(\lambda,\mathsfit{T\/})  =
i_{\scriptscriptstyle D} + \mathcal{F} \left[
\mathcal{R}_{\scriptscriptstyle F}(\lambda,\mathsfit{T\/})\;(\mathcal{T\/}
\overline{P_0})
+ \mathcal{L}_{\scriptscriptstyle C}(\lambda,\mathsfit{T\/})\;(\mathcal{T\/}
\overline{P_0})^2\, \Gamma/A \right].
\end{equation}
The photoelectron count rate $\,\overline{\mu}_{\scriptscriptstyle TC}\,$, the digital
counterpart of $\overline{\imath}_{\scriptscriptstyle TC}$, is obtained via
\begin{equation}\label{firstiTOmu}
  \overline{\mu}_{\scriptscriptstyle TC} =
  \frac{\overline{\imath}_{\scriptscriptstyle TC}}{e}\,.
\end{equation}
When $i_{\scriptscriptstyle D}$ includes an
electronic baseline offset rather than physical
dark counts, $i_{\scriptscriptstyle D}/e$
represents an electron-equivalent background
rate, rather than an actual dark-count rate.
Using the relation \,$\mathcal{R}_{\scriptscriptstyle [\cdot]} =
(e/h\nu)\,\eta_{\scriptscriptstyle [\cdot]}$\, provided in Eq.~(\ref{iSRESPeta}), and
setting $\mathcal{F} = 1$ for digital-counting implementations in which lock-in
detection is not employed, Eq.~(\ref{totalcurrentfullCoherent}) becomes
\begin{equation}\label{imathTOmu}
\overline{\mu}_{\scriptscriptstyle TC}(\lambda,\mathsfit{T\/})  =
\frac{i_{\scriptscriptstyle D}}{e} +
\eta_{\scriptscriptstyle F}(\lambda, \mathsfit{T\/})\;\frac{(\mathcal{T\/}
\overline{P_0})}{h\nu }
+ \mathcal{L}_{\scriptscriptstyle C}(\lambda,\mathsfit{T\/})\;(\mathcal{T\/}
\overline{P_0})^2 \;\frac{\Gamma}{eA}\,.
\end{equation}

\subsubsection{Entangled-Photon-Pair Illumination}\label{fosteringent}

When the sample is illuminated by a CW or pulsed source
of entangled-photon pairs instead, the
photoemission terms analogous to those in Eq.~(\ref{totalcurrentfullCoherent}) are
augmented by the entangled-two-photon contribution specified in
Eq.~(\ref{iepcurrentETA}), whereupon the total mean photocurrent
$\,\overline{\imath}_{\scriptscriptstyle TE}\,$ becomes
\begin{equation}\label{totalcurrentfullE}
\overline{\imath}_{\scriptscriptstyle TE}(\lambda,\mathsfit{T\/})  =
i_{\scriptscriptstyle D} + \mathcal{F} \left[
\mathcal{R}_{\scriptscriptstyle F}(\lambda,\mathsfit{T\/})\;(\mathcal{T\/}
\overline{P_{\!\scriptscriptstyle E}}) +
\mathcal{R}_{\scriptscriptstyle E}(\lambda,\mathsfit{T\/})\;(\mathcal{T\/}^2
\overline{P_{\!\scriptscriptstyle E}})
+ \mathcal{L}_{\scriptscriptstyle C}(\lambda,\mathsfit{T\/})\;(\mathcal{T\/}
\overline{P_{\!\scriptscriptstyle E}})^2\,\Gamma/A\right],
\end{equation}
where $\overline{P_{\!\scriptscriptstyle E}}$ represents the mean optical power
emitted by the entangled-photon source,
$\mathcal{T\/}
\overline{P_{\!\scriptscriptstyle E}}$ indicates the total incident optical power,
and $\mathcal{T\/}^2
\overline{P_{\!\scriptscriptstyle E}}$ signifies the intact-pair incident optical power.

By using the same symbols in writing Eqs.~(\ref{totalcurrentfullE}) and
(\ref{totalcurrentfullCoherent}), it is implicitly assumed that:
\begin{enumerate}
\item Individual SPDC photons and
coherent photons generate Fermi-tail photoemission in the same way (as
validated in Fig.~\ref{fig23}).
\item The magnitude of the optical-system transmittance
$\mathcal{T\/}$ is the same whether a photon is coherent or a member of an
entangled-photon pair (as borne out in Fig.~\ref{fig23}).
\item SPDC cousins and singleton pairs generate
two-photon photoemission in the same way as pairs of coherent photons
(as corroborated in Fig.~\ref{fig26}).
\item Cousins are independently absorbed (as confirmed in Fig.~\ref{fig26}).
\end{enumerate}

Moreover, the total mean optical power incident
on the sample is $\mathcal{T\/}
\overline{P_{\!\scriptscriptstyle E}}$, whereas
the power associated with surviving intact pairs
is represented by $\mathcal{T\/}^2
\overline{P_{\!\scriptscriptstyle E}}$,
highlighting the different dependencies on
$\mathcal{T\/}$ of the terms in
Eq.~(\ref{totalcurrentfullE}).
 The parameter $\Gamma$ accommodates the use of pulsed
entangled-photon sources. Again, the presence of the factors $\mathcal{F}$,
$\mathcal{T\/}$, $\Gamma$, and $A$ in Eq.~(\ref{totalcurrentfullE}) render the
responsivities as CW-equivalent values.

Although the terms involving $\mathcal{R}_{\scriptscriptstyle F}$ and
$\mathcal{L}_{\scriptscriptstyle C}$ in Eq.~(\ref{totalcurrentfullE}) have the same
form as those in Eq.~(\ref{totalcurrentfullCoherent}), they have different
interpretations. In the context of entangled-photon excitation, the contributions
involving $\mathcal{R}_{\scriptscriptstyle F}$, $\mathcal{R}_{\scriptscriptstyle E}$,
and $\mathcal{L}_{\scriptscriptstyle C}$ represent, respectively, Fermi-tail
photoemission induced by individual SPDC photons, entangled-two-photon
photoemission induced by twins, and two-photon photoemission induced by
independent cousins and singleton pairs.

The digital counterpart of the total mean photocurrent
$\,\overline{\imath}_{\scriptscriptstyle TE}\,$ given in Eq.~(\ref{totalcurrentfullE}) is
\begin{equation}\label{muimathTOmue}
  \overline{\mu}_{\scriptscriptstyle TE}(\lambda,\mathsfit{T\/}) =
  \mu_{\scriptscriptstyle D} + \mu_{\scriptscriptstyle F} + \mu_{\scriptscriptstyle E} +
  \mu_{\scriptscriptstyle C}\,,
\end{equation}
representing the total mean photoelectron count rate.
For frequency-degenerate photon pairs, both twins have photon energy $h\nu$.
On expansion,
Eq.~(\ref{muimathTOmue}) becomes
\begin{eqnarray}
\overline{\mu}_{\scriptscriptstyle TE}(\lambda,\mathsfit{T\/}) & =
& \frac{i_{\scriptscriptstyle D}}{e} +
\eta_{\scriptscriptstyle F}(\lambda, \mathsfit{T\/})\;\frac{(\mathcal{T\/}
\overline{P_{\!\scriptscriptstyle E}})}{h\nu } +
\eta_{\scriptscriptstyle E}(\lambda, \mathsfit{T\/})\;\frac{\mathcal{T\/} (\mathcal{T\/}
\overline{P_{\!\scriptscriptstyle E}})}{h\nu }
+ \mathcal{L}_{\scriptscriptstyle C}(\lambda,\mathsfit{T\/})\;(\mathcal{T\/}
\overline{P_{\!\scriptscriptstyle E}})^2\,\frac{\Gamma}{eA} \label{imathTOmue1} \\[0.6mm]
 & = & \frac{i_{\scriptscriptstyle D}}{e} +
\eta_{\scriptscriptstyle F}(\lambda, \mathsfit{T\/})\;\frac{(\mathcal{T\/}
\overline{P_{\!\scriptscriptstyle E}})}{h\nu } +
\eta_{\scriptscriptstyle E}(\lambda, \mathsfit{T\/})\;\frac{\mathcal{T\/} (\mathcal{T\/}
\overline{P_{\!\scriptscriptstyle E}})}{h\nu }
+ \mathcal{L}_{\scriptscriptstyle C}(\lambda,\mathsfit{T\/})\;(\mathcal{T\/}
\overline{I_{\scriptscriptstyle E}})^2\,\frac{\Gamma A}{e}\,.\qquad \label{imathTOmue2}
\end{eqnarray}
In the ETPP term, one factor of $\mathcal{T\/}$ accounts
for photon transmission, while the second accounts for
the conditional probability that the transmitted photon's
twin also survives, yielding an intact-pair survival
probability of $\mathcal{T\/}^2$. Equation~(\ref{imathTOmue1}) for
entangled-photon-pair illumination parallels Eq.~(\ref{imathTOmu}) for
coherent-light illumination.

\subsection{Linear-Quadratic Crossover Intensities} \label{measidencross}

Expressions for the \textbf{crossover intensities},
defined as the mean incident intensities at which
the quadratic two-photon photocurrent equals the
sum of the relevant linear photocurrent
contributions, are calculated for classical,
coherent, and entangled-photon-pair illumination.
 While these crossover intensities are
operational, experiment-specific quantities that depend on the illumination statistics
and measurement configuration, they can be useful for determining parameter
values and for identifying the form of subthreshold photoemission.

\subsubsection{Classical-Light Illumination}\label{fosteringcohCROSS}

As the classical light power increases, the Fermi-tail term, which scales linearly with
power, is increasingly dominated by the two-photon photocurrent, which scales
quadratically with power. A useful benchmark is the crossover intensity at the
sample, $I_{\scriptscriptstyle FC}$, where the two contributions are equal,
representing the boundary between linear and quadratic scaling. Making use of
Eq.~(\ref{totalcurrentfullC}), and using $\mathcal{T\/} \overline{P_0}/A =
\overline{P}/A = \overline{I}$,\, we obtain
\begin{equation}\label{IFCcross}
  I_{\scriptscriptstyle FC} =  \frac{\mathcal{R}_{\scriptscriptstyle F}}
  {\mathcal{L}_{\scriptscriptstyle C} g_2\Gamma}\,. \!\!\qquad \text{(classical light)}
\end{equation}

\subsubsection{Coherent-Light Illumination}\label{fosteringcoherentCROSS}

For coherent-light illumination $g_2 = 1$, so that Eq.~(\ref{IFCcross}) becomes
\begin{equation}\label{IFCcrosscoh}
  I_{\scriptscriptstyle FC} =  \frac{\mathcal{R}_{\scriptscriptstyle F}}
  {\mathcal{L}_{\scriptscriptstyle C} \Gamma}\,.\;\;\;\qquad \text{(coherent light)}
\end{equation}
For illumination with CW coherent light, Eq.~(\ref{IFCcrosscoh}) is modified
by setting $\Gamma = 1$, whereupon
\begin{equation}\label{IFCcrosscohCW}
  \qquad I_{\scriptscriptstyle FC} =  \frac{\mathcal{R}_{\scriptscriptstyle F}}
  {\mathcal{L}_{\scriptscriptstyle C}}\,.\;\;\;\qquad \,\, \text{(CW coherent light)}
\end{equation}

\subsubsection{Entangled-Photon-Pair Illumination}\label{fosteringentCROSS}

As the incident entangled-photon power increases, the quadratic two-photon
photoemission term in Eq.~(\ref{totalcurrentfullE}) increasingly dominates the other
two terms, which scale linearly. The useful benchmark in this case is the crossover
intensity $I_{\scriptscriptstyle EC}$, which represents the boundary between linear
and quadratic scaling, i.e., where the sum of the Fermi-tail and
entangled-two-photon photocurrents equals the two-photon photocurrent. This
quantity, obtained from Eq.~(\ref{totalcurrentfullE}), is
\begin{equation}\label{IFEcross}
  I_{\scriptscriptstyle EC} = \frac{\mathcal{R}_{\scriptscriptstyle F} +
  \mathcal{T\/} \mathcal{R}_{\scriptscriptstyle E} }{\mathcal{L}_{\scriptscriptstyle C}
  \Gamma }.
\end{equation}

In the course of carrying out experiments to observe entangled-two-photon
photoemission, it is often useful to conduct controlled companion experiments that
make use of coherent light or eliminate the twin photons. It then proves convenient
to relate $I_{\scriptscriptstyle EC}$ for the entangled-photon experiment to
$I_{\scriptscriptstyle FC}$ for the control experiment. If the experimental parameters
are the same for both, combining Eqs.~(\ref{IFCcrosscoh}) and (\ref{IFEcross}) leads
to
\begin{equation}\label{IFEcrossFC}
  I_{\scriptscriptstyle EC}  = I_{\scriptscriptstyle FC}\left( 1 +
  \frac{\mathcal{T\/}\mathcal{R}_{\scriptscriptstyle E}}
  {\mathcal{R}_{\scriptscriptstyle F}}\right)\,.
\end{equation}
The crossover intensity for entangled-photon pairs $I_{\scriptscriptstyle EC}$ is
larger than the crossover intensity $I_{\scriptscriptstyle FC}$ because of the
additional contribution from entangled-two-photon photoemission.

\begin{figure}[htb!]
\centering\includegraphics[width=3in]{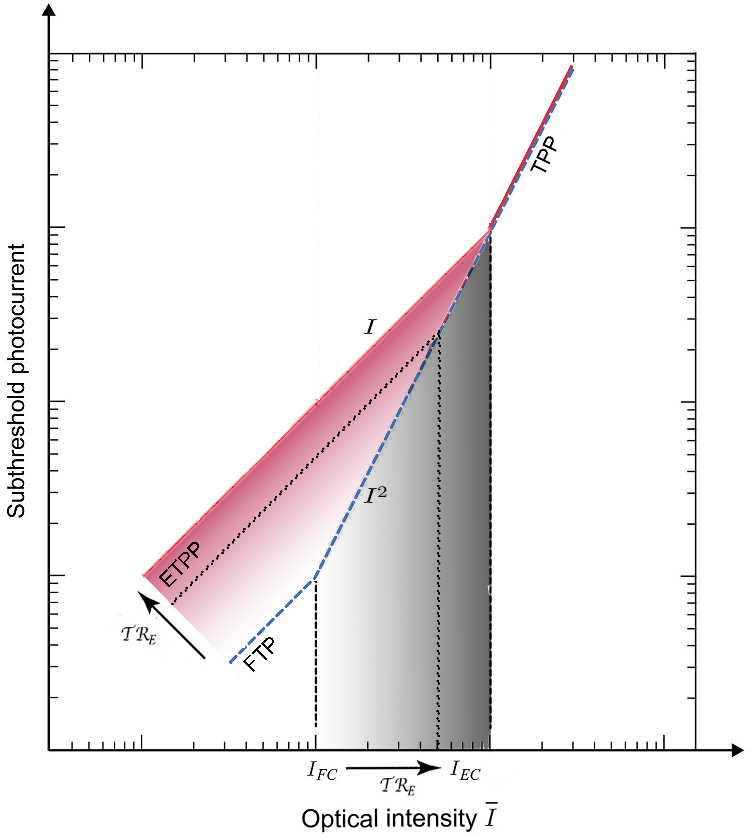}
\caption{Total subthreshold photocurrent vs. optical intensity under coherent-light
illumination (dashed blue lines) and under
entangled-photon-pair illumination (graded red and shaded gray regions).
As specified in Eq.~(\ref{totalcurrentfullE}), the photocurrent elicited by
entangled-photon pairs comprises three contributions: 1)~Fermi-tail
photoemission (FTP) induced by individual SPDC photons $(\propto \mathcal{R}_{\scriptscriptstyle F}
\overline{I})$; 2)~entangled-two-photon photoemission (ETPP) induced by twins
$(\propto \mathcal{T\/} \mathcal{R}_{\scriptscriptstyle E} \overline{I})$; and
3)~two-photon photoemission (TPP) induced by independent cousins and/or
singleton pairs $\smash{(\propto \mathcal{L}_{\scriptscriptstyle C}\Gamma \overline{I}^2)}$.
The degree of ETPP admixture is quantified by the magnitude of $\mathcal{T\/}
\mathcal{R}_{\scriptscriptstyle E}$ and is represented visually by the saturation
of the red shading.
The intensity region $I_{\scriptscriptstyle FC} <  \overline{I}  <  I_{\scriptscriptstyle EC}$
has a width $I_{\scriptscriptstyle EC}-I_{\scriptscriptstyle FC} = \mathcal{T\/}
\mathcal{R}_{\scriptscriptstyle E} /(\mathcal{L}_{\scriptscriptstyle C} \Gamma)$
that increases with increasing $\mathcal{T\/} \mathcal{R}_{\scriptscriptstyle E}$
and is represented visually by the increasing darkness of the gray shading.
A representative value of $\mathcal{T\/} \mathcal{R}_{\scriptscriptstyle E}$, along
with the corresponding value of $I_{\scriptscriptstyle EC}$, are indicated by the
dotted lines. This figure is a refined version of Fig.~\ref{fig2} that incorporates optical loss and
an arbitrary level of ETPP.} \label{fig12}
\end{figure}

Equation~(\ref{IFEcrossFC}) teaches that the scaling signatures of two-photon and
entangled-two-photon photoemission differ in the intensity region
\begin{equation}\label{IFCEC}
  I_{\scriptscriptstyle FC} <  \overline{I}  <  I_{\scriptscriptstyle EC}\,.
\end{equation}
As displayed in Fig.~\ref{fig12}, in this
region the linear contribution (dotted black line) dominates
the photocurrent under entangled-photon-pair
illumination, whereas the quadratic contribution (dashed blue line)
dominates under coherent-light illumination. The ratio of the two crossover
parameters, obtained from  Eq.~(\ref{IFEcrossFC}), is
\begin{equation}\label{ICEratio}
  \frac{I_{\scriptscriptstyle EC}}{I_{\scriptscriptstyle FC}} =
  1 +\frac{\mathcal{T\/}\mathcal{R}_{\scriptscriptstyle E}}
  {\mathcal{R}_{\scriptscriptstyle F}}\,.
\end{equation}

The absolute width of this intensity region,
$I_{\scriptscriptstyle EC}-I_{\scriptscriptstyle FC}
=\mathcal{T\/}\mathcal{R}_{\scriptscriptstyle E}/
(\mathcal{L}_{\scriptscriptstyle C}\Gamma)$,
increases with increasing
$\mathcal{R}_{\scriptscriptstyle E}$ and
$\mathcal{T\/}$, while decreasing the
Fermi-tail responsivity
$\mathcal{R}_{\scriptscriptstyle F}$ increases
the ratio $I_{\scriptscriptstyle EC}/
I_{\scriptscriptstyle FC}$ and hence the
logarithmic span of the region.
These results are foundational for the analyses
presented in Secs.~\ref{entPMT} and \ref{entchannel}.

\subsection{Parameter Estimation for Subthreshold Photoemission}\label{paramestPE}

In carrying out subthreshold photoemission measurements, it is often desired to
estimate from the data the value of a parameter that characterizes the underlying
photoemission process of interest. Carrying out this task in an optimal manner relies
on the use of statistical estimation
theory~\cite{fisher25,frechet43,rao45,cramer46,kay93,helstrom76,holevo82,braunsteincaves94,paris09,wiseman09}.
The parameter of interest might be one of the subthreshold quantum efficiencies or
responsivities, such as $\,\eta_{\scriptscriptstyle E}\,$ or
$\,\mathcal{R}_{\scriptscriptstyle E}$, or possibly one of the crossover intensities,
$\,I_{\scriptscriptstyle EC}\,$ or $\,I_{\scriptscriptstyle FC}$. These parameters may
be inferred from the natural primary observables, the subthreshold photocurrent or
photoelectron count rate, via the calibrated relations provided in
Sec.~\ref{measideniandmu}.

In classical estimation theory, an unknown parameter $\,\theta\,$ is inferred from
data $\,x\,$ that is distributed according to a probability law $p_\theta(x)$.
Under the usual regularity conditions, the
variance of any locally unbiased estimator
$\tilde{\theta}$ is bounded from below by the
Cram{\'e}r--Rao inequality~\cite{frechet43,rao45,cramer46},
\begin{equation}
\mathrm{Var}(\tilde{\theta}) \ge \frac{1}{F(\theta)},
\end{equation}
where $F(\theta)$ is the Fisher information \cite{fisher25,kay93}. In a quantum
setting, further optimization over the measurement itself leads to the quantum
Fisher information and the corresponding quantum Cram{\'e}r--Rao
bound~\cite{helstrom76,holevo82,braunsteincaves94,paris09,wiseman09}. The
precision achievable in the quantum domain depends not only on the observable
being measured, but also on the statistical character of the illuminating
field~\seccite{9.4}{saleh25}. Researchers have developed various estimation
techniques for diverse applications (Sec.~\ref{EP-Metrology}).

\subsection{Identifying the Form of Subthreshold Photoemission} \label{measideniden}

The phenomenological signatures serialized in Secs.~\ref{idenidenclascoh} and
\ref{idenidenepp} facilitate the identification of the specific form of subthreshold
photoemission induced by classical/coherent and   entangled-photon-pair
illumination, respectively. However, the behaviors cited should be regarded as
diagnostic signatures of the underlying photoemission mechanism, rather than as
logically definitive identifications in every experimental circumstance.

\subsubsection{Classical- and Coherent-Light Illumination} \label{idenidenclascoh}

Based on Eqs.~(\ref{totalcurrentfullC})--(\ref{imathTOmu})
and (\ref{IFCcross})--(\ref{IFCcrosscohCW}), and in conjunction with Fig.~\ref{fig12},
we can make several inferences regarding the origin of subthreshold
photocurrents induced by classical or coherent light.
The following scaling signatures presume
subtraction of the background contribution
and fixed values of $g_2$, $\Gamma$, and $A$
as the incident intensity is varied.
\begin{enumerate}
  \item In regions where the measured mean photocurrent
      $\overline{\imath}_{\scriptscriptstyle TC}$ or photoelectron count rate
      $\,\overline{\mu}_{\scriptscriptstyle TC}\,$ scales linearly with the incident
      optical intensity $\overline{I}$, the current represents Fermi-tail photoemission.
  \item If the measured photocurrent $\overline{\imath}_{\scriptscriptstyle TC}$ or
      photoelectron count rate $\,\overline{\mu}_{\scriptscriptstyle TC}\,$ scales
      linearly over the full range of $\overline{I}$, two-photon photoemission is either
      absent or negligible ($I_{\scriptscriptstyle FC}$ lies well above the
      measured intensity range).  (This situation is encountered in
      Figs.~\ref{fig14} and \ref{fig15}.)
  \item In regions where the measured mean photocurrent
      $\overline{\imath}_{\scriptscriptstyle TC}$ or photoelectron count rate
      $\,\overline{\mu}_{\scriptscriptstyle TC}\,$  scales quadratically with the
      incident optical intensity $\overline{I}$, the current represents two-photon
      photoemission. (This situation is encountered over portions of Figs.~\ref{fig20},
      \ref{fig21}, and \ref{fig23}.)
  \item If the measured photocurrent $\overline{\imath}_{\scriptscriptstyle TC}$ or
      photoelectron count rate $\,\overline{\mu}_{\scriptscriptstyle TC}\,$  behaves
      quadratically over the full range of $\overline{I}$,\, Fermi-tail photoemission is
      either absent or negligible ($I_{\scriptscriptstyle FC}$ lies well below
      the measured intensity range).   (This situation is encountered for the thick Na
      sample illuminated by pulsed classical light in Fig.~\ref{fig18} and for the
      CsK$_2$Sb sample illuminated by CW coherent light in Fig.~\ref{fig26}.)
  \item In regions where the measured mean photocurrent
      $\overline{\imath}_{\scriptscriptstyle TC}$ or photoelectron count rate
      $\,\overline{\mu}_{\scriptscriptstyle TC}\,$  depends on $g_2$, $\Gamma$,
      and/or $A$ in the manner expected, this is indicative of a quadratic
      contribution from two-photon photoemission.
\end{enumerate}

\subsubsection{Entangled-Photon-Pair Illumination} \label{idenidenepp}

Based on Eqs.~(\ref{totalcurrentfullE})--(\ref{imathTOmue2}) and
(\ref{IFEcross})--(\ref{ICEratio}), and with the help of Fig.~\ref{fig12}, we can make a
number of inferences regarding the origin of empirical subthreshold photocurrents
induced by entangled-photon pairs. Comparisons with coherent-light illumination
presume matched incident photon spectra, illumination areas, and sample and detection
conditions, unless otherwise specified.
The intensity-scaling tests described below refer to varying
the SPDC generation rate at fixed optical transmittance.
The attenuation tests instead refer to varying the
transmittance of filters placed after the SPDC source
while holding its generation rate fixed.
\begin{enumerate}
  \item If the measured mean photocurrent $\overline{\imath}_{\scriptscriptstyle
      TE}$ or photoelectron count rate $\,\overline{\mu}_{\scriptscriptstyle TE}\,$
       scales linearly with the incident entangled-photon intensity~$\overline{I}$ over
      a given range of values, and has the same magnitude as the linear photocurrent
      $\,\overline{\imath}_{\scriptscriptstyle TC}$ or photoelectron count rate
      $\,\overline{\mu}_{\scriptscriptstyle TC}\,$ measured using a coherent source of
      light over that same range of values, the current represents
      individual-SPDC-photon induced
      Fermi-tail photoemission. In accordance with Eqs.~(\ref{totalcurrentfullE}) and
      (\ref{imathTOmue1}), the attenuation provided by filters inserted in the
      entangled-photon beam path then follows the filter transmittance
      $\mathcal{T\/}$ in that region. (This situation is encountered over a portion of
      Fig.~\ref{fig23}.)
  \item If the measured mean photocurrent $\overline{\imath}_{\scriptscriptstyle
      TE}$ or photoelectron count rate $\,\overline{\mu}_{\scriptscriptstyle TE}\,$
      scales linearly with the incident entangled-photon intensity~$\overline{I}$ over a
      given range of values, and has a magnitude greater than the linear photocurrent
      $\,\overline{\imath}_{\scriptscriptstyle TC}$ or photoelectron count rate
      $\,\overline{\mu}_{\scriptscriptstyle TC}\,$ measured using a coherent source of
      light over that same range of values, the current represents a combination of
      individual-SPDC-photon induced Fermi-tail photoemission and entangled-two-photon
      photoemission. In accordance with  Eqs.~(\ref{totalcurrentfullE}) and
      (\ref{imathTOmue1}), the attenuation provided by filters inserted in the
      entangled-photon beam path then behaves as a mixture of $\mathcal{T\/}$ and
      $\mathcal{T\/}^2$ in that region.
  \item If the measured mean photocurrent $\overline{\imath}_{\scriptscriptstyle
      TE}$ or photoelectron count rate $\,\overline{\mu}_{\scriptscriptstyle TE}\,$
      scales linearly with the incident entangled-photon intensity~$\overline{I}$ over a
      given range of values, and has a magnitude greater than the quadratic
      photocurrent $\,\overline{\imath}_{\scriptscriptstyle TC}$ or photoelectron
      count rate $\,\overline{\mu}_{\scriptscriptstyle TC}\,$ measured using a
      coherent source of light over that same range of values,
     the current indicates the presence of an additional pair-induced
     contribution when
     one-photon backgrounds are independently constrained.
    In accordance with Eqs.~(\ref{totalcurrentfullE})
    and (\ref{imathTOmue1}), when the
    entangled-two-photon contribution dominates,
    attenuation by filters inserted in the
    entangled-photon beam path follows
    $\mathcal{T\/}^2$ rather than $\mathcal{T\/}$
    in that region.  (This situation is encountered over a portion of
      Fig.~\ref{fig26}.)
  \item If the measured mean photocurrent $\overline{\imath}_{\scriptscriptstyle
      TE}$ or photoelectron count rate $\,\overline{\mu}_{\scriptscriptstyle TE}\,$
       scales quadratically with the incident entangled-photon intensity~$\overline{I}$
      over a given range of values, and has the same magnitude as the quadratic
      photocurrent $\,\overline{\imath}_{\scriptscriptstyle TC}$ or photoelectron
      count rate $\,\overline{\mu}_{\scriptscriptstyle TC}\,$ measured using a
      coherent source of light over that same range of values, the current represents
      two-photon photoemission induced by cousins or singleton pairs. In accordance
      with Eqs.~(\ref{totalcurrentfullE}) and (\ref{imathTOmue1}), the attenuation
      provided by filters inserted in the entangled-photon beam path then behaves as
      $\mathcal{T\/}^2$ in that region. (This situation is encountered over a portion of
      Fig.~\ref{fig26}.)
        \item As \,$\mathcal{T\/} \mathcal{R}_{\scriptscriptstyle
        E}/\mathcal{R}_{\scriptscriptstyle F} \to 0$,
        indicating that the entangled-two-photon
        contribution becomes negligible relative
        to Fermi-tail photoemission, the two
        crossover intensities coincide in
        accordance with Eq.~(\ref{ICEratio}).
         Individual-SPDC-photon induced Fermi-tail
      photoemission then transitions directly into cousin-/singleton-pair-induced
      two-photon photoemission at $I_{\scriptscriptstyle EC} = I_{\scriptscriptstyle
      FC}$, reproducing the results obtained for coherent light.
\end{enumerate}

\subsection{Enhancing a Selected Form of Subthreshold Photoemission}\label{enhancingf}

Qualitative guidance for enhancing a selected form of subthreshold photoemission at
the expense of others is sketched in Table~\ref{tab:PEchar}. Techniques include
manipulating the characteristics of the incident light; the properties of the sample
such as its temperature, area, and thickness; and/or the structure and specifications
of the measurement system.
Optical loss suppresses entangled-two-photon photoemission relative to
Fermi-tail photoemission, since the former scales as $\mathcal{T\/}^2$
and the latter as $\mathcal{T\/}$ (Sec.~\ref{optlosssys}).
\begin{table}[htb!]
\centering
\begin{small}
\caption{Characteristics of the incident light and sample temperature
that, for a given material, can enhance a selected form of subthreshold photoemission
at the expense of competing forms: 1)~Fermi-tail, 2)~two-photon, and
3)~entangled-two-photon. The characteristics considered are the wavelength region,
incident optical power ($\overline{P}$), illumination type and photon statistics,
modulation format ($\Gamma$), illumination area ($A$), incident photon
wavelength ($\lambda$); and sample temperature ($\mathsfit{T\/}$).
Here, $\lambda_{\texttt{W}} = hc/\texttt{W}\,$ is the material's ionization-wavelength
and $\lambda_g = hc/E_{\!g}$ is its bandgap wavelength.
The wavelength ranges shown for two-photon processes assume degenerate
illumination with photon energies above the bandgap; more  generally,
subthreshold two-photon photoemission requires a combined photon
energy exceeding the ionization energy, and transitions via virtual intermediate
states need not satisfy $\lambda<\lambda_g$.
For ETPP, the preference for a large illumination area
presumes an unchanged entanglement area.}
\renewcommand{\arraystretch}{1.05}
\label{tab:PEchar}
\begin{tabular}{@{}l|c|c|c|c|c|c|c}
\multicolumn{1}{c}{} & \multicolumn{7}{c}{\textbf{Characteristics of Incident Light
and Sample Temperature}} \\ [0.7mm]\cline{2-8}
\noalign{\vskip 0.5mm}
\textbf{Photoemission Form} & Wavelength region & $\overline{P}$ & Light type &
$\Gamma$ & $A$ & $\lambda$ & $\mathsfit{T\/}$ \\[0.3mm]
\hline
\noalign{\vskip 0.3mm}
Suprathreshold  & $\lambda < \lambda_\texttt{W}$  &   &  &   &  &  & \\
Fermi-tail    & $\lambda > \lambda_\texttt{W}$  & low &  coherent & CW & large
& short & warm     \\
Two-photon       & $\lambda/2 < \lambda_{\texttt{W}} < \lambda < \lambda_g$
&  high  &  bunched &   pulsed  & small &   long  & cool  \\
Entangled-two-photon   & $\lambda/2 < \lambda_{\texttt{W}} < \lambda < \lambda_g$
&  low   & entangled &  CW  & large &  long  & cool \\[0.3mm]
\hline
\end{tabular}
\end{small}
\end{table}

Two examples are spelled out for the sake of illustration: 1)~A cooled
sample of small area, illuminated by a high-power, pulsed laser focused to a small
spot, favors two-photon photoemission over Fermi-tail photoemission.
2)~A cooled sample illuminated by a low-power CW source of entangled-photon
pairs, with a large illumination area but an unchanged entanglement area,
favors entangled-two-photon photoemission over independent-photon
two-photon photoemission, while cooling suppresses Fermi-tail photoemission.

\section{OBSERVATION OF FERMI-TAIL PHOTOEMISSION UNDER
COHERENT\\ ILLUMINATION}\label{subthreshold}

At finite temperatures  ($\mathsfit{T\/}>0$~K), the idealized intrinsic-semiconductor
picture introduced in Sec.~\ref{theory} predicts a small population of electrons in the
high-energy tail of the Fermi--Dirac occupation distribution whose energies extend
into the conduction band, with the preponderance near its lower
edge~\chapcite{17}{saleh2019}. As illustrated in  Fig.~\ref{fig7}($b$) and discussed
in Sec.~\ref{subthreshtheory}, single subthreshold photons can impart sufficient
energy to these \textbf{Fermi-tail electrons} to allow them to escape to the vacuum.
Fermi-tail photoemission can therefore occur when
\begin{equation}\label{rangemetals}
  h\nu < \texttt{W} = E_{\!g} + \chi.
\end{equation}

The physical origin of this form of photoemission makes it clear that the magnitude
of the Fermi-tail photocurrent depends on the temperature of the sample as well as
on the wavelength of the illumination. Since the carrier concentration available in the
conduction band increases with increasing temperature, so too does the Fermi-tail
photocurrent. Moreover, the photocurrent increases with increasing photon energy
(decreasing wavelength) as the incident photons can promote emission from
progressively more heavily occupied states nearer the lower portion of the
conduction-band manifold. These expectations have been borne out in the
alkali-antimonide semiconductor experiments reported by Booth~\emph{et
al.}~\cite{booth2006}.

In this section, we review \textbf{Fermi-tail photoemission} from CsK$_2$Sb and
from two other alkali-antimonide materials, Na$_2$KSb and Cs$_3$Sb, under
coherent illumination. Similar experiments can be carried out with other materials by
constructing photomultiplier tubes with vacuum-deposited custom photocathodes
(Sec.~\ref{sub:earlyNa}). Fermi-tail photoemission from CsK$_2$Sb stimulated by
entangled-photon pairs is analyzed in Sec.~\ref{entPMT}. It is important to
thoroughly understand Fermi-tail photoemission in the context of examining the two
other forms of subthreshold photoemission addressed in this work: two-photon
photoemission and especially entangled-two-photon photoemission.

\subsection{Experimental Arrangement} \label{sub:SPSTCsK2Sb}

Fermi-tail photoemission yields a far smaller photocurrent than suprathreshold
photoemission. This weak photocurrent can be
amplified, however, by using a photomultiplier-tube (PMT) configuration
in which the photocathode comprises the material under study.
 This approach engages the high-gain and comparatively
low-noise current amplification offered by the PMT's internal electron-multiplier
structure (Fig.~\ref{fig8}).

\begin{figure}[htb!]
\centering\includegraphics[width=3.25in]{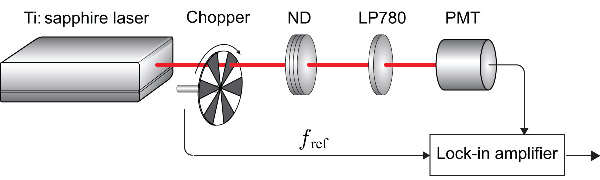}
\caption{Experimental arrangement for observing Fermi-tail photoemission from
PMTs with CsK$_2$Sb, Na$_2$KSb, and Cs$_3$Sb semiconductor photocathodes.
The light from a  mode-locked Ti:sapphire laser, operating at a wavelength of
800, 830, or 845~nm, was incident on the photocathode, which was maintained
at temperature $\mathsfit{T\/}$. Neutral-density (ND) filters were used to
attenuate the laser power. A long-pass (LP780) filter isolated the detector from
stray light with wavelengths shorter than 780~nm. The PMT output was fed to a
lock-in amplifier referenced to the frequency of a rotating light
chopper ($f_\mathrm{ref} = 150$~Hz). The 3-mm-diameter laser beam
remained unfocused for these experiments.
(Adapted from Fig.~2.3 of~\cite{Booth04}.)
} \label{fig13}
\end{figure}

Figure~\ref{fig13} portrays the experimental arrangement used by Booth~\emph{et
al.}~\cite{booth2006} to observe Fermi-tail photoemission from several
alkali-antimonide semiconductors. A mode-locked Spectra-Physics (Tsunami)
Ti:sapphire laser~\chapcite{16}{saleh2019} was operated at one of three
wavelengths (photon energies): \,1)~$\lambda= 800$~nm ($h\nu = 1.55$~eV),
\,2)~830~nm ($h\nu = 1.50$~eV), or \,3)~845~nm ($h\nu = 1.47$~eV). The laser
output comprised a sequence of $\tau_0 = 120$-fs duration light pulses of
approximately Gaussian shape, emitted at a rate $f_\mathrm{rep} \equiv 1/\tau_1 =
82$~MHz. The average power emitted by the laser was $\overline{P_0} \approx
1$~W and the peak power for each pulse was $\widehat{P}_0 \approx 90$~kW.
Approximating the light pulses by an equivalent square-pulse model, the duty cycle
defined in Eq.~(\ref{eq:deltadef}) is  estimated to be $\Delta = \tau_0/\tau_1 =
\tau_0 f_\mathrm{rep} \approx 0.98 \times 10^{-5}$.

The mean optical power $\overline{P}$ incident on the faceplate of the PMT was
measured using a laser power meter (Model 818-SL, Newport Corp., Irvine, CA)
placed directly in front of the faceplate.  The mean optical power was restricted to a
value $\leqslant 20~\mu$W to preclude local photocathode heating. Calibrated
neutral-density (ND) filters (Wratten, Eastman Kodak, Rochester, NY) were used to
attenuate the optical power for the various sequences of measurements. A long-pass
(LP780) optical filter was used to isolate the detector from stray light at wavelengths
shorter than 780~nm.

To extract the weak Fermi-tail photocurrent from various sources of noise, lock-in
detection~\cite{dicke46,stutt1949} was implemented, and a rotating optical light
chopper imposed intensity modulation. The output of the PMT was fed to a lock-in
amplifier (LIA) (Stanford Research Systems 850, Sunnyvale, CA), for which the light
chopper provided the reference signal at $f_\mathrm{ref} = 150$~Hz. The
experiments reported in this section were carried out at Boston University in the
time frame
1998--2004~\cite{booth1998ETPA,Lissandrin99,Booth01,Booth04,booth2006}.
The original experimental reports do not fully specify
the optical-power averaging and lock-in amplitude
calibration conventions; the resulting uncertainty
in the absolute responsivity normalization is
mentioned in Sec.~\ref{sub:Exp2pCsKSb}.

\subsection{Fermi-Tail Photoemission from Cesium-Potassium Antimonide}
\label{sub:SPSTCsK2Sb-results}

Detailed studies of the Fermi-tail photocurrent observed from CsK$_2$Sb have been
carried out using the experimental arrangement depicted in Fig.~\ref{fig13}.
CsK$_2$Sb is a widely used bialkali-antimonide semiconductor
(Sec.~\ref{singletheoryCsK2Sb}) that is often modeled, for phenomenological
purposes, as an intrinsic direct-bandgap material whose Fermi level lies near the
center of the bandgap~\cite{Sommer68,Nathan70,Varma78,Ghosh80}, so that its
energy-band diagram is reasonably represented by Fig.~\ref{fig7}. After describing
the PMT used in these experiments, we review the Fermi-tail photoemission data
obtained at different temperatures and wavelengths.

\subsubsection{PMT with Semitransparent CsK$_2$Sb Photocathode}
\label{subsub:CsPMT}

The Fermi-tail photoemission experiments made use of a Hamamatsu R464 PMT
enclosed in a Hamamatsu C4877 thermoelectric housing that allowed the
photocathode temperature $\mathsfit{T\/}$ to be reduced. The R464 is a head-on,
51-mm-diameter photomultiplier tube featuring a semi-transparent CsK$_2$Sb
bialkali-antimonide photocathode deposited on the front borosilicate window. The
thickness of the photocathode is on the order of tens of nm while the escape depth $d$
for CsK$_2$Sb has been estimated to be about 40~nm, the value used in
Table~\ref{tab:photoemissiveparams}.

The (one-photon) spectral response of this device extends over the wavelength range
300--650~nm, with the peak sensitivity at 420~nm. The tube employs a 12-stage
box-and-grid dynode structure. When operated at an anode voltage of 1000~V
(maximum 1500~V), it offers a mean anode gain $\overline{G} \approx 6.0 \times
10^6$ and a gain--responsivity product $\overline{G}\mathcal{R}_{\scriptscriptstyle
S} \approx 3.0 \times 10^5\ \mathrm{A/W}$. Designed for photon-counting
applications, the R464 has a low dark-count rate $\approx 5$~s$^{-1}$ (maximum
$15$~s$^{-1}$), risetime $\approx 13$~ns, and transit time $\approx 70$~ns. In the
experiments described in this section, the PMT was operated at 1400~V and in
analog mode.

The statistical properties of the gain process and pulse-height distribution for the
particular R464 PMT used in these experiments were measured and carefully
characterized by Lissandrin~\cite{Lissandrin99}; an important effect that is often
ignored, electron backscattering at the dynodes, was incorporated in the analysis.

\subsubsection{Fermi-Tail Photocurrent from CsK$_2$Sb at Different Temperatures}
\label{sub:SPSTCsK2Sb-FerTemp}

\begin{figure}[htb!]
\centering\includegraphics[width=3.25in]{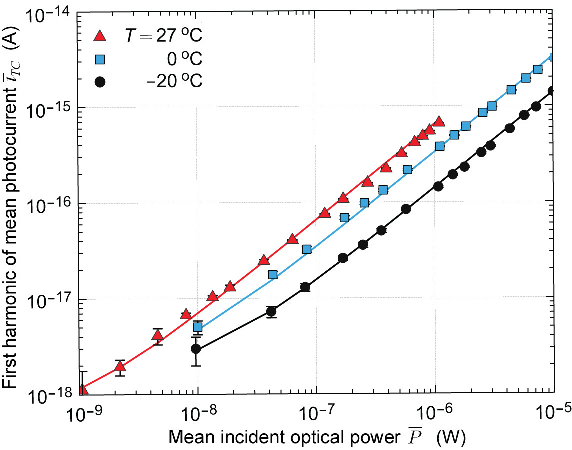}
\caption{Fermi-tail photoemission from CsK$_2$Sb at several temperatures.
Doubly logarithmic plots of the fundamental component of the photocurrent at the
photocathode vs. the mean optical power incident on the PMT faceplate.
Data are reported for three temperatures: $\mathsfit{T\/} =
27~^\circ$C (triangles), $0~^\circ$C (squares), and $-20~^\circ$C (circles).
Error bars indicate standard deviations for individual data points. The curves
incorporate the measured PMT background currents and represent near-linear
fits to the three data sets. (Adapted with permission from Fig.~2.4 of~\cite{Booth04}
and Fig.~6 of~\cite{booth2006}, \textcopyright 2006 AIP Publishing.)
} \label{fig14} \end{figure}

The Fermi-tail photocurrent from the CsK$_2$Sb photocathode of the R464 PMT is
displayed in Fig.~\ref{fig14} for three values of the temperature: $\mathsfit{T\/} =
27~^\circ$C (triangles), $0~^\circ$C (squares), and $-20~^\circ$C (circles). The
measurements were made using the unfocused light beam from the mode-locked
Ti:sapphire laser operated at a central wavelength $\lambda = 800$~nm ($h\nu =
1.55$~eV).

The data are displayed in the form of doubly logarithmic plots of the fundamental
Fourier component of the photocurrent at the photocathode
$\overline{\imath}_{\scriptscriptstyle TC}$ (A) vs. the mean optical power incident on
the PMT faceplate $\overline{P} = \mathcal{T\/}\overline{P_0}$ \,(W). The
photocurrent at the PMT anode was measured over a series of 3-min acquisition
periods with the low-pass-filter of the lock-in amplifier set to a bandwidth
$\mathsfit{B\/} = 26$~mHz.
The fundamental Fourier component of the photocathode current was calculated
by dividing the corresponding lock-in-detected anode-current component by the
mean PMT current gain $\,\overline{G} \approx 10^8$ at the
operating voltage of 1400~V.  The standard deviations of the data
points were also recorded.

The curves are near-linear fits to the three data sets; the slight departures from strict
linearity at low photocurrent are primarily attributable to additive dark/circuit
background currents. The Fermi-tail photocurrent increases with increasing
temperature, consistent with the increasing thermal population of electronically
accessible high-energy states in the semiconductor. The associated values of the
Fermi-tail responsivity $\mathcal{R}_{\scriptscriptstyle F}(\lambda,\mathsfit{T\/})$
are displayed in Table~\ref{tab:PMT}.

\subsubsection{Fermi-Tail Photocurrent from CsK$_2$Sb at Different Wavelengths}
\label{sub:SPSTCsK2Sb-results-Ferlambda}

\begin{figure}[htb!]
\centering\includegraphics[width=3.25in]{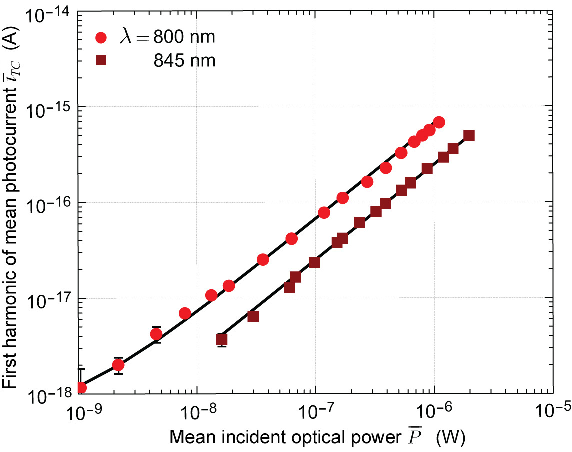}
\caption{Fermi-tail photoemission from CsK$_2$Sb at different wavelengths.
Doubly logarithmic plots of the fundamental component of the photocurrent at
the photocathode vs. the mean optical power incident on the PMT faceplate at
two wavelengths: $\lambda = 800$~nm (circles) and $845$~nm (squares).
The photocathode is maintained at $\mathsfit{T\/} = 27~^\circ$C. Error bars
indicate standard deviations for individual data points.  The curves incorporate
the measured PMT background currents and represent near-linear fits to the two
data sets.  (Adapted from Fig.~2.5 of~\cite{Booth04} and Fig.~7
of~\cite{booth2006}, \textcopyright 2006 AIP Publishing.)}
\label{fig15}
\end{figure}

The wavelength dependence of the Fermi-tail photocurrent from the same
CsK$_2$Sb photocathode is depicted in Fig.~\ref{fig15}. The data are again exhibited
in the form of doubly logarithmic plots of the fundamental Fourier component of the
photocurrent at the photocathode $\overline{\imath}_{\scriptscriptstyle TC}$ (A) vs.
the mean optical power incident on the PMT faceplate $\overline{P} =
\mathcal{T\/}\overline{P_0}$ \,(W). The source was the same unfocused light beam
from the mode-locked Ti:sapphire laser, which was operated at two different
wavelengths: $\lambda =800$~nm ($h\nu = 1.55$~eV, circles) and 845~nm ($h\nu =
1.47$~eV, squares). The photocathode temperature was maintained at
$27~^\circ$C.

The curves are near-linear fits to the data sets; again, the slight curvatures at low
values of the photocurrent arise from additive dark/circuit currents. The Fermi-tail
photocurrent declines with increasing wavelength (decreasing photon energy),
consistent with the reduced ability of the incident photons to induce emission from
the more strongly populated states. Again, the associated values of the Fermi-tail
responsivity $\mathcal{R}_{\scriptscriptstyle F}(\lambda,\mathsfit{T\/})$ are
displayed in Table~\ref{tab:PMT}.

\subsection{Fermi-Tail Responsivity and Quantum Efficiency} \label{sub:FTailRespon}

As discussed in Sec.~\ref{enhancingf}, Fermi-tail photoemission can be enhanced
relative to competing forms of subthreshold photoemission by using large-area,
low-power illumination and a large-area sample maintained at an elevated
temperature. In the absence of two-photon photoemission, the total photocurrent
displayed in Eq.~(\ref{totalcurrentfullCoherent}) reduces to the Fermi-tail
photocurrent, augmented by the dark/circuit current:
\begin{equation}\label{Fphcurrent}
\overline{\imath}_{\scriptscriptstyle TC}(\lambda,\mathsfit{T\/})  =
i_{\scriptscriptstyle D} +  \mathcal{F} \, \mathcal{R}_{\scriptscriptstyle F}
(\lambda,\mathsfit{T\/})\,(\mathcal{T\/} \overline{P_0}).
\end{equation}
The Fermi-tail photocurrent thus depends on the mean optical power present at the
sample, and on the fundamental Fourier component $\mathcal{F}$ associated with
the lock-in detection, but not on $\Gamma$, $g_2$, or $A$. Reordering
Eq.~(\ref{Fphcurrent}) leads to an expression for the CW-equivalent, Fermi-tail
responsivity:
\begin{equation} \label{eq:yield1p}
\mathcal{R}_{\scriptscriptstyle F}(\lambda,\mathsfit{T\/}) =
\frac{\overline{\imath}_{\scriptscriptstyle TC}(\lambda,\mathsfit{T\/}) -
i_{\scriptscriptstyle D}}{\mathcal{F}\, (\mathcal{T\/} \overline{P_0})}\,.
\end{equation}
Finally, in accordance with Eq.~(\ref{iFRESPeta}), the associated CW-equivalent,
Fermi-tail quantum efficiency is
\begin{equation}\label{RestoQE}
  \eta_{\scriptscriptstyle F}(\lambda,\mathsfit{T\/}) =
  \frac{h c}{e\lambda}\,\mathcal{R}_{\scriptscriptstyle F}(\lambda,\mathsfit{T\/}).
\end{equation}

\subsubsection{Three Alkali-Antimonide Photocathodes:
CsK$_2$Sb, Na$_2$KSb, and Cs$_3$Sb} \label{sub:FTailResponPMT}

Fermi-tail responsivities and quantum efficiencies have been determined for the
alkali-antimonide photocathodes in three commercially available Hamamatsu
photomultiplier tubes, at several temperatures and wavelengths:
\begin{enumerate}
  \item The CsK$_2$Sb (cesium-potassium antimonide) photocathode in the
      Hamamatsu \textbf{R464} PMT, with mean current gain $\overline{G} \approx
      10^8$ and representative phenomenological parameter values $\,E_{\!g} =
      1.0$~eV,
    $\,\chi = 1.1$~eV,
    and \,$\texttt{W} = 2.1$~eV~\cite{Ghosh80}.
  \item The Na$_2$KSb (sodium-potassium antimonide) photocathode in the
      Hamamatsu \textbf{R2557} PMT, with mean current gain $\overline{G} \approx
      5 \times 10^6$ and representative phenomenological parameter values
      $\,E_{\!g} = 1.0$~eV, $\,\chi =  1.0$~eV, and \,$\texttt{W} =
      2.0$~eV~\cite{Ghosh80}.
  \item The Cs$_3$Sb (cesium antimonide) photocathode in the Hamamatsu
      \textbf{1P28} PMT, with mean current gain $\overline{G} \approx 10^7$ and
      representative phenomenological parameter values $\,E_{\!g} = 1.6$~eV, $\,\chi
      =  0.4$~eV, and \,$\texttt{W} =  2.0$~eV~\cite{Hattori00}.
\end{enumerate}

The measurements reported by Booth~\emph{et
al.}~\cite{booth2006} were carried out using
the experimental arrangement depicted in Fig.~\ref{fig13}. For these measurements,
the fixed system parameters are taken to be:
\begin{itemize}
  \item Dark/circuit current $i_{\scriptscriptstyle D} \approx 10^{-18}$~A.
  \item Intrinsic optical-system transmittance $\mathcal{T\/}_{\!\!0} \approx 0.7$.
  \item First-harmonic component $\mathcal{F} = 2/\pi$.
\end{itemize}
The Fermi-tail responsivity $\mathcal{R}_{\scriptscriptstyle
F}(\lambda,\mathsfit{T\/})$ for a given data set is calculated by inserting these
system parameters, together with the mean optical power at the sample
$\overline{P} = \mathcal{T\/}\overline{P_0}$ and its associated photocurrent
$\overline{\imath}(\lambda,\mathsfit{T\/})$, into Eq.~(\ref{eq:yield1p}). The
resulting responsivities, and  associated quantum efficiencies calculated using
Eq.~(\ref{RestoQE}), are compiled in Table~\ref{tab:PMT}.
\begin{table}[htb!]
\centering
\begin{small}
\caption{Experimental values of the (CW-equivalent) Fermi-tail
responsivities $\mathcal{R}_{\scriptscriptstyle F}(\lambda,\mathsfit{T\/})$ (A/W),
and  quantum efficiencies $\eta_{\scriptscriptstyle F}(\lambda,\mathsfit{T\/}) =
(h c/e\lambda)\,\mathcal{R}_{\scriptscriptstyle F}$, for Hamamatsu PMT types
R464, R2557, and 1P28, along with a supplementary data set for a Soviet-era CsK$_2$Sb
PMT-130 device. The associated photocathode materials are identified in
parentheses. The wavelength of the incident light $\,\lambda\,$ and the photocathode
temperature $\,\mathsfit{T\/}\,$ at which the measurements were conducted
are indicated. The entries from Booth~\emph{et al.}~\cite{booth2006} were
calculated using Eq.~(\ref{eq:yield1p}), which corrects for additive dark/circuit
current, mechanical light chopping, and system optical loss.
 (Data adapted from Table~I of
Booth \emph{et al.}~\cite{booth2006}, except for the CsK$_2$Sb PMT-130
data$^{\text{a}}$ extracted from Chernov~\cite{chernov03} and the R2557 PMT
data$^{\text{b}}$ drawn from Kobayashi \emph{et al.}~\cite{kobayashi2006}.)}
\label{tab:PMT}
\renewcommand{\arraystretch}{1.1}
\begin{tabular}{lcccc}
\hline
\noalign{\vskip 0.7mm}
  \textbf{PMT}  & $\lambda$ (nm)&  $\mathsfit{T\/}$~($^\circ$C)&
  $\mathcal{R}_{\scriptscriptstyle F}(\lambda,\mathsfit{T\/})$ \,(A/W) &
  $\eta_{\scriptscriptstyle F}(\lambda,\mathsfit{T\/})$  \\[0.7mm]
  \hline
\noalign{\vskip 0.7mm}
  \textbf{R464}  & 800  & 27  & $\!\!1.50 \times 10^{-9}$ & $\!\!2.32 \times 10^{-9}$ \\
   (CsK$_2$Sb)               &        & 0  & $7.40 \times 10^{-10}$ & $\!\!1.15 \times 10^{-9}$ \\
        &                  & $\!\!\!\!-$20  & $3.10 \times 10^{-10}$ & $4.80 \times 10^{-10}$ \\
        &           830  & 27  & $\!\!1.30 \times 10^{-9}$ & $\!\!1.94 \times 10^{-9}$ \\
        &                  & 0  & $5.70 \times 10^{-10}$ & $8.51 \times 10^{-10}$ \\
        &           845  & 27  & $5.20 \times 10^{-10}$ & $7.63 \times 10^{-10}$ \\
        &                  & $\!\!\!\!-$20  & $2.00 \times 10^{-11}$ & $2.93 \times 10^{-11}$ \\
  \textbf{PMT-130}& 1080$^{\text{a}}$  & ---  & $1.04 \times 10^{-10}$ &
  $1.20 \times 10^{-10}$ \\[0.7mm]
  \hline
\noalign{\vskip 0.7mm}
  \textbf{R2557} &  800  & 27  & $\!\!5.30 \times 10^{-7}$ & $\!\!8.21 \times 10^{-7}$                                  \\
  (Na$_2$KSb)                & 845  & 27  & $\!\!6.80 \times 10^{-8}$ & $\!\!9.98 \times 10^{-8}$\\
  & 1064$^{\text{b}}$  & 27  & $5.65 \times 10^{-11}$ & $6.58 \times 10^{-11}$ \\[0.7mm]
  \hline
\noalign{\vskip 0.7mm}
  \textbf{1P28}  &  800  & 27  & $\!\!6.00 \times 10^{-8}$ &  $\!\!9.30 \times 10^{-8}$                              \\
   (Cs$_3$Sb)                & 845  & 27  & $\!\!6.20 \times 10^{-9}$ &
   $\!\!9.10 \times 10^{-9}$\\[0.7mm]
   \hline
\end{tabular}
\end{small}
\end{table}

\begin{framed}\noindent
For the range of parameters investigated, the entries in
Table~\ref{tab:PMT} reveal that $\mathcal{R}_{\scriptscriptstyle F}$ generally decreases as
the photocathode temperature is lowered and as the illumination wavelength is
increased, for the reasons suggested in the introduction to this section.
\end{framed}

As an example, $\mathcal{R}_{\scriptscriptstyle F}$ for the R464 decreased by a factor of
$\approx 5$ when the PMT temperature was lowered from $\mathsfit{T\/} =
27~^\circ$C to $-20~^\circ$C; additional cooling would have further reduced
$\mathcal{R}_{\scriptscriptstyle F}$. Note that
$\mathcal{R}_{\scriptscriptstyle F}(\lambda,\mathsfit{T\/})$ is
smaller for the R464
than for the R2557 and 1P28 PMTs, which is why CsK$_2$Sb
was chosen for the entangled-two-photon photoemission experiments
discussed in Secs.~\ref{entPMT} and \ref{entchannel}.

\section{OBSERVATION OF TWO-PHOTON PHOTOEMISSION
UNDER CLASSICAL\\ ILLUMINATION}\label{twophoton}

\textbf{Two-photon photoemission (TPP)} is a broadly-adopted tool for studying
electronic structure at surfaces, interfaces, and image potential states in both metals
and semiconductors; it is often used for the spectroscopic investigation of
unoccupied electronic states between the Fermi and vacuum levels. Time-resolved
TPP employing pump-probe schemes allows carrier lifetimes and relaxation pathways
to be characterized with femtosecond resolution. Two-photon photoemission has
been used to measure the relaxation of hot carriers in bulk materials, the lifetimes of
adsorbate-modified electronic states, and electron localization in thin molecular
films~\cite{Schuppler92,Fann1992a,Wang95,petek1997,Fauster02,Kentsch02,
Weinelt2002, Toeben03,Weinelt04,Lisowski04,ferrini2009,Lerch2020}. The
technique has been widely applied to semiconductor surfaces such as GaAs, Si, and
perovskites to resolve ultrafast carrier dynamics.

As explained in Sec.~\ref{twophotontheory}, for the real-conduction-band
intermediate-state configuration illustrated in Fig.~\ref{fig7}($c$)
two-photon photoemission can take place when
\begin{equation}\label{2Prange}
  E_{\!g} <  h\nu <  \texttt{W} < 2h\nu, \qquad\quad \mbox{(semiconductors)}
\end{equation}
where $h\nu$ is the energy of the individual photons (energy-degenerate photons
are assumed for simplicity). This formula can also be written in terms of
wavelength-based parameters as
\begin{equation}\label{2Pwave}
\lambda/2 < \lambda_{\texttt{W}} < \lambda < \lambda_g, \qquad\quad
\mbox{(semiconductors)}
\end{equation}
where \,$\lambda = hc/h\nu$\, is the wavelength of the individual photons,
\,$\lambda_{\texttt{W}} = hc/\texttt{W}$\, is defined as the ionization-energy
wavelength, and \,$\lambda_g = hc/E_{\!g}$\, is the bandgap wavelength. When
$h\nu$, $\texttt{W}$, and $E_{\!g}$ are expressed in electron volts (eV), and
$\lambda$, $\lambda_{\texttt{W}}$, and $\lambda_g$ are expressed in nm, the
following approximate, but convenient, relationships ensue (see Eqs.~(5.1-1) and
(5.1-2) of~\cite{teich25}):
\begin{equation}\label{2PwaveConvert}
\lambda \approx \frac{1240}{h\nu} \qquad\quad \lambda_{\texttt{W}}
\approx \frac{1240}{\texttt{W}} \qquad\quad \lambda_g \approx \frac{1240}{E_{\!g}}\,.
\end{equation}

For metals, we make use of the work function \,${\scriptstyle \mathcal{W}}$\, in
place of the ionization energy \,\texttt{W}, and the notion of bandgap energy is
inapplicable, so that Eqs.~(\ref{2Prange}) and (\ref{2Pwave}) become
\begin{equation}\label{2PrangeMetal}
  h\nu <  {\scriptstyle \mathcal{W}} < 2h\nu, \qquad\quad \mbox{(metals)}
\end{equation}
and
\begin{equation}\label{2PwaveMetal}
\lambda/2 < \lambda_{{\scriptstyle \mathcal{W}}} < \lambda, \qquad\quad \mbox{(metals)}
\end{equation}
respectively.

In Sec.~\ref{TPP-photocurrent}, we indicate several approaches that are useful for
enhancing the strength of the two-photon photocurrent. Section~\ref{sub:earlyNa}
is devoted to discussing how to incorporate vacuum-depositable materials into
photomultiplier tubes to harness the high-gain and comparatively low-noise current
amplification of these devices. Section~\ref{sub:Nametal} is dedicated to reviewing the technical
aspects of carrying out two-photon photoemission experiments with sodium metal
and reporting the results under classical light illumination.
Section~\ref{sub:TPP-CsK2Sb} considers two-photon photoemission from CsK$_2$Sb
illuminated with coherent light, providing a detailed analysis of its temperature and
wavelength dependencies.

\subsection{Facilitating the Observation of Two-Photon Photoemission}
\label{TPP-photocurrent}

As set forth in Eq.~(\ref{totalcurrentfullC}), the observed mean total photocurrent
$\,\overline{\imath}_{\scriptscriptstyle TC}\,$, including the additive dark/circuit
current $i_{\scriptscriptstyle D}$ and the Fermi-tail and two-photon subthreshold
contributions specified in Eqs.~(\ref{iFETAresp}) and (\ref{i2pcurrentETA}),
respectively, can be written as
\begin{eqnarray}
\overline{\imath}_{\scriptscriptstyle TC}(\lambda,\mathsfit{T\/}) & = &
i_{\scriptscriptstyle D} + \mathcal{F} \left[
\mathcal{R}_{\scriptscriptstyle F}(\lambda,\mathsfit{T\/})\,( \mathcal{T\/} \overline{P_0})
+ \mathcal{L}_{\scriptscriptstyle C}(\lambda,\mathsfit{T\/})\,( \mathcal{T\/}
\overline{P_0})^2\,g_2\,\Gamma/A\right]  \label{eq:current1corr} \\
 & = & i_{\scriptscriptstyle D} +  \mathcal{F}\left[\mathcal{R}_{\scriptscriptstyle F}
 (\lambda,\mathsfit{T\/})\,( \mathcal{T\/}  \overline{P_0})
+ \mathcal{L}_{\scriptscriptstyle C}(\lambda,\mathsfit{T\/}) \,( \mathcal{T\/}
\overline{I_0}) \,( \mathcal{T\/} \overline{P_0})\,g_2\,\Gamma \right]
\label{eq:current2corr} \\
 & = & i_{\scriptscriptstyle D} +  \mathcal{F}\left[\mathcal{R}_{\scriptscriptstyle F}
 (\lambda,\mathsfit{T\/})\,( \mathcal{T\/} \overline{P_0})
+ \mathcal{R}_{\scriptscriptstyle C}(\lambda,\mathsfit{T\/}) \,( \mathcal{T\/}
\overline{P_0})\,g_2\,\Gamma \right].
\label{eq:current3corr}
\end{eqnarray}
Here, $\mathcal{R}_{\scriptscriptstyle C}  =\mathcal{L}_{\scriptscriptstyle C}
(\mathcal{T\/}\overline{I_0})$ is evaluated at the mean intensity incident on the
sample, $\mathcal{T\/}\overline{I_0} =\mathcal{T\/}\overline{P_0}/A$,
and therefore remains intensity-dependent.
As summarized in Sec.~\ref{measidenaux}, $\mathcal{F}$ is the calibrated factor
associated with extracting a Fourier component of the modulated photocurrent
by lock-in detection;  $\overline{P_0}$ is the optical
power emitted by the source; $A$ is the illumination area at the sample;
$\mathcal{T\/}$ is the fraction of the source power that reaches the sample;
$\Gamma$ is the TPP pulsed-source enhancement factor; and $g_2$ is the
intensity-fluctuation enhancement factor. Accommodating these factors assures
that the extracted responsivities are CW-equivalent values.

Referring to Eqs.~(\ref{eq:current1corr})--(\ref{eq:current3corr}) suggests that the
two-photon photocurrent may be enhanced by implementing the  approaches
outlined below.

\subsubsection{Detection Paradigm (Maximize $\mathcal{F}$)} \label{sub:Fmodloss}

Using direct digital counting rather than chopper-based lock-in detection eliminates
the need for the light chopper, thereby avoiding its optical-throughput loss and
the associated Fourier-component normalization. In this case, $\mathcal{F}=1$.

\subsubsection{Beam Power and Focusing (Maximize $P$ and Minimize $A$)}
\label{sub:TPPfocus}

Focusing a high-power light beam on the sample offers the simplest way of
enhancing the TPP photocurrent relative to the Fermi-tail photocurrent. The
advantage arises because, as specified in Eq.~(\ref{eq:current1corr}), the
two-photon photocurrent is inversely proportional to the illumination area  $A$
while the Fermi-tail photocurrent is independent of $A$. As is clear from
Eq.~(\ref{eq:current1corr}), two-photon photoemission dominates Fermi-tail
photoemission when $A/\overline{P} \ll \mathcal{L}_{\scriptscriptstyle
C}g_2\,\Gamma/\mathcal{R}_{\scriptscriptstyle F}$.

\subsubsection{Optical Loss (Maximize $\mathcal{T\/}$)} \label{sub:TPPloss}

Utilizing an optical path that is as direct as possible, and that contains as few optical
components as possible, maximizes the  optical-system transmittance
$\mathcal{T\/}$, which enters the TPP photocurrent as $\mathcal{T\/}^2$.

\subsubsection{Pulsed Source (Maximize $\Gamma$)} \label{sub:TPPpulse}

Since the two-photon photocurrent scales quadratically with the optical power, as
indicated in Eq.~(\ref{eq:current1corr}), a sequence of brief but powerful optical
pulses gives rise to two-photon photoemission more effectively than does a CW
optical beam of the same mean power. Stated differently, two-photon
photoemission is governed by the mean-square rather than by the square-mean
optical power. Increasing $\Gamma$, which is defined in Eq.~(\ref{eq:intenfactor})
as the ratio of these two quantities, therefore enhances the two-photon
photocurrent. Two-photon photoemission can be observed using a CW source,
however~\cite{roth2002}.

\subsubsection{Intensity Fluctuations (Maximize $g_2$)} \label{sub:TPPintfluct}

In analogy with the results for pulsed sources discussed in Sec.~\ref{sub:TPPpulse},
the presence of intensity fluctuations provides an increased mean-square to
square-mean power ratio $g_2$, as provided in Eq.~(\ref{g2def}), thereby enhancing
the two-photon photocurrent.

\subsection{Two-Photon Photoemission from Vacuum-Depositable Materials}
\label{sub:earlyNa}

\begin{figure}[htb!]
\centering\includegraphics[width=3.25in]{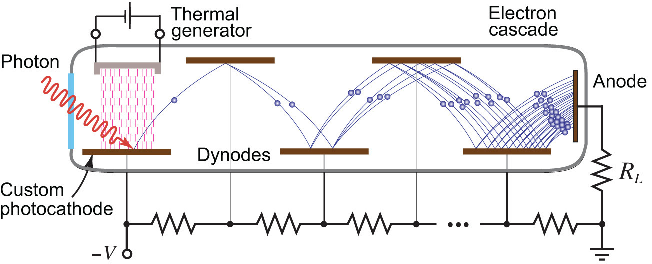}
\caption{Sketch of the interior of a custom PMT designed to allow two-photon
photoemission to be studied for suitable vacuum-depositable materials.
Passage of a current through the thermal generator initiates a chemical reaction
that  releases a vapor of the desired material (e.g., Na). The vapor condenses on
a cold metallic plate (e.g., Ni) that serves as the substrate for the custom
photocathode.} \label{fig16}
\end{figure}

The method initially used to prepare a sample of sodium metal suitable for
conducting two-photon photoemission
experiments~\cite{Teich64,teich66PhD,Teich68} provides a useful framework for
suitable vacuum-depositable photoemissive materials. As depicted in
Fig.~\ref{fig16}, the envelope of a specially designed PMT can be fitted with a
thermal generator that deposits a custom photocathode of selected material and
thickness. The merit of this approach is that it allows the high-gain, low-noise current
amplification offered by the internal electron-multiplication structure in a PMT to be
exploited~\cite{teich66PhD}.

\subsection{Two-Photon Photoemission from Sodium Metal} \label{sub:Nametal}

\begin{figure}[htb!]
\centering\includegraphics[width=3.25in]{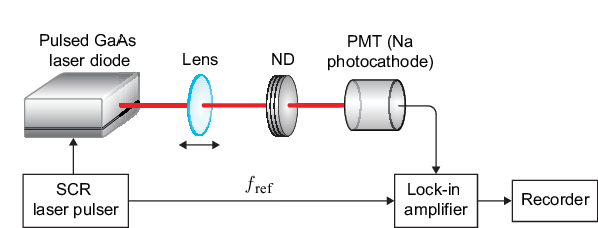}
\caption{Experimental arrangement for measurements of two-photon photoemission
from Na metal conducted in the 1960s~\cite{Teich64,teich66PhD}.
The light source, a 1963-vintage pulsed gallium arsenide (GaAs) laser diode driven by
a silicon-controlled-rectifier (SCR) laser pulser, emitted temporally and spatially
multimode light with a central wavelength of 845~nm. The light was focused by a
50-mm-diameter lens of focal length $\,\mathsfit{f}\, = 57$~mm onto a
vapor-deposited Na-metal photocathode in a specially constructed photomultiplier
tube operated at room temperature ($\mathsfit{T\/} = 27~^\circ$C).
For the recoated tube, the transmittance through the faceplate and residual sodium
coating was estimated to be approximately 0.4, including a coarse allowance for
faceplate reflection. Decrements of the light intensity incident
on the photocathode were implemented by inserting thin gelatin (Kodak Wratten)
neutral-density (ND) filters in the beam path (conventional glass filters were
unsuitable because refraction in the glass caused the imaged illumination area
and location to change randomly from trial-to-trial).
The PMT output was fed to a lock-in amplifier whose reference signal
($f_\mathrm{ref} = 2.2$~kHz) was provided by the SCR driver. The output of the
lock-in amplifier was fed to a recorder operated in voltage mode. Signal strengths
were enhanced by inserting low-noise preamplifiers just before the lock-in amplifier,
one in the signal arm and the other in the reference arm (not shown). (Adapted
with permission from Fig.~1 of~\cite{Teich64}, \textcopyright 1964 American
Physical Society, and Fig.~6 of~\cite{teich66PhD}.)
} \label{fig17}
\end{figure}

Early photoemission experiments often made use of metallic sodium because it was a
well-studied, simple alkali metal with a nearly spherical Fermi surface, thereby
facilitating the comparison of theory and experiment. As a result, sodium was chosen
as the first metal in which two-photon photoemission was investigated. A block
diagram of the experimental arrangement for these studies is displayed in
Fig.~\ref{fig17}. Details pertaining to the constituent optical components are
presented in Secs.~\ref{subsub:NaPMT}--\ref{subsubPhaseSens}, and analyses of
the observed two-photon photocurrents, responsivities, and quantum efficiencies
follow in Secs.~\ref{subsubTPPMeas} and \ref{sub:TPRespon}.
Section~\ref{subsub:VolSurf} contains a discussion of volume and surface
two-photon photoemission in the context of thick and thin Na films. The experiments
reported in this section were conducted at Cornell University in the time frame
1963--1965~\cite{Teich64,teich65,teich66PhD,teich1966PRL,Teich1967,Teich68}.

\subsubsection{Custom PMT with Na-Metal Photocathode} \label{subsub:NaPMT}

A Na metal film, vapor-deposited on a Ni substrate, served as the photocathode in a
specially constructed photomultiplier tube that functionally resembles the sketch
portrayed in Fig.~\ref{fig16}. This Na-photocathode PMT contained an 11-stage
electron multiplier that was operated at 2000~V and exhibited a gain of 50\,000. The
light was normally incident on the photocathode surface.

Fowler-plot analyses~\cite{Fowler1931} of the original, unrecoated Na film
yielded work-function estimates of ${\scriptstyle \mathcal{W}}=1.96$~eV
from measurements at $h\nu=1.48$~eV, and ${\scriptstyle \mathcal{W}}=2.3$~eV
from measurements involving a two-photon energy $2h\nu=2.96$~eV.
 Inserting the lower of these values into Eq.~(\ref{2PrangeMetal}) provides
\begin{equation}\label{2PenergyNa}
 h\nu < 1.96~\mbox{eV} < 2h\nu, \qquad\quad \mbox{(Na metal)}
\end{equation}
and expressing this in the wavelength domain via Eqs.~(\ref{2PwaveConvert}) and
(\ref{2PwaveMetal}) yields
\begin{equation}\label{2PwavelengthreverseNa}
\lambda/2 < 633~\mbox{nm} < \lambda. \qquad\quad\,\,\,\, \mbox{(Na metal)}
\end{equation}
The Na metal TPP experiments reported in this section rely on a GaAs laser diode
operating at $\lambda = 845$~nm (Sec.~\ref{subsub:GaAs}), so that
Eq.~(\ref{2PwavelengthreverseNa}) is obeyed, i.e., \,$423 < 633 < 845$~nm.
The two-photon energy condition is also satisfied for the larger work-function
estimate of 2.3~eV.

\subsubsection{Controlling and Estimating Sodium Film Thickness}
\label{subsub:NaThickness}

Thick and thin Na films were used as samples. The film
thickness could be increased at will by making use of a sodium-vapor generator
internal to the PMT that consisted of a 15-mm-long/2-mm-diameter Ni collar
containing reagent-grade Na$_2$CrO$_4$ and Si. As illustrated in Fig.~\ref{fig16},
external leads fed through the glass envelope of the PMT allowed a current to be
passed through the generator. Applied for 90~s, this current heated the reagents
within the Ni collar sufficiently to cause the Si to reduce the Na$_2$CrO$_4$ and to
liberate Na vapor, which condensed on the relatively cold Ni first dynode that served
as the photocathode substrate. Some condensation also appeared on the inside
faceplate of the PMT, most of which was subsequently evaporated by heating it in
the yellow flame of a Bunsen burner. About 40\% of the light incident on the
faceplate was estimated to reach the photocathode after recoating. Hence, a peak
power of approximately 400~mW at the faceplate corresponded to approximately
160~mW incident on the thick Na film. This external transmission loss is distinct
from the factor $\beta$ that describes optical coupling into the Na material in
the photoemission model.

The thickness of the thick Na metal sample, which was deposited on the
photocathode some two hours prior to collecting data, was estimated
to be $l \approx 50$~nm, slightly greater than the adopted electron escape depth
$d \approx 40$~nm~\cite{teich66PhD}. These are rough estimates rather than
independent thickness measurements. The thickness of the thin Na metal
sample, which was deposited on the photocathode about six months prior to use, is
unknown but apparently small. Indeed, most of the residual film on the faceplate of
the PMT when it initially arrived from the manufacturer appeared to have
evaporated by the time the thin-sample experiments were conducted. For that
sample, the power incident on the photocathode was taken to be approximately
equal to the faceplate power, neglecting glass reflection.
Additional details are disclosed in~\cite{teich66PhD}.

\subsubsection{Pulsed GaAs Laser Diode} \label{subsub:GaAs}

The experiments were conducted using a GaAs laser diode, which was developed
contemporaneously at GE, IBM, and MIT Lincoln Laboratory in
1962~\cite{quist62,LLhistory2011,teich25}~\chapcite{18}{saleh2019}. Several of
these new devices were graciously donated to Teich in 1963 by their coinventors at
MIT Lincoln Laboratory, which enabled his doctoral project to
proceed~\cite{teich2025RG}. The GaAs photon energy, $h\nu = 1.48$~eV, was ideal for
investigating two-photon photoemission from Na metal, whose work function is
${\scriptstyle \mathcal{W}} = 1.96$~eV. The experiments reviewed here were
initiated in 1963, just months after this new device was developed, so only
pulsed, multimode operation at 77~K was possible. The emission comprised multiple
Fabry--Perot modes that, on average, were $\approx75\%$ polarized with the
electric-field vector principally in the plane of the junction. The temporal mode
structure varied from pulse to pulse while the spatial modal
structure varied from experiment to experiment.
Hence, the light emitted by this device was classical, rather than the stable
coherent illumination considered elsewhere in this review.

Mounted snugly in a copper heat sink, and housed in a styrofoam dewar fitted with
an antireflection-coated optical window, the laser diode delivered a peak optical
power of approximately 400~mW. It was driven by a custom silicon-controlled-rectifier
(SCR) long-pulse driver that provided a train of current pulses (peak current 21~A)
of roughly semicircular shape with a full width of 35~$\mu$s and a repetition rate of
2.2~kHz~\cite{teich65}. The SCR current driver was controlled by a unit pulser
(General Radio 1217A).

By way of comparison, modern discrete laser diodes ---
including those fabricated from the ternary and quaternary
semiconductors AlGaAs, InGaAsP, AlInGaP, and AlInGaN ---
can operate CW at room temperature, with single-temporal-
and single-spatial-mode operation available in suitable
designs, while offering compositional wavelength tuning
and output powers ranging from mW to W~\chapcite{18}{saleh2019}. A
device of this type was used as a pump in the experiments conducted by Kobayashi
\emph{et al.}~\cite{kobayashi2007} to observe entangled-two-photon
photoemission, as discussed in Sec.~\ref{entchannelexp}.

\subsubsection{Optical Pulse Train and its Square} \label{subsub:ltpt}

The pulse train $P(t)$ for the optical-power envelope incident on the photocathode
can be approximated as a periodic sequence of semicircular functions,
\begin{equation}\label{persemirect}
  P(t)=\sum_{n=-\infty}^{\infty} p(t-n\tau_1),
  \qquad
  p(t)=
  \begin{cases}
    \widehat{P}\sqrt{1-(2t/\tau_0)^2}, & |t|<\tau_0/2,\\
    0, & \text{otherwise}.
  \end{cases}
\end{equation}
Here, $\widehat{P}$ is the peak power at the photocathode, $\tau_0$ is the full
pulse width, and $\tau_1$ is the pulse-train period. This envelope follows the
current pulse train provided by the SCR driver reasonably well. The rapid
multimode fluctuations are treated separately through the factor $g_2$.

Since the two-photon photocurrent $i_{\scriptscriptstyle C}(t)$ is proportional to
the square of the optical power, in accordance with Eq.~(\ref{i2phiA}), the square
of the optical pulse train is of central importance:
\begin{equation}\label{2Pptsq}
  P^2(t)=\sum_{n=-\infty}^{\infty}p^2(t-n\tau_1),
  \qquad
  p^2(t)=
  \begin{cases}
    \widehat{P}^{\,2}\left[1-(2t/\tau_0)^2\right],
       & |t|<\tau_0/2,\\
    0, & \text{otherwise}.
  \end{cases}
\end{equation}
The pulses do not overlap because $\tau_0<\tau_1$. The squared pulse train is
therefore a sequence of parabolic caps of base width $\tau_0$ and height
$\widehat{P}^{\,2}$. The semicircular and parabolic functions
that appear in Eqs.~(\ref{persemirect}) and (\ref{2Pptsq}) reasonably approximate
the corresponding experimental oscilloscope traces presented in Figs.~8b and 8a
of~\cite{teich66PhD}, respectively. Departures of the observed photocurrent
waveform from the square of the measured optical envelope were discussed
in \cite{teich66PhD} in connection with the changing spatial and
temporal modal structure.

\subsubsection{Lock-In Detection}\label{subsubPhaseSens}

The PMT anode current, proportional to the two-photon photocathode current
$i_{\scriptscriptstyle C}(t)$ and hence to the squared optical-power envelope
$P^2(t)$ under the adopted model, was fed to a lock-in amplifier (Princeton
Applied Research JB5, Princeton, NJ).  Lock-in detection, which makes use
of phase-sensitive demodulation (in this case at the fundamental frequency
$f_{\mathrm{ref}}=2.2~\mathrm{kHz}$) followed by narrowband low-pass filtering,
was employed to enhance sensitivity to the small two-photon photocurrent and to
reject out-of-band noise~\cite{dicke46,stutt1949}. The SCR pulser, in addition to
driving the laser diode, provided the $2.2~\mathrm{kHz}$ reference signal to the
lock-in amplifier via the ``synchronization pulse'' output of the unit pulser driving the
SCR circuitry (General Radio 1217A).

The measured lock-in gain was calibrated from sinusoidal RMS input to dc output.
In converting the recorded voltage to the photocathode current, the $\sqrt{2}$
factor required to express the recovered
fundamental component as a peak amplitude was explicitly included
[Eq.~(2.6) of~\cite{teich66PhD}]. The photocurrents reproduced here therefore
already represent fundamental peak amplitudes; no additional RMS-to-peak
conversion is required.

For the two-photon contribution, the appropriate waveform normalization is the
ratio $\mathcal{F}_{\scriptscriptstyle C}$ of the fundamental peak amplitude to the
cycle-averaged photocurrent. Since
$i_{\scriptscriptstyle C}(t)\propto P^2(t)$, this ratio can be calculated from the
squared optical envelope,
\begin{equation} \label{defnFC}
\smash{\mathcal{F}_{\scriptscriptstyle C}  \equiv  {a_1}/{\overline{P^2}}} \approx 1.988,
\end{equation}
where $a_1$ is the fundamental cosine coefficient of $P^2(t)$, as calculated in
Sec.~\ref{subsubTPPMeas}.  The final stage in the experimental
arrangement portrayed in Fig.~\ref{fig17}
is the recorder (Bausch \& Lomb, VOM6), which was driven by the differential
output of the lock-in amplifier and operated in voltage mode.

\subsubsection{Two-Photon Photocurrent from Samples of Different Thicknesses}
\label{subsubTPPMeas}

\begin{figure}[htb!]
\centering\includegraphics[width=4.50in]{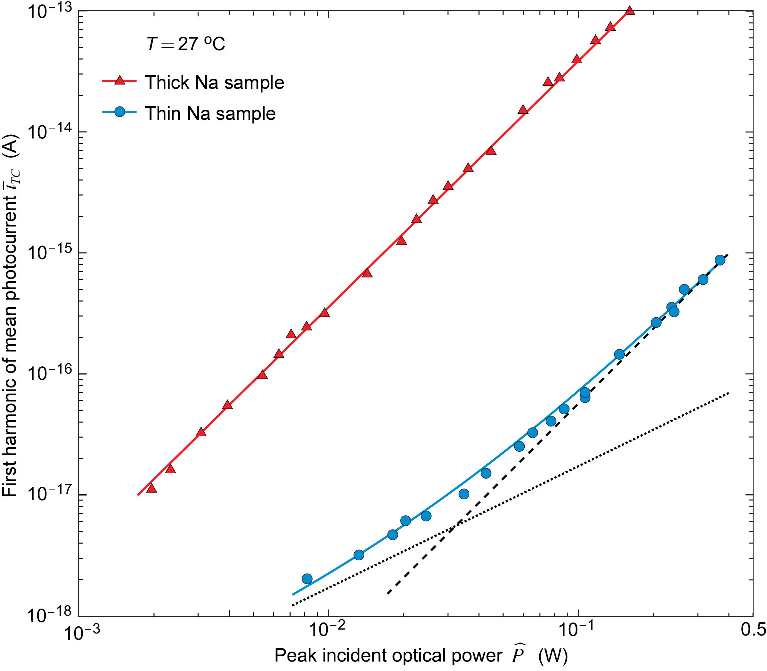}
\caption{Two-photon photocurrent from Na metal in a specially constructed
photomultiplier tube. Data are presented for thick (red triangles) and thin (blue circles)
Na metal films. This doubly logarithmic plot displays the peak amplitude of the
fundamental Fourier component of the mean photoelectric current at the
photocathode vs. the peak optical power incident on the photocathode.
For the thick-film data, the faceplate power has been multiplied by the estimated
faceplate transmittance of 0.4; for the thin-film data, faceplate and photocathode
powers were taken to be approximately equal, neglecting glass reflection.
The dashed and dotted lines, with slopes of two and one, represent
quadratic and linear scaling, respectively.
The photocurrent from the thick film (red) exhibits quadratic scaling over some four
orders of magnitude and is nearly three orders of magnitude larger than that from
the thin film (blue) at comparable incident powers.
(Data adapted from Figs.~7 and 9 of~\cite{teich66PhD}.)
} \label{fig18}
\end{figure}

The observed room-temperature, two-photon photocurrents from thick and thin Na
metal samples are presented in Fig.~\ref{fig18}. The thick-film thickness was
estimated to exceed the electron escape depth $d$, whereas the thin-film thickness
was unknown but apparently small. The ordinate displays the peak amplitude of
the fundamental Fourier component of the mean photoelectric current at the
photocathode, denoted here by
$\overline{\imath}_{\scriptscriptstyle TC}$, and the abscissa
represents the peak optical power $\widehat{P}$ incident on the sample. The sodium
photocathode was evaporated onto the first dynode of a specially constructed
photomultiplier tube, as described in Secs.~\ref{subsub:NaPMT} and
\ref{subsub:NaThickness}. The illumination, generated by the pulsed GaAs laser
diode characterized in Secs.~\ref{subsub:GaAs} and \ref{subsub:ltpt}, was normally
incident on the photocathode. The fundamental photocurrent component was
extracted using the lock-in detection scheme described in
Sec.~\ref{subsubPhaseSens}. The magnitude of the photocurrent was found to
depend only weakly on the polarization of the incident light.

The dashed and dotted lines in Fig.~\ref{fig18}, with slopes of two and one, represent
quadratic and linear scaling, respectively. The thick-film (red) data points indicate a
quadratic photocurrent for all measured values of the peak incident optical power
$\widehat{P}$, a hallmark of two-photon photoemission. The thin-film (blue) data
points indicate a substantially smaller photocurrent, with quadratic scaling
only at the higher reaches of $\widehat{P}$.

The parameters of the incident light that give rise to the two-photon photocurrent
for the thick film illustrated in Fig.~\ref{fig18} are presented below. The numerical
values provided in items \#4--7 and \#10 are calculated explicitly for the
largest-current/highest-peak-power data point --- the uppermost/rightmost red
triangle. The waveform calculations refer to the optical envelope; the
multimode-fluctuation enhancement is included separately through $g_2$.
\begin{enumerate}
  \item Laser-pulse full width:\, $\tau_0 = 35~\mu$s.

  \item Pulse-train period:\, $\tau_1 = 1/f_{\mathrm{ref}} =
  1/(2.2 \times 10^3~\text{Hz}) \approx 0.455$~ms.

  \item Duty cycle (ratio of full pulse width to pulse-train
  period):\newline
  $\Delta \equiv \tau_0/\tau_1 = \tau_0 f_{\mathrm{ref}} =
  (35 \times 10^{-6})\cdot(2.2 \times 10^3) \approx 0.077$ \,(7.7\%).

  \item Highest peak laser power at sample:\,
  $\widehat{P}=160~\mathrm{mW}$.

  \item Highest mean laser power at sample:\,
  $\overline{P} =
  (\pi/4)\,\widehat{P}\,(\tau_0/\tau_1)
  =(\pi/4)\,\widehat{P}\,\Delta \approx 9.68$~mW.

  \item Highest square-mean laser power at sample:\,
  ${\overline{P}}^2 =
  [(\pi/4)\,\widehat{P}\,\Delta]^2
  \approx 9.36 \times 10^{-5}$~W$^2$.

  \item Highest mean-square laser power at sample:\,
  $\overline{P^2} =
  (2/3)\,\widehat{P}^{\,2}\,\Delta
  \approx 1.31 \times 10^{-3}$~W$^2$.

  \item Mean-square to square-mean power ratio associated with laser pulses:\,
  $\Gamma = \overline{P^2}/\overline{P}^{\,2}
  =32/(3\pi^2\Delta)\approx14.0$.
  This result, $\Gamma\approx1.08/\Delta$, is close to the value
  $\Gamma=1/\Delta$ obtained for a rectangular pulse train, as specified in
  Eq.~(\ref{eq:equivfactor5}).

  \item Mean-square to square-mean power ratio associated with laser multimode
  fluctuations:\, $g_2\approx2$, as provided in Eq.~(\ref{g2=2}).
  Its use as a separate multiplier assumes that this enhancement can be
  approximated as constant over the optical envelope.

  \item Fundamental cosine coefficient of $P^2(t)$:
  $a_1 =
  (2/\tau_1)\!\int_{-\tau_0/2}^{\tau_0/2}
  \!\widehat{P}^{\,2}(1-4t^2/\tau_0^2)
  \cos(2\pi t/\tau_1)\,dt$.\newline
  Writing $x=\pi\Delta$, this integral yields
 $ a_1 =
  4\widehat{P}^{\,2}\Delta
  (\sin x-x\cos x)/x^3
  \approx2.61\times10^{-3}~\mathrm{W^2}$.
  The coefficient normalized to the squared peak power is therefore
  $a_1/\widehat{P}^{\,2}\approx0.1021$.

  \item Fundamental peak-amplitude to cycle-average normalization for the
  two-photon photocurrent:
  $\mathcal{F}_{\scriptscriptstyle C}
  \equiv {a_1}/{\overline{P^2}}
  = 6\,(\sin x-x\cos x)/{x^3}   \approx1.988$. This is the amplitude factor required when the photocurrent is expressed in
  terms of $\overline{P}^{\,2}$ and $\Gamma$.
\end{enumerate}

\subsubsection{Two-Photon Responsivity and Quantum Efficiency} \label{sub:TPRespon}

The CW-equivalent two-photon responsivity is obtained by relating the measured
fundamental peak amplitude to the mean incident optical power. For a general
mixture of linear and quadratic photoemission, the two contributions require
their respective waveform-amplitude factors.
In addition to $\mathcal{F}_{\scriptscriptstyle C}$ defined above, we introduce
\begin{equation}
\mathcal{F}_{\scriptscriptstyle F}
\equiv  {b_1}/{\overline{P}},
\qquad
b_1=\frac{2}{\tau_1}
\int_{-\tau_0/2}^{\tau_0/2}
P(t)\cos(2\pi t/\tau_1)\,dt,
\end{equation}
where $b_1$ is the fundamental cosine coefficient of the optical-power envelope.
The linear photocurrent is proportional to $P(t)$, whereas the two-photon
photocurrent is proportional to $P^2(t)$.

With $\overline{P}$ and $\overline{I}=\overline{P}/A$ evaluated at the sample,
the measured fundamental peak amplitude can consequently be written as
\begin{equation}
\overline{\imath}_{\scriptscriptstyle TC}
=
i_{\scriptscriptstyle D}
+\mathcal{F}_{\scriptscriptstyle F}
 \mathcal{R}_{\scriptscriptstyle F}\overline{P}
+\mathcal{F}_{\scriptscriptstyle C}
 \mathcal{L}_{\scriptscriptstyle C}
 \frac{\overline{P}^{\,2}}{A}\,g_2\Gamma.
\end{equation}
Here, $i_{\scriptscriptstyle D}$ denotes an effective measurement baseline,
rather than the unmodulated dc dark current, which is rejected by ideal lock-in
detection. The resulting responsivity is
\begin{equation}\label{eq:yield}
\begin{split}
\mathcal{R}_{\scriptscriptstyle C}(\lambda,\mathsfit{T\/})
&=
\mathcal{L}_{\scriptscriptstyle C}(\lambda,\mathsfit{T\/})
\,\overline{I}\\
&=
\frac{\overline{\imath}_{\scriptscriptstyle TC}
      (\lambda,\mathsfit{T\/})-i_{\scriptscriptstyle D}}
     {\overline{P}\,
      \mathcal{F}_{\scriptscriptstyle C}\,g_2\Gamma}
-\frac{\mathcal{F}_{\scriptscriptstyle F}}
       {\mathcal{F}_{\scriptscriptstyle C}}
 \frac{\mathcal{R}_{\scriptscriptstyle F}
       (\lambda,\mathsfit{T\/})}{g_2\Gamma}.
\end{split}
\end{equation}

The thick Na metal sample, represented by the red triangular data points in
Fig.~\ref{fig18}, exhibits quadratic scaling over some four orders of magnitude
of photocurrent. Linear Fermi-tail photoemission and an additive measurement
baseline are therefore neglected in extracting its two-photon coefficient.
Setting $\mathcal{R}_{\scriptscriptstyle F}=i_{\scriptscriptstyle D}=0$ in
Eq.~(\ref{eq:yield}) yields
\begin{equation}\label{eq:yieldthick}
\mathcal{R}_{\scriptscriptstyle C}(\lambda,\mathsfit{T\/})
=
\mathcal{L}_{\scriptscriptstyle C}(\lambda,\mathsfit{T\/})
\,\overline{I}
\approx
\frac{\overline{\imath}_{\scriptscriptstyle TC}
      (\lambda,\mathsfit{T\/})}
     {\overline{P}\,
      \mathcal{F}_{\scriptscriptstyle C}\,g_2\Gamma},
\end{equation}
and hence
\begin{equation}\label{eq:yieldthickL}
\mathcal{L}_{\scriptscriptstyle C}(\lambda,\mathsfit{T\/})
\approx
\frac{\overline{\imath}_{\scriptscriptstyle TC}
      (\lambda,\mathsfit{T\/})}
     {\overline{P}\,\overline{I}\,
      \mathcal{F}_{\scriptscriptstyle C}\,g_2\Gamma}
=
\frac{A\,\overline{\imath}_{\scriptscriptstyle TC}
      (\lambda,\mathsfit{T\/})}
     {g_2a_1}.
\end{equation}
This result generalizes Eq.~(\ref{i2pcurrentETA}) by accommodating the
fundamental-amplitude normalization and the enhancement factors $g_2$ and
$\Gamma$, discussed in Secs.~\ref{optpowvar} and
\ref{2Qphotstatspulse}, respectively. The coefficient extracted in this way
refers to CW coherent illumination, with the pulse-envelope and multimode
enhancements removed.

We now proceed to use Eq.~(\ref{eq:yieldthickL}) to obtain a numerical estimate of
$\mathcal{L}_{\scriptscriptstyle C}$ for the thick Na metal sample, using the
following parameters:
\begin{enumerate}
  \item It is convenient to anchor Eq.~(\ref{eq:yieldthickL}) to the
  largest-current/highest-peak-power data point in Fig.~\ref{fig18}: the
  uppermost/rightmost red triangle. This datum shows that the measured
  fundamental peak-amplitude photocurrent is
  $\overline{\imath}_{\scriptscriptstyle TC}
  =1.0\times10^{-13}$~A at a peak incident optical power
  $\widehat{P}=160$~mW.

  \item Referring to item \#5 in Sec.~\ref{subsubTPPMeas}, the highest mean laser
  power $\overline{P}$ at the sample is related to the highest peak laser power
  $\widehat{P}$ via
  $\overline{P}=(\pi/4)\Delta\widehat{P}\approx9.68$~mW.

  \item Given $\overline{P}\approx9.68$~mW and an illumination area
  $A=2.0\times10^{-9}$~m$^2$, the highest mean laser intensity at the sample is
  $\overline{I}=(9.68\times10^{-3})/(2.0\times10^{-9})
  =4.84\times10^6$~W/m$^2$.

  \item Referring to item \#11 in Sec.~\ref{subsubTPPMeas}, the fundamental
  peak-amplitude to cycle-average normalization is
  $\mathcal{F}_{\scriptscriptstyle C}\approx1.988$.

  \item Referring to item \#9 in Sec.~\ref{subsubTPPMeas}, the multimode
  enhancement factor is $g_2\approx2$.

  \item Referring to item \#8 in Sec.~\ref{subsubTPPMeas}, the pulse-envelope
  enhancement factor is $\Gamma\approx14.0$.
\end{enumerate}

Inserting these quantities into Eq.~(\ref{eq:yieldthickL}), using the
unrounded waveform factors, yields
$\mathcal{L}_{\scriptscriptstyle C}
\approx 3.83\times10^{-20}~\mathrm{A\,m^2/W^2}$.
The resulting two-photon responsivity
$\mathcal{R}_{\scriptscriptstyle C}
=\mathcal{L}_{\scriptscriptstyle C} I$
and quantum efficiency
$\eta_{\scriptscriptstyle C}
=(hc/e\lambda)\mathcal{R}_{\scriptscriptstyle C}$,
calculated via Eq.~(\ref{iCRESPeta}), are displayed in the top row of
Table~\ref{respqeadNa}. Here and in the table, $I$ denotes the intensity of
the corresponding CW coherent illumination at the sample.

\begin{table}[htb!]
\caption{Experimental and theoretical CW-equivalent two-photon responsivities
$\mathcal{R}_{\scriptscriptstyle C}$ and quantum efficiencies
$\eta_{\scriptscriptstyle C}$ for subthreshold photoemission from Na metal
at $\mathsfit{T\/}=27~^\circ$C.
The relationship between $\mathcal{R}_{\scriptscriptstyle C}$
and $\eta_{\scriptscriptstyle C}$ is provided in Eq.~(\ref{iCRESPeta}).
Experimental values for thick and thin Na metal films illuminated with classical
light at $\lambda=845$~nm have been extracted from Fig.~\ref{fig18}, using
the fundamental peak-amplitude calibration and removing the pulse-envelope
and multimode enhancement factors. The entries therefore refer to CW coherent
illumination. The volume-theoretical coefficient lies within a factor of about two
of the thick-film experimental coefficient, whereas the surface-theoretical
coefficient lies within a factor of about four of the thin-film experimental
coefficient. The CW intensity $I$ at the sample is expressed in W/m$^2$.}
\label{respqeadNa}
\centering
\small
\renewcommand{\arraystretch}{1.15}
\begin{tabular}{@{}lcc@{}}
\cline{2-3}
\noalign{\vskip 1.3mm}
 & \shortstack{Responsivity\\$\mathcal{R}_{\scriptscriptstyle C}$ (A/W)}
 & \shortstack{Quantum efficiency\\$\eta_{\scriptscriptstyle C}$ (electrons/photon)}\\[0.7mm]
\hline
\noalign{\vskip 0.7mm}
\begin{tabular}[c]{@{}l@{}}{\sc \textbf{thick sodium sample}}\\[-0mm]\end{tabular}\\
Experiment & $3.83\times10^{-20}\,I$ & $5.62\times10^{-20}\,I$\\[1.8mm]
\begin{tabular}[c]{@{}l@{}} \\[-1mm]\end{tabular}\\[-6.3mm]
Volume theory & $1.80\times10^{-20}\,I$ & $2.65\times10^{-20}\,I$\\[0.7mm]
\hline
\noalign{\vskip 0.7mm}
\begin{tabular}[c]{@{}l@{}}{\sc \textbf{thin sodium sample}}\\[-0mm]\end{tabular}\\
Experiment & $6.44\times10^{-23}\,I$ & $9.45\times10^{-23}\,I$\\[1.8mm]
\begin{tabular}[c]{@{}l@{}} \\[-1mm]\end{tabular}\\[-6.3mm]
Surface theory & $2.40\times10^{-22}\,I$ & $3.53\times10^{-22}\,I$\\
\noalign{\vskip 0.7mm}
\hline
\end{tabular}
\end{table}

An analogous calculation has been carried out for the thin Na metal sample, in which
Fermi-tail photoemission and the measurement baseline are ignored for simplicity
at the selected high-power point. In this case, it is convenient to anchor
Eq.~(\ref{eq:yieldthickL}) to the uppermost/rightmost blue circle in
Fig.~\ref{fig18}, which provides a measured fundamental peak-amplitude
photocurrent of
$\overline{\imath}_{\scriptscriptstyle TC}=9.0\times10^{-16}$~A
at a peak incident optical power $\widehat{P}=370$~mW.
In accordance with item \#5 in Sec.~\ref{subsubTPPMeas}, the maximum mean
laser power is
$\overline{P}_{\max} =(\pi/4)\Delta\widehat{P}\approx22.4$~mW.
Given $\overline{P}$ and the illumination area
$A=2.0\times10^{-9}$~m$^2$, the maximum mean laser intensity at the sample is
$\overline{I}_{\max}=(22.4\times10^{-3})/(2.0\times10^{-9})
=1.12\times10^7$~W/m$^2$.
The parameters
$\mathcal{F}_{\scriptscriptstyle C}\approx1.988$,
$g_2\approx2$, and $\Gamma\approx14.0$ are the same for both samples.
The fundamental cosine coefficient for the thin-film optical envelope is
$a_1\approx0.1021(0.370)^2\approx1.40\times10^{-2}$~W$^2$.
Inserting these quantities into Eq.~(\ref{eq:yieldthickL}) gives
$\mathcal{L}_{\scriptscriptstyle C}
\approx6.44\times10^{-23}~\mathrm{A\,m^2/W^2}$.
The associated responsivity and quantum efficiency are displayed in the third
row of Table~\ref{respqeadNa}. At a common CW intensity, the responsivity
(and quantum efficiency) for the thick Na metal sample is approximately
600~times greater than that for the thin sample.

The responsivities cited in the original reports were $\mathcal{R}_{\scriptscriptstyle C}\approx
8.0\times10^{-20}\,I~\text{A/W}$ for the thick film
and approximately $2.8\times10^{-22}\,I~\text{A/W}$ for the thin
film~\cite{Teich64,teich66PhD,Teich68}.
The original experimental analysis explicitly corrected for the
fundamental Fourier component of the squared optical-power
waveform using an estimated coefficient of 0.1 (Sec.~IV.B-1 of~\cite{teich66PhD}).
The present semicircular-envelope calculation gives the  closely related coefficient
$a_1/\widehat{P}^{\,2}\approx0.1021$.
The original theoretical calculation also explicitly included
the multimode enhancement factor $g_2\approx2$  (Eq.~(3.37) of~\cite{teich66PhD}).
The historical experimental coefficients retained this
enhancement, whereas the present values remove it to
represent CW coherent illumination.
Dividing the historical thick-film experimental coefficient
by two gives approximately  $4.0\times10^{-20}~\mathrm{A\,m^2/W^2}$,
close to the present value of  $3.83\times10^{-20}~\mathrm{A\,m^2/W^2}$.
The corresponding theoretical coefficient likewise changes
from $3.6\times10^{-20}$ to   $1.8\times10^{-20}~\mathrm{A\,m^2/W^2}$.
The values extracted here remain approximate because of
uncertainties in the optical waveform, illumination area,
faceplate transmission, and experimental calibration;
for the thin film, neglect of the residual linear contribution constitutes an
additional approximation.

\subsubsection{Volume vs. Surface Effects} \label{subsub:VolSurf}

As summarized in Sec.~\ref{2pvpt}, the generally accepted theoretical approaches
for calculating the two-photon surface and volume photocurrents were initiated by
Smith in 1962~\cite{Smith62} and by Bloch in 1964~\cite{Bloch64}, respectively.

The suitability of the two-photon volume model for Na metal can be appraised by
considering the two-photon photoemission experiments using sodium films of
different thicknesses, as described in Secs.~\ref{subsubTPPMeas} and
\ref{sub:TPRespon}. The photocurrent generated in the thick Na metal film, shown as
the red line transecting the triangular data points in Fig.~\ref{fig18}, is quadratic
(slope~2) for all measured values of the incident optical power, the hallmark of
two-photon photoemission specified in Eq.~(\ref{i2phiA}). When the illumination
is a CW coherent beam of light in a single temporal and spatial mode, the
theoretical two-photon volume photocurrent and responsivity are expressed in
Eqs.~(\ref{i2pcurrent})$/$(\ref{matrixelement}) and
(\ref{i2pcurrentETA})$/$(\ref{eqRLI}), respectively.
The numerical estimate in Eq.~(3.37) of~\cite{teich66PhD},
$3.6\times10^{-20}P^2/A$, includes the multimode enhancement
$g_2\approx2$. Removing this enhancement gives the coherent-light coefficient
$\mathcal{L}_{\scriptscriptstyle C}\approx
1.8\times10^{-20}~\mathrm{A\,m^2/W^2}$, presented in row~2 of
Table~\ref{respqeadNa} and labeled ``Volume theory''.
This value is smaller than the revised experimental coefficient for the thick
Na metal film by a factor of approximately 2.1, indicating encouraging
agreement in view of the uncertainties associated with the model and
experimental parameters.

The thin film is estimated to have a significantly smaller thickness (and therefore
volume). The photocurrent generated in the thin sample, plotted as the curve
passing through the blue circles in Fig.~\ref{fig18}, is substantially smaller than that
of the thick sample (red triangles). It exhibits quadratic scaling only at higher levels
of the optical power and scales linearly at lower levels~\cite{Teich64}. The linear
behavior is found to be independent of the illumination area and is plausibly
attributable to Fermi-tail photoemission (Sec.~\ref{subthreshtheory}). The origin of
the quadratic component of the photocurrent is more enigmatic. It could arise from
a volume photoeffect, as for the thick sample, but exhibit lower responsivity because
of its reduced volume.

A surface contribution provides another possible interpretation of the weak
quadratic photocurrent. As reported in Sec.~\ref{2pvpt}, the theoretical expression
for the two-photon surface responsivity at $\mathsfit{T\/}=0$~K, as derived by
Smith~\cite{Smith62} and corrected by Bowers~\cite{bowers64} and
Marinchuk~\cite{marinchuk66,marinchuk71}, was conveniently included as Eq.~(2.5)
in Barashev's review paper~\cite{barashev1972I}. Barashev also used his Eq.~(2.5)
to estimate the surface responsivity expected for Na metal, which he reported in
Table~2 of his paper to be
$\mathcal{R}_{\scriptscriptstyle C}\approx2.4\times10^{-22}\,I$,
and which has been inserted in row~4 of Table~\ref{respqeadNa} and labeled
``Surface theory''.
This theoretical coefficient is larger than the revised experimental coefficient
for the thin Na metal sample by a factor of approximately 3.7, rather than
smaller than it. The agreement remains encouraging in view of the simplicity
of the model and the uncertainty of the experimental parameters.
\begin{framed}\noindent
Taken together, these comparisons support a substantial volume contribution
to two-photon photoemission from the thick Na sample. The much weaker
quadratic response of the thin Na sample is compatible with a surface
contribution, although a reduced-volume contribution cannot be excluded
on the basis of these measurements alone.
\end{framed}

\subsection{Two-Photon Photoemission from Cesium-Potassium Antimonide
Semiconductor} \label{sub:TPP-CsK2Sb}

As discussed at the very beginning of this section, two-photon photoemission is a
useful tool for investigating electronic structure at metallic and semiconductor
surfaces, interfaces, and image potential states. Techniques for facilitating the
observation of two-photon photoemission were described in
Sec.~\ref{TPP-photocurrent}, and methods for preparing vacuum-depositable
photoemissive samples were outlined in Sec.~\ref{sub:earlyNa}. This was followed, in
Sec.~\ref{sub:Nametal}, by a detailed description of two-photon photoemission from
Na metal. In this section, we examine two-photon photoemission from CsK$_2$Sb,
Na$_2$KSb, and Cs$_3$Sb photocathodes under coherent illumination, using light
at several wavelengths and samples at several temperatures.  The experiments
reported in this section were carried out at Boston University in the time frame
1998--2004~\cite{booth1998ETPA,Lissandrin99,Booth01,Booth04,booth2006}.

\subsubsection{Experimental Arrangement Using Analog Detection}
\label{sub:Exp2pCsKSb}

As with Fermi-tail photoemission, two-photon photoemission gives rise to a
photocurrent that is significantly smaller than that obtained with suprathreshold
photoemission. Its observation once again makes use of the comparatively low-noise,
high current gain of a PMT in which the sample serves as the photocathode.

\begin{figure}[htb!]
\centering\includegraphics[width=3.25in]{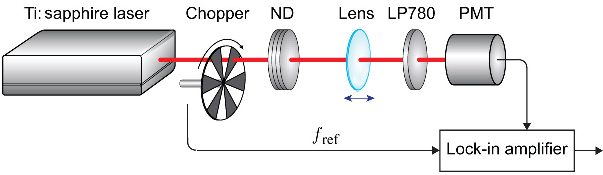}
\caption{Analog-detection experimental arrangement for observing two-photon
photoemission (TPP) from PMTs with CsK$_2$Sb, Na$_2$KSb, and Cs$_3$Sb
semiconductor photocathodes. The light from a  mode-locked Ti:sapphire laser,
operating at a wavelength of 800, 830, or 845~nm, was focused by a lens of focal
length $\mathsfit{f}\, = 88$~mm onto the photocathode, which was maintained at
temperature $\mathsfit{T\/}$. The illuminated area at the photocathode was maintained
at $A \approx 2.5 \times 10^{-9}$~m$^2$, as estimated by the properties
of the focusing lens. Wratten neutral-density (ND) filters were used to attenuate the
laser power. A long-pass (LP780) filter isolated the detector from stray light with
wavelengths shorter than 780~nm. The PMT output was fed to a lock-in amplifier
referenced to the frequency of a rotating light chopper ($f_\mathrm{ref} = 150$~Hz).
The experimental configuration shown here is the same as that displayed in Fig.~\ref{fig13},
except that the latter lacks a lens as it was designed to observe Fermi-tail photoemission.
A digital-detection system is discussed in Sec.~\ref{2Pcounting}.
(Adapted from Fig.~2.6 of~\cite{Booth04}.)
}\label{fig19}
\end{figure}

The experimental arrangement used by Booth~\emph{et al.}~\cite{booth2006} for
observing two-photon photoemission from CsK$_2$Sb, and two other
alkali-antimonide materials, is depicted in Fig.~\ref{fig19}. As with the Fermi-tail
photoemission experiments described in Sec.~\ref{subthreshold}, a mode-locked
Spectra-Physics (Tsunami) Ti:sapphire laser~\chapcite{16}{saleh2019} was operated
at one of three wavelengths (photon energies): \,1)~$\lambda= 800$~nm ($h\nu =
1.55$~eV), \,2)~830~nm ($h\nu = 1.50$~eV), or \,3)~845~nm ($h\nu = 1.47$~eV).
The laser output comprised a sequence of $\tau_0 = 120$-fs duration light pulses
emitted at a rate $f_\mathrm{rep} \equiv 1/\tau_1 = 82$~MHz, which, when
modeled as square, yield $\Delta = \tau_0 f_\mathrm{rep} \approx 0.98 \times
10^{-5}$.

The setup in Fig.~\ref{fig19} is the same as that used to observe Fermi-tail
photoemission (Fig.~\ref{fig13}), except that a lens is inserted in the beam path to
allow the illumination area on the photocathode to be adjusted. As pointed out in
Sec.~\ref{sub:TPPfocus}, for a fixed mean optical power, reducing the illumination
area allows the TPP photocurrent to be enhanced relative to the Fermi-tail
photocurrent.

The mean optical power $\overline{P}$ incident on the faceplate of the PMT was
measured with a laser power meter; the power
was restricted to a value $\overline{P}
\leqslant 20~\mu$W to minimize the possibility of local photocathode heating.
Calibrated Wratten neutral-density (ND) filters were used to attenuate the optical
power; these filters were deliberately chosen because they are sufficiently thin that
their insertion caused minimal changes to the area and location of the focused light
spot. A long-pass (LP780) optical filter served to isolate the detector from stray light
at wavelengths shorter than 780~nm.

As described in Sec.~\ref{subsubPhaseSens} in connection with the two-photon
photocurrent from Na metal, lock-in detection was employed to mitigate the
deleterious effects of various sources of noise. In this case, the output of the PMT
was fed to the lock-in amplifier  described in Sec.~\ref{sub:SPSTCsK2Sb} for the
Fermi-tail photoemission experiments, and the $f_\mathrm{ref} = 150$~Hz reference
signal was provided by the rotating light chopper. The presence of the chopper
reduced the duty cycle reported in Sec.~\ref{sub:SPSTCsK2Sb} by a factor of 2,
yielding an effective duty cycle $\Delta/2 \approx 0.5 \times 10^{-5}$. In accordance
with Eq.~(\ref{eq:equivfactor5}), therefore, the chopper-adjusted
pulse-enhancement factor is
$\Gamma \equiv (\Delta/2)^{-1} \approx 2\times10^{5}$.
The square-bladed chopper imposed square-wave modulation of unity
depth. In extracting the tabulated responsivities, we retain the
first-harmonic factor $\mathcal{F}=2/\pi$ used in the analysis
of Booth~\emph{et al.}~\cite{booth2006}.
 However, the published experimental descriptions
do not expressly indicate whether the optical power was measured with the
chopper operating or with uninterrupted
transmission, or whether the lock-in amplitudes  were reported as rms or peak values,
thereby introducing  possible uncertainties of up to a factor
of about three in the responsivities. As
indicated in the caption to Table~\ref{tab:PMT2}, however,
the responsivities are only accurate to within a
factor of about ten, owing principally to uncertainties in the
optical losses, illumination area, and mean current gain.

\subsubsection{Choice of CsK$_2$Sb} \label{subsub:2PCsPMT}

As discussed in Sec.~\ref{singletheoryCsK2Sb}, CsK$_2$Sb is a widely used
bialkali-antimonide semiconductor that is often modeled, for phenomenological
purposes, as an intrinsic direct-bandgap material whose Fermi level lies near the
center of the bandgap, so that its energy-band diagram is reasonably represented by
Fig.~\ref{fig7}. The Hamamatsu R464 semitransparent CsK$_2$Sb-photocathode
PMT used in these experiments has been described in detail in
Sec.~\ref{subsub:CsPMT}.

As shown in Fig.~\ref{fig9}($a$), the energy bandgap and ionization energy for this
material are $E_{\!g} = 1.0$~eV \,and\, $\,\texttt{W} = E_{\!g} + \chi = 1.0 + 1.1 =
2.1$~eV, respectively. Inserting these values into Eq.~(\ref{2Prange}) yields
\begin{equation}\label{2PenergyCs}
1.0~\mbox{eV} <  h\nu <  2.1~\mbox{eV} < 2h\nu, \qquad\quad \mbox{(CsK$_2$Sb)}
\end{equation}
which, when translated into the wavelength domain via Eqs.~(\ref{2Pwave}) and
(\ref{2PwaveConvert}), provides
\begin{equation}\label{2PwavelengthreverseCs}
\lambda/2 < 590~\mbox{nm} < \lambda < 1240~\mbox{nm}. \qquad\quad
\mbox{(CsK$_2$Sb)}
\end{equation}
The experiments reported in this section use light
of wavelength $\lambda=800$, 830, or 845~nm,
all of which satisfy Eq.~(\ref{2PwavelengthreverseCs}).
For $\lambda=800$~nm, for example,
\begin{equation}\label{2Pwavelengthdetail}
  \;\;\; 400 < 590 < 800 < 1240~\text{nm}, \qquad\quad \mbox{(CsK$_2$Sb)}
\end{equation}
confirming that the wavelength condition for two-photon photoemission,
Eq.~(\ref{2Pwave}), is satisfied. The observation of two-photon and
entangled-two-photon photoemission from CsK$_2$Sb under entangled-photon-pair
excitation is reported in Sec.~\ref{entchannel}.

\subsubsection{Two-Photon Photocurrent at Different Temperatures}
\label{subsubTPPMeasCs}

\begin{figure}[htb!]
\centering\includegraphics[width=3.25in]{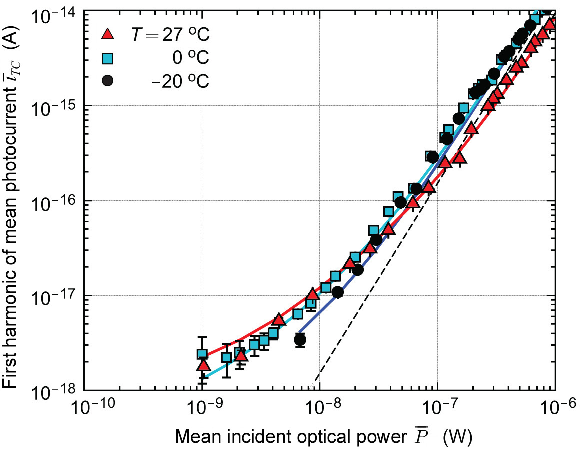}
\caption{Measured subthreshold photocurrents, comprising
both Fermi-tail and two-photon contributions, from
the  CsK$_2$Sb photocathode of a
Hamamatsu R464 PMT at different temperatures. The device was enclosed in a
Hamamatsu C4877 thermoelectric housing that allowed the PMT
temperature~$\mathsfit{T\/}$ to be varied. The doubly logarithmic plots display the
fundamental components of the photoelectric currents at the photocathode vs. the
mean optical power incident on the PMT faceplate.
Data are presented for three temperatures:  $\mathsfit{T\/} = 27~^\circ$C (triangles),
$0~^\circ$C (squares), and $-20~^\circ$C (circles). Error bars indicate standard deviations
for individual data points. The curves represent best polynomial fits.
The dashed line of slope two portrays quadratic scaling. The radiation source was a
Ti:sapphire mode-locked laser operating at $\lambda = 800$~nm.
(Adapted from Fig.~2.7 of~\cite{Booth04} and Fig.~3
of~\cite{booth2006}, \textcopyright 2006 AIP Publishing.)} \label{fig20}
\end{figure}

The measured subthreshold photocurrents, comprising
both Fermi-tail and two-photon contributions, from
the CsK$_2$Sb photocathode of a Hamamatsu R464 PMT
are displayed in Fig.~\ref{fig20} for three values of the temperature:
$\mathsfit{T\/} = 27~^\circ$C (triangles),  $0~^\circ$C (squares), and
$-20~^\circ$C (circles).   The measurements were made at a
wavelength $\lambda = 800$~nm ($h\nu = 1.55$~eV) using the experimental
arrangement portrayed in Fig.~\ref{fig19}.

As with the Fermi-tail photocurrent plots presented in Fig.~\ref{fig14}, these data are
displayed in the form of doubly logarithmic plots of the fundamental Fourier
component of the photocurrent at the photocathode
$\overline{\imath}_{\scriptscriptstyle TC}$ (A) vs. the mean optical power incident on
the PMT faceplate $\overline{P}$ (W). Again, the photocurrent at the PMT anode
was measured over a series of 3-min acquisition periods with the low-pass-filter of
the lock-in amplifier set to a bandwidth $\mathsfit{B\/} = 26$~mHz.  The
photocathode current was calculated by dividing the anode current by the mean
PMT current gain $\,\overline{G} \approx 10^8$ at the operating voltage of 1400~V.
The standard deviations of the data points were also recorded.

It is clear from the data displayed in Fig.~\ref{fig20} that the slopes of the curves
asymptote toward a value of two as the incident optical power increases and the
quadratic term in Eq.~(\ref{totalcurrentfullCoherent}) increasingly dominates the
linear term. For the data at $\mathsfit{T\/} = -20~^\circ$C and $0~^\circ$C, the
slope ultimately reaches a value of two. At $\mathsfit{T\/} = 27~^\circ$C, on the
other hand, the temperature is sufficiently high that the Fermi-tail photocurrent is
more robust (see Fig.~\ref{fig14}) and the quadratic contribution fails to fully
dominate at the maximum optical power available at the PMT faceplate.

The associated values of the two-photon responsivity
$\mathcal{R}_{\scriptscriptstyle C}(\lambda,\mathsfit{T\/})$ are displayed in
Table~\ref{tab:PMT2}. Although the scales on the ordinates and abscissas of
Figs.~\ref{fig20} and \ref{fig14} are nearly the same, the associated responsivities,
$\mathcal{R}_{\scriptscriptstyle C}$ and $\mathcal{R}_{\scriptscriptstyle F}$,
respectively, nevertheless differ in both magnitude and character, as is apparent
when comparing Table~\ref{tab:PMT2} with Table~\ref{tab:PMT}. The two-photon
component of the photocurrent is relatively insensitive to temperature for these
semiconductors, as it is for metals~\cite{Fowler1931,DuBridge33}.

\subsubsection{Two-Photon Photocurrent at Different Wavelengths}
\label{subsubTPPMeasCsLambda}

\begin{figure}[htb!]
\centering\includegraphics[width=3.25in]{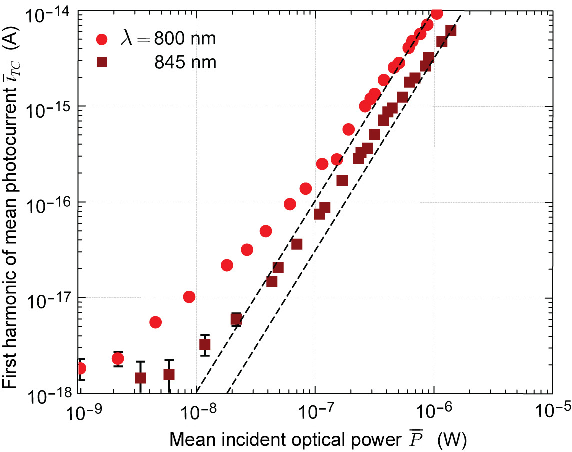}
\caption{Subthreshold photocurrent, comprising both Fermi-tail and
two-photon contributions, from the  CsK$_2$Sb photocathode of a
Hamamatsu R464 PMT for two different wavelengths. The doubly logarithmic
plots display the fundamental components of the photoelectric current at the
photocathode vs. the
mean optical power incident on the PMT faceplate.
The photocathode was maintained at $\mathsfit{T\/} = 27~^\circ$C.
The radiation source was
a Ti:sapphire mode-locked laser operating at $\lambda = 800$~nm (circles) or
at $845$~nm (squares). Error bars indicate standard deviations for individual
data points. Dashed lines of slope two portray quadratic scaling.
(Adapted from Fig.~2.8 of~\cite{Booth04} and Fig.~4
of~\cite{booth2006}, \textcopyright 2006 AIP Publishing.)
}\label{fig21}
\end{figure}

The wavelength dependence of the total subthreshold
photocurrent, comprising both Fermi-tail and
two-photon contributions, from the same
CsK$_2$Sb bialkali-antimonide semiconductor
is displayed in Fig.~\ref{fig21}.
 The
data are again exhibited in the form of doubly logarithmic plots of the fundamental
Fourier component of the photocurrent at the photocathode
$\overline{\imath}_{\scriptscriptstyle TC}$~(A) vs. the mean optical power incident
on the PMT faceplate $\overline{P}$ (W).

The data were recorded at $\mathsfit{T\/} = 27~^\circ$C for two wavelengths:
$\lambda =800$~nm ($h\nu = 1.55$~eV, circles) and 845~nm ($h\nu = 1.47$~eV,
squares). Again, the slopes increase toward a value of two as the optical power
increases, indicating that the quadratic term increasingly dominates the linear term.
At a given value of the optical power, the longer-wavelength photons at 845~nm
($h\nu = 1.47$~eV) carry less energy per photon than the shorter-wavelength
photons at 800 nm ($h\nu = 1.55$~eV), but there are more of them.
Because of the exponential dependence of the
Fermi-tail occupation probability on energy,
however, the Fermi-tail photocurrent is larger
at 800~nm. The two-photon responsivity is also
larger at 800~nm (Table~\ref{tab:PMT2}), so
both contributions favor a greater overall
photocurrent at this wavelength.

Again, the associated values of the two-photon responsivity
$\mathcal{R}_{\scriptscriptstyle C}(\lambda,\mathsfit{T\/})$ are displayed in
Table~\ref{tab:PMT2}.

\subsubsection{Responsivity and Quantum Efficiency for Alkali-Antimonide
Photocathodes} \label{sub:2PResponPMT}

As with the Fermi-tail responsivities reported in Sec.~\ref{sub:FTailResponPMT}
and Table~\ref{tab:PMT},
 two-photon responsivities have been established at several temperatures
and wavelengths for three alkali-antimonide
photocathodes (CsK$_2$Sb, Na$_2$KSb, and Cs$_3$Sb), all of which are
commercially available in Hamamatsu photomultiplier tubes, and tabulated
in Table~\ref{tab:PMT2}:
\begin{enumerate}
  \item CsK$_2$Sb (cesium-potassium antimonide) photocathode in the Hamamatsu
      R464 PMT, with mean current gain $\overline{G} \approx 10^8$, bandgap
      energy $\,E_{\!g} = 1.0$~eV, electron affinity$\,\chi = 1.1$~eV, and ionization
      energy \,$\texttt{W} = 2.1$~eV~\cite{Ghosh80}.
  \item Na$_2$KSb (sodium-potassium antimonide) photocathode in the
      Hamamatsu R2557 PMT, with mean current gain $\overline{G} \approx 5 \times
      10^6$, bandgap energy $\,E_{\!g} = 1.0$~eV,
    electron affinity$\,\chi =  1.0$~eV,
    and ionization energy \,$\texttt{W} =  2.0$~eV~\cite{Ghosh80}.
  \item Cs$_3$Sb (cesium antimonide) photocathode in the Hamamatsu 1P28 PMT,
      with mean current gain $\overline{G} \approx 10^7$, bandgap energy $\,E_{\!g}
      = 1.6$~eV,
    electron affinity$\,\chi =  0.4$~eV,
    and ionization energy \,$\texttt{W} =  2.0$~eV~\cite{Hattori00}.
\end{enumerate}

The measurements reported by Booth~\emph{et al.}~\cite{booth2006}
were carried out using the experimental arrangement depicted
in Fig.~\ref{fig19}. For those measurements, the fixed
system parameters are taken to be:
\begin{itemize}
  \item Dark/circuit current $i_{\scriptscriptstyle D} \approx 10^{-18}$~A.
  \item Intrinsic optical-system transmittance $\mathcal{T\/}_{\!\!0} \approx 0.7$.
  \item Fourier component $\mathcal{F} = 2/\pi$.
  \item Pulsed-source photocurrent enhancement factor $\Gamma \approx 2\times10^{5}$.
  \item Illumination area $A \approx 2.5 \times 10^{-9}$~m$^2$.
\end{itemize}
The additional measurements included in Table~\ref{tab:PMT2}
were obtained using different experimental arrangements,
as indicated by the references in the table caption.

For the measurements reported by Booth~\emph{et al.}~\cite{booth2006}, the
two-photon responsivity $\mathcal{R}_{\scriptscriptstyle
C}(\lambda,\mathsfit{T\/})$ for a given data set is calculated by inserting the mean
optical power $\overline{P} = \mathcal{T\/}\overline{P_0}$, and its associated
photocurrent $\overline{\imath}_{\scriptscriptstyle TC}(\lambda,\mathsfit{T\/})$,
into Eq.~(\ref{eq:yield}), along with the Fermi-tail responsivity
$\mathcal{R}_{\scriptscriptstyle F}(\lambda,\mathsfit{T\/})$ drawn from
Table~\ref{tab:PMT}. The resulting responsivities, and the associated quantum
efficiencies established via Eq.~(\ref{iCRESPeta}), are compiled in
Table~\ref{tab:PMT2}. Since these values  are determined using
Eq.~(\ref{eq:yield}), they are corrected for the effects of dark/circuit current, optical loss,
the mechanical light chopper, the pulsed nature of the source, and
focusing of the optical beam. The values presented in Table~\ref{tab:PMT2} are in
reasonably good agreement with those reported earlier for similar materials;
variations are expected given the uncertainties involved in the measurements. It is
worth emphasizing that the responsivities $\mathcal{R}_{\scriptscriptstyle F}$ and
$\mathcal{R}_{\scriptscriptstyle C}$, displayed in Tables~\ref{tab:PMT} and
\ref{tab:PMT2}, respectively, differ in both magnitude and in character.

\begin{table}[htb!]
\centering
\begin{small}
\caption{Experimental values of the (CW-equivalent) two-photon
responsivities $\mathcal{R}_{\scriptscriptstyle C}(\lambda,\mathsfit{T\/})$ (A/W),
along with the associated quantum efficiencies
$\eta_{\scriptscriptstyle C}(\lambda,\mathsfit{T\/}) = (h c/e\lambda)\,
\mathcal{R}_{\scriptscriptstyle C}$, for Hamamatsu PMT types R464, R2557,
and 1P28, along with a supplementary data set for a Soviet-era CsK$_2$Sb
PMT-130 device. The associated photocathode materials are identified in parentheses.
The wavelength of the incident light $\,\lambda\,$ and the photocathode
temperature $\,\mathsfit{T\/}\,$ at which the measurements were conducted are
indicated. The intensity of the illuminated area at the photocathode $I$ is expressed
in units of W/m$^2$.
The entries derived from Booth~\emph{et al.}~\cite{booth2006} are corrected for
the effects of dark/circuit current, optical loss,
the mechanical light chopper, the pulsed nature of the
source, and focusing of the optical beam.
They are precise to within 10\% but are accurate only to within a
factor of about ten, owing principally to uncertainties in the
optical losses, illumination area, and mean current gain.
The supplementary entries were obtained from the
independent experiments referenced below.  (Data adapted from Table~I of
Booth \emph{et al.} (2006)~\cite{booth2006}, except for the CsK$_2$Sb CPM module
data$^{\text{a}}$ extracted from Kobayashi \emph{et al.} (2007)~\cite{kobayashi2007},
the CsK$_2$Sb PMT-130 data$^{\text{b}}$ drawn from Chernov (2003)~\cite{chernov03},
and the R2557 PMT data$^{\text{c}}$ extracted from
Kobayashi \emph{et al.} (2006)~\cite{kobayashi2006}.)}
\label{tab:PMT2}
\renewcommand{\arraystretch}{1.1}
\begin{tabular}{lcccc}
\hline
\noalign{\vskip 0.7mm}
  {\small \textbf{PMT/CPM}}  & {\small $\lambda$ (nm}) &
  {\small $\mathsfit{T\/}$~($^\circ$C)} &
  {\small $\mathcal{R}_{\scriptscriptstyle C}(\lambda,\mathsfit{T\/}) =
  \mathcal{L}_{\scriptscriptstyle C}(\lambda,\mathsfit{T\/}) \, I$ \,(A/W)} &
  {\small $\eta_{\scriptscriptstyle C}(\lambda,\mathsfit{T\/})$}
  \\[0.7mm] \hline \\[-3.5mm]
\renewcommand{\arraystretch}{1}%
\textbf{R464}  & 800  & 27  & $2.60 \times 10^{-16} \,I$ & $4.03 \times 10^{-16} \,I$ \\
(CsK$_2$Sb)               &        & 0  & $6.70 \times 10^{-16} \,I$ & $1.04 \times 10^{-15} \,I$ \\
        &                  & $\!\!\!\!-$20  & $6.60 \times 10^{-16} \,I$ & $1.02 \times 10^{-15} \,I$ \\
        &           830  & 27  & $2.00 \times 10^{-16} \,I$ & $2.99 \times 10^{-16} \,I$ \\
        &                  & 0  & $5.20 \times 10^{-16} \,I$ & $7.77 \times 10^{-16} \,I$ \\
        &           845  & 27  & $0.98 \times 10^{-16} \,I$ & $1.44 \times 10^{-16} \,I$ \\
        &                  & $\!\!\!\!-$20  & $2.70 \times 10^{-16} \,I$ & $3.96 \times 10^{-16} \,I$ \\
 \textbf{CPM MP942}$^{\text{a}}$  &  1064  & 27  & $5.80 \times 10^{-16} \,I$ &
 $6.76 \times 10^{-16} \,I$ \\
 \textbf{PMT-130}$^{\text{b}}$  &  1080  & ---  & $2.37 \times 10^{-17} \,I$ &
 $2.72 \times 10^{-17} \,I$  \\[0.7mm]
 \hline  \\[-3.5mm]
  \textbf{R2557} &  800  & 27  & --- & --- \\
  (Na$_2$KSb)                & 845  & 27  & $7.10 \times 10^{-16} \,I$ &
  $1.04 \times 10^{-15} \,I$\\
 &           1064$^{\text{c}}$  & 27  & $2.10 \times 10^{-14} \,I$ &
 $2.45 \times 10^{-14} \,I$ \\[0.7mm] \hline \\[-3.5mm]
  \textbf{1P28}  &  800  & 27  & --- & --- \\
   (Cs$_3$Sb)                & 845  & 27  & $1.80 \times 10^{-18} \,I$ &
   $2.64 \times 10^{-18} \,I$\\[0.7mm]
    \hline
\end{tabular}
\end{small}
\end{table}

The results displayed in Table~\ref{tab:PMT2} confirm that the two-photon
responsivity $\mathcal{R}_{\scriptscriptstyle C}(\lambda,\mathsfit{T\/})$ for the
Hamamatsu R464 PMT depends only weakly on operating wavelength and
temperature for the parameter ranges and devices considered. It was previously
established in Table~\ref{tab:PMT} that the Fermi-tail responsivity
$\mathcal{R}_{\scriptscriptstyle F}(\lambda,\mathsfit{T\/})$ for the R464 is
substantially smaller than that for the R2557 and 1P28. Moreover,
$\mathcal{R}_{\scriptscriptstyle F}$ for the R464 decreases with decreasing
temperature and increasing wavelength. Since Fermi-tail photoemission takes
the form of a current that hinders the observation of both two-photon and entangled-two-photon
photoemission, employing a cooled R464 PMT, and making use of longer-wavelength
light, appear to be good strategies for conducting experiments to observe these
effects.

Not all PMTs and photocathode materials are suitable for exhibiting two-photon
photoemission. While the Na$_2$KSb photocathode of the R2557 PMT offers a
two-photon responsivity that exceeds those of the R464's CsK$_2$Sb and the 1P28's
Cs$_3$Sb photocathodes, for example, this is only true for sufficiently long
wavelengths. Experiments conducted with  these PMTs at wavelengths shorter than
845~nm gave rise to a residual Fermi-tail photocurrent that masked much of the
two-photon photocurrent; attempts to elicit an appreciable two-photon effect by
increasing the incident optical power above $20~\mu$W resulted in signal
saturation, probably as a result of local photocathode heating. Values of the
two-photon responsivities for the R2557 and 1P28 are therefore not displayed at a
wavelength of $\lambda = 800$~nm in Table~\ref{tab:PMT2}. An interesting parallel
is observed between one-photon and two-photon photoisomerization in human
foveal
vision~\cite{goodeve36,vasilenko1965,sliney76,palczewska2014,gil2023,kaczkos2024}.
\begin{framed}\noindent
A comparison of the semiconductor responsivities compiled in
Table~\ref{tab:PMT2} with those listed in Table~\ref{respqeadNa} for Na metal is
consistent with the observation offered in the introduction to
Sec.~\ref{twophotontheory}, where it was noted that two-photon responsivities
for semiconductors are typically several orders of magnitude larger than those for
metals~\cite{Teich68,logothetis69,barashev1972I,ferrini2009,li2016}.
\end{framed}

An exception to this observation would appear to be the small value of the two-photon
responsivity for the 1P28 PMT shown in Table~\ref{tab:PMT2}, but this is most likely
a result of the fact that the bandgap energy of its Cs$_3$Sb photocathode exceeds
the photon energy at $\lambda = 845$~nm ($E_{\!g} = 1.6~\text{eV} >
1.47~\text{eV} = h\nu$), signifying that a virtual, rather than real, intermediate state
may be involved in the transition.

\subsubsection{Linear-Quadratic Crossover Intensity}\label{CsCrossInt}

The linear-quadratic crossover intensity for coherent light specified in
Eq.~(\ref{IFCcrosscoh}),\, $I_{\scriptscriptstyle FC} = \mathcal{R}_{\scriptscriptstyle
F}/(\mathcal{L}_{\scriptscriptstyle C}\Gamma)$,\, may be extracted from
Fig.~\ref{fig20} or \ref{fig21}. Increasing $\mathcal{R}_{\scriptscriptstyle F}$
signifies an increasing Fermi-tail contribution, which moves $I_{\scriptscriptstyle
FC}$ higher, whereas increasing $\mathcal{L}_{\scriptscriptstyle C}$ and/or
$\Gamma$ signifies an increasing two-photon contribution, which moves
$I_{\scriptscriptstyle FC}$ lower.

\subsubsection{Volume vs. Surface Effects} \label{subsub:VolSurfCs}

As discussed in Sec.~\ref{singletheoryCsK2Sb}, CsK$_2$Sb is a highly efficient
photoemitter, with a suprathreshold quantum efficiency that can reach
$\eta_{\scriptscriptstyle S} \approx 0.25$. Single-photon photoemission from
CsK$_2$Sb is generally considered to be primarily a volume effect by virtue of the
material's relatively long electron mean free path at relevant excitation energies
(Sec.~\ref{volvssurf}).
A volume contribution is therefore plausible for two-photon photoemission
from CsK$_2$Sb as well.
Surface photoemission could also contribute;  however, the stronger near-surface
weighting
of quadratic optical excitation does not, by itself, distinguish a surface-specific
photoemission mechanism from volume photoemission.

\subsubsection{Experimental Arrangement Using Digital Detection}\label{2Pcounting}

A disadvantage of the analog-detection arrangements depicted
in Figs.~\ref{fig13} and \ref{fig19} for observing subthreshold
photoemission is the presence of the mechanical light chopper.
For a chopper duty cycle of 50\%, the time-averaged
transmitted optical power is reduced by a factor of two.
The lock-in amplifier extracts the fundamental Fourier
component of the modulated photocurrent, with the
associated amplitude normalization represented by
$\mathcal{F}$.
Digital photon and electron counting techniques allow
measurements to be carried out without this chopping loss.

In both the analog and digital arrangements, the signal of interest is integrated over
time and low-pass filtered: in the analog case this takes place via phase-sensitive
detection in a lock-in amplifier while in the digital case it occurs by averaging the
counts over a fixed integration time. For the case at hand, the counting window of
the digital counter is synchronized to the 12.2-ns repetition period (82-MHz
repetition rate) of the mode-locked Ti:sapphire laser and the integration time is set
to 10~s. The measurements made using the two approaches are consistent, as will be
shown in Sec.~\ref{FTQERsame} and Table~\ref{respqead}.

\section{OBSERVATION OF FERMI-TAIL PHOTOEMISSION
UNDER\\ ENTANGLED-PHOTON-PAIR ILLUMINATION}\label{entPMT}

Fermi-tail photoemission (FTP) and two-photon photoemission (TPP) elicited from
the photocathodes of several photomultiplier tubes illuminated by classical light
were considered in Secs.~\ref{subthreshold} and \ref{twophoton}, respectively.
Using light at multiple wavelengths, measurements were reported for several
materials at a number of temperatures, most prominently metallic
sodium and the semiconductor cesium-potassium antimonide.

In the  experiments described in Secs.~\ref{entPMTexar} and
\ref{ssec:expPMT}, carried out at Boston University during
1998--2004~\cite{booth1998ETPA,Lissandrin99,Booth01,Booth04,booth2006},
entangled-photon pairs were the principal source of illumination, with
coherent light serving only as a control.
 The initial intent of these experiments was to explore the feasibility of
observing entangled-two-photon photoemission (ETPP) from the cooled CsK$_2$Sb
photocathode in a commercially available Hamamatsu R464
PMT~\cite{booth1998ETPA,Lissandrin99,teich00capri,Teich00MIT,Booth01,Booth04,lissandrin2004,booth2006}.
It turned out, however, that any ETPP that might have been present was
masked by Fermi-tail photoemission from the  photocathode, despite its being
cooled. The experiments were then redirected toward examining Fermi-tail
photoemission \emph{per se} in this PMT. Contemporaneous attempts to observe
ETPP, also unsuccessful, were made by Chernov ~\cite{chernov03} and
Strekalov \emph{et al.}~\cite{strekalov02}, as reported in
Secs.~\ref{ssec:expContemp} and \ref{ssec:expStrek}, respectively.

As illustrated in Fig.~\ref{fig7}($b$) and summarized in Sec.~\ref{subthreshtheory},
Fermi-tail photoemission (FTP) is a form of subthreshold photoemission that can
occur when $h\nu < \texttt{W} = E_{\!g} + \chi$. Since \,$\texttt{W}$ is taken to be
2.1~eV for CsK$_2$Sb, as shown in Fig.~\ref{fig9}($a$), this form of photoemission
can take place when
\begin{equation}\label{Fermisubtd}
h\nu <  2.1~\mbox{eV}
\end{equation}
or, equivalently, when
\begin{equation}\label{Fermisubtdvalue}
\lambda > 590~\mbox{nm},
\end{equation}
where $\lambda$ is the wavelength of the downconverted photons. The Fermi-tail
photoemission experiments reported here made use of  degenerate photon pairs
of wavelength $\lambda = 800$~nm,
generated by spontaneous parametric downconversion (SPDC),
so that Eq.~(\ref{Fermisubtdvalue}) is satisfied. The two other
forms of subthreshold photoemission that can be elicited by entangled-photon pairs
in this wavelength range --- two-photon and entangled-two-photon photoemission
--- are addressed in Sec.~\ref{entchannel}.

\subsection{Booth's Experimental Arrangement}\label{entPMTexar}

\begin{figure}[htb!]
\centering\includegraphics[width=4.25in]{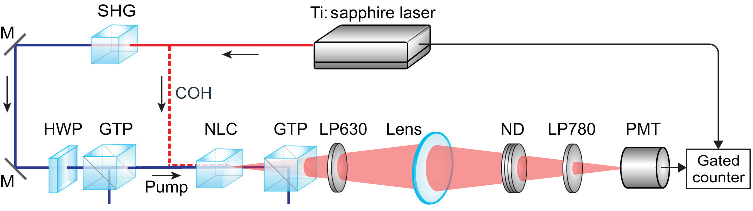}
\caption{Digital-detection experimental arrangement for observing Fermi-tail
photoemission (FTP) using spontaneous parametric downconversion (SPDC)
generated in a bulk LiIO$_3$ nonlinear optical crystal (NLC). Individual SPDC photons
elicit FTP from the cooled CsK$_2$Sb photocathode of a Hamamatsu R464 PMT.
The principal subsystems of the apparatus --- the SPDC pump/generator and
PMT/counter --- along with the intervening elements depicted in the beam path
around the periphery of the figure, are described in the text. The dashed vertical line
labeled COH identifies the alternative beam path used for carrying out experiments
with coherent light in the same apparatus. The digital configuration portrayed here is
to be contrasted with the analog-detection arrangement for studying FTP displayed
in Fig.~\ref{fig13}. (Adapted from Fig.~3.3 of~\cite{Booth04}.)
}
\label{fig22}
\end{figure}

The experimental arrangement used by Booth~\cite{Booth04} for investigating
entangled-photon-pair-induced Fermi-tail photoemission from CsK$_2$Sb is
displayed in Fig.~\ref{fig22}. The pairs were generated in a lithium-iodate (LiIO$_3$)
second-order nonlinear optical crystal (NLC).

As discussed in Sec.~\ref{sub:SPSTCsK2Sb}, detection of the small photocurrent
generated by Fermi-tail photoemission often makes use of the relatively low-noise,
high current gain offered by internal electron multiplication in a photomultiplier tube
(Fig.~\ref{fig8}).

\subsubsection{SPDC Pump and Generator}\label{ssec:SPDCpump}
As displayed in Fig.~\ref{fig22}, the heart of the SPDC pump was a Spectra-Physics
(Tsunami) Ti:sapphire laser whose output was frequency-doubled in a nonlinear
optical crystal optimized for second-harmonic generation
(SHG)~\cite{franken1961}~\chapcite{22}{saleh2019}. The laser, which was itself
pumped by a CW diode-pumped solid-state (DPSS) laser operating at
532~nm~\chapcite{16}{saleh2019}, produced mode-locked optical pulses of duration
$\tau_0 = 120$~fs at a wavelength of 800~nm and a repetition rate $f_\mathrm{rep}
= 1/\tau_1 = 82$~MHz. The pulsed pump for the SPDC, centered at $\lambda_p =
400$~nm and with a duty cycle $\Delta = \tau_0/\tau_1 = \tau_0 f_\mathrm{rep}
\approx 0.98 \times 10^{-5}$, delivered a maximum mean optical pump power
$\overline{P_{\!\scriptscriptstyle P}} \approx 750$~mW. Entangled-photon pairs
were generated via type-I collinear degenerate SPDC in a 25-mm-long LiIO$_3$
nonlinear downconversion crystal aligned for maximal
conversion~(Sec.~\ref{sssec:EPPs-prop-birefring}).
Degenerate entangled-photon pairs of center wavelength
800~nm were estimated to emerge from the LiIO$_3$
crystal with a mean source power
$\overline{P_{\!\scriptscriptstyle E}} \approx 135$~nW,
corresponding to an estimated downconversion efficiency
$\eta_\mathrm{\scriptscriptstyle SPDC}
\approx 1.8\times10^{-7}$.
Values for the entanglement time $T_{\scriptscriptstyle E}$
and entanglement area $A_{\scriptscriptstyle E}$ defined in
Sec.~\ref{sssec:TETA} were not independently established.

\subsubsection{PMT and Gated Photoelectron Counter}\label{ssec:photodetector}

The Hamamatsu R464 photomultiplier tube depicted in Fig.~\ref{fig22}, along with
the C4877 thermoelectric housing that allowed the photocathode temperature
$\mathsfit{T\/}$ to be reduced, is described in detail in Sec.~\ref{subsub:CsPMT}.
The experiments considered in this section used a digital experimental configuration
so the PMT was operated in photon-counting mode, again at a voltage of 1400~V.
A discriminator was used to reject low-amplitude
electronic noise and register PMT output pulses
as discrete counting events.
 The PMT output was fed to a
single-photon counting unit (Becker \& Hickl GmbH) configured with a gated
counting window that could be synchronized to the mode-locked laser's 82-MHz
repetition rate. An integration time of 10~s was used for each data point.

\subsubsection{Entangled-Photon Beam Path}\label{ssec:beampath}

The 400-nm SHG pump light was directed by two mirrors (M) to the LiIO$_3$ NLC
through a half-wave plate (HWP) and Glan--Taylor prism (GTP) (bottom-left of
Fig.~\ref{fig22}) that optimized the pump polarization and provided attenuation.
Residual pump light was removed by making use of a second GTP rotated at an angle
of 90$^\circ$ with respect to the first (since the pump and entangled-photon beams
are orthogonally polarized in type-I SPDC) followed by a long-pass filter (LP630). The
intrinsic optical-system transmittance in the beam path following the NLC was
estimated to be $\mathcal{T\/}_{\!\!0} \approx 0.5$. The entangled-photon beam
was attenuated by means of calibrated Wratten neutral-density (ND) filters inserted
following the $\mathsfit{f}\, = 88.3$-mm lens that focused the light onto the PMT
photocathode.
The  illumination area (which is immaterial for FTP) was estimated to
be $A \approx 2.5 \times 10^{-9}$~m$^2$ (the estimated diffraction-limited
focal area was $\approx 7.0 \times 10^{-10}$~m$^2$). The PMT was isolated from stray
room light by a long-pass filter (LP 780). The parameters associated with these
experiments are summarized in Table~\ref{compare}; additional information is
available in~\cite{Booth01,Booth04,booth2006}.

\subsubsection{Coherent-Light Beam Path}\label{ssec:beampathCOH}

The beam path for coherent light (labeled COH in
Fig.~\ref{fig22}) bypassed the SHG stage and SPDC
pump optics. The 800-nm mode-locked Ti:sapphire
laser beam passed through the transparent LiIO$_3$
NLC and followed the remainder of the
entangled-photon beam path to the photocathode.
 Calibrated Wratten ND filters provided the adjustable
attenuation.

\subsection{Subthreshold Photoemission from Cesium-Potassium Antimonide
Under Coherent and Entangled-Photon-Pair Illumination}\label{ssec:expPMT}

\begin{figure}[htb!]
\centering\includegraphics[width=3.25in]{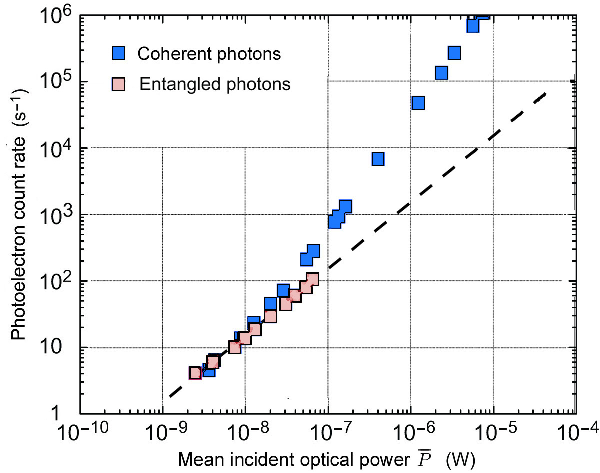}
\caption{Fermi-tail photoemission elicited by coherent light and entangled-photon light.
Doubly logarithmic plot of the photoelectron count rate vs. the mean incident optical
power for experiments using a Hamamatsu R464 PMT with a CsK$_2$Sb photocathode
cooled to $\mathsfit{T\/} = -20~^\circ$C. The blue squares represent data points
collected using coherent photons generated by a Ti:sapphire mode-locked laser (beam
path labeled COH in Fig.~\ref{fig22}).
The light red squares represent data points collected using entangled-photon
illumination  generated
via SPDC (full beam path in Fig.~\ref{fig22}). The dashed line of unity slope represents
linear scaling. The maximum attainable incident optical power under
entangled-photon illumination was
$\overline{P}_{\!\max} \approx 67$~nW, which comprised a
mixture of intact entangled-photon pairs ($\overline{P}_\mathrm{\!intact}
= \mathcal{T\/}_{\!\!0} \overline{P}_{\!\max} \approx 34$~nW) and singletons.
(Adapted from Fig.~3.6 of~\cite{Booth04}.) } \label{fig23}
\end{figure}

The observed photoelectron count rate for coherent photons
(\begin{tikzpicture}[baseline={(0,0)}] \filldraw[draw=black, fill={rgb,255:red,29;
green,121; blue,254}] (0,0) rectangle + (1.2ex,1.2ex); \end{tikzpicture}) and for
entangled-photon pairs (\begin{tikzpicture}[baseline={(0,0)}] \filldraw[draw=black,
fill={rgb,255:red,238; green,188; blue,187}] (0,0) rectangle +(1.2ex,1.2ex);
\end{tikzpicture}) vs. the mean incident optical power \,$\overline{P}$\, at the PMT
faceplate is plotted on doubly logarithmic coordinates in Fig.~\ref{fig23}.
The maximum attainable mean incident optical power
for entangled-photon illumination was
$\overline{P}_{\!\max} =\mathcal{T\/}_{\!\!0} \overline{P_{\!\scriptscriptstyle E}}
\approx 67$~nW
(the estimated source power at the output of the NLC
was $\overline{P_{\!\scriptscriptstyle E}} \approx 135$~nW
and the intrinsic optical-system transmittance was
$\mathcal{T\/}_{\!\!0} \approx 0.5$).    The incident entangled-photon-pair power was
$\overline{P}_\mathrm{\!intact} =\mathcal{T\/}_{\!\!0}^2
\overline{P_{\!\scriptscriptstyle E}} \approx 34$~nW.
The data
were collected by inserting different neutral-density Wratten filters (ND in
Fig.~\ref{fig22}) in the entangled-photon beam, which allowed the mean incident
optical power $\overline{P}$ to be adjusted --- its value was determined by
multiplying $67$~nW by the transmittance of the Wratten filters. The data were
collected using the CsK$_2$Sb photocathode in the R464 PMT, cooled to
$\mathsfit{T\/} = -20~^\circ$C.

\subsubsection{Photoelectron Count Rates}\label{countrate-20}

The data represented by the coherent photons (\begin{tikzpicture}[baseline={(0,0)}]
\filldraw[draw=black, fill={rgb,255:red,29; green,121; blue,254}] (0,0) rectangle
+(1.2ex,1.2ex); \end{tikzpicture}) in Fig.~\ref{fig23} reveal both linear and quadratic
components. These features represent Fermi-tail and two-photon photoemission,
respectively, as is readily understood from the results presented in
Sec.~\ref{sub:TPP-CsK2Sb}. The absence of a zero-slope component in the data
indicates that there is negligible dark/circuit current. An expression for the coherent
photoelectron count rate \,$\overline{\mu}_{\scriptscriptstyle TC}$\, may thus be
obtained directly from Eq.~(\ref{imathTOmu}) with $i_{\scriptscriptstyle D} = 0$:
\begin{equation}\label{imathTOmuCOH}
\overline{\mu}_{\scriptscriptstyle TC}(\lambda,\mathsfit{T\/})  \approx
\eta_{\scriptscriptstyle F}(\lambda, \mathsfit{T\/})\;\frac{(\mathcal{T\/}
\overline{P_0})}{h\nu }
+ \mathcal{L}_{\scriptscriptstyle C}(\lambda,\mathsfit{T\/})\;(\mathcal{T\/}
\overline{P_0})^2 \;\frac{\Gamma}{eA}\,,
\end{equation}
where $\overline{P_0}$ represents the mean optical power provided by the coherent
source.

The data represented by the entangled-photon pairs
(\begin{tikzpicture}[baseline={(0,0)}] \filldraw[draw=black, fill={rgb,255:red,238;
green,188; blue,187}] (0,0) rectangle +(1.2ex,1.2ex); \end{tikzpicture}) in
Fig.~\ref{fig23}, on the other hand, fall squarely along the unit-slope dashed line over
their full range of values.
Neither a power-independent background nor a quadratic
contribution is resolved in these data, so both are
neglected in the following approximation.
 The analog of Eq.~(\ref{imathTOmuCOH}) suitable for
characterizing the photoelectron count rate for entangled-photon-pair illumination is
 obtained from the expression for $\mu_{\scriptscriptstyle TE}$ provided in
Eq.~(\ref{imathTOmue1}) by setting $i_{\scriptscriptstyle D} =
\mathcal{L}_{\scriptscriptstyle C} = 0$. This leads to a mean photoelectron count
rate given by
\begin{equation}\label{imathTOmueNONE}
\overline{\mu}_{\scriptscriptstyle TE}(\lambda,\mathsfit{T\/})  \approx
\eta_{\scriptscriptstyle F}(\lambda, \mathsfit{T\/})\;\frac{(\mathcal{T\/}
\overline{P_{\!\scriptscriptstyle E}})}{h\nu } +
\eta_{\scriptscriptstyle E}(\lambda, \mathsfit{T\/})\;\frac{\mathcal{T\/} (\mathcal{T\/}
\overline{P_{\!\scriptscriptstyle E}})}{h\nu }\,,
\end{equation}
where $\overline{P_{\!\scriptscriptstyle E}}$ represents the mean optical power
emitted by the entangled-photon source.

Figure~\ref{fig23} shows that the count rates and incident optical power levels are
similar for the coherent and entangled-photon data over nearly the full range of
transmittances imposed by the Wratten filters. (This situation is well-described by
statement \#1 in Sec.~\ref{idenidenepp}.) This indicates that entangled-two-photon
photoemission is not observable in these data.
The Fermi-tail photocurrent appears to constitute
a dominant background that masks any entangled-two-photon
contribution, consistent with
$\mathcal{T\/}\eta_{\scriptscriptstyle E}
\ll \eta_{\scriptscriptstyle F}$.
Since the attenuation was applied after the NLC,
the entangled-two-photon term in
Eq.~(\ref{imathTOmueNONE}) scales quadratically
with $\mathcal{T\/}$, whereas the Fermi-tail term
scales linearly. The absence of a resolved quadratic
contribution is therefore consistent with
entangled-two-photon photoemission lying below
the experimental detection limit.

Since \,$\overline{\mu}_{\scriptscriptstyle TE} \approx \overline{\mu}_{\scriptscriptstyle
TC}$\,, and \,$\mathcal{T\/}\overline{P_0} = \mathcal{T\/}
\overline{P_{\!\scriptscriptstyle E}} = \overline{P}$\,, in the region where the
coherent-photon data (\begin{tikzpicture}[baseline={(0,0)}] \filldraw[draw=black,
fill={rgb,255:red,29; green,121; blue,254}] (0,0) rectangle +(1.2ex,1.2ex);
\end{tikzpicture}) and entangled-photon data (\begin{tikzpicture}[baseline={(0,0)}]
\filldraw[draw=black, fill={rgb,255:red,238; green,188; blue,187}] (0,0) rectangle
+(1.2ex,1.2ex); \end{tikzpicture}) overlap, we infer from Eqs.~(\ref{imathTOmuCOH})
and (\ref{imathTOmueNONE}) that the Fermi-tail quantum efficiency
$\eta_{\scriptscriptstyle F}$ is the same for both forms of light, and therefore so too
is the Fermi-tail responsivity $\mathcal{R}_{\scriptscriptstyle F}$, by virtue of
Eq.~(\ref{iFRESPeta}). These observations confirm the first two of the three assumptions
stated in  Sec.~\ref{fosteringent}; the third is confirmed in Sec.~\ref{Kobacompare}.
\begin{framed}\noindent
Coherent photons and individual SPDC photons
generate Fermi-tail photoemission in the same manner, and manifest the same optical
transmittance $\mathcal{T\/}\!$.
\end{framed}

\subsubsection{Fermi-Tail Responsivity and Quantum Efficiency} \label{FTQERsame}

Since entangled-two-photon photoemission was
not experimentally resolved, its contribution
may be neglected in Eq.~(\ref{imathTOmueNONE}),
which then reduces to
\begin{equation}\label{imathTOmueNONEreduce}
\overline{\mu}_{\scriptscriptstyle TE}(\lambda,\mathsfit{T\/})  \approx
\eta_{\scriptscriptstyle F}(\lambda,\mathsfit{T\/}) \,\frac{(\mathcal{T\/}
\overline{P_{\!\scriptscriptstyle E}})}{h\nu }
= \eta_{\scriptscriptstyle F}(\lambda,\mathsfit{T\/}) \,\frac{\overline{P}}{h\nu }\,,
\end{equation}
where \,$\overline{P}  = \mathcal{T\/} \overline{P_{\!\scriptscriptstyle E}}$ is the
mean incident optical power. Rearranging Eq.~(\ref{imathTOmueNONEreduce}) leads
to an expression for the Fermi-tail quantum efficiency:
\begin{equation}\label{imathTOmueRedFi}
\eta_{\scriptscriptstyle F} (\lambda,\mathsfit{T\/}) \approx
\frac{\overline{\mu}_{\scriptscriptstyle TE}}
{\overline{P}/ h\nu } \,.
\end{equation}
The experimental values of $\overline{\mu}_{\scriptscriptstyle TE}$ and
$\overline{P}$ can be read directly from the ordinate and abscissa of Fig.~\ref{fig23},
respectively. Inserting the pair of values associated with any point along the
unit-slope line in Fig.~\ref{fig23} into Eq.~(\ref{imathTOmueRedFi}) yields the values
for \,$\eta_{\scriptscriptstyle F}$\, and \,$\mathcal{R}_{\scriptscriptstyle F}$\,
entered in Table~\ref{respqead}. Also listed in Table~\ref{respqead} are the values
for \,$\eta_{\scriptscriptstyle F}$\, and \,$\mathcal{R}_{\scriptscriptstyle F}$\,
obtained by Booth \emph{et al.}~\cite{booth2006} using coherent light in
conjunction with the analog-detection experimental configuration displayed in
Fig.~\ref{fig13} (the data for which are reported in Table~\ref{tab:PMT}).
The digital and analog entries in
Table~\ref{respqead} are in reasonably good agreement,
supporting the consistency of the
results obtained using the two experimental
approaches despite their different detection
and normalization procedures.
\begin{table}[htb!]
\caption{Responsivity $\mathcal{R}_{\scriptscriptstyle F}$ and quantum efficiency
$\eta_{\scriptscriptstyle F}$ for Fermi-tail photoemission from the CsK$_2$Sb
photocathode of a Hamamatsu R464 PMT cooled to $\mathsfit{T\/}=-20~^\circ$C.
The relationship between \,$\mathcal{R}_{\scriptscriptstyle F}$ and
\,$\eta_{\scriptscriptstyle F}$\, is given in Eq.~(\ref{iFRESPeta}). Experimental results
are reported for two systems, both of which operate at 800~nm: 1)~digital detection
using individual SPDC photons from a LiIO$_3$ NLC, as depicted in Fig.~\ref{fig22} (data
plotted in Fig.~\ref{fig23}); and 2)~analog detection using coherent photons from
a Ti:sapphire laser, as illustrated in Fig.~\ref{fig13} (data reported in Table~\ref{tab:PMT}).
The close agreement between the two sets of data provides evidence that
individual SPDC photons and coherent photons yield consistent values,
within experimental uncertainty, for the Fermi-tail
quantum efficiency and responsivity; and that the results obtained using digital and
analog detection systems are consistent.}
\label{respqead}
\centering
\small
\renewcommand{\arraystretch}{1.15}
\begin{tabular}{@{}lcc@{}}
\cline{2-3}
\noalign{\vskip 1.1mm}
 & \shortstack{Responsivity\\$\mathcal{R}_{\scriptscriptstyle F}$ (A/W)}
 & \shortstack{Quantum efficiency\\$\eta_{\scriptscriptstyle F}$ (electrons/photon)}\\[0.7mm]
\hline
\noalign{\vskip 0.7mm}
\begin{tabular}[c]{@{}l@{}}Individual SPDC photons\\[-1mm](digital)\end{tabular} & $2.4\times10^{-10}$ &
$3.7\times10^{-10}$\\[2.8mm]
\hline
\noalign{\vskip 0.7mm}
\begin{tabular}[c]{@{}l@{}}Coherent\\[-1mm](analog)\end{tabular}    &
$3.1\times10^{-10}$ & $4.8\times10^{-10}$\\
\noalign{\vskip 0.7mm}
\hline
\end{tabular}
\end{table}

\subsection{Chernov's Experiment}\label{ssec:expContemp}

Chernov~\cite{chernov03} reported in 2003 an experiment
conducted during 2001--2002 that was similar to the one  carried out by Booth and
discussed above, with the same initial goal: to observe
entangled-two-photon photoemission from a CsK$_2$Sb
photocathode in a PMT.
 As we now proceed to
demonstrate, however, some of Chernov's parameters, a hybrid of experiment and
calculation, are implausible.

\subsubsection{Experimental Arrangement}\label{chernovexp}

Chernov's experimental design was similar to Booth's
configuration~\cite{Booth01,Booth04}. However, Chernov made use of a pulsed
Nd:YAlO$_3$ laser and a bulk LiNbO$_3$ NLC rather than the mode-locked
Ti:sapphire laser and bulk LiIO$_3$ NLC used by Booth (Fig.~\ref{fig22}). Both
researchers employed: 1)~NLC pumping by SHG derived from a pulsed solid-state
laser; 2)~type-I collinear degenerate SPDC generated in the NLC; 3)~a PMT with a
CsK$_2$Sb photocathode; and 4)~photoelectron-counting detection. The
parameters associated with Chernov's experiment~\cite{chernov03}, along with
those for Booth's experiments~\cite{Booth01,Booth04,booth2006}, are displayed in
Table~\ref{compare}. Chernov's experiment was carried out at Moscow State
University in the time frame 2001--2002.

Chernov used 1080-nm laser pulses of duration $\tau_0 = 34$~ns and a pulse
repetition rate $f_\mathrm{rep} = 1/\tau_1 = 50$~Hz. The pulsed-laser output,
frequency-doubled via SHG, served as the SPDC pump. The pump had a central
wavelength $\lambda_p = 540$~nm, a duty cycle $\Delta = \tau_0 f_\mathrm{rep}
\approx 1.7 \times 10^{-6}$, and mean power $\overline{P_{\!\scriptscriptstyle P}}
\approx 20$~mW.
Entangled-photon pairs were generated via type-I
collinear degenerate SPDC in a LiNbO$_3$ NLC,
whose length was estimated to be 10~mm, aligned for maximal conversion
(Sec.~\ref{sssec:EPPs-prop-birefring}).
The 1080-nm entangled-photon pairs emerging from
the NLC were estimated to have a mean power
$\overline{P_{\!\scriptscriptstyle E}} \approx 3.9$~pW,
corresponding to a calculated downconversion
efficiency $\eta_\mathrm{\scriptscriptstyle SPDC} =
\overline{P_{\!\scriptscriptstyle E}}/
\overline{P_{\!\scriptscriptstyle P}}
\approx 2.0 \times 10^{-10}$.
 The CsK$_2$Sb photocathode of a Soviet-era PMT-130
photomultiplier tube served as the sample.
The mean optical power carried by intact
entangled-photon pairs incident on the photocathode
was estimated to be
$\overline{P}_\mathrm{\!intact} =
\mathcal{T\/}_{\!\!0}^2 \overline{P_{\!\scriptscriptstyle E}} \approx 0.35$~pW.
Values for the entanglement time $T_{\scriptscriptstyle E}$
and entanglement area $A_{\scriptscriptstyle E}$, defined in
Sec.~\ref{sssec:TETA}, were not provided in the account of this
experiment.

\subsubsection{Fermi-Tail and Two-Photon Quantum Efficiencies}\label{chernovFT2P}

The area-specific residual one-photon response
reported by Chernov~\cite{chernov03}, interpreted
here as Fermi-tail photoemission, was characterized
by a coefficient   $\sigma_{\!\scriptscriptstyle F}
\approx 1.5 \times 10^{-17}\ \text{m}^2$, corresponding to a Fermi-tail quantum efficiency
$\eta_{\scriptscriptstyle F}   \approx \sigma_{\!\scriptscriptstyle F}/A
\approx 1.2 \times 10^{-10}$ electrons/photon.
This is in line with values measured by others in PMTs with CsK$_2$Sb
photocathodes (Table~\ref{tab:PMT}). The (area-specific) two-photon quantum
efficiency, given by Chernov as $\delta \approx 5 \times 10^{-36}\
\text{m}^2\,\text{s}$\,, corresponds to a two-photon photoemission quantum
efficiency \,$\eta_{\scriptscriptstyle C}(1080~\text{nm}) \equiv
\delta\,(I/h\nu) \approx 2.7 \times 10^{-17}\,I$,\, where $I$ is the optical intensity
at the photocathode in W/m$^2$. This result is also in line with values measured by
others for CsK$_2$Sb (Table~\ref{tab:PMT2}).

\subsubsection{Implausible Entangled-Two-Photon Quantum Efficiency}
\label{chernovE2PP}

Chernov specified that the photon-pair flux density during a pump pulse, calculated
from Klyshko's  theory of entangled-two-photon processes~\cite{klyshko82}, was
$\phi_{\mathrm{pair}}\approx5\times10^{19}\ \text{m}^{-2}\,\text{s}^{-1}$,
corresponding to an in-pulse optical intensity
$I=2h\nu\phi_{\mathrm{pair}}\approx18.4$~W/m$^2$.
 Using a beam area $A = 1.26 \times
10^{-7}$~m$^2$ (obtained from the specified beam diameter $D = 0.4$~mm) and a
pump-pulse duration $\tau_0 = 34$~ns, and incorporating the square of the
optical-system transmittance $\mathcal{T}^2 = 0.09$, results in a calculated
entangled-photon population per gate $N_\gamma =
2 \mathcal{\/T}^2 \phi_{\mathrm{pair}}A
\tau_0 \approx 3.9 \times 10^{4}$ photons at the photocathode.

Chernov reported 127 counted events, of which
24 were attributed to noise. After allowing for
an estimated residual one-photon contribution,
approximately 88 events were attributed to
entangled-two-photon photoemission.
Using the total photon population at the photocathode
($N_\gamma/\mathcal{T}\approx1.3\times10^5$ photons per gate,
including singletons) and the one-photon quantum efficiency
$\eta_{\scriptscriptstyle F}\approx1.2\times10^{-10}$ from
Sec.~\ref{chernovFT2P}, we estimate about 110 one-photon events
over the 40-h observation, compared with Chernov's estimate of
about 14; the attribution of 88 events to ETPP therefore depends
on a background subtraction that our reconstruction does not reproduce.
Using Chernov’s attribution of 88 events to ETPP, the inferred mean number
of ETPP photoelectrons per pump pulse was
$N_e=88/[50~\text{s}^{-1}\times
(40\times3600~\text{s})]
\approx1.22\times10^{-5}$.
Dividing by the theoretically estimated photon
population per gate gives
$\eta_{\scriptscriptstyle E}(1080~\text{nm})
\approx N_e/N_\gamma
\approx3.2\times10^{-10}$.
This result depends on both the attribution of
the residual photoelectron counts to ETPP and
the theoretical estimate of the incident
photon-pair population; it therefore does not
constitute an independently measured ETPP
quantum efficiency.

\subsubsection{Methodological and Experimental Shortcomings}\label{chernovLimit}

Chernov's report does not establish a definitive
observation of entangled-two-photon photoemission
because of the following methodological and
experimental limitations:
\begin{itemize}
  \item The reported entangled-two-photon quantum
  efficiency was obtained by a hybrid procedure.
  Chernov attributed approximately 88 of the 127 recorded events to ETPP
  after subtracting 24 measured noise events and an estimated
  one-photon contribution. Our reconstruction in Sec.~\ref{chernovE2PP}
  gives a one-photon contribution comparable to the entire
  noise-subtracted count. The quoted quantum efficiency was obtained by dividing the mean
number of photoelectrons per gate inferred from those 88 events
by the theoretically estimated number of photons in surviving
intact pairs per gate at the photocathode.

  \item The experiment required approximately 200~h, including about 40~h of effective
  observation. Although several control  configurations were employed, the publication
  does not provide a quantitative assessment of  long-term instrumental stability or the
  uncertainties associated with the background  subtraction.

  \item The SPDC source was too weak to permit a post-crystal attenuation experiment capable
  of distinguishing ETPP from residual one-photon photoemission (Sec.~\ref{measideniden}).
  The estimated mean optical power carried by   surviving intact photon pairs was only
  $\overline{P}_\mathrm{\!intact}\approx0.35$~pW,    corresponding to
  $\overline{I}_\mathrm{intact}\approx  2.8\times10^{-6}$~W/m$^2$.
\end{itemize}

\subsection{Strekalov et al.'s Experiment}\label{ssec:expStrek}

Another early attempt at observing entangled-two-photon photoemission was
undertaken by Strekalov~\emph{et al.}~\cite{strekalov02} at Caltech's Jet Propulsion
Laboratory, as reported in 2002. Their experiment employed an Ar-ion
laser operated at 351~nm pumping a $\beta$-BaB$_2$O$_4$ (BBO) NLC that
produced type-I near collinear degenerate SPDC at 702~nm. The entangled-photon
pairs were focused onto the Cs$_2$Te photocathode of a PerkinElmer MH922P
channel photomultiplier (CPM) module (with a channel similar to that displayed in
Fig.~\ref{fig25}).

Strekalov~\emph{et al.} reported a nonlinear CPM response that peaked when the
photocathode was placed in the image plane. However, a similar peak was obtained
with attenuated CW laser light, leading them to conclude that the observed effect
was not evidence of entangled-two-photon photoemission, but rather reflected a
slow, nonlinear response of the Cs$_2$Te photocathode, likely involving long-lived
intermediate or trap states.
Supporting this interpretation, separate measurements
using attenuated CW light revealed two-photon
self-sensitization of the photocathode, with the
sensitivity increasing by more than an order of
magnitude over approximately 100~s. Subsequent
relaxation in darkness exhibited two characteristic
decay times, approximately 5 and 100~s,
consistent with the possible involvement of
long-lived trapped states.

In a recent private communication (2026),  Dmitry Strekalov suggested
that an additional limitation, not discussed in the original publication,
might have arisen from birefringent walk-off
associated with the tilted optical axis used for type-I BBO phase matching.
This effect could have
broadened the effective SPDC source, degraded the imaging of
photon-pair spatial correlations
onto the photocathode, and thereby reduced the expected
entangled-two-photon enhancement.

\section{OBSERVATION OF TWO-PHOTON AND ENTANGLED-TWO-PHOTON\\
PHOTOEMISSION UNDER ENTANGLED-PHOTON-PAIR ILLUMINATION}\label{entchannel}

Whereas Sec.~\ref{entPMT} was directed to characterizing Fermi-tail photoemission
(FTP) from CsK$_2$Sb when illuminated by entangled-photon pairs, this section is
devoted to examining two-photon photoemission (TPP) and entangled-two-photon
photoemission (ETPP) from  CsK$_2$Sb under entangled-photon-pair illumination.
TPP is a subthreshold photoeffect that can be driven either by classical photons (as
discussed in Sec.~\ref{twophoton}) or by effectively independent photons, such as
singleton pairs or cousins originating from an entangled-photon-pair source. ETPP,
on the other hand, is two-photon photoemission induced by twins, which are
characterized by a nonseparable two-photon state.

As explained in Secs.~\ref{twophotontheory} and \ref{enttheory}, and illustrated in
Figs.~\ref{fig7}($c$) and \ref{fig7}($d$), respectively, TPP and ETPP can occur
for the real-conduction-band     intermediate-state configuration when
\begin{equation}\label{2Prange2}
  E_{\!g} <  h\nu <  (E_{\!g} + \chi) < 2h\nu,
\end{equation}
where $h\nu$ is the energy of a member of a degenerate entangled-photon pair. The
parameter values for CsK$_2$Sb, displayed in Fig.~\ref{fig9}($a$), are \,$E_{\!g} =
1.0$~eV,\, $\chi = 1.1$~eV,\, and \,$\texttt{W} = E_{\!g} + \chi = 2.1$~eV.\, Inserting
these values into Eq.~(\ref{2Prange2}) leads to
\begin{equation}\label{2EPenergy}
1.0~\mbox{eV} <  h\nu <  2.1~\mbox{eV} < 2h\nu,
\end{equation} or equivalently
\begin{equation}\label{2PwavelengthrevETPP}
\lambda/2 < 590~\mbox{nm} < \lambda < 1240~\mbox{nm}\,,
\end{equation}
where $\lambda$ is the entangled-photon wavelength.

The entangled-two-photon experiments described in this section
make use of degenerate entangled-photon pairs of wavelength $\lambda =
1064$~nm, generated by spontaneous parametric downconversion in a periodically
poled lithium niobate waveguide structure pumped by a laser diode operating at
$\lambda_p =\lambda/2 =  532$~nm. Equation~(\ref{2PwavelengthrevETPP}) is
therefore satisfied.

\subsection{Kobayashi et al.'s Experimental Arrangement}\label{entchannelexp}

\begin{figure}[htb!]
\centering\includegraphics[width=4.25in]{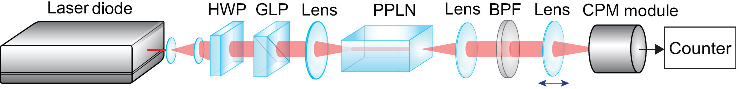}
\caption{Experimental arrangement for observing entangled-two-photon
photoemission (ETPP) using spontaneous parametric downconversion (SPDC)
generated in a periodically poled lithium niobate (PPLN) waveguide structure.
The entangled-photon pairs impinged on the CsK$_2$Sb photocathode of a
PerkinElmer MP942 channel photomultiplier (CPM) module. A 532-nm CW laser
diode served as the SPDC pump. The principal subsystems of the apparatus --- the
SPDC pump/generator and the CPM module/counter --- along with the intervening
elements in the beam path, are described in the text. (Adapted with permission from Fig.~3 of~\cite{kobayashi2007},
\copyright 2007 IEICE.)}
\label{fig24}
\end{figure}

The experimental arrangement used for investigating entangled-two-photon
photoemission induced by entangled-photon pairs, as implemented by
Kobayashi~\emph{et al.}~\cite{kobayashi2007}, is displayed in Fig.~\ref{fig24}.
An earlier paper reported coherent-light
two-photon photoemission measurements using
a Hamamatsu R2557 PMT, along with the
generation of entangled-photon pairs using
a PPLN waveguide~\cite{kobayashi2006}.
Both studies were conducted at Kyoto University
and published in 2006 and 2007, respectively.

\subsubsection{SPDC Pump and Generator}\label{ssec:SPDCcrystalKob}

As illustrated in Fig.~\ref{fig24}, the focused light from a 532-nm CW laser
diode~\chapcite{18}{saleh2019} served as the pump for a periodically poled lithium
niobate (PPLN) waveguide structure (Sec.~\ref{sssec:EPPs-prop-waveguide}) of
estimated length $L \approx 30$~mm. Time--frequency-entangled
collinear degenerate photon
pairs of central wavelength $\lambda = 1064$~nm were generated in the PPLN
(which was held at 52~$^\circ$C) via type-I quasi-phase-matched (QPM) SPDC.
The use of PPLN permitted quasi-phase matching, while the waveguide
geometry served to enhance the efficiency of the downconversion and
collection processes.
The pump power and the generated
entangled-photon power, which were confirmed to be linearly related, were
$P_{\!\scriptscriptstyle P} = 175$~mW and $P_{\!\scriptscriptstyle E} = 900$~nW,
respectively. This corresponds to a downconversion efficiency
$\eta_\mathrm{\scriptscriptstyle SPDC}  \approx 5 \times 10^{-6}$, a value very
close to that reported by Tanzilli~\emph{et al.} in 2001~\cite{Tanzilli01-EL}.
The PPLN waveguide in this experiment acts as a single-pass $\chi^{(2)}$ source that
generates entangled-photon pairs via SPDC.

\subsubsection{CPM Module and Photoelectron Counter}\label{ssec:photodetectorKob}

\begin{figure}[htb!]
\centering\includegraphics[width=3.25in]{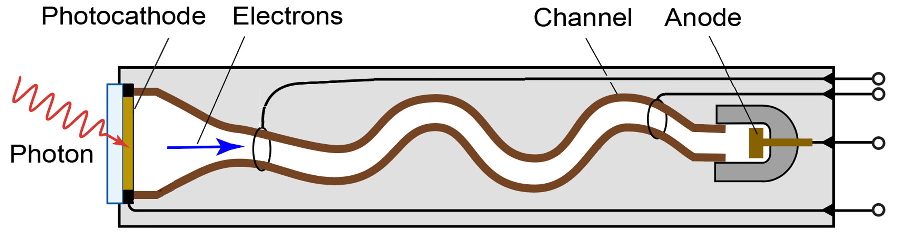}
\caption{The principal elements of a channel photomultiplier (CPM) are
a semitransparent  photocathode, an electron-multiplier channel, and an anode.
The channel typically has a diameter of 1--2~mm
and a length of several cm; it is usually curved to minimize ion feedback.
The entangled-two-photon  photoemission experiment depicted
in Fig.~\ref{fig24} made use of a PerkinElmer MP942 CPM module
with a  bialkali photocathode (treated as CsK$_2$Sb) operated at nominal room
temperature (assumed to be $27~^\circ$C).}\label{fig25}
\end{figure}

A PerkinElmer MP942 channel-photomultiplier (CPM) module served
as the  photoelectron-counting detector for the
entangled-two-photon photoemission experiment portrayed in Fig.~\ref{fig24}.
As sketched in Fig.~\ref{fig25}, the CPM contains a semitransparent bialkali
photocathode (treated here as CsK$_2$Sb) coated on the inside of
a quartz entrance window
and a small low-noise electron multiplier that achieves gain.
The interior wall of the channel is coated with a material
that enables secondary electron emission and behaves as a continuous
dynode in the presence of an accelerating electric field (in contrast to the
discrete dynodes of a photomultiplier tube, as illustrated in Fig.~\ref{fig8}).

At 1064~nm, the CPM module used by Kobayashi~\emph{et al.}
exhibited no resolvable Fermi-tail contribution under coherent-light illumination
over the measured intensity range, in contrast to
the cooled-PMT measurements at 800~nm made by Booth and
described in Sec.~\ref{ssec:expPMT}.
According to the manufacturer's specifications, the suprathreshold quantum
efficiency of the MP942 CPM module is $\eta_{\scriptscriptstyle S} \approx 0.2$, the
dark-count rate is typically $i_{\scriptscriptstyle D}/e \approx 10$~counts/s, the (TTL-pulse)
count rate increases linearly with incident intensity up to 5~MHz, and the output
pulse width is adjustable between 15~ns and 1~$\mu$s.

The experiment employed CW illumination, and
the photoelectron pulses were registered using
a photon counter without the need for synchronization to pump pulses.

\subsubsection{Entanglement Time and Entanglement Area}\label{EntTA-1064}
The entanglement time $T_{\scriptscriptstyle E}$ and entanglement
area $A_{\scriptscriptstyle E}$, defined in
Sec.~\ref{sssec:TETA}, were not reported by
Kobayashi \emph{et al.}~\cite{kobayashi2006,kobayashi2007}.
In the regime where the entangled-two-photon photoemission contribution
was observed, the illumination area at the photocathode
was estimated to be  $A \approx 1600~\mu\mathrm{m}^2 =
1.6 \times 10^{-9}$~m$^2$.   In the calculations that follow, we identify the
transverse correlation-area scale with the illumination-area
scale, adopting $A_{\scriptscriptstyle E} \approx A$ as an effective
single-spatial-mode approximation. The accuracy of this approximation depends
on the collected  spatial modes and their propagation to the sample.
Using this approximation, an entanglement time
$T_{\scriptscriptstyle E} \approx 340$~fs can be inferred  from the
linear-quadratic crossover
photon-flux density $\phi_{\scriptscriptstyle EC}$ specified in Eq.~(\ref{phiaete}),
as shown in Eq.~(\ref{TE340fs}) and explained in Sec.~\ref{Kobacomparecrosss}.

For comparison, G{\"a}bler \emph{et al.}~\cite{graefe2023APR}
obtained an entanglement time of approximately 409~fs
from a reconstructed joint temporal intensity for
type-0 collinear degenerate SPDC at 810~nm, generated in a 20-mm-long PPLN waveguide.
Their estimate includes the effects of dispersion in the optical fibers, however, and therefore
does not independently validate the value inferred here for the experiment of
Kobayashi \emph{et al.}   The estimated area--time product, and its assumed
decomposition into $A_{\scriptscriptstyle E}$ and
$T_{\scriptscriptstyle E}$, are used in Table~\ref{kobaliss} and
Sec.~\ref{sssec:Kobalissandrin} to compare the results of
Lissandrin \emph{et al.}'s entangled-two-photon photoemission theory with those of
Kobayashi~\emph{et al.}'s entangled-two-photon photoemission experiment.

\subsubsection{Entangled-Photon Beam Path} \label{kobeampath}

After passing through a half-wave plate (HWP)/Glan--Laser prism (GLP) that served
to attenuate the pump beam and allow its power to be varied, the 532-nm output of
the laser diode shown in Fig.~\ref{fig24} was focused on the PPLN waveguide.
Residual pump light was removed by a bandpass filter (BPF). A lens mounted on a
translation stage then directed the entangled-photon beam to the CPM
photocathode, while allowing the illumination area to be adjusted over the range $A
\approx 26~\mu\text{m}^2$ ($2.6 \times 10^{-11}~\text{m}^2$) to $A \approx
1600~\mu\text{m}^2$ ($1.6 \times 10^{-9}~\text{m}^2$), depending on the
experiment being conducted.
The beam cross sections were measured by scanning
a 5-$\mu$m slit across the beam. Kobayashi \emph{et al.}~\cite{kobayashi2006} had previously
identified uncertainties in beam-area measurements
as a significant source of experimental error.

For the four optical elements traversed by the entangled photons between the PPLN and the
photocathode (two lenses, one bandpass filter, and one quartz window), we estimate the
single-photon optical-system transmittance to be $\mathcal{T\/}_{\!\!0} \approx 0.7$,
corresponding to an intact-pair survival fraction $\mathcal{T\/}_{\!\!0}^2 \approx 0.5$.
The experimental parameters are summarized in Table~\ref{compare}; the photocathode temperature
is assumed to be nominal room temperature ($\mathsfit{T\/} = 27~^\circ$C).
Further details of the experimental arrangement
can be found in~\cite{kobayashi2006,kobayashi2007}.

\subsubsection{Coherent-Light Beam Path}
The coherent-light comparison reported by
Kobayashi~\emph{et al.} in 2007~\cite{kobayashi2007}
used illumination at 1064~nm, rather than the 532-nm pump employed for SPDC.
The earlier (2006) coherent-light arrangement using a Hamamatsu R2557 PMT
is presented in
Fig.~3 of~\cite{kobayashi2006}. Although the 2007 paper does not explicitly specify
the details of the optical arrangement used for the coherent-light comparison,
we infer that the MP942 CPM was also used for the coherent-light comparison.
This deduction follows from  the
substantially different coherent-light responses reported in
Fig.~6 of~\cite{kobayashi2006} and Fig.~6 of~\cite{kobayashi2007}.
As reported in Table~\ref{tab:PMT2}, the two-photon
responsivity coefficient $\mathcal{L}_{\scriptscriptstyle C}$
obtained from the 2006 measurements using the Hamamatsu R2557 PMT is about
a factor of 36 greater than that for the 2007 coherent-light data collected using the
MP942 CPM. Moreover, the 2006 measurements exhibit a linear
one-photon contribution at low intensities and quadratic two-photon behavior at higher
intensities. In contrast, the 2007 coherent-light data exhibit essentially
quadratic behavior, with no resolvable linear contribution, over the full measured
intensity range.

\subsection{Subthreshold Photoemission from Cesium-Potassium Antimonide
Under Coherent and Entangled-Photon-Pair Illumination}\label{ssec:expCPM}

\begin{figure}[htb!]
\centering\includegraphics[width=3.25in]{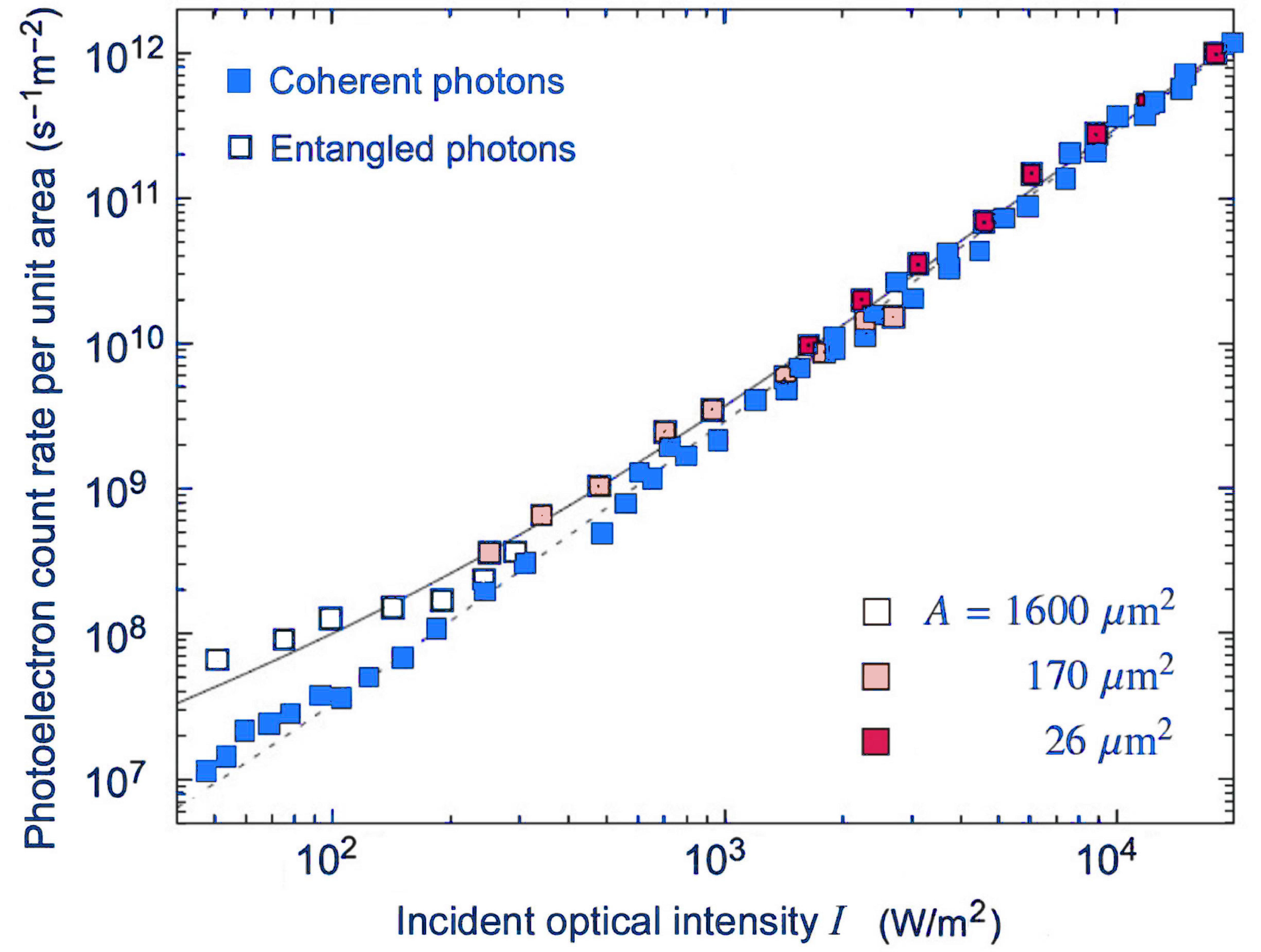}
\caption{Entangled-two-photon photoemission (ETPP) and two-photon
photoemission (TPP) from the bialkali photocathode (taken to be
CsK$_2$Sb) of a PerkinElmer MP942 channel photomultiplier (CPM) module,
operated at nominal room temperature (assumed to be $27~^\circ$C).
The open squares, and those filled with shades of red, represent data points for entangled-two-photon
beams at $\lambda = 1064$~nm focused to spots of different areas on the
photocathode, as specified in the legend. The blue squares represent data points for coherent
photons generated by a CW laser diode operated at 1064~nm. The ordinate
represents the photoelectron count rate per unit area, $\mu_{\scriptscriptstyle TC}/A$
or  $\mu_{\scriptscriptstyle TE}/A$. The abscissa represents the optical intensity incident on
the photocathode, \,$I = P/A$, which was adjusted via the
HWP/GLP optic (see Fig.~\ref{fig24}) and the entangled-photon-beam area.
This figure reveals the transition of $\mu_{\scriptscriptstyle TE}/A$ from linear to
quadratic behavior. (Adapted
with permission from Fig.~6 of~\cite{kobayashi2007}, \copyright~2007 IEICE.)
\label{fig26}}
\end{figure}

In Fig.~\ref{fig26}, the observed photoelectron count rates per unit area,
$\mu_{\scriptscriptstyle TC}/A$ and $\mu_{\scriptscriptstyle TE}/A$,
are plotted on doubly logarithmic coordinates
against the optical intensity \,$I$\, incident on the CsK$_2$Sb photocathode of the
CPM, for coherent photons (\begin{tikzpicture}[baseline={(0,0)}]
\filldraw[draw=black, fill={rgb,255:red,38; green,122; blue,254}] (0,0) rectangle
+(1.2ex,1.2ex); \end{tikzpicture}) and for entangled-photon pairs
(\begin{tikzpicture}[baseline={(0,0)}] \filldraw[draw=black, fill={rgb,255:red,255;
green,255; blue,255}] (0,0) rectangle +(1.2ex,1.2ex); \end{tikzpicture}), respectively.

For coherent light, the optical power was
varied while the illumination area was
maintained at $A = 26~\mu\text{m}^2$.
For entangled-photon pairs, as reported
in Table~\ref{compare}, the power generated
in the PPLN was
$P_{\!\scriptscriptstyle E} \approx 900$~nW.
With an estimated intrinsic optical-system
transmittance $\mathcal{T\/}_{\!\!0} \approx 0.7$,
the maximum total optical power reaching
the photocathode was $P_{\max} = \mathcal{T\/}_{\!\!0} P_{\!\scriptscriptstyle E}
\approx 630$~nW,
with approximately $P_\mathrm{intact} \approx 450$~nW carried
by surviving intact photon pairs.
While the incident optical intensity could
be decreased by adjusting the HWP/GLP optic
that governs the SPDC pump power
$P_{\!\scriptscriptstyle P}$
(Fig.~\ref{fig24}), attaining higher
intensities with the available SPDC power
required reducing the illumination area.
Three  values of the area were used to do so, as specified in
the insert to Fig.~\ref{fig26}: \newline \hspace{4em}
\begin{tikzpicture}[baseline={(0,0)}]
\filldraw[draw=black, fill={rgb,255:red,255; green,255; blue,255}] (0,0) rectangle
+(1.2ex,1.2ex);
\end{tikzpicture} \, {\small $A=1600$~$\mu$m$^2$}
\quad
\begin{tikzpicture}[baseline={(0,0)}]
\filldraw[draw=black, fill={rgb,255:red,238; green,188; blue,187}] (0,0) rectangle
+(1.2ex,1.2ex);
\end{tikzpicture} \, {\small $A=170$~$\mu$m$^2$}
\quad
\begin{tikzpicture}[baseline={(0,0)}]
\filldraw[draw=black, fill={rgb,255:red,221; green,5; blue,88}] (0,0) rectangle
+(1.2ex,1.2ex);
\end{tikzpicture} \, {\small $A=26$~$\mu$m$^2$}
\quad
\begin{tikzpicture}[baseline={(0,0)}]
\filldraw[draw=black, fill={rgb,255:red,38; green,122; blue,254}] (0,0) rectangle
+(1.2ex,1.2ex);
\end{tikzpicture} \, {\small $A = 26~\mu \text{m}^2$ (coherent).}

\subsubsection{Photoelectron Count Rates}

The mean photoelectron count rate \,$\overline{\mu}_{\scriptscriptstyle TC}$\, for
coherent photons provided in Eq.~(\ref{imathTOmu}), modified for CW operation by
removing the averaging overlines and setting $\Gamma = 1$, is written as
\begin{equation}\label{imathTOmu2}
\mu_{\scriptscriptstyle TC}(\lambda,\mathsfit{T\/})  =  \frac{i_{\scriptscriptstyle D}}{e} +
\eta_{\scriptscriptstyle F}(\lambda, \mathsfit{T\/})\;\frac{(\mathcal{T\/} P_0)}{h\nu }
+  \mathcal{L}_{\scriptscriptstyle C}(\lambda,\mathsfit{T\/})\;(\mathcal{T\/} I_0)^2\,
\frac{A}{e}\,.
\end{equation}
The data representing coherent photons (\begin{tikzpicture}[baseline={(0,0)}]
\filldraw[draw=black, fill={rgb,255:red,38; green,122; blue,254}] (0,0) rectangle
+(1.2ex,1.2ex); \end{tikzpicture}) in Fig.~\ref{fig26}, which is a plot of
\,$\mu_{\scriptscriptstyle TC}/A$\, vs. \,$I$,\, scale approximately quadratically
over the measured intensity range, indicating the dominance of
two-photon photoemission (see Sec.~\ref{sub:TPP-CsK2Sb}).
No zero-slope or unit-slope contributions
are resolved in these data.
We therefore neglect dark/circuit current
and Fermi-tail photoemission, setting
$i_{\scriptscriptstyle D} =
\eta_{\scriptscriptstyle F} = 0$
as an approximation in
Eq.~(\ref{imathTOmu2}), to arrive at
\begin{equation}\label{imathTOmuCOHnoF} \frac{\mu_{\scriptscriptstyle TC}}{A}
\approx \frac{\mathcal{L}_{\scriptscriptstyle C}(\mathcal{T\/} I_0)^2}{e} =
\frac{\mathcal{L}_{\scriptscriptstyle C}I^2}{e}\,,
\end{equation}
where $I = \mathcal{T\/}I_0$ is the CW coherent light intensity at the photocathode.
The original publication does not report an
independent dark-count measurement or specify
whether a dark-count background was subtracted;
the neglect of this contribution is thus
an assumption based on the observed scaling.

When the sample is illuminated by entangled-photon pairs instead,
it is evident that at the higher reaches of the optical intensity
(\begin{tikzpicture}[baseline={(0,0)}]
\filldraw[draw=black, fill={rgb,255:red,221; green,5; blue,88}] (0,0) rectangle
+(1.2ex,1.2ex);
\end{tikzpicture})
in Fig.~\ref{fig26}, the data also scale quadratically. In that region, therefore,
Eq.~(\ref{imathTOmue2}) can be approximated as
 \begin{equation}\label{imathTOmuCOHnoFEP}
\frac{\mu_{\scriptscriptstyle TE}}{A} \approx \frac{\mathcal{L}_{\scriptscriptstyle C}
(\mathcal{T\/} I_{\scriptscriptstyle E})^2}{e} = \frac{\mathcal{L}_{\scriptscriptstyle C}I^2}{e}\,,
\end{equation}
where $I = \mathcal{T\/}I_{\scriptscriptstyle E}$ is the entangled-photon intensity at
the photocathode. (This situation is well-described by statement \#4 in
Sec.~\ref{idenidenepp}.)

\subsubsection{Two-Photon Responsivity and Quantum Efficiency} \label{Kobacompare}

In the high-intensity region of Fig.~\ref{fig26}, the entangled-photon data
(\begin{tikzpicture}[baseline={(0,0)}] \filldraw[draw=black, fill={rgb,255:red,221;
green,5; blue,88}] (0,0) rectangle +(1.2ex,1.2ex); \end{tikzpicture}) are seen to
coincide with the coherent-photon data (\begin{tikzpicture}[baseline={(0,0)}]
\filldraw[draw=black, fill={rgb,255:red,38; green,122; blue,254}] (0,0) rectangle
+(1.2ex,1.2ex); \end{tikzpicture}) over a substantial range of incident intensities.
The approximate coincidence of these data,
together with Eqs.~(\ref{imathTOmuCOHnoF})
and (\ref{imathTOmuCOHnoFEP}), indicates the consistency of
the two-photon responsivity coefficients
$\mathcal{L}_{\scriptscriptstyle C}$, and therefore also the values of
$\mathcal{R}_{\scriptscriptstyle C}$ and $\eta_{\scriptscriptstyle C}$ at a
given intensity, being the same for both forms
of illumination.
This supports the third and fourth assumptions stated
in Sec.~\ref{fosteringent}; the first two
were supported in Sec.~\ref{countrate-20}.
\begin{framed}\noindent
Two-photon photoemission is generated in the same manner
by pairs of coherent
photons, pairs of singletons, and cousins derived from different entangled pairs.
\end{framed}

The values of the two-photon responsivity \,$\mathcal{R}_{\scriptscriptstyle C}
\text{(CsK}_2\text{Sb, 1064~nm,}~27~^\circ \text{C})$ and quantum efficiency
\,$\eta_{\scriptscriptstyle C} \text{(CsK}_2\text{Sb, 1064~nm,}~27~^\circ \text{C})$
extracted from the data in the fully quadratic region of Fig.~\ref{fig26}, collected
using the digital detection system illustrated in Fig.~\ref{fig24}, are tabulated in
Table~\ref{respqeadKob}. Also listed in Table~\ref{respqeadKob} are the values for
\,$\mathcal{R}_{\scriptscriptstyle C} \text{(CsK}_2\text{Sb, 800~nm,}~27~^\circ
\text{C})$ and \,$\eta_{\scriptscriptstyle C} \text{(CsK}_2\text{Sb,
800~nm,}~27~^\circ\text{C})$ obtained by Booth \emph{et al.}~\cite{booth2006}
and reported in Table~\ref{tab:PMT2}, collected using coherent light and the analog
detection system shown in Fig.~\ref{fig19}.  The digital and analog results in
Table~\ref{respqeadKob} agree within a factor of about two, despite
the different wavelengths, detectors, and measurement methods employed.
This level of agreement is consistent with the experimental uncertainties
discussed earlier.

\begin{table}[htb!]
\caption{Experimental values for the responsivity \,$\mathcal{R}_{\scriptscriptstyle C}$\,
and quantum efficiency \,$\eta_{\scriptscriptstyle C}$\, for two-photon photoemission
from CsK$_2$Sb at a nominal room temperature. The relationship
between \,$\mathcal{R}_{\scriptscriptstyle C} = \mathcal{L}_{\scriptscriptstyle C} I$
and \,$\eta_{\scriptscriptstyle C}$\, is provided in Eq.~(\ref{iCRESPeta}). Results are
shown for
a digital-detection setup that employed coherent photons or SPDC cousins/singleton
pairs at 1064~nm using a PerkinElmer MP942 CPM module (the experimental
arrangement is depicted in Fig.~\ref{fig24} and the results are plotted in Fig.~\ref{fig26});
and for
an analog-detection setup that employed coherent photons at 800 nm using a
Hamamatsu R464 PMT (the experimental arrangement is depicted in Fig.~\ref{fig19}
and the results are reported in Table~\ref{tab:PMT2}).
The entries are comparable in magnitude, although the 1064-nm coherent and SPDC
values are inferred from the overlap of the same high-intensity data sets, rather
than from independent determinations.}
\label{respqeadKob}
\centering
\small
\renewcommand{\arraystretch}{1.15}
\begin{tabular}{@{}lcc@{}}
\cline{2-3}
\noalign{\vskip 1.3mm}
 & \shortstack{Responsivity\\$\mathcal{R}_{\scriptscriptstyle C} =
 \mathcal{L}_{\scriptscriptstyle C} I$ \,\, (A/W)} \hspace{3mm}
 & \shortstack{Quantum efficiency\\$\eta_{\scriptscriptstyle C}$ (electrons/photon)}
 \\[0.7mm]
\hline
\noalign{\vskip 0.7mm}
\begin{tabular}[c]{@{}l@{}}Coherent\\[-1mm](digital)\end{tabular} &
$\!\!\!\!\!5.80\times10^{-16}\,I$ & $6.76\times10^{-16}\,I$\\[1.8mm]
\begin{tabular}[c]{@{}l@{}}Cousins/singleton pairs\\[-1mm](digital)\end{tabular}   &
$\!\!\!\!\!5.80\times10^{-16}\,I$ & $6.76\times10^{-16}\,I$\\[2.5mm]
\hline
\noalign{\vskip 0.7mm}
\begin{tabular}[c]{@{}l@{}}Coherent\\[-1mm](analog)\end{tabular} &
$\!\!\!\!\!2.60\times10^{-16}\,I$ & $4.03\times10^{-16}\,I$\\
\noalign{\vskip 0.7mm}
\hline
\end{tabular}
\end{table}

\subsubsection{Entangled-Two-Photon Responsivity and Quantum Efficiency}
\label{KobalissSec} Focusing on the entangled-photon (red and open) data points in
Fig.~\ref{fig26} as the incident intensity \,$I$\, decreases from its maximum value of
20,000~W/m$^2$ toward 200~W/m$^2$, the data are seen to follow a trajectory
that gradually moves from slope-2 (\begin{tikzpicture}[baseline={(0,0)}]
\filldraw[draw=black, fill={rgb,255:red,221; green,5; blue,88}] (0,0) rectangle
+(1.2ex,1.2ex); \end{tikzpicture}) to slope-1 (\begin{tikzpicture}[baseline={(0,0)}]
\filldraw[draw=black, fill={rgb,255:red,255; green,255; blue,255}] (0,0) rectangle
+(1.2ex,1.2ex); \end{tikzpicture}) behavior, representing a shift from
cousin-/singleton-pair-induced two-photon photoemission to entangled-two-photon
photoemission.
Over the range $50$--$240~\mathrm{W/m^2}$, the response evolves
from predominantly linear behavior at the lowest measured intensity
to the estimated crossover at which the linear and quadratic
contributions are equal (Sec.~\ref{Kobacomparecrosss}).
In this region, the photoelectron count rate per unit area
\,$\mu_{\scriptscriptstyle TE}/A$\, for entangled-photon pairs
(\begin{tikzpicture}[baseline={(0,0)}] \filldraw[draw=black, fill={rgb,255:red,255;
green,255; blue,255}] (0,0) rectangle +(1.2ex,1.2ex); \end{tikzpicture}) exceeds that
for coherent photons (\begin{tikzpicture}[baseline={(0,0)}] \filldraw[draw=black,
fill={rgb,255:red,38; green,122; blue,254}] (0,0) rectangle +(1.2ex,1.2ex);
\end{tikzpicture}) (by about a factor of 4 at the lowest intensity).

The absence of a resolvable linear contribution under coherent-light
illumination at 1064~nm argues against a comparable one-photon
contribution to the entangled-photon data.
At sufficiently low intensities, the quadratic TPP contribution is subdominant.
Even at the lowest intensity examined, however, the
quadratic contribution is estimated to be approximately
20\% of the linear ETPP contribution. Neglecting
the dark-count, one-photon, and quadratic contributions in Eq.~(\ref{imathTOmue2})
then leaves the entangled-two-photon term:

\begin{equation}\label{imathTOmueR}
\mu_{\scriptscriptstyle TE}  \approx
\eta_{\scriptscriptstyle E}\frac{\mathcal{T\/}_{\!\!0} P}{h\nu }\,.
\end{equation}
Reordering Eq.~(\ref{imathTOmueR}) yields the entangled-two-photon quantum
efficiency  $\eta_{\scriptscriptstyle E}$,
\begin{equation}\label{imathTOmueRed}
\eta_{\scriptscriptstyle E} \approx
\frac{\mu_{\scriptscriptstyle TE}}
{\mathcal{T\/}_{\!\!0} P/ h\nu } =
\frac{\mu_{\scriptscriptstyle TE}/A}
{\mathcal{T\/}_{\!\!0} I/ h\nu }\,.
\end{equation}
Values of $\mu_{\scriptscriptstyle TE}/A$ and $I$ may be read directly from the
ordinate and abscissa of Fig.~\ref{fig26}, respectively. Division by the intrinsic
optical-system transmittance $\mathcal{T\/}_{\!\!0}$,
taken to be approximately 0.7 as discussed in Sec.~\ref{kobeampath}, corrects for
the additional loss associated with intact-pair survival under the assumed
independent-loss model. (This situation is well-described by statement \#3
in Sec.~\ref{idenidenepp}.)

We estimated $\eta_{\scriptscriptstyle E}$ by reading
$\mu_{\scriptscriptstyle TE}/A$ and the corresponding intensity $I$
from the low-intensity portion of the curve drawn by
Kobayashi~\emph{et al.}~\cite{kobayashi2007} through the
open data points (\begin{tikzpicture}[baseline={(0,0)}] \filldraw[draw=black,
fill={rgb,255:red,255; green,255; blue,255}] (0,0) rectangle +(1.2ex,1.2ex);
\end{tikzpicture}) in Fig.~\ref{fig26}. Inserting these values
and $\mathcal{T\/}_{\!\!0}\approx0.7$ into
Eq.~(\ref{imathTOmueRed}) gives
$\eta_{\scriptscriptstyle E}\approx2.3\times10^{-13}$
electrons/photon and a corresponding responsivity
$\mathcal{R}_{\scriptscriptstyle E}\approx2.0\times10^{-13}$~A/W
at 1064~nm (Table~\ref{respqeadKobEP}). Because the ordinate
includes the residual quadratic contribution, this procedure
tends to overestimate the linear ETPP response.

\begin{table}[htb!]
\caption{Experimentally inferred values for the responsivity
$\mathcal{R}_{\scriptscriptstyle E}$ and quantum efficiency
$\eta_{\scriptscriptstyle E}$ for entangled-two-photon photoemission
from CsK$_2$Sb at $\mathsfit{T\/}=27~^\circ$C with an entanglement
area--time product estimated to be
$A_{\scriptscriptstyle E} T_{\scriptscriptstyle E} \approx
5.4 \times 10^{-22} \text{ m}^2\,\text{s}$ using the
heuristic crossover model [Eq.~(\ref{AETEestimate})].
The relationship
between \,$\mathcal{R}_{\scriptscriptstyle E}$ and \,$\eta_{\scriptscriptstyle E}$\,
is given in Eq.~(\ref{ieRESPeta}).
The normalized entangled-two-photon cross section per primitive cell
$\delta_{\scriptscriptstyle E} =
\sigma_{\!\scriptscriptstyle E} A_{\scriptscriptstyle E} T_{\scriptscriptstyle E}$,
defined in Eq.~(\ref{sigAETE}), facilitates comparison with entangled-two-photon
molecular absorption experiments.
The 1064-nm entangled-photon pairs were generated in the PPLN waveguide
structure sketched in Fig.~\ref{fig24} and the parameter values were extracted
from the data displayed in Fig.~\ref{fig26}.}
\label{respqeadKobEP}
\centering
\small
\renewcommand{\arraystretch}{1.15}
\begin{tabular}{@{}lccc@{}}
\cline{2-4}
\noalign{\vskip 1.3mm}
 & \shortstack{Responsivity\\$\mathcal{R}_{\scriptscriptstyle E}$ (A/W)} \hspace{4.5mm}
 & \shortstack{Quantum efficiency\\$\eta_{\scriptscriptstyle E}$ (electrons/photon)}
 \hspace{1.5mm}
 & \shortstack{Normalized cross section\\per primitive cell
 $\delta_{\scriptscriptstyle E}$ (m$^4$\,s)}
 \\[0.7mm]
\hline
\noalign{\vskip 0.7mm}
\begin{tabular}[c]{@{}l@{}}Entangled-\\[-1mm]photon pairs\end{tabular} &
$\!\!\!\!\!\!\!\!\!2.0\times10^{-13}$ & $2.3\times10^{-13}$  & $2.1\times10^{-54}$ \\
\noalign{\vskip 0.7mm}
\hline
\end{tabular}
\end{table}

The plausibility of these values is supported by the following reasoning. In Booth's
data, illustrated in Fig.~\ref{fig23}, the presence of Fermi-tail photoemission and the
absence of entangled-two-photon photoemission implies that the Fermi-tail
response establishes a noise floor indicating, in accordance with
Eq.~(\ref{imathTOmue2}), that the dominance of entangled-two-photon
photoemission over Fermi-tail photoemission requires
$\mathcal{T\/}_{\!\!0}\eta_{\scriptscriptstyle E} >
\eta_{\scriptscriptstyle F}$. Using  the entangled-two-photon quantum efficiency for
the data of Kobayashi~\emph{et al.} provided in Table~\ref{respqeadKobEP}, along
with the intrinsic transmittance for his experiment indicated in
Sec.~\ref{kobeampath} ($\mathcal{T\/}_{\!\!0} \approx 0.7$), yields
\,$\mathcal{T\/}_{\!\!0}\eta_{\scriptscriptstyle E} \approx  1.6 \times 10^{-13}$.
This
value does indeed exceed the (unmeasured) Fermi-tail quantum efficiency associated
with the experiment of Kobayashi~\emph{et al.}, which is \,$\eta_{\scriptscriptstyle
F} <3.7 \times 10^{-14}$,\, as may be inferred from Fig.~\ref{fig26}. At the same
time, the value \,$\mathcal{T\/}_{\!\!0}\eta_{\scriptscriptstyle E} \approx  1.6 \times
10^{-13}$\,  lies far below the Fermi-tail quantum efficiency for Booth's data
presented in Fig.~\ref{fig23}, \,$\eta_{\scriptscriptstyle F} \approx  3.7 \times
10^{-10}$\, (Table~\ref{respqead}), in which entangled-two-photon photoemission
was not discerned.

\begin{framed}\noindent
For the experimental configuration used by
Kobayashi \emph{et al.}~\cite{kobayashi2007}, a plausible value for the
entangled-two-photon photoemission quantum efficiency of
CsK$_2$Sb is \,$\eta_{\scriptscriptstyle E} \approx  2.3 \times 10^{-13}.$
\end{framed}

The parameters extracted from the experiment of Kobayashi~\emph{et
al.}~\cite{kobayashi2006,kobayashi2007}, as well as those associated with the
experiments of Booth~\cite{Booth01,Booth04,booth2006} and
Chernov~\cite{chernov03}, are displayed in Table~\ref{compare} for ready
comparison.

\begin{table}[ht!]
\centering
\begin{small}
\caption{Experimental parameters for three subthreshold-photoemission experiments
using: 1)~isolated entangled-photon pairs from type-I collinear degenerate
SPDC; \,2)~a CsK$_2$Sb photocathode;
and \,3)~photoelectron-counting detection. Booth's experiments (Sec.~\ref{entPMTexar})
reported individual-SPDC-photon induced Fermi-tail photoemission while Kobayashi \emph{et al.}'s
experiment (Sec.~\ref{entchannelexp}) reported entangled-two-photon photoemission
(and also cousin-/singleton-pair-induced two-photon photoemission). Chernov's
value for $\eta_{\scriptscriptstyle E}$ is implausibly large (Sec.~\ref{ssec:expContemp}).
Parameters associated with coherent light are denoted by an asterisk *.
\,Estimated/calculated parameters are denoted by a dagger $\dagger$.
 \vspace{-1.5mm} }
\label{compare}
\renewcommand{\arraystretch}{1.05}
\begin{tabular}{@{}llll@{}
}
\hline
\noalign{\vskip 0.5mm}
{\sc \textbf{subsystem$/$}}      & Booth~\cite{Booth01,Booth04,booth2006} &
Chernov~\cite{chernov03} & \hspace{-2mm}Kobayashi \emph{et al.} \cite{kobayashi2006}\\
[-.5mm]
Parameter& (2001, 2004, 2006) & (2003) & \hspace{-2mm}\cite{kobayashi2007} \,\,
(2006, 2007) \\[0.5mm]
\hline
\noalign{\vskip 0.5mm}
{\sc \textbf{coherent source (control)}} & \hspace{-3.9mm}* &  &  \\
Laser & Ti:sapphire & Nd:YAlO$_3$ & laser diode \\
Wavelength $\lambda$ & 800~nm & 1080~nm & 1064~nm \\
Pulse duration $\tau_0$ & 120~fs & CW & CW \\
Repetition rate $f_\mathrm{rep} = 1/\tau_1$  & 82~MHz & CW  & CW \\[0.5mm]
\hline
\noalign{\vskip 0.5mm}
{\sc \textbf{spdc pump}*} &  &  &  \\
Source & Ti:sapphire SHG & Nd:YAlO$_3$ SHG & laser diode \\
Wavelength $\lambda_p$ & 400~nm & 540~nm & 532~nm \\
Pulse duration $\tau_0$ & 120~fs & 34~ns  & CW \\
Repetition rate $f_\mathrm{rep} = 1/\tau_1$  & 82~MHz & 50~Hz  & CW \\
Mean pump power $\overline{P_{\!\scriptscriptstyle P}}$  & 750~mW    & 20~mW
& 175~mW\\[0.5mm]
\hline
\noalign{\vskip 0.5mm}
{\sc \textbf{spdc source}} &  &  &  \\
Nonlinear crystal & LiIO$_3$ & LiNbO$_3$  & PPLN waveguide \\
Nonlinear crystal length $L$ & 25~mm & $\dagger 10$~mm  & $\dagger 30$~mm \\
Wavelength $\lambda_{1,2}$ & 800~nm & 1080~nm & 1064~nm \\
Pulse enhancement $\Gamma = \tau_1/\tau_0$  &
$1.0 \times 10^5$ & $5.9 \times 10^5$  & 1 \\
Entangled-photon power $\overline{P_{\!\scriptscriptstyle E}}$ & 135~nW    &
$\dagger$3.9~pW & 900~nW \\
Efficiency $\eta_\mathrm{\scriptscriptstyle SPDC} =
\overline{P_{\!\scriptscriptstyle E}}/\overline{P_{\!\scriptscriptstyle P}}$ &
$1.8 \times 10^{-7}$ &  $2.0 \times 10^{-10}$  & $5.1 \times 10^{-6}$  \\
Max. incident pair power $\overline{P}_\mathrm{\!intact}$ & 34~nW    & $\dagger$0.35~pW
& 450~nW \\
Illumination area $A$ & $2.5 \times 10^{-9}~\text{m}^2$ &
$1.3 \times 10^{-7}~\text{m}^2$ & $1.6 \times 10^{-9}~\text{m}^2$ \\
Max. incident pair inten. $\overline{I}_\mathrm{intact}$ & 14~W$/$m$^2$  &
$\dagger$$2.8 \times 10^{-6}$~W$/$m$^2$ & 280~W$/$m$^2$\\
Entanglement time $T_{\scriptscriptstyle E}$ & ---
& ---           &  $\dagger$340~fs  \\
Entanglement area $A_{\scriptscriptstyle E}$ & ---
 & ---    & $\dagger$$1.6 \times 10^{-9}$~m$^2$
 \\[0.5mm]
\hline
\noalign{\vskip 0.5mm}
{\sc \textbf{system transmittance\,/}}&\hspace{-10.mm}{\sc \textbf{photodetector}} &  &
\\
Intrinsic pair-survival prob. $\mathcal{T\/}_{\!\!0}^2$ &
$\dagger$\,$0.25$ & $ 0.09$ &
$\dagger$\,$0.5$ \\
CsK$_2$Sb photocathode & PMT R464 & PMT-130 & CPM MP942 mod.\\
Photocathode temperature $\mathsfit{T\/}$  & $-20~^\circ$C & --- &
$27~^\circ$C \\
Counter gate time$/$rep. rate &  --- \, $/$82~MHz  & 50~ns$/$50~Hz & free-running\\ [0.5mm]
\hline
\noalign{\vskip 0.5mm}
{\sc \textbf{responsivities\,$/$\,quantum}} & \hspace{-7.5mm}{\sc \textbf{efficiencies}} &
&  \\
Responsivity $\mathcal{R}_{\scriptscriptstyle F}$ &$2.4 \times 10^{-10}$~A/W
& *$1.0 \times 10^{-10}$~A/W   & *$<3.2 \times 10^{-14}$~A/W\\
Responsivity $\mathcal{R}_{\scriptscriptstyle C}$ & *$6.6 \times 10^{-16} \,I$~A/W
&  *$2.4 \times 10^{-17} \,I$~A/W   & $5.8 \times 10^{-16} \,I$~A/W \\
Responsivity $\mathcal{R}_{\scriptscriptstyle E}$ & --- & $\dagger$$2.
8 \times 10^{-10}$~A/W  & $2.0 \times 10^{-13}$~A/W \\
Quantum efficiency $\eta_{\scriptscriptstyle F}$ & $3.7 \times 10^{-10}$
& *$1.2 \times 10^{-10}$ &  $*<3.7 \times 10^{-14}$ \\
Quantum efficiency $\eta_{\scriptscriptstyle C}$
& *$1.0 \times 10^{-15} \,I$& *$2.7 \times 10^{-17} \,I$
& $6.8 \times 10^{-16} \,I$ \\
Quantum efficiency $\eta_{\scriptscriptstyle E}$ & --- &
$\dagger$$3.2 \times 10^{-10}$ & $2.3 \times 10^{-13}$ \\ [0.5mm]
\hline
\noalign{\vskip 0.5mm}
{\sc \textbf{crossover intensities}} &  &  &  \\
Crossover intensity $I_{\scriptscriptstyle FC}$& *3.6 W$/$m$^2$
& *$4.4 \times 10^6$~W$/$m$^2$ &  *$\lesssim 55$~W$/$m$^2$ \\
Crossover intensity $I_{\scriptscriptstyle EC}$&--- &  ---  & 240~W$/$m$^2$\\[0.5mm]
\hline
\end{tabular}
\end{small}
\end{table}

As a final note, we mention that it would have been desirable to have additional
confirmation of the presence of
entangled-photon pairs at the sample in the experiment
carried out by Kobayashi \emph{et al.}~\cite{kobayashi2006,kobayashi2007}. It
would have been useful, for example, to insert attenuating filters in the
entangled-photon beam to confirm the expected quadratic ($\mathcal{T\/}^2$)
scaling. There is no mention of such a measurement in their reports, possibly
because of the difficulty of carrying them out given the limited range of optical
intensities over which linear behavior was present.
The presence of photon-pair correlations at the photocathode
could also have been verified by temporarily replacing the
CPM with a coincidence-counting arrangement, although
establishing entanglement would require additional measurements.
It would also have been helpful to have an experimental measurement of the entanglement
time and entanglement area at the sample. We are not
aware of any attempts to repeat, or expand upon, Kobayashi \emph{et al.}'s  experiments.

\subsubsection{Linear-Quadratic Crossover Intensity and Photon-Flux Density}
\label{Kobacomparecrosss}

For a CW source of entangled-photon pairs ($\Gamma = 1$),
with Fermi-tail photoemission neglected
($\mathcal{R}_{\scriptscriptstyle F}=0$),
Eq.~(\ref{IFEcross}) for the linear-quadratic crossover
intensity reduces to
\begin{equation}
\label{IEC}
I_{\scriptscriptstyle EC}
=
\frac{
\mathcal{T\/}_{\!\!0}\,
\mathcal{R}_{\scriptscriptstyle E}
}{
\mathcal{L}_{\scriptscriptstyle C}
}.
\end{equation}
Inserting the experimentally inferred values of
$\mathcal{L}_{\scriptscriptstyle C}$ and
$\mathcal{R}_{\scriptscriptstyle E}$ from
Tables~\ref{respqeadKob} and \ref{respqeadKobEP},
respectively, together with the estimate
$\mathcal{T\/}_{\!\!0} \approx 0.7$ provided in
Sec.~\ref{kobeampath}, yields
\begin{equation}
\label{IEC2}
I_{\scriptscriptstyle EC}
\approx 240~\mathrm{W/m^2}.
\end{equation}
This value is consistent with the crossover region in
Fig.~\ref{fig26}, but does not provide an independent confirmation  of the value
since the responsivity parameters were inferred from the same
photoemission data.
Since the data in Fig.~\ref{fig26} were acquired using
different illumination areas, the possibility that the
entanglement area varied with the focusing conditions
introduces an additional uncertainty into the interpretation
of the crossover intensity.
The corresponding crossover photon-flux density at the
sample is
\begin{equation}
\label{phiEC2}
\phi_{\scriptscriptstyle EC}
=
I_{\scriptscriptstyle EC}/h\nu
\approx
1.3 \times 10^{21}~\mathrm{photons/m^2\,s}.
\end{equation}

Within the heuristic framework set forth in
Sec.~\ref{EPparticlemodel}, which is applicable for entangled-two-photon
photoemission as well as entangled-two-photon absorption, the crossover photon-flux
density can be related to the entanglement area--time
product via Eq.~(\ref{sigdelATtophiEC}).
For CW illumination ($\Gamma = 1$),  this relation becomes
\begin{equation}
\label{phiaete}
\phi_{\scriptscriptstyle EC}
=
\frac{
\mathcal{T\/}_{\!\!0}
}{
A_{\scriptscriptstyle E} T_{\scriptscriptstyle E}
}.
\end{equation}
Equation~(\ref{phiaete}) incorporates the cross-section relation used in
the heuristic framework and is thus not a general, source-only identity.
Its use assumes the applicability of that relation
to the photoemission response and neglects departures
arising from the detailed two-photon amplitude and
material dynamics.

Using the values of $\mathcal{T\/}_{\!\!0}$ and  $\phi_{\scriptscriptstyle EC}$
specified above leads to the estimate
\begin{equation}\label{AETEestimate}
A_{\scriptscriptstyle E} T_{\scriptscriptstyle E}
=
\frac{
\mathcal{T\/}_{\!\!0}
}{
\phi_{\scriptscriptstyle EC}
}
\approx
5.4 \times 10^{-22}~\mathrm{m^2\,s}.
\end{equation}
The estimate provided in Eq.~(\ref{AETEestimate}) depends on
the prefactor adopted in the heuristic cross-section relation
[Eq.~(\ref{sigdelAT})]; we have selected this prefactor to correspond
to $\alpha=1$ in the
more general formulation presented in Eq.~(\ref{sigdelATqalpha}).
Had we used the prefactor $\slfrac{1}{2}$ employed by
Fei~\emph{et al.}~\cite{Fei97}, the estimated
entanglement area--time product would have been reduced by a factor of 2.
While the crossover relation
constrains the product $A_{\scriptscriptstyle E} T_{\scriptscriptstyle E}$,
rather than either factor separately, adopting the effective
single-spatial-mode approximation  $A_{\scriptscriptstyle E} \approx A
\approx 1.6 \times 10^{-9}$~m$^2$ discussed in  Sec.~\ref{EntTA-1064} leads to
\begin{equation}\label{TE340fs}
T_{\scriptscriptstyle E}
\approx
\frac{
5.4 \times 10^{-22}~\mathrm{m^2\,s}
}{
1.6 \times 10^{-9}~\mathrm{m^2}
}
\approx 340~\mathrm{fs}.
\end{equation}
Again, this value is a crossover-inferred estimate
conditioned on the applicability of both the heuristic model and the assumed
entanglement area.

The intermediate-state lifetime
$T_{\scriptscriptstyle A} \approx 270$~fs cited by
Kobayashi \emph{et al.}~\cite{kobayashi2006} was drawn
from the measurements of
Hattori \emph{et al.}~\cite{Hattori00} for the
Na$_2$KSb photocathode in a Hamamatsu R2557 PMT
(Table~\ref{tab:PMT2}), rather than from a direct measurement of the CsK$_2$Sb photocathode.
The same lifetime was adopted for CsK$_2$Sb in the calculations of
Lissandrin \emph{et al.}~\cite{lissandrin2004} ($\tau_j \equiv T_{\scriptscriptstyle A}$ in
Table~\ref{tab:photoemissiveparams}).  The inferred value of $T_{\scriptscriptstyle E}$ is
nominally greater than $T_{\scriptscriptstyle A}$, but the
two times are comparable. Hence, the temporal condition
$T_{\scriptscriptstyle E} \geqslant  T_{\scriptscriptstyle A}$ underlying the heuristic
relation is provisionally consistent with these estimates, but the spatial condition
$A_{\scriptscriptstyle E} \geqslant \sigma_1$
has not been independently established. Had we adopted the prefactor of $\slfrac12\,$, the inferred
entanglement time would have been approximately 170~fs for the same assumed entanglement area, thereby violating
the temporal condition of the heuristic crossover relation for the adopted value of $T_{\scriptscriptstyle A}$.

We make use of the estimated area--time product, together with
the assumed values of its individual factors, in
Table~\ref{kobaliss} and
Sec.~\ref{sssec:Kobalissandrin} to compare the entangled-two-photon photoemission
theory   of Lissandrin \emph{et al.}~\cite{lissandrin2004} with the
experimental results obtained by
Kobayashi \emph{et al.}~\cite{kobayashi2007}.

\subsubsection{Facilitating the Observation of Entangled-Two-Photon Photoemission}
\label{kobabooth}

The experiments conducted by Booth, the data for which are presented in
Fig.~\ref{fig23} and Tables~\ref{respqead} and \ref{compare}, reported Fermi-tail
photoemission induced by individual photons derived from entangled-photon pairs. The
complementary experiment of Kobayashi \emph{et al.}, the data for which are
presented in Fig.~\ref{fig26} and Tables~\ref{respqeadKob}--\ref{compare}, reported
two-photon photoemission induced by entangled-photon cousins or singleton pairs,
and entangled-two-photon photoemission induced by entangled-photon twins. Both
sets of experiments shared the following features:
\begin{enumerate}
  \item Entangled-photon pairs generated via type-I collinear degenerate SPDC.
  \item SPDC photons with wavelengths capable of inducing ETPP.
 \item A bialkali photocathode (CsK$_2$Sb confirmed for Booth and presumed for
 Kobayashi~\emph{et al.}).
  \item Photoelectron-counting detection.
\end{enumerate}

Techniques for enhancing a selected form of subthreshold photoemission were
discussed in Sec.~\ref{enhancingf}, and a qualitative guide for doing so was set forth
in Table~\ref{tab:PEchar}. In the context of facilitating the observation of
entangled-two-photon photoemission, the experimental configuration used by
Kobayashi~\emph{et al}. was superior to that used by Booth.
As detailed below, this superiority rests on two principal features (see
Table~\ref{compare}): 1)~an increase in entangled-photon power and
2)~a decrease in Fermi-tail photoemission:

\begin{enumerate}
  \item Kobayashi~\emph{et al.}'s use of a CW laser-diode pump in combination with
      a PPLN waveguide delivered greater entangled-photon power with a simpler
      experimental configuration (compare Figs.~\ref{fig22} and \ref{fig24}). The PPLN
      waveguide NLC, which made use of quasi-phase-matching, offered substantially
      higher conversion efficiency.
As a result, the estimated entangled-photon
power at the output of Kobayashi~\emph{et al.}'s
PPLN waveguide was     $\overline{P_{\!\scriptscriptstyle E}} \approx 900$~nW,
about 6.7 times the estimated value
$\overline{P_{\!\scriptscriptstyle E}} \approx 135$~nW
at the output of Booth's nonlinear crystal.
The corresponding maximum power carried by
surviving intact pairs at Kobayashi~\emph{et al.}'s
photocathode was  $\overline{P}_\mathrm{\!intact} \approx 450$~nW,
a factor of 13.4 greater than that at Booth's photocathode, which was
$\overline{P}_\mathrm{\!intact} \approx 34$~nW.
 Moreover, the CW nature of the pump ($\Gamma = 1$)
 eliminated the pulsed-source enhancement of TPP, which is a source
 of noise for the observation of ETPP.

 \item The CPM module used in the experiment of Kobayashi~\emph{et al.}
      exhibited negligible Fermi-tail photoemission for the full range of incident
      intensities over which ETPP was observed ($50 \leqslant I \leqslant
      240$~W/m$^2$). The coherent-light data of  Kobayashi~\emph{et al.} suggest an
upper bound for the Fermi-tail quantum  efficiency of
$\eta_{\scriptscriptstyle F}  \lesssim 3.7\times10^{-14}$,
as inferred from Fig.~\ref{fig26}.  This is at least four orders of magnitude
below Booth's measured value of  $\eta_{\scriptscriptstyle F}
\approx 3.7\times10^{-10}$.  This difference cannot be
attributed solely to the CPM architecture, however, because the two experiments
had different configurations and different illumination wavelengths
(1064 and 800~nm, respectively). Indeed, it is possible that the longer wavelength light
used by Kobayashi~\emph{et al.} could have served to reduce any putative Fermi-tail
photoemission.

\end{enumerate}
\noindent The data in Table~\ref{compare} can be further scrutinized for more
granular insights.

\subsection{Lissandrin et al.'s Theory of Entangled-Two-Photon Photoemission:
Comparison with Experiment} \label{sssec:Kobalissandrin}

We bring Sec.~\ref{entchannel} to a close by comparing  the results of
the 2004 entangled-two-photon photoemission theory for
CsK$_2$Sb developed by  Lissandrin \emph{et al.}~\cite{lissandrin2004}
(Sec.~\ref{enttheory}) with the entangled-two-photon quantum efficiency
$\eta_{\scriptscriptstyle E}$ and responsivity
$\mathcal{R}_{\scriptscriptstyle E}$ for
CsK$_2$Sb inferred from the 2007 experiment of
Kobayashi \emph{et al.}~\cite{kobayashi2007} (Sec.~\ref{ssec:expCPM}).

Table~\ref{kobaliss} brings together the parameters relevant
to this comparison. The original model parameters
are carried over from Sec.~\ref{enttheory} and
Tables~\ref{tab:spdcparams} and
\ref{tab:photoemissiveparams}.
The parameters associated
with Kobayashi \emph{et al.}'s experiment are carried over
from Sec.~\ref{ssec:expCPM} and Table~\ref{compare}.
As discussed in Secs.~\ref{EntTA-1064} and
\ref{Kobacomparecrosss}, the value
$A_{\scriptscriptstyle E}T_{\scriptscriptstyle E}
\approx 5.4 \times 10^{-22}$~m$^2$\,s reported in Table~\ref{kobaliss} is a
heuristic crossover inference [Eq.~(\ref{AETEestimate})].
Its decomposition into
$A_{\scriptscriptstyle E}\approx 1.6 \times 10^{-9}$~m$^2$
and $T_{\scriptscriptstyle E}\approx 340$~fs additionally
assumes that the effective entanglement area is
approximately equal to the illumination area.

\begin{table}[htb!]
\centering
\small
\caption{
Parameters used for comparing
entangled-two-photon photoemission theory and experiment.
Column~2 lists parameters of the original model
for CsK$_2$Sb by Lissandrin \emph{et al.}~\cite{lissandrin2004},
carried over from Tables~\ref{tab:spdcparams} and
\ref{tab:photoemissiveparams}.
Column~3 lists parameters  associated with the experiment of
Kobayashi \emph{et al.}~\cite{kobayashi2007}, carried over from
Sec.~\ref{entchannelexp} and Table~\ref{compare}.
In column~3, the entanglement area is an effective
single-spatial-mode approximation, the area--time product
is inferred using the heuristic crossover model, and
the entanglement time follows from their ratio.
The lower portion of the table reports the original theoretical values of
$\eta_{\scriptscriptstyle E}$ and $\mathcal{R}_{\scriptscriptstyle E}$,
along with scaled estimates assuming that the remaining material-response
factor is unchanged between the two parameter sets ($\mathcal{Q}=1$).}
\label{kobaliss}
\renewcommand{\arraystretch}{1.1}
\begin{tabular}{lcc}
\hline
\noalign{\vskip 0.7mm}
 & Model parameters & Experimental parameters \\[-0.5mm]
 & Lissandrin \emph{et al.} & Kobayashi \emph{et al.} \\[-0.5mm]
Parameter (units)
 & 2004~\cite{lissandrin2004}
 & 2007~\cite{kobayashi2007} \\[0.7mm]
\hline
\noalign{\vskip 0.7mm}
Pump wavelength $\lambda_p$ (nm)
 & 406 & 532 \\
Pump photon energy $\hbar\omega_p$ (eV)
 & 3.05 & 2.33 \\
Nondegeneracy ratio $\omega_1^0/\omega_p$
 &  ½ &  ½ \\
SPDC wavelength $\lambda_{1,2}$ (nm)
 & 812 & 1064 \\
SPDC photon energy $\hbar\omega_{1,2}$ (eV)
 & 1.53 & 1.17 \\
Entanglement time $T_{\scriptscriptstyle E}$ (fs)
 & 10 & 340 \\
Entanglement area $A_{\scriptscriptstyle E}$ (m$^2$)
 & $1.0 \times 10^{-10}$
 & $1.6 \times 10^{-9}$ \\
Entanglement area--time product
$A_{\scriptscriptstyle E}T_{\scriptscriptstyle E}$
(m$^2$\,s)
 & $1.0 \times 10^{-24}$
 & $5.4 \times 10^{-22}$ \\
Intrinsic transmittance squared
$\mathcal{T\/}_{\!\!0}^2$
 & $1.0$ & $0.5$ \\
Illumination area $A$ (m$^2$)
 & $1.0 \times 10^{-10}$
 & $1.6 \times 10^{-9}$ \\[0.7mm]
\cline{2-3}
\noalign{\vskip 0.7mm}
 & Original calculation
 & Scaled estimate ($\mathcal{Q}=1$) \\[0.7mm]
\cline{2-3}
\noalign{\vskip 0.7mm}
Quantum efficiency $\eta_{\scriptscriptstyle E}$
(electrons/photon)
 & $1.6 \times 10^{-9}$
 & $3.0 \times 10^{-12}$ \\
Responsivity $\mathcal{R}_{\scriptscriptstyle E}$ (A/W)
 & $1.0 \times 10^{-9}$
 & $2.6 \times 10^{-12}$ \\[0.7mm]
\hline
\end{tabular}
\end{table}

A direct comparison of the theory with experiment would entail recomputing
Eqs.~(\ref{iepcurrent})--(\ref{ieRESPeta}) using
the source and material parameters
appropriate to Kobayashi \emph{et al.}'s experiment. These equations represent the
photocurrent $i_{\scriptscriptstyle E}$ and its link to the quantum efficiency
$\eta_{\scriptscriptstyle E}$ and responsivity $\mathcal{R}_{\scriptscriptstyle E}$.
However, an approximate comparison can be constructed far more simply --- by
separating the explicit entanglement area--time factor
from the remaining material-response factor in
Eq.~(\ref{iepcurrent}) and assuming that the
CsK$_2$Sb material parameters (including $\beta$, $d$,
$\xi$, and $T_{\scriptscriptstyle A}$)
remain fixed at the values used in the original model.
For degenerate SPDC, the explicit frequency combination
$\omega_p^2/(4\omega_1^0\omega_2^0)$ in
Eq.~(\ref{iepcurrent}) equals unity.  Still, the material-response integral
in Eq.~(\ref{iepcurrent}), which relies on $F(k,T_{\scriptscriptstyle E})$ in
Eq.~(\ref{iepcurrentF}), retains a dependence on frequency through the intermediate-state
denominators, the accessible electronic states, and the integration limits; it also
depends on  $T_{\scriptscriptstyle E}$.

To accommodate this frequency dependence as the excitation changes
from 812~nm for Lissandrin \emph{et al.}'s model to 1064~nm
for Kobayashi \emph{et al.}'s experiment,
we employ Eqs.~(\ref{iepcurrent}) and (\ref{efluxtopflux}) to calculate
the ratio \,$\eta_{\scriptscriptstyle
E}^\text{\sc kob}/\eta_{\scriptscriptstyle E}^\text{\sc lis}$ [the individual efficiencies are
normalized to an optical-system transmittance of unity,
as defined in Eq.~(\ref{imathTOmueRed})].
Removing  the explicit
$1/A_{\scriptscriptstyle E}T_{\scriptscriptstyle E}$
factors, holding the common material parameters fixed, and defining
$\mathcal{Q}$ as the dimensionless ratio of the
remaining response factors for the Kobayashi and
Lissandrin parameter sets then leads to
\begin{equation}
\label{koblisratio}
\frac{
\eta_{\scriptscriptstyle E}^{\textsc{kob}}
}{
\eta_{\scriptscriptstyle E}^{\textsc{lis}}
}
=
\frac{
(A_{\scriptscriptstyle E}T_{\scriptscriptstyle E})
^{\textsc{lis}}
}{
(A_{\scriptscriptstyle E}T_{\scriptscriptstyle E})
^{\textsc{kob}}
}
\,\mathcal{Q}\,.
\end{equation}
Equation~(\ref{koblisratio}) effectively renormalizes the parameters in
Lissandrin~\emph{et al.}'s  original calculations to the parameters in
Kobayashi~\emph{et al.}'s experiment.
The factor $\mathcal{Q}$ accommodates changes in the
response integral that arise from both the excitation
frequency and the adopted entanglement time; still, determining its explicit value
requires recalculation of the full model.

As an example of the use of Eq.~(\ref{koblisratio}), we make use of the parameters from
Table~\ref{kobaliss} and set
$\mathcal{Q}=1$, which leads to a scaled estimate  for the quantum efficiency given by
\begin{equation} \label{qescaledest}
\eta_{\scriptscriptstyle E}^{\textsc{kob}}\:(\mathcal{Q}=1)
\approx
1.6 \times 10^{-9} \times
\frac{1.0 \times 10^{-24}}{5.4 \times 10^{-22}}  \times 1
\approx 3.0 \times 10^{-12}.
\end{equation}
At $\lambda=1064$~nm, the corresponding scaled estimate for the responsivity is then
\begin{equation} \label{respscaledest}
\mathcal{R}_{\scriptscriptstyle E}^{\textsc{kob}}\:(\mathcal{Q}=1)
=
\frac{e\lambda}{hc}\,
\eta_{\scriptscriptstyle E}^{\textsc{kob}}
\approx 2.6 \times 10^{-12}~\mathrm{A/W}.
\end{equation}
These illustrative estimates, obtained by
rescaling the explicit entanglement area--time
factor in Lissandrin~\emph{et al.}'s theory and assuming $\mathcal{Q}=1$, are presented
in Table~\ref{thexpetpp}; they have also been inserted in column~3 of
Table~\ref{kobaliss}.  Experimentally inferred values for
these quantities, provided in Tables~\ref{respqeadKobEP} and \ref{compare}, have also
been carried over to Table~\ref{thexpetpp} to facilitate the comparison
between theory and experiment.

\begin{table}[htb!]
\caption{
Experimentally inferred responsivity
$\mathcal{R}_{\scriptscriptstyle E}$ and quantum efficiency
$\eta_{\scriptscriptstyle E}$ for entangled-two-photon
photoemission from CsK$_2$Sb at
$\mathsfit{T\/}=27~^\circ$C, compared with
scaled estimates of these quantities ($\mathcal{Q}=1$) calculated via
Eqs.~(\ref{respscaledest}) and (\ref{qescaledest}), respectively.
The experimental values are carried over from
Tables~\ref{respqeadKobEP} and \ref{compare}.   }
\label{thexpetpp}
\centering
\small
\renewcommand{\arraystretch}{1.15}
\begin{tabular}{@{}lcc@{}}
\cline{2-3}
\noalign{\vskip 1.3mm}
 & \shortstack{Responsivity\\
 $\mathcal{R}_{\scriptscriptstyle E}$ (A/W)}
 & \shortstack{Quantum efficiency\\
 $\eta_{\scriptscriptstyle E}$ (electrons/photon)}
 \\[0.7mm]
\hline
\noalign{\vskip 0.7mm}
Experimentally inferred
 & $2.0 \times 10^{-13}$
 & $2.3 \times 10^{-13}$ \\
Scaled estimates ($\mathcal{Q}=1$)
 & $2.6 \times 10^{-12}$
 & $3.0 \times 10^{-12}$ \\
\noalign{\vskip 0.7mm}
\hline
\end{tabular}
\end{table}

\begin{framed}\noindent
The scaled estimate of the entangled-two-photon
quantum efficiency calculated from the theory of
Lissandrin \emph{et al.}~\cite{lissandrin2004}  with $\mathcal{Q}=1$
is about a factor of 13 larger than the inferred value measured
in the experiment of Kobayashi \emph{et al.}~\cite{kobayashi2007}.
\end{framed}

The disparity could be the result of a number of factors,
including: 1)~a response ratio $\mathcal{Q} \lessgtr 1$ (agreement with experiment
would require
$\mathcal{Q}\approx0.077$, if the other assumed parameter values are retained);\,
2)~uncertainties in the source and photocathode parameters;\, and 3)~experimental error.
Moreover, the area--time product for the
experiment was inferred using the heuristic crossover model (Sec.~\ref{measidencrossPFD}),
rather than  being independently measured.
While the comparison of theory and experiment  presented here thus offers
a valuable indication of consistency, it does not provide a fully independent
validation of either the experimental interpretation or the theoretical prediction,
which would require an independent  characterization of the
two-photon field at the photocathode and an evaluation of the full material response
at 1064~nm (the wavelength at which the experiment was conducted).

\section{SUBTHRESHOLD ABSORPTION AND FLUORESCENCE
MICROSCOPY\\ UNDER ENTANGLED-PHOTON-PAIR ILLUMINATION}\label{entabsfluor}

The three forms of entangled-photon-pair initiated subthreshold photoemission
examined in Secs.~\ref{theory}, \ref{measiden}, \ref{entPMT}, and \ref{entchannel}
have direct counterparts in entangled-photon-pair initiated subthreshold absorption:
    \begin{enumerate}
      \item Individual-SPDC-photon induced Fermi-tail photoemission
      (FTP)~[\begin{tikzpicture}[baseline={(0,0)}]
\filldraw[draw=black, fill={rgb,255:red,238; green,188; blue,187}] (0,0) rectangle +(1.2ex,1.2ex);
\end{tikzpicture} Fig.~\ref{fig23}] $\Leftrightarrow$ Individual-SPDC-photon induced Boltzmann-tail
absorption (BTA)~\cite{jimenez2022}.
      \item Twin-induced entangled-two-photon photoemission
          (ETPP)~[\begin{tikzpicture}[baseline={(0,0)}] \filldraw[draw=black,
          fill={rgb,255:red,255; green,255; blue,255}] (0,0) rectangle +(1.2ex,1.2ex);
\end{tikzpicture} Fig.~\ref{fig26}] $\Leftrightarrow$ Twin-induced entangled-two-photon
absorption (ETPA)~\cite{georgiades1995PRL}.
      \item Cousin and singleton-pair induced two-photon photoemission
          (TPP)~[\begin{tikzpicture}[baseline={(0,0)}] \filldraw[draw=black,
          fill={rgb,255:red,221; green,5; blue,88}] (0,0) rectangle +(1.2ex,1.2ex);
\end{tikzpicture} Fig.~\ref{fig26}] $\Leftrightarrow$ Cousin and singleton-pair induced
two-photon absorption (TPA)~\cite{raymer2024PRA}.
    \end{enumerate}

As articulated in earlier sections of this review, subthreshold photoemission is readily
characterized by measures that relate the flux of emitted photoelectrons to the flux
of incident photons, i.e., the quantum efficiencies $\eta_{\scriptscriptstyle F,E,C}$
and the corresponding responsivities $\mathcal{R}_{\scriptscriptstyle F,E,C}$\,. [The
relation between the dimensionless quantum efficiency and the responsivity (A/W) is
provided in Eq.~(\ref{iSRESPeta}).] These quantities were used in
Sec.~\ref{fosteringent} to craft expressions for the total mean photocurrent
$\,\overline{\imath}_{\scriptscriptstyle TE}\,$ and its digital counterpart, the total
mean photoelectron count rate \,$\overline{\mu}_{\scriptscriptstyle TE}$. [The
relation between the two-photon responsivity coefficient
$\mathcal{L}_{\scriptscriptstyle C}$ \,(A\,m$^2$/W$^2$)\, and the two-photon
responsivity $\mathcal{R}_{\scriptscriptstyle C}$ \,(A/W)\, is provided in
Eq.~(\ref{eqRLI}).]

Subthreshold absorption, on the other hand, is typically characterized by measures
that relate the rate of absorption events $R_{\scriptscriptstyle TE}$ to the incident
photon-flux density, i.e., the cross sections $\sigma_{\!\scriptscriptstyle B,E}$
\,[m$^2$]\, that do so linearly and the cross section $\sigma^{(2)}$ \,[m$^4$\,s]\,
that does so quadratically. Subthreshold fluorescence microscopy, a form of
fluorescence-mode subthreshold absorption, is characterized similarly. Weaker forms
of subthreshold absorption, such as higher-order multiphoton absorption,
may also occur but are beyond the scope of this review.

Section~\ref{EPparticlemodel} provides a review of the commonly used heuristic
framework of Fei \emph{et al.}~\cite{Fei97}, which is extended by incorporating
single-photon Boltzmann-tail absorption and optical loss.
Section~\ref{ETPAquantummodel} discusses quantum models of subthreshold
absorption and highlights the conditions under which the quantum outcomes
resemble, or differ from, those obtained from the heuristic model.
Section~\ref{ssec:ETFAFM} chronicles multiple
experimental reports of purported ETPA and ETPFM
and examines the unsuccessful efforts to replicate
these observations independently. Finally,
Section~\ref{impabmic} outlines a number of methodologies that could prove useful
for enhancing the observability of ETPA and for implementing ETPFM.

\subsection{Fei et al.'s Heuristic Model of Entangled-Two-Photon
Absorption}\label{EPparticlemodel}

We review and expand a simple probabilistic framework for phenomenologically
describing entangled-two-photon absorption at the level of a single absorber.
Introduced by Hong-Bing Fei \emph{et al.}~\cite{Fei97} in 1997, this approach treats
photons as particles and ignores interference effects.
Two-photon absorption is represented schematically
as a transition from the ground state to the final
state via an intermediate state, which may be real
or virtual; a virtual intermediate state does not
correspond to a separately observable absorption event.
 This framework is suitable for modeling both
transmission- and fluorescence-mode ETPA, as well as entangled-two-photon
fluorescence microscopy~\cite{teich1997}. It is also useful for ETPP.
The heuristic model provides a useful phenomenological
description but is not universally quantitative: the effective cross sections depend on the
spatiotemporal properties of the incident light and the material response, and the simple
linear--quadratic decomposition can require modification, particularly in the high-gain
regime~\cite{Schlawin2024}.

We begin by writing simple formulas that relate the absorption rate \,(s$^{-1}$)\, to
the incident photon-flux density \,(photons/m$^{2}$\,s)\,  in terms of the cross
sections for the three forms of subthreshold absorption discussed above. We do this
first in the absence of photon loss described by Bernoulli random deletion, and then
in its presence. We conclude by cataloging a number of experimental procedures that
can be used to identify the particular form of subthreshold absorption being
observed.

\subsubsection{Individual-SPDC-Photon Induced Boltzmann-Tail
Absorption (BTA)}\label{EPparticlemodel-01} The absorption rate
\,$R_{\scriptscriptstyle B}$\, for \textbf{Boltzmann-tail absorption (BTA)}, a
one-photon process, is directly proportional to the photon-flux density of the
entangled-photon source $\phi_{\scriptscriptstyle E}$. The BTA cross section
\,$\sigma_{\!\scriptscriptstyle B}$\, serves as the constant of proportionality, so that
\begin{equation}\label{RsubB}
  R_{\scriptscriptstyle B} = \sigma_{\!\scriptscriptstyle B} \phi_{\scriptscriptstyle E}.
\end{equation}
Here, $\phi_{\scriptscriptstyle E}$ denotes the photon-flux density, counting both photons
of each SPDC pair, rather than the pair-flux density $\phi_{\mathrm{pair}}$
($\phi_{\scriptscriptstyle E}=2\phi_{\mathrm{pair}}$ for a lossless source).
The effective cross section $\sigma_{\!\scriptscriptstyle B}$ depends on
the thermal population of the relevant initial
states and the illumination wavelength.

\subsubsection{Twin-Induced Entangled-Two-Photon
Absorption (ETPA)}\label{EPparticlemodel-02}

For the simplified version of \textbf{entangled-two-photon absorption (ETPA)} in
which the temporal and spatial aspects of the process are treated as independent,
the absorption rate is proportional to a product of two functions:
\begin{enumerate}
  \item The probability $\xi(T_{\scriptscriptstyle E})$
  that the relative arrival-time separation of the
  twin photons at the absorber falls within the
  characteristic response time
  $T_{\scriptscriptstyle A}$ of the absorber.

  \item The probability $\zeta(A_{\scriptscriptstyle E})$
  that the twin photons interact within the same
  effective transverse absorption area
  $\sigma_1$, given their characteristic
  entanglement area $A_{\scriptscriptstyle E}$.
\end{enumerate}
The effective absorption cross section under entangled-photon illumination then
becomes
\begin{equation}\label{xisigmae}
  \sigma_{\!\scriptscriptstyle E}
  = \sigma_{1}\,\xi(T_{\scriptscriptstyle E})\,\zeta(A_{\scriptscriptstyle E}),
\end{equation}
and the entangled-two-photon absorption rate takes the form
\begin{equation}\label{RsubE}
  R_{\scriptscriptstyle E} = \sigma_{\!\scriptscriptstyle E} \phi_{\scriptscriptstyle E}.
\end{equation}
Here, $\sigma_{\!\scriptscriptstyle E}$ is an effective
cross section referenced to the individual-photon flux density.
The normalization adopted in Eq.~(\ref{xisigmae}) constitutes
a heuristic modeling assumption, as discussed in
Sec.~\ref{EPparticlemodel-03}.

\subsubsection{Cousin and Singleton-Pair Induced Two-Photon
Absorption (TPA)}\label{EPparticlemodel-03}

Photons belonging to different SPDC pairs can
participate in two-photon absorption even when
the twins within each pair remain correlated.
Within the heuristic model, the characteristic
interaction volume is represented by
\begin{equation}\label{intvol}
  V_\mathrm{int} = cT_{\scriptscriptstyle A}\sigma_{1}\,.
\end{equation}
As the photon-flux density increases, accidental
coincidences involving photons from different pairs become increasingly important.
Such events, involving cousins or surviving singleton photons, contribute to conventional
\textbf{two-photon absorption (TPA)}, represented  phenomenologically by
\begin{equation}\label{deltasubrRr}
  R_{\scriptscriptstyle C} = \sigma^{(2)} \Gamma \phi_{\scriptscriptstyle E}^2\,.
\end{equation}
Here, $\sigma^{(2)}$ is the two-photon absorption cross section, usually specified in
Goeppert-Mayer (GM) units (1 GM $\equiv 10^{-58}\,\text{m}^4\,\text{s}/$photon),
and $\Gamma$ is the enhancement factor associated with a pulsed
entangled-photon source. As described in Sec.~\ref{2Qphotstatspulse}, a pulsed
source comprising a periodic sequence of rectangular optical pulses of duration
$\tau_0$ and repetition rate $f_\mathrm{rep} = 1/\tau_1$ has a duty cycle $\Delta
= \tau_0 f_\mathrm{rep}$. In accordance with Eq.~(\ref{eq:equivfactor5}), the
pulse-enhancement factor is then $\Gamma = 1/\Delta$.
Equation~(\ref{deltasubrRr}) adopts the conventional independent-photon quadratic
contribution, including the pulse-enhancement factor $\Gamma$.
For non-Poisson illumination,
the coefficient can also depend on intensity correlations, as will be discussed in
Sec.~\ref{ETPAquantummodelBSV}.
 In the special case of CW entangled-photon illumination, we have $\Gamma =
1$, whereupon Eq.~(\ref{deltasubrRr}) reduces to
\begin{equation}\label{deltasubrRrNoGamma}
  R_{\scriptscriptstyle C} = \sigma^{(2)} \phi_{\scriptscriptstyle E}^2.
\end{equation}

Within the particlelike heuristic model, the
two-photon absorption cross section is estimated
as the product of two effective one-photon
cross sections and a characteristic absorber
response time $T_{\scriptscriptstyle A}$~\cite{mainfray1984},
giving
\begin{equation}\label{deltasubr}
  \sigma^{(2)} = \sigma_{1}^2\, T_{\scriptscriptstyle A}\,.
\end{equation}
In the regime where $T_{\scriptscriptstyle E} < T_{\scriptscriptstyle A}$ and
$A_{\scriptscriptstyle E} < \sigma_{1}$, the arrival probabilities set forth in
Sec.~\ref{EPparticlemodel-02} behave as $\xi(T_{\scriptscriptstyle E}) \to 1$ and
$\zeta(A_{\scriptscriptstyle E}) \to 1$, so that Eqs.~(\ref{xisigmae}) and (\ref{deltasubr})
yield an effective entangled-photon absorption cross section given by
\begin{equation}\label{deltasubralt}
  \sigma_{\!\scriptscriptstyle E} = \frac{\sigma^{(2)}}{\sigma_{1} T_{\scriptscriptstyle A}}
  \qquad \quad T_{\scriptscriptstyle E} < T_{\scriptscriptstyle A},\,\,\,
  A_{\scriptscriptstyle E} <
  \sigma_{1}.
\end{equation}

On the other hand, in the (usually experimentally relevant) regime where
  $T_{\scriptscriptstyle E} \geqslant T_{\scriptscriptstyle A}$ and
  $A_{\scriptscriptstyle E} \geqslant \sigma_{1}$, the arrival probabilities stipulated in
  Sec.~\ref{EPparticlemodel-02} are estimated to be $\xi(T_{\scriptscriptstyle E}) =
  T_{\scriptscriptstyle A}/T_{\scriptscriptstyle E}$ and $\zeta(A_{\scriptscriptstyle E})
  = \sigma_{1}/A_{\scriptscriptstyle E}$, whereupon Eqs.~(\ref{xisigmae}) and
  (\ref{deltasubr}) yield the cross-section relation \begin{equation}\label{sigdelAT}
  \sigma_{\!\scriptscriptstyle E}
   = \frac{\sigma^{(2)}}{A_{\scriptscriptstyle E} T_{\scriptscriptstyle E}}
   \qquad \quad T_{\scriptscriptstyle E} \geqslant T_{\scriptscriptstyle A},\,\,\,
   A_{\scriptscriptstyle E} \geqslant \sigma_{1}.
\end{equation}
Equations~(\ref{deltasubralt}) and (\ref{sigdelAT}) relate the effective
entangled-two-photon absorption cross section per absorber
\,$\sigma_{\!\scriptscriptstyle E}$\, to the conventional two-photon absorption
cross section per absorber \,$\sigma^{(2)}$; a large value of $\sigma^{(2)}$ implies a
large value of $\sigma_{\!\scriptscriptstyle E}$ in both cases.
Equation~(\ref{sigdelAT}) is algebraically analogous to Eq.~(\ref{sigAETE}) for the
normalized entangled-two-photon cross section $\delta_{\scriptscriptstyle E}$.
However, Eq.~(\ref{sigdelAT}) is designed to enable $\sigma_{\!\scriptscriptstyle E}$
to be determined from an established value of $\sigma^{(2)}$, whereas
Eq.~(\ref{sigAETE}) is designed to simply provide a complete description of
$\sigma_{\!\scriptscriptstyle E}$, using the symbol $\delta_{\scriptscriptstyle E}$ to
do so. A generalization of these results suitable for the description of
entangled-multiphoton absorption was set forth by Pe{\v r}ina \emph{et al.} in
1998~\cite{Perina98}.

It is useful to highlight three distinctions between the calculations presented here
and those employed in the earlier version of this model reported by Fei \emph{et
al.}~\cite{Fei97}: \,1)~the definition of \,$\sigma_{\!\scriptscriptstyle E}$\, used in
Eq.~(\ref{xisigmae}) is a factor of two greater than that employed by Fei \emph{et
al.}, so that the right-hand side of Eq.~(\ref{sigdelAT}) is a factor of two greater than
that of Eq.~(2) in~\cite{Fei97}; \,2)~the absorption rate $R_{\scriptscriptstyle C}$
and the two-photon absorption cross section $\sigma^{(2)}$ specified in
Eqs.~(\ref{deltasubrRr}) and (\ref{deltasubr}) refer to illumination by
independent  entangled-photon cousins and singleton pairs, whereas the absorption rate $R_r$
and the two-photon absorption cross section $\delta_r$ presented in Eq.~(1)
of~\cite{Fei97} represent illumination by randomly arriving pairs of
randomly arriving photons; \,3)~the absorption rate
$R_{\scriptscriptstyle C}$ specified in Eq.~(\ref{deltasubrRr}) accommodates a pulsed
source of entangled-photon pairs via the parameter $\Gamma$, whereas the
absorption rate $R_r$ in Eq.~(1) of~\cite{Fei97} represents randomly arriving photons
without an explicit pulse-enhancement factor.
Since the photon-flux density counts individual photons in both
formulations, the factor-of-two difference is not attributable to
the distinction between photon and photon-pair flux densities.
Rather, the prefactor adopted here represents a modeling assumption
that corresponds to $\alpha=1$ in the more general expression
presented in Sec.~\ref{ETPAquantummodelCrossRate}.

\subsubsection{Overall Subthreshold Absorption Rate}\label{EPparticlemodel-04}

We now consider the overall \textbf{subthreshold
absorption rate} for an individual absorber
illuminated by a nondepleted source of pulsed
or CW entangled photons. To facilitate comparison
with measured signals, we initially include
an effective background rate
$R_{\scriptscriptstyle D}$, expressed in
equivalent absorption-event rate units per
absorber. The resulting rate is
\begin{equation}\label{totalrate}
   R_{\scriptscriptstyle TE} = R_{\scriptscriptstyle D} +
  R_{\scriptscriptstyle B} + R_{\scriptscriptstyle E} + R_{\scriptscriptstyle C}\,,
\end{equation}
where the four terms on the right-hand side of Eq.~(\ref{totalrate}) represent,
respectively, contributions to the rate from dark/circuit noise, Boltzmann-tail
absorption (BTA), entangled-two-photon absorption (ETPA), and two-photon
absorption (TPA). Equation~(\ref{totalrate}) is analogous to
Eq.~(\ref{muimathTOmue}) for the ETPP photoelectron count rate
$\overline{\mu}_{\scriptscriptstyle TE}$.

The background rate $R_{\scriptscriptstyle D}$
is independent of $\phi_{\scriptscriptstyle E}$
and can arise from detector dark counts,
electronic noise, or unrelated fluorescence.
Unlike the other terms in Eq.~(\ref{totalrate}),
it does not represent absorption by the sample.
Its magnitude depends on the experimental
configuration and need not be negligible.
For the following analysis, we assume that
this background has been subtracted or is
sufficiently small as to be neglected.
Using Eqs.~(\ref{RsubB}), (\ref{RsubE}),
and (\ref{deltasubrRr}) then yields
\begin{eqnarray}\label{fei}
 R_{\scriptscriptstyle TE} & = & \,\,\,R_{\scriptscriptstyle B}\,\,\,\, +\,\,\,
 R_{\scriptscriptstyle E} \,\,\,+ \,\,\, R_{\scriptscriptstyle C}\\
   &=& \sigma_{\!\scriptscriptstyle B} \phi_{\scriptscriptstyle E} +
   \sigma_{\!\scriptscriptstyle E} \phi_{\scriptscriptstyle E} + \sigma^{(2)}
   \Gamma \phi_{\scriptscriptstyle E}^2.  \label{fei3}
\end{eqnarray}

Still, the results provided in Eq.~(\ref{fei3}) assume that there is no loss between the
source of entangled-photon pairs and the absorber, i.e., that all of the intervening
optical components have unity optical transmittance ($\mathcal{T\/}_{\!\!0} = 1$).
Since optical loss affects the three forms of subthreshold absorption differently,
however, and since it is unavoidable in real systems, it is essential to explicitly
incorporate it into the formula for the overall absorption rate. Relying on the
discussion provided in Sec.~\ref{optlosssys}, and guided by the entries in
Table~\ref{tab:optlosssys}, incorporating the intrinsic optical transmittance
$\mathcal{T\/}_{\!\!0}$ in Eq.~(\ref{fei3}) converts it to
\begin{eqnarray}\label{feiLOSS}
  R_{\scriptscriptstyle TE} &=& \sigma_{\!\scriptscriptstyle B} (\mathcal{T\/}_{\!\!0}\,
  \phi_{\scriptscriptstyle E}) + \sigma_{\!\scriptscriptstyle E} (\mathcal{T\/}_{\!\!0}^2
  \phi_{\scriptscriptstyle E}) + \sigma^{(2)} \Gamma (\mathcal{T\/}_{\!\!0}\,
  \phi_{\scriptscriptstyle E})^2 \label{feiLOSS01}\\
  &=& \sigma_{\!\scriptscriptstyle B} \phi + \mathcal{T\/}_{\!\!0}\,
  \sigma_{\!\scriptscriptstyle E} \phi + \sigma^{(2)}  \Gamma \phi^2, \label{feiLOSS02}
\end{eqnarray}
where $\phi  = \mathcal{T\/}_{\!\!0}\, \phi_{\scriptscriptstyle E}$ is the photon-flux
density at the absorber.

\subsubsection{Detectability Limits Imposed by Photon Loss and Absorption}
\label{degradation}

Focusing on the observation of entangled-two-photon absorption, we see from
Eq.~(\ref{feiLOSS02}) for the total absorption rate $R_{\scriptscriptstyle TE}$ that
the detection threshold for ETPA is limited not only by optical loss characterized by
the transmittance $\mathcal{T\/}_{\!\!0}$, but also by the additive one- and
two-photon absorption contributions $\sigma_{\!\scriptscriptstyle B} \phi$ and
$\sigma^{(2)}  \Gamma \phi^2$, respectively:
\begin{itemize}

\item At fixed source photon-flux density
$\phi_{\scriptscriptstyle E}$, optical loss reduces
ETPA and TPA by the same factor
$\mathcal{T\/}_{\!\!0}^2$, whereas BTA is
reduced only by $\mathcal{T\/}_{\!\!0}$.
At fixed photon-flux density $\phi$ at the
absorber, on the other hand, the ETPA contribution carries
an additional factor $\mathcal{T\/}_{\!\!0}$
relative to both other contributions,
as shown in Eq.~(\ref{feiLOSS02})
 ($\sigma_{\!\scriptscriptstyle E}$ is the
      analog of $\mathcal{R}_{\scriptscriptstyle E}$).

\item One-photon Boltzmann-tail absorption (BTA),
which is analogous to FTP in subthreshold
photoemission, produces a competing linear
absorption contribution
$\sigma_{\!\scriptscriptstyle B}\phi$
that can mimic ETPA
($\sigma_{\!\scriptscriptstyle B}$ is the
analog of $\mathcal{R}_{\scriptscriptstyle F}$).

\item Independent two-photon absorption (TPA),
which is analogous to TPP in subthreshold
photoemission, produces a competing quadratic
absorption contribution
$\sigma^{(2)}\Gamma\phi^2$
that dominates ETPA at sufficiently high values of
$\phi$ within the heuristic model
($\sigma^{(2)}$ is the analog of
$\mathcal{L}_{\scriptscriptstyle C}$).

\end{itemize}
Within the heuristic model, the detectability of ETPA is optimized by maximizing
$\sigma_{\!\scriptscriptstyle E}$ and $\mathcal{T\/}_{\!\!0}~(\leqslant 1)$, while
minimizing $\sigma_{\!\scriptscriptstyle B}$ and $\Gamma$
\,[$\sigma_{\!\scriptscriptstyle E} \propto \sigma^{(2)}$, in accordance with
Eq.~(\ref{sigdelAT})].

\subsubsection{Linear-Quadratic Crossover Photon-Flux Densities}
\label{measidencrossPFD}

When the photon-flux density at the absorber is sufficiently large, it is clear from
Eq.~(\ref{feiLOSS02}) that the two-photon absorption rate, which scales
quadratically with $\phi$, dominates the Boltzmann-tail and entangled-two-photon
absorption rates, which scale linearly with $\phi$. The \textbf{crossover photon-flux
density} $\phi_{\scriptscriptstyle EC}$, at which the linear and quadratic
contributions to the rate are equal, is therefore established by setting
\,$(\sigma_{\!\scriptscriptstyle B} + \mathcal{T\/}_{\!\!0}\,
\sigma_{\!\scriptscriptstyle E}) \phi_{\scriptscriptstyle EC} = \sigma^{(2)} \Gamma
\phi_{\scriptscriptstyle EC}^2$\,, which yields
\begin{equation}\label{IFEcross+EC}
  \phi_{\scriptscriptstyle EC} = \frac{\sigma_{\!\scriptscriptstyle B} +
  \mathcal{T\/}_{\!\!0}\, \sigma_{\!\scriptscriptstyle E}}{\sigma^{(2)} \Gamma}\,.
\end{equation}
Equation~(\ref{IFEcross+EC}) is analogous to Eq.~(\ref{IFEcross}) for subthreshold
photoemission (the crossover photon-flux density and the crossover intensity are
related by $\phi_{\scriptscriptstyle EC} = I_{\scriptscriptstyle EC}/h\nu$). In fact, all
of the formulas presented in this section are analogous to those provided in
Sec.~\ref{fosteringentCROSS}.

In the absence of Boltzmann-tail absorption ($\sigma_{\!\scriptscriptstyle B} \to 0$),
 Eq.~(\ref{IFEcross+EC}) reduces to
\begin{equation}\label{IFEcross+ECnoB}
  \phi_{\scriptscriptstyle EC} = \frac{\mathcal{T\/}_{\!\!0}\,
  \sigma_{\!\scriptscriptstyle E}}{\sigma^{(2)} \Gamma },
\end{equation}
which, when combined with Eq.~(\ref{sigdelAT}), yields
\begin{equation}\label{sigdelATtophiEC}
  \phi_{\scriptscriptstyle EC}
    = \frac{\mathcal{T\/}_{\!\!0} / \Gamma}{A_{\scriptscriptstyle E} T_{\scriptscriptstyle E}}
    \qquad \quad T_{\scriptscriptstyle E} \geqslant T_{\scriptscriptstyle A},\,\,\,
    A_{\scriptscriptstyle E} \geqslant \sigma_{1}.
\end{equation}
(This relation is also applicable for entangled-two-photon photoemission in the
absence of Fermi-tail photoemission. Indeed, it is carried over to
Sec.~\ref{Kobacomparecrosss} as Eq.~(\ref{phiaete}), where it is used to provide an
estimate for the entanglement time $T_{\scriptscriptstyle E}$ associated
with a particular entangled-two-photon photoemission experiment.)

In the absence of entangled-two-photon absorption, we have
$\sigma_{\!\scriptscriptstyle E} \to 0$, so that Eq.~(\ref{IFEcross+EC}) reduces to
\begin{equation}\label{IFEcross+BC}
  \phi_{\scriptscriptstyle BC} = \frac{\sigma_{\!\scriptscriptstyle B}}{\sigma^{(2)} \Gamma}.
\end{equation}
Within the present heuristic model, Eq.~(\ref{IFEcross+BC}) applies to both
coherent and entangled-photon-pair illumination,
provided that the effective quadratic coefficients and other relevant experimental
parameters are matched. Quantum-statistical
corrections are considered in Sec.~\ref{ETPAquantummodel}.

The crossover photon-flux densities $\phi_{\scriptscriptstyle EC}$ and
$\phi_{\scriptscriptstyle BC}$, specified in Eqs.~(\ref{IFEcross+EC}) and
(\ref{IFEcross+BC}), respectively, are useful measures for determining parameter
values and for identifying the form of subthreshold absorption under observation.
Indeed, in the course of carrying out experiments to observe entangled-two-photon
absorption, it is often useful to conduct controlled companion experiments that
make use of coherent light or samples that are unresponsive to entangled-photon
pairs. It is then convenient to relate $\phi_{\scriptscriptstyle EC}$ for the
entangled-photon experiment to $\phi_{\scriptscriptstyle BC}$ for the control
experiment. If the operative experimental parameters are the same for both,
combining Eqs.~(\ref{IFEcross+EC}) and (\ref{IFEcross+BC})  leads to
\begin{equation}\label{IFEcrossFCphi}
  \phi_{\scriptscriptstyle EC}  = \phi_{\scriptscriptstyle BC}\left( 1 +
  \frac{\mathcal{T\/}_{\!\!0}\,\sigma_{\!\scriptscriptstyle E}}
  {\sigma_{\!\scriptscriptstyle B}}\right)\,,
\end{equation}
which is analogous to Eq.~(\ref{IFEcrossFC}) for subthreshold photoemission. The
crossover flux density for entangled-photon pairs $\phi_{\scriptscriptstyle EC}$ is
larger than the crossover flux density $\phi_{\scriptscriptstyle BC}$ because of the
additional contribution from entangled-two-photon absorption.

\begin{figure}[htb!]
\centering\includegraphics[width=3in]{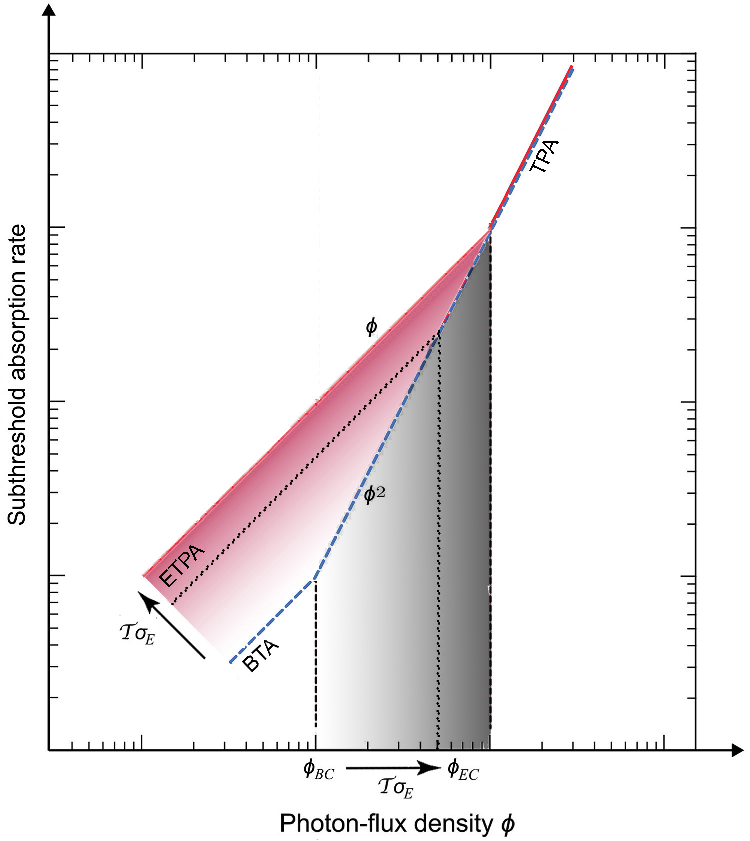}
\caption{Subthreshold absorption rate $R_{\scriptscriptstyle TE}$ vs. photon-flux
density $\phi$ at an absorber illuminated by entangled-photon pairs (graded red
and shaded gray regions).
As specified in Eq.~(\ref{feiLOSS02}), the total absorption rate comprises three
contributions: 1)~Boltzmann-tail absorption (BTA) induced by individual SPDC photons
$(\propto \sigma_{\!\scriptscriptstyle B} \phi$); 2)~entangled-two-photon
absorption (ETPA) induced by twins $(\propto \mathcal{T\/}_{\!\!0}\,
\sigma_{\!\scriptscriptstyle E} \phi)$; and 3)~two-photon absorption (TPA) induced
by independent cousins and/or singleton pairs $\smash{(\propto \sigma^{(2)}
\Gamma \phi^2)}$.
The degree of ETPA admixture is quantified by the magnitude of $\mathcal{T\/}_{\!\!0}\,
\sigma_{\!\scriptscriptstyle E}$ and is visually represented by increasing
saturation of the  red shading.
The photon-flux-density region $\phi_{\scriptscriptstyle BC} <  \phi  <
\phi_{\scriptscriptstyle EC}$ has a width
$\phi_{\scriptscriptstyle EC}-\phi_{\scriptscriptstyle BC} =
\mathcal{T\/}_{\!\!0}\,\sigma_{\!\scriptscriptstyle E} /(\sigma^{(2)}\Gamma)$ that increases with
increasing $\mathcal{T\/}_{\!\!0}\, \sigma_{\!\scriptscriptstyle E}$
 and is represented visually by the increasing darkness of the gray shading.
A representative value of $\mathcal{T\/}_{\!\!0}\,\sigma_{\!\scriptscriptstyle E}$, along with
the corresponding value of $\phi_{\scriptscriptstyle EC}$, are indicated by the
dotted lines.  This figure is a refined version of Fig.~\ref{fig2} that incorporates optical loss and an
arbitrary level of ETPA.
It mimics Fig.~\ref{fig12} for subthreshold photoemission.} \label{fig27}
\end{figure}

Equation~(\ref{IFEcrossFCphi}) informs us that the distinction between
entangled-two-photon and Boltzmann-tail absorption is perhaps most clearly
manifested in the photon-flux-density region
\begin{equation}\label{IFCECphi}
  \phi_{\scriptscriptstyle BC} <  \phi  <  \phi_{\scriptscriptstyle EC}\,.
\end{equation}
Within this region, the linear contribution
dominates the total absorption rate when ETPA
is present, whereas the quadratic contribution
dominates when ETPA is absent, as illustrated
in Fig.~\ref{fig27}. The corresponding
log--log slopes approach unity and two,
respectively, away from the crossover points.

The difference between the crossover parameters, calculated by combining
Eqs.~(\ref{IFEcross+EC}) and  (\ref{IFEcross+BC}), is
\begin{equation} \label{ICEdiffphi}
\phi_{\scriptscriptstyle EC}-\phi_{\scriptscriptstyle BC} =
\frac{\mathcal{T\/}_{\!\!0}\,\sigma_{\!\scriptscriptstyle E}}{\sigma^{(2)}\Gamma}\,,
\end{equation}
while their ratio, obtained from Eq.~(\ref{IFEcrossFCphi}),
is given by
\begin{equation}\label{ICEratiophi}
  \frac{\phi_{\scriptscriptstyle EC}}{\phi_{\scriptscriptstyle BC}} = 1 +
  \frac{\mathcal{T\/}_{\!\!0}\,\sigma_{\!\scriptscriptstyle E}}
  {\sigma_{\!\scriptscriptstyle B}}\,.
\end{equation}
With $\sigma_{\!\scriptscriptstyle E}$, $\mathcal{T\/}_{\!\!0}$,
and $\Gamma$ held fixed, Eq.~(\ref{ICEdiffphi}) shows that the
width of the crossover interval varies inversely with $\sigma^{(2)}$.
If the heuristic relation in Eq.~(\ref{sigdelAT}) applies, however,
the width reduces to
$\mathcal{T\/}_{\!\!0}/(\Gamma A_{\scriptscriptstyle E}T_{\scriptscriptstyle E})$
and is independent of $\sigma^{(2)}$ for fixed illumination parameters.
In either case, $\sigma_{\!\scriptscriptstyle B}$ does not enter
the width specified in Eq.~(\ref{ICEdiffphi}). On the other hand, at fixed $\mathcal{T\/}_{\!\!0}\,\sigma_{\!\scriptscriptstyle E}$,
decreasing $\sigma_{\!\scriptscriptstyle B}$ increases the ratio of
the two crossover flux densities, as is clear from Eq.~(\ref{ICEratiophi}).

\subsubsection{Identifying the Form of Subthreshold Absorption}
\label{measidenidenABS}

Equations~(\ref{feiLOSS01})--(\ref{ICEratiophi}) and Fig.~\ref{fig27}, along with the templates
prescribed in Sec.~\ref{idenidenepp} for subthreshold photoemission, provide guides for
identifying the form of experimentally observed subthreshold absorption.

The photon-flux-scaling tests described below refer to varying the SPDC generation rate at
fixed optical transmittance. The attenuation tests instead refer to varying the transmittance
of filters placed after the SPDC source while holding its generation rate fixed.
Comparisons with coherent illumination presume matched photon spectra, illumination areas,
sample conditions, and detection procedures.
For the attenuation tests, the total optical transmittance
is $\mathcal{T\/}=\mathcal{T\/}_{\!\!0}\mathcal{T\/}_{\!\!f}$,
where $\mathcal{T\/}_{\!\!f}$ is the transmittance of the
inserted filters.
\begin{enumerate}
  \item If the measured absorption rate $R_{\scriptscriptstyle TE}$ scales linearly
      with the incident entangled photon-flux density $\phi$ over a given range of
      values, and has the same magnitude as the linear absorption rate measured
      using a coherent source of light over that same range of values,
      the response is consistent with individual-SPDC-photon induced
      Boltzmann-tail absorption being dominant.  In accordance with
      Eq.~(\ref{feiLOSS01}), the attenuation provided by filters inserted in the
      entangled-photon beam path then follows the filter transmittance
      $\mathcal{T\/}_{\!\!f}$ in that region.

  \item If the measured absorption rate $R_{\scriptscriptstyle TE}$ scales linearly
      with the incident entangled photon-flux density $\phi$ over a given range of
      values, and has a magnitude greater than the linear absorption rate measured
      using a coherent source of light over that same range of values,
      the excess response is consistent with an ETPA contribution in addition to BTA,
      provided that differences in one-photon absorption and instrumental backgrounds
      have been excluded.  In accordance with Eq.~(\ref{feiLOSS01}), the
      attenuation provided by filters inserted in the entangled-photon beam path then
      behaves as a mixture of $\mathcal{T\/}_{\!\!f}$ and  $\mathcal{T\/}_{\!\!f}^2$
      in that region.

\item If the measured absorption rate
$R_{\scriptscriptstyle TE}$ scales linearly with the SPDC generation rate but quadratically
with the transmittance of filters placed after the source, the response exhibits the
characteristic signatures expected for entangled-two-photon absorption in the
isolated-pair regime. Identification of ETPA additionally requires
independent constraints on BTA and other competing absorption and background processes.
Comparison with the magnitude of coherent-light quadratic absorption alone is insufficient
to establish ETPA.

\item If the measured absorption rate $R_{\scriptscriptstyle TE}$ scales quadratically
with the SPDC generation rate, the response is consistent with TPA dominance within the
heuristic model. Attenuation by filters placed after the source also produces quadratic scaling.
The quadratic coefficient need not equal that obtained with coherent illumination, since
quantum-statistical and modal correlations can modify the response, as discussed in
Sec.~\ref{ETPAquantummodel}.

  \item As \,$\mathcal{T\/}_{\!\!0}\, \sigma_{\scriptscriptstyle
      E}/\sigma_{\scriptscriptstyle B} \to 0$, indicating that the effective entangled-two-photon
      contribution becomes negligible relative to BTA, the two crossover densities coincide, as is
      clear from Eq.~(\ref{ICEratiophi}).  Individual-SPDC-photon induced Boltzmann-tail absorption
      then transitions directly into two-photon absorption at $\phi_{\scriptscriptstyle
      EC} = \phi_{\scriptscriptstyle BC}$, reproducing the results obtained for coherent
      light.
\end{enumerate}

\subsection{Quantum Models of Entangled-Two-Photon
Absorption}\label{ETPAquantummodel}

We now review the key results that emerge from the quantum theory of
entangled-two-photon absorption (ETPA), whose conceptual antecedent reaches
back to David Klyshko's 1982 analysis of two-photon processes using spontaneous
parametric downconversion (SPDC)~\cite{klyshko82}.
The emphasis in this review is on ETPA driven by
\textbf{isolated photon pairs} that exhibit
strong temporal correlations and, when generated
using a narrowband SPDC pump, frequency
anticorrelations. The combination of short
relative arrival times and a narrow sum-frequency
bandwidth has motivated proposals for low-flux
two-photon spectroscopy with enhanced temporal
and spectral selectivity and reduced photodamage.
However, the predicted ETPA rates for typical
molecular systems generally turn out to be too small for
practical observation.

Vacuum-seeded parametric downconversion at
\textbf{low gain} produces \textbf{weak squeezed
vacuum}, characterized by low photon occupation
per spatiotemporal mode and, in the isolated-pair
regime, negligible overlap between distinct
photon pairs. At \textbf{high gain}, the output
becomes \textbf{bright squeezed vacuum}, with
many photons per mode and strong signal--idler
photon-number correlations. We begin by examining
the low-gain predictions of the quantum models
developed by Dayan and Raymer~\emph{et al.}
(Sec.~\ref{ETPAquantummodelRaymer}),
Drago and Sipe (Sec.~\ref{ETPAquantummodelSipe}),
and Schlawin~\emph{et al.}
(Sec.~\ref{ETPAquantummodelSchlawin}).
 In Sec.~\ref{ETPAquantummodelBSV}, we
turn to ETPA quantum models suitable for the bright-squeezed-vacuum domain.
Section~\ref{ETPAquantummodelCrossRate} is devoted to comparing the
quantum-theoretical formulas for the subthreshold absorption rate and
entangled-two-photon cross section with the corresponding expressions obtained via
the heuristic model presented in Sec.~\ref{EPparticlemodel}. Finally, in
Sec.~\ref{paramestABS}, we discuss procedures for estimating the values of
subthreshold absorption parameters of interest.

\begin{framed}\noindent
In certain limits, the predictions of the in-depth quantum-theoretic models reviewed
here turn out to mirror those of the heuristic
model described in Sec.~\ref{EPparticlemodel}.
\end{framed}

\subsubsection{Models of Dayan and Raymer \emph{et al.}}
\label{ETPAquantummodelRaymer}

In the early 2020s, Michael Raymer, together with his students and colleagues, wrote
a series of papers that advanced the quantum theory of entangled-two-photon
absorption, with an emphasis on in-solution molecular systems of biological and
chemical interest.
Most of their calculations considered excitation
far from one-photon resonances with intermediate
molecular states, although their 2021 tutorial~\cite{raymer2021JCP}
also examined additional density-matrix excitation
pathways, including those enabled by dephasing.
Their calculations extended from the isolated-photon-pair (low-gain)
domain~\cite{raymer2020arxiv,raymer2021optica,raymer2021OE,raymer2021JCP}
to the bright-squeezed-vacuum (high-gain) domain~\cite{raymer2022PRA}. Their
approach is closely related to the analysis spearheaded by Dayan in
2007~\cite{dayan2007PRA}.

In the isolated-photon-pair domain, Raymer and Landes' theoretical treatment
reveals that it is possible, in principle, to significantly enhance the two-photon
absorption rate by making use of entangled-photon pairs generated by SPDC in place
of coherent light (of the same spectral bandwidth and photon flux). The quantum
enhancement emerges because the frequencies of the paired photons sum to the
frequency of the SPDC pump laser.
The frequency anticorrelations provide enhanced
spectral selectivity when the narrow sum-frequency
distribution overlaps the two-photon transition,
even though the individual photons occupy much
broader marginal spectra. This enhancement does
not necessarily imply a large absolute absorption
rate, which remains constrained by the molecular
two-photon cross section. The
near-simultaneous arrival of the twins leads to an absorption rate that scales linearly
with photon flux.

It follows that the quantum enhancement can be attained when: 1)~the SPDC
photons are generated using a narrowband pump, 2)~the two-photon transition is
resonant with the sum of the paired photon frequencies, and 3)~the two-photon
absorption linewidth of the transition $\gamma_{fg}$ is narrow in comparison with
the marginal entangled-photon-pair bandwidth $B$.
The pronounced enhancement associated with
linear photon-flux scaling pertains to the
isolated-pair, low-gain regime, in which the
mean photon occupation per spatiotemporal mode
is small. For typical molecular absorbers,
the corresponding absolute ETPA rates are
extremely low. At high gain, correlation-dependent
enhancements may persist, but their magnitude
depends on the molecular linewidth and the
spatiotemporal properties of the illumination.

\subsubsection{Model of
Drago and Sipe}\label{ETPAquantummodelSipe}
While the theoretical models discussed by Dayan and Raymer \emph{et al.}
in Sec.~\ref{ETPAquantummodelRaymer} focused on systems in which near-resonant
intermediate states were excluded, theoretical models that accommodate such states
have been considered by
others~\cite{Fei97,schlawin2016,schlawin2017NJP,schlawin2021arxiv,nakanishi2009,%
oka2018,sipe2022PRA,portela2026}. In particular, Drago and
Sipe~\cite{sipe2022PRA} permitted their excitation spectrum to overlap with
intermediate-state transition wavelengths. These authors analyzed ETPA by
determining the total energy removed from the incident field rather than by
following the density-operator/final-state population approach favored by Raymer
\emph{et al.}  Their theoretical formalism accommodates
near-resonant and far-detuned intermediate states, low- and high-intensity illumination,
and both CW and pulsed excitation. The explicit calculations presented in this
paper, however, focus on CW illumination.

Drago and Sipe considered limiting cases for two-photon molecular absorption
linewidths that were narrow and broad in comparison with the bandwidth of the
entangled-photon pairs. They found that broadband squeezed-light
excitation could substantially enhance two-photon absorption through near-resonant
intermediate states. However, for the  molecular parameters they examined,
one-photon absorption dominated in the regimes where the two-photon enhancement
was appreciable, thereby obscuring its experimental observability.
(An experimental investigation of resonance-enhanced ETPA in indocyanine green (ICG)
found no detectable ETPA signal, suggesting that a near-resonant intermediate state
does not necessarily provide a useful enhancement in molecular
systems~\cite{2024-RES-ENHANCED-TRICK-Cushing-ETPEF-JCP}.)
Drago and Sipe's  analysis in which near-resonant intermediate states were
excluded in the isolated-photon-pair domain led to results that were in
essential agreement with those obtained by Raymer \emph{et al.}

\subsubsection{Models of Schlawin \emph{et al.}}\label{ETPAquantummodelSchlawin}

Schlawin and his collaborators have extensively examined entangled-two-photon
absorption and spectroscopy from the isolated-photon-pair domain through to the
bright-squeezed-vacuum
domain~\cite{schlawin2013,schlawin2016,schlawin2017JPB,schlawin2017NJP,
schlawin2018,schlawin2021arxiv,schlawin2021PRR,schlawin2022PRA,Schlawin2024}.
In 2024, Schlawin generalized the earlier analyses carried out by
Klyshko~\cite{klyshko82}, Dayan~\cite{dayan2007PRA}, Raymer \emph{et
al.}~\cite{raymer2020arxiv,raymer2021optica,raymer2021OE,raymer2021JCP,raymer2022PRA},
and Drago and Sipe~\cite{sipe2022PRA} by analyzing molecular ETPA cross sections
for pulsed broadband parametric-downconversion
light spanning the low- and high-gain regimes in the context of a unified formalism
that accommodates both spatial and spectral degrees of freedom~\cite{Schlawin2024}.

The spatial entanglement of downconverted photons
has been investigated using Schmidt-mode methods
and near- and far-field quantum correlations~\cite{law2004,brambilla2004}.
More generally, the joint spatiotemporal amplitude
need not factorize into spatial and spectral
components, giving rise to high-dimensional
$\mathsf{X\/}$-entanglement~\cite{gatti2009,gatti2012}.
Related spatial and temporal correlations in
pulsed twin beams have also been modeled
across different gain regimes~\cite{gatti2023}.
Schlawin's 2024 calculations, however, employ
a factorized bigaussian model, leaving the nonseparable case for another day.
Within this model, he recovered Fei \emph{et al.}'s
cross-section relation in the low-gain regime, with a correction factor dependent on the
two-photon linewidth and pump bandwidth, and demonstrated that the correlated
contribution can itself scale quadratically with photon flux at high gain.
In the limit of a single spatial mode and a narrowband pump, Schlawin's results
largely agree with those derived by Raymer and Landes~\cite{raymer2022PRA}.

\subsubsection{Models for the Bright-Squeezed-Vacuum (BSV) Domain}
\label{ETPAquantummodelBSV}

Although the preceding subsections have emphasized the isolated-photon-pair regime,
the quantum models also describe parametric downconversion at higher gain, where
the same
second-order $\chi^{(2)}$ interaction produces high-flux parametric downconversion
with zero mean field, i.e., bright squeezed vacuum. In the generic
(nondegenerate) case, the output then comprises a macroscopic twin-beam
quantum field with strong photon-number correlations between the signal and idler
modes, which is conveniently represented by a Schmidt (singular-value)
decomposition of the joint spectral amplitude (JSA), within
a fixed spatial-mode description,
\begin{equation}\label{Schmidt}
f(\omega_1,\omega_2)=\sum_{k}\sqrt{\lambda_k}\,u_k(\omega_1)\,v_k(\omega_2),
\end{equation}
where $\{u_k\}$ and $\{v_k\}$ are orthonormal signal and idler mode functions,
$\{\lambda_k\}$ are Schmidt weights ($\sum_k\lambda_k=1$), and each mode pair
$(u_k,v_k)$ undergoes an independent two-mode squeezing transformation. The
degenerate case, used for illustrative purposes below, corresponds to equal central
angular frequencies, $\omega_1^0=\omega_2^0=\omega_p/2$.

A general analysis of downconversion-induced two-photon absorption (TPA) and the
related two-photon interactions spanning the low- to high-flux domains was given by
Dayan in 2007 for broadband downconversion driven by a spectrally narrow
pump~\cite{dayan2007PRA}. In 2013, Schlawin and Mukamel~\cite{schlawin2013}
treated the case of ultrashort-pulse pumping using a Schmidt decomposition to
obtain the time--frequency correlation functions governing absorption --- scaling
behavior consistent with that found by Dayan and Raymer \emph{et al.} was
recovered. In 2022, Raymer and Landes~\cite{raymer2022PRA} theoretically
investigated ETPA in the  far-off-resonance regime for an in-solution molecular model
that stretched from the isolated-photon-pair to the BSV domains. Their
single-spatial-, multi-temporal-mode formalism relied on fourth-order perturbation
theory applied to the molecular-state density operator. Also in 2022, a framework
accommodating near-resonant intermediate states was investigated by Drago and
Sipe~\cite{sipe2022PRA} --- in the far-off-resonance limit, their results were similar
to those obtained by Raymer \emph{et al.}

\begin{figure}[htb!]
\centering\includegraphics[width=3.25in]{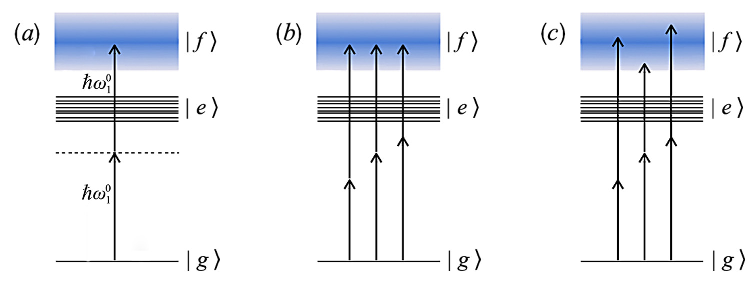}
\caption{Energy-level diagrams
depicting two-photon absorption induced by
(degenerate) parametric-downconversion light
of central angular frequency $\omega_1^0$ in
two-photon transitions from the molecular ground state $|g\rangle$ to the final
state $|f\rangle$ via far-off-resonance intermediate states $|e\rangle$.
The two-photon linewidth is represented in blue. ($a$)~Idealized monochromatic
case. ($b$)~Correlated (coherent) contribution: broadband frequency-anticorrelated
photon pairs with $\omega_1+\omega_2\approx\omega_p=2\omega_1^0$ (three
examples are shown). ($c$)~Uncorrelated (incoherent) contribution: broadband
frequency-noncorrelated photon combinations with widely varying sum frequencies
(three examples are shown); these transitions can contribute appreciably only if the
two-photon linewidth accommodates the resulting sum-frequency band. Although
illustrated for the degenerate case, the same distinction applies for the nondegenerate
case with \,$\omega_1^0\neq\omega_2^0$\,
and \,$\omega_p=\omega_1^0+\omega_2^0$.}
\label{fig28}
\end{figure}

Across the various quantum treatments, the BSV-induced TPA rate naturally
separates into two physically distinct contributions: 1)~a correlated (sometimes
termed coherent) contribution that depends on the phase-sensitive signal--idler
correlations, and 2)~an uncorrelated (sometimes termed incoherent) contribution
associated with accidental photon combinations. Operationally, the correlated term
is dominated by frequency-anticorrelated pairs with frequencies that satisfy the
energy constraint
\begin{equation}
\omega_1 + \omega_2  \approx \omega_p
\end{equation}
within the pump bandwidth, whereas the uncorrelated term samples a broad
distribution of sum frequencies governed primarily by the marginal spectra of the
individual beams. The characteristic timescale of the correlated term is set by the
inverse bandwidth of the downconverted photons, while that of the uncorrelated
term is set by the duration of the pump-intensity envelope (or by any imposed
intensity modulation or shuttering). For a narrow final-state linewidth (a spectrally
sharp two-photon resonance), the material response effectively postselects the
near-$\omega_p$ sum-frequency component, which strongly favors the correlated
contribution (even at high photon flux). These concepts are illustrated in
Fig.~\ref{fig28} for the degenerate case ($\omega_1^0=\omega_2^0=\omega_p/2$).
\par

The ideal monochromatic case is sketched in panel~($a$). Panel~($b$) depicts the
correlated contribution arising from broadband frequency-anticorrelated photon
pairs whose sum frequency remains locked to the narrowband pump
$\omega_p=2\omega_1^0$, while panel~($c$) depicts the uncorrelated contribution
in which the sum frequencies span a wide range and contribute significantly only
when the two-photon linewidth is sufficiently broad. The same interpretation applies
in the general nondegenerate case when \,$\omega_1^0 \neq \omega_2^0$\, and
\,$\omega_p=2\omega_1^0$\, is replaced by
\,$\omega_p=\omega_1^0+\omega_2^0$.

The heuristic model considered in
Sec.~\ref{EPparticlemodel} represents the
quadratic absorption rate by a single term,
associated with accidental two-photon
absorption involving cousins and/or
surviving singleton photons
[Eq.~(\ref{feiLOSS02})].
The quantum models, however, distinguish
correlated and uncorrelated contributions
to the quadratic rate, since the correlated
contribution can itself acquire a quadratic
photon-flux dependence at high gain.

\subsubsection{Absorption Rates and Cross Sections for Quantum and
Heuristic Models}\label{ETPAquantummodelCrossRate}

To compare the quantum models with the heuristic
rate law, we use a phenomenological decomposition
into linear and quadratic photon-flux contributions,
\begin{equation}\label{feiQgam}
  R_{\scriptscriptstyle TE}  = \sigma_{\!\scriptscriptstyle B} \phi + \mathcal{T\/}_{\!\!0} \,
  \sigma_{\!\scriptscriptstyle E} \, \phi + \gamma \, \sigma^{(2)} \Gamma \phi^2.
\end{equation}
This expression differs from the heuristic
rate law in Eq.~(\ref{feiLOSS02}) by the presence of the parameter
$\gamma$, a dimensionless coefficient for the effective quadratic contribution
after the explicitly displayed linear pair contribution has been separated.
The decomposition is not universally exact, since the coefficients, and even the separation
of contributions, can depend on the photon flux, quantum state, and material response.

The first term on the right-hand side of Eq.~(\ref{feiQgam}) denotes
Boltzmann-tail single-photon absorption. The second term is the quantum
contribution representing twin-photon absorption that can manifest in the
isolated-photon-pair (low-flux) domain. The third term is a quadratic
contribution that may contain both correlated and accidental components,
with their relative weights determined by the state and material response.

We proceed to discuss, in turn, the form of the entangled-two-photon cross section
$\sigma_{\!\scriptscriptstyle E}$ (along with its normalized counterpart
$\delta_{\scriptscriptstyle E}$) and the prefactor $\gamma$ that appears in the
third term on the right-hand side of Eq.~(\ref{feiQgam}).

In the isolated-photon-pair domain, the quantum expression for
$\sigma_{\!\scriptscriptstyle E}$ has the same form as Eq.~(\ref{sigdelAT}) for the
heuristic model, the distinction being the presence of a dimensionless prefactor
\,$\alpha$\,~\cite{raymer2021optica,raymer2021OE,raymer2021JCP,jimenez2021,Schlawin2024}:
\begin{equation}\label{sigdelATqalpha}
  \sigma_{\!\scriptscriptstyle E}
    = \frac{\alpha \sigma^{(2)}}{A_{\scriptscriptstyle E} T_{\scriptscriptstyle E}}.
\end{equation}
The parameter $\alpha$ quantifies the efficiency with which the excitation activates
the full two-photon response of the absorber. Equation~(\ref{sigdelATqalpha}),
which is tightly bounded by the physics of photon entanglement and the nature of
the nonlinear response~\cite{raymer2021OE}, is valid when the wavefunction of the
entangled-photon system can be factorized into spatial and spectral
components~\cite{Schlawin2024}. Schlawin has reiterated that
\begin{equation}\label{linewidthsymbol}
  \sigma^{(2)} \propto \frac{1}{\gamma_{fg}}\,,
\end{equation}
where $\gamma_{fg}$ is the linewidth of the two-photon transition, and established
that $\sigma_{\!\scriptscriptstyle E}(\gamma_{fg})$ decreases monotonically with
increasing $\gamma_{fg}$~\cite{Schlawin2024}.

Comprehensive analyses of entangled-two-photon absorption carried out by
Raymer \emph{et al.}~\cite{raymer2022PRA} and
Schlawin~\cite{Schlawin2024} distinguish the marginal bandwidth $B$
of the downconverted photons from the pump bandwidth $\mathcal{B}$,
which governs their sum-frequency distribution. For the correlated
contribution in the isolated-photon-pair domain, Schlawin's resonant
bigaussian model yields an efficiency factor governed by the ratio
of the two-photon transition linewidth to the pump bandwidth.
When $\gamma_{fg}\gg\mathcal{B}$, including
the narrowband CW-pump limit, Schlawin's
efficiency factor approaches
$\alpha=\sqrt{2/\pi}\approx0.80$.
Neglecting this order-unity coefficient,
Eq.~(\ref{sigdelATqalpha}) reduces to
the heuristic approximation

\begin{equation}\label{sigdelATqalpha=1}
  \sigma_{\!\scriptscriptstyle E}
    \approx \frac{\sigma^{(2)}}
    {A_{\scriptscriptstyle E} T_{\scriptscriptstyle E}}
    \qquad \gamma_{fg} \gg \mathcal{B}.
\end{equation}
ETPA cross sections in this domain thus behave as
$\sigma_{\!\scriptscriptstyle E} \propto 1/\gamma_{fg}$,
with the other material and illumination parameters held fixed.

\begin{framed}\noindent
Though molecules in solution are difficult to model because of their many degrees
of freedom, they typically have relatively broad absorption linewidths and minuscule
entangled-two-photon cross sections. These features lead to
entangled-two-photon absorption rates that lie below the
detection limits of many current experimental systems.
\end{framed}

In the opposite limit, when the two-photon transition linewidth is
much narrower than the pump bandwidth, Schlawin's model yields
$\alpha \approx \gamma_{fg}/\mathcal{B}$, with $\mathcal{B}$
identified with the pump bandwidth parameter used in that model.
Equation~(\ref{sigdelATqalpha}) then becomes
\begin{equation}\label{sigdelATqalpha1}
  \sigma_{\!\scriptscriptstyle E}
    \approx \frac{\sigma^{(2)}}
    {A_{\scriptscriptstyle E} T_{\scriptscriptstyle E}}
    \frac{\gamma_{fg}}{\mathcal{B}}
    \qquad \gamma_{fg} \ll \mathcal{B}.
\end{equation}
In this limit, the dependence
$\sigma^{(2)} \propto 1/\gamma_{fg}$
cancels the linewidth dependence of $\alpha$, yielding an
entangled-two-photon cross section that is approximately
independent of $\gamma_{fg}$ and behaves as
$\sigma_{\!\scriptscriptstyle E}   \propto
1/(A_{\scriptscriptstyle E}  T_{\scriptscriptstyle E}\mathcal{B})$.
\begin{framed}\noindent
As specified in Eq.~(\ref{sigdelATqalpha}), the entangled-two-photon cross section
$\sigma_{\!\scriptscriptstyle E}$ is determined not only by the properties of the
absorber via $\sigma^{(2)}\!$, but also by the entanglement area--time product
$A_{\scriptscriptstyle E} T_{\scriptscriptstyle E}$ characterizing the
illumination.  As such, the absorption of entangled-photon pairs is
more fully described by $ \delta_{\scriptscriptstyle E}$,
the normalized form of $\sigma_{\!\scriptscriptstyle E}$
defined in Eq.~(\ref{sigAETE}):
\begin{equation}\label{sigAETE2}
  \delta_{\scriptscriptstyle E} = \sigma_{\!\scriptscriptstyle E}
  A_{\scriptscriptstyle E} T_{\scriptscriptstyle E}.
\end{equation}
\end{framed}

Representative theoretical estimates of $\sigma_{\!\scriptscriptstyle E}$ and
$\delta_{\scriptscriptstyle E}$ (per absorber) for a hydrogen atom and a rhodamine
6G molecule in solution are provided in Table~\ref{sigevarious}.
Also shown are values inferred from the entangled-two-photon photoemission
experiment
of Kobayashi~\emph{et al.} for CsK$_2$Sb, expressed per primitive cell
using the model
developed in Sec.~\ref{entxsect}. The inferred value of $\delta_{\scriptscriptstyle E}$
for CsK$_2$Sb is contingent on the entanglement area--time
product estimated using the heuristic crossover model
[Eq.~(\ref{AETEestimate})]. The correspondence between a condensed-matter
primitive cell and a molecular absorber is necessarily approximate.
The entries in the table reveal that the heuristically estimated
value of $\delta_{\scriptscriptstyle E}$ for ETPA in rhodamine 6G is approximately
400 times smaller than the inferred value of ETPP for CsK$_2$Sb. Both
are vastly smaller than the theoretical value of $\delta_{\scriptscriptstyle E}$ for the
1S$\to$2S transition in atomic hydrogen, which benefits from a far smaller value of
$\gamma_{fg}$. As explained in Sec.~\ref{EPphoto}, it is far easier to observe ETPP
than fluorescence-mode ETPA, principally because the former relies on the emission
of electrons rather than photons.

\begin{table}[htb!]
\centering
\caption{Degenerate entangled-photon pairs, generated by SPDC in the isolated-
photon-pair domain, are characterized by their wavelengths $\lambda_{1,2}$ and
entanglement area--time products $A_{\scriptscriptstyle E} T_{\scriptscriptstyle E}$,
which serve as normalization factors whose values are specified in the
References column below.
Representative theoretical estimates for the ETPA cross sections
$\sigma_{\!\scriptscriptstyle E}$, and their normalized counterparts
$\delta_{\scriptscriptstyle E} = \sigma_{\!\scriptscriptstyle E} A_{\scriptscriptstyle E}
T_{\scriptscriptstyle E}$, are provided for an atomic system and for an organic
molecule in solution.
For comparison, experimentally inferred estimates of $\sigma_{\!\scriptscriptstyle E}$
and $\delta_{\scriptscriptstyle E}$ are provided for ETPP from CsK$_2$Sb.}
\label{sigevarious}
\centering
\small
\renewcommand{\arraystretch}{1.10}
\begin{tabular}{@{}lccccc@{}}
\cline{2-6}
\noalign{\vskip 1.1mm}
& \shortstack{$\lambda_{1,2}$\\(nm)}
& \shortstack{Normalization\\$A_{\scriptscriptstyle E} T_{\scriptscriptstyle E}$ (m$^2$\,s)}
& \shortstack{{\sc etpa/etpp} cross\\section $\sigma_{\!\scriptscriptstyle E}$ (m$^2$)}
 & \shortstack{Normalized $\delta_{\scriptscriptstyle E}$ \\$\sigma_{\!\scriptscriptstyle E}
 A_{\scriptscriptstyle E} T_{\scriptscriptstyle E}$ (m$^4$\,s)}
& References\\[0.7mm]
\hline
\noalign{\vskip 0.7mm}
\begin{tabular}[c]{@{}l@{}}Hydrogen atom\\[-1mm]1S$\to$2S transition\end{tabular} &
243 & \begin{tabular}[c]{@{}l@{}}$\approx 6.0\times10^{-24}$
\\[-1mm]\phantom{X}{\sc (theory)}\end{tabular} &
\begin{tabular}[c]{@{}l@{}}$\approx 3.0\times10^{-17}$
\\[-1mm]\phantom{X}{\sc (theory)}\end{tabular} &
\begin{tabular}[c]{@{}l@{}}$\approx 1.8\times10^{-40}$
\\[-1mm]\phantom{X}{\sc (theory)}\end{tabular}
& \cite{Fei97}, Fig.~\ref{fig4}\\[2.8mm]
\hline
\noalign{\vskip 0.7mm}
\begin{tabular}[c]{@{}l@{}}Rhodamine 6G\\[-1mm](in methanol)\end{tabular}  & 810
& \begin{tabular}[c]{@{}l@{}}$\approx 3.4\times10^{-24}$
\\[-1mm]\phantom{X}{\sc (theory)}\end{tabular} &
\begin{tabular}[c]{@{}l@{}}$\approx 1.5\times10^{-33}$
\\[-1mm]\phantom{X}{\sc (theory)}\end{tabular}   &
\begin{tabular}[c]{@{}l@{}}$\approx5.1\times10^{-57}$
\\[-1mm]\phantom{X}{\sc (theory)}\end{tabular}
 &  \begin{tabular}[c]{@{}c@{}}\cite{jimenez2021,rebane2016OE,raymer2021PRR}
 \\[-1mm]using Eq.~(\ref{sigdelAT})\end{tabular} \\[2.8mm]
\hline
\noalign{\vskip 0.7mm}
\begin{tabular}[c]{@{}l@{}}CsK$_2$Sb \,  (ETPP\\[-1mm]in semiconductor)\end{tabular}
& 1064 & \begin{tabular}[c]{@{}l@{}}\phantom{|}$\approx 5.4\times10^{-22}$
\\[-1mm]\phantom{M}{\sc (inferred)}\end{tabular}&
\begin{tabular}[c]{@{}l@{}}\phantom{|}$\approx 3.8\times10^{-33}$
\\[-1mm]\phantom{Xf}{\sc (inferred)}\end{tabular}&
\begin{tabular}[c]{@{}l@{}}\phantom{|}$\approx 2.1\times10^{-54}$
\\[-1mm]\phantom{Xf}{\sc (inferred)}\end{tabular}
&
\begin{tabular}[c]{@{}c@{}}\cite{kobayashi2006,kobayashi2007,lissandrin2004}\\
[-1mm]$\!\!\!\!\!$Eq.~(\ref{iEsigma2})/Tables~\ref{tab:photoemissiveparams},
\ref{compare}\end{tabular}\\
\noalign{\vskip 0.7mm}
\hline
\end{tabular}
\end{table}

\begin{framed}\noindent
The entangled-photon-pair illumination and the parameters of the medium are
generally intertwined in a nonseparable way by quantum interference so the
cross section
$\sigma_{\!\scriptscriptstyle E}$ can be a nonmonotonic function of
$T_{\scriptscriptstyle E}$.
\end{framed}

Strong nonmonotonicity of the entangled-two-photon cross section
plotted as a function of $T_{\scriptscriptstyle E}$
was illustrated  by Fei \emph{et al.}~\cite{Fei97} for
the exactly calculable case of the 1S$\to$2S narrowband electronic transition in
atomic hydrogen, as displayed in Fig.~\ref{fig4}. Similar nonmonotonicity has also recently been
theoretically predicted for cesium~\cite{portela2026}. The molecular
quantum calculations carried out by Fu \emph{et
al.}~\cite{THEORY-2023-Thew-formula-JPCL} for a number of model chromophores
similarly display strong periodic nonmonotonicities in the entangled-two-photon
cross section. Calculations by
Lissandrin~\emph{et al.}~\cite{lissandrin2004} for entangled-two-photon
photoemission from the semiconductor CsK$_2$Sb and the metal Na
exhibit vestiges of this
nonmonotonicity, by virtue of the  transition being broadband, as
displayed in Figs.~\ref{fig10} and \ref{fig11}, respectively.

Finally, we compare the absorption rates and cross sections for the quantum and
heuristic models in the high-gain BSV domain. The two-photon
quantum-theoretical absorption rate dominant in this domain, denoted
$R_{\scriptscriptstyle C}$, is represented by the third term on the right-hand side of
Eq.~(\ref{feiQgam}) --- this contribution exhibits quadratic scaling and contains the
dimensionless prefactor $\gamma$. An in-depth derivation by Raymer and
Landes~\cite{raymer2022PRA} demonstrated that when the molecular
two-photon-absorption final-state linewidth is much greater than the marginal
bandwidth of the incident entangled photons ($\gamma_{fg} \gg B$), the absorption
rate is given by
\begin{equation}\label{deltasubrRrqg2}
  R_{\scriptscriptstyle C} = g_2\, \sigma^{(2)} \Gamma \phi^2,
\end{equation}
so that $\gamma=g_2$ in Eq.~(\ref{feiQgam}) in this limit (at high gain).
For collinear type-I or type-0 phase-matched
SPDC with indistinguishable degenerate photons,
the ideal normalized second-order correlation function is given by
$g_2\approx 3+1/\bar n$, which approaches 3 at high gain ($\bar n\gg1$).
For distinguishable photons, on the other hand, the corresponding
expression is $g_2 \approx 2+1/\bar n$, which approaches 2 at high gain.
Dispersion and averaging over multiple
statistically independent modes can reduce
the observed value of $g_2$.
Drago and Sipe's
results~\cite{sipe2022PRA} agree with those of Raymer and
Landes~\cite{raymer2022PRA} and Cutipa \emph{et al.}~\cite{cutipa2020} in this
domain.

In the opposite limit, when the two-photon-absorption linewidth is narrow in
comparison with the  marginal entangled-photon-pair bandwidth ($\gamma_{fg} \ll
B$), Raymer and Landes~\cite{raymer2022PRA} have shown that BSV has the same
effectiveness in inducing TPA as a quasi-monochromatic coherent pulse of the same
temporal shape and mean photon number. Drago and Sipe's~\cite{sipe2022PRA}
calculations, in the absence of near-resonant intermediate states, match those of
Raymer and Landes when $\gamma_{fg}$ is very much smaller than $B$. In the
presence of near-resonant intermediate states, on the other hand, Drago and Sipe's
calculations predict a significant enhancement in the ETPA rate. Just as in
the isolated-photon-pair domain discussed in Sec.~\ref{ETPAquantummodelSipe},
however, the one-photon absorption dominates in regimes where the
two-photon enhancement is appreciable, thereby obscuring its  observability.

\subsubsection{Parameter Estimation for Subthreshold Absorption}\label{paramestABS}
In carrying out subthreshold absorption measurements, it is often desired to
estimate from the data the value of a parameter that characterizes the underlying
absorption process of interest. This might be one of the subthreshold molecular
cross sections, such as $\,\sigma_{\!\scriptscriptstyle E}$, or  a crossover photon-flux
density, such as $\,\phi_{\scriptscriptstyle EC}\,$ or $\,\phi_{\scriptscriptstyle BC}$.
As discussed in Sec.~\ref{paramestPE} for subthreshold photoemission, carrying out
this task in an optimal way involves the use of statistical estimation
theory~\cite{fisher25,frechet43,rao45,cramer46,helstrom76,holevo82,kay93,braunsteincaves94,paris09,wiseman09}.

The parameter of interest is inferred by fitting calibrated measurements of the
sample optical transmittance or fluorescence count rate, obtained over a range of incident
photon-flux densities, to an appropriate absorption model, such as
Eq.~(\ref{feiQgam}).  The conversion from these observables to
the absorption-event rate requires accounting for sample concentration and geometry,
optical losses, and detection efficiency; fluorescence measurements additionally
require the fluorescence quantum yield and collection efficiency.
Since BTA and ETPA both exhibit linear photon-flux dependence, independent control
measurements or downstream attenuation  tests are needed to distinguish their
contributions, as discussed in Sec.~\ref{measidenidenABS}.

\subsection{Experimental Studies of Subthreshold Absorption and
Fluorescence Microscopy} \label{ssec:ETFAFM}

Unadorned introductions to subthreshold entangled-two-photon absorption (ETPA)
and entangled-two-photon fluorescence microscopy (ETPFM) were provided in
Secs.~\ref{EPabsorp} and \ref{EPmicroscopy}, respectively. In
Sec.~\ref{irreproducibilityETPA}, we review a number of studies in which these two
effects were purportedly observed in the isolated-photon-pair domain.
As discussed in Sec.~\ref{irreproducibilityFAIL},
however, independent efforts have not convincingly
replicated the reported molecular ETPA and ETPFM signals in the isolated-photon-pair domain.
Subsequent investigations have demonstrated that Boltzmann-tail absorption (BTA), including
hot-band absorption in molecular systems, and other one-photon processes can mimic experimental
signatures attributed to ETPA (Sec.~\ref{BTA}).   These mechanisms provide plausible
explanations for many earlier observations, but the origins of all purported ETPA signals
have not been conclusively established.

\subsubsection{Experimental Reports of ETPA and ETPFM} \label{irreproducibilityETPA}

In 2004, French and Goodson examined the
angular dependence of biphoton statistics
and estimated the expected entangled-two-photon
absorption and fluorescence rates in organic
materials~\cite{goodson2004}.
Beginning with the 2006 experimental report
by Lee and Goodson~\cite{ETPA-2006-goodson-JPCBL},
   numerous publications have reported
   entangled-two-photon absorption under
   isolated-photon-pair illumination,
   using transmission- or fluorescence-based
   measurements.
These effects have been purportedly observed across a wide
range of molecular chromophores, polymers, dendrimers, and biological samples.
The preponderance of these reports originated from laboratories in Ann
Arbor~\cite{Harpham2009Entangled,
Guzman2010Spatial,Upton2013Illuminate,goodson-ETPF-2017-JPCL,
VillabonaMonsalve2018Flavins,ETPA+ETPF-2020-goodson-JPCC,goodson2024JPPC,goodson2025JACS,goodson2025JPCA},
Bogotá~\cite{villabona2017JPC}, and Geneva~\cite{tabakaev2021PRA,thew22},
with a more recent contribution from Yokohama~\cite{hashimoto2026}. Reports of the implementation of
entangled-two-photon fluorescence microscopy also emerged from the laboratory in
Ann
Arbor~\cite{ETPFM-2020-goodson-JACS,goodson2022ETPFM,goodson2024ETPFM,goodson2026ETPFM}.

\subsubsection{Failures to Replicate Experimental Reports of
ETPA and ETPFM} \label{irreproducibilityFAIL}

Notwithstanding the breadth of the claims
reported in Sec.~\ref{irreproducibilityETPA},
no convincing independent replication of
molecular ETPA has been achieved in the
isolated-photon-pair domain, and the reported
implementations of ETPFM remain without
independent confirmation.
Sustained experimental efforts to investigate
molecular ETPA have been undertaken at
laboratories in
Boulder~\cite{jimenez2020SPIE,jimenez2021,jimenez2022,jimenez2025},
Eugene~\cite{raymer2021PRR,raymer2024PRA}, Pasadena~\cite{cushing2022JPCL},
Mexico City~\cite{corona2022JPC},
Le{\'o}n~\cite{2023-HOM-TRICK-Triana-JPCA,2024-HOM-TRICK-Triana-JPCA,2025-HOM-TRICK-RamirezRamos-JPCA},
Bern~\cite{stefanov2021JCP}, Jena~\cite{graefe2023APR,graefe2025APLQ}, and
Hangzhou~\cite{hangzhou2024}.
In particular, Krsti{\'c} et al.~\cite{graefe2025APLQ} reported absorption
measurements in Nile Red that exhibited partial agreement with theoretical ETPA
predictions, but their measurements did not establish
an unambiguous two-photon contribution,
and no ETPA-induced fluorescence was detected.

Many of these reports have also established upper
experimental bounds for the magnitudes of the associated cross sections for these
molecular samples. For molecules in solution, conventional
two-photon absorption cross sections $\sigma^{(2)}$ are typically of order
1--1000~GM, where $1\,\mathrm{GM}=10^{-58}\,
\mathrm{m}^4\,\mathrm{s}/\mathrm{photon}$.
Quantum-theoretical estimates of the corresponding ETPA cross sections can
lie well below experimentally established upper bounds.
Under the illumination conditions and sample geometries of current experiments,
the predicted fluorescence count rates are generally below achievable detection
thresholds~\cite{dayan2007PRA,raymer2021optica,raymer2021OE,
sipe2022PRA,jimenez2021,raymer2024PRA}.

\begin{framed}\noindent
Independent experimental investigations have not established a detectable
ETPA enhancement for common molecular fluorophores in solution under
isolated-photon-pair illumination. Consequently, ETPFM in this regime
remains experimentally unverified.
\end{framed}

\subsubsection{Conflating BTA and Other One-Photon Processes with ETPA} \label{BTA}

A number of the experiments cited in Sec.~\ref{irreproducibilityFAIL} explicitly
demonstrated that one-photon effects can masquerade as ETPA.
In particular, one-photon absorption from thermally populated initial states can produce
subthreshold signals that scale linearly with photon flux.
We coin the term \textbf{Boltzmann-tail absorption (BTA)} for this mechanism, by analogy with
Fermi-tail photoemission (FTP) discussed in Sec.~\ref{entPMT}.
In organic materials, BTA can take the form of \textbf{hot-band absorption (HBA)}, whereby
one-photon excitation proceeds from thermally populated vibrational levels of the ground
electronic state~\cite{jimenez2022}. Mikhaylov et al.~\cite{jimenez2022}
demonstrated HBA in LDS798, showing that the fluorescence signal could mimic the
linear photon-flux dependence expected for ETPA; its dependence on post-source
attenuation and sample temperature supported its one-photon origin.

Other effects that can be conflated with ETPA include single-photon scattering of
the downconverted light, leakage or scattering of residual pump light,
Fresnel reflection, and fluorescence from optical components.
In particular, Hickam et al.~\cite{cushing2022JPCL} demonstrated that
single-photon scattering in rhodamine 6G could produce a signal
that could be misinterpreted as ETPA.  Additional complications include one-photon
absorption into real intermediate states and aggregation-dependent changes in the
optical properties of the sample \cite{corona2022JPC,sipe2022PRA,
2024-HOM-TRICK-Triana-JPCA}.

\subsection{Methodologies for Enhancing ETPA and
Implementing ETPFM}\label{impabmic}

As reported in Sec.~\ref{entchannel}, ETPP has been observed from CsK$_2$Sb and a
theoretical model constructed on the basis of the band structure of this material
can reproduce the experimentally inferred quantum efficiency for a suitable
value of the material response ratio $\mathcal{Q}$. Techniques for enhancing
the observation of ETPP were discussed in Secs.~\ref{enhancingf} and
\ref{kobabooth}.  Although ETPA and ETPP involve different
material transitions, both depend on photon-pair correlations and compete with
one-photon and conventional two-photon processes.
We therefore recount a number of techniques that may be useful for enhancing the
observability of ETPA and for implementing ETPFM, neither of which has yet been
independently replicated in molecular solution under isolated-photon-pair
illumination. We also
highlight a number of other approaches, drawn from recent technical reports, that
could prove useful for this purpose. The approaches discussed in this section
focus on increasing the strength of the ETPA signal, reducing background
contributions and measurement noise,  and distinguishing ETPA from competing
one-photon processes.

\subsubsection{Sample Selection}\label{impabmic-03-XSect}

As discussed in Sec.~\ref{measidencrossPFD}, the most straightforward means for
elevating the observability of ETPA is to choose a sample with a large value of the
entangled-two-photon cross section $\sigma_{\!\scriptscriptstyle E}$. Studies of
conventional two-photon absorption can serve as a guide~\cite{pawlicki2009}.
A number of candidate substances have emerged
from theoretical calculations \cite{bochenkova2025e2pa,bochenkova2025KR2}.
Mart\'{i}nez-Tapia and Le{\'o}n-Montiel~\cite{2025-ANN-TRICK-LeonMontiel-AVSQSci}
developed an artificial-neural-network approach for classifying the number of
intermediate states contributing to ETPA, using theoretically generated,
delay-dependent absorption signals. Their simulations suggest that this
approach may facilitate the characterization and selection of candidate absorbers,
although experimental validation remains necessary.

Atomic species~\cite{georgiades1995PRL,drago2025arxiv} or color centers with
narrow two-photon transitions and large matrix elements may be suitable samples.
In particular, ultracold Rb atoms confined in a magneto-optical trap (MOT), which
greatly suppresses Doppler broadening, appear promising for observing ETPA.
McLean et al.~\cite{jimenez2026arxiv} recently demonstrated conventional
two-photon-excited fluorescence from $^{85}$Rb and $^{87}$Rb in a MOT at
excitation powers as low as approximately $1\,\mu$W, with measured two-photon
cross sections of order $10^{12}$~GM. Although ETPA was not observed, these
results establish a sensitive atomic platform for future investigations.
Condensed-matter platforms, such as CsK$_2$Sb semiconductor, which has been
shown to exhibit ETPP (Sec.~\ref{ssec:expCPM}), and suitably
prepared biological samples can offer large numbers of potential absorbers
within the illuminated volume.However, increasing absorber density
does not in and of itself increase the entangled-two-photon cross section per
absorber and may introduce additional one-photon absorption and scattering.

As pointed out in Sec.~\ref{2PEinst}, the induction of entangled-two-photon
photoemission in CsK$_2$Sb involves a real intermediate state (in the conduction
band). Hence, it might be thought that the presence of a resonant intermediate state
would enhance the ETPA transition probability in molecular samples. However, ETPA
experiments carried out with the organic molecular dye indocyanine green (ICG),
which does possess such an intermediate state, did not lead to an observable signal.
Nor was the conventional two-photon ICG cross section $\sigma^{(2)}$ significantly
enhanced by the presence of the resonant intermediate
state~\cite{2024-RES-ENHANCED-TRICK-Cushing-ETPEF-JCP}.

\subsubsection{One-Photon Interactions}\label{impabmic-01}
As discussed in Secs.~\ref{measidencrossPFD} and \ref{BTA}, it is important to
maximize the optical-system transmittance $\mathcal{T\/}_{\!\!0}$\, by using a
direct optical path that contains as few optical components as possible. It is also
crucial to minimize Boltzmann-tail absorption (BTA), which, like Fermi-tail
photoemission (FTP), can be mitigated by cooling the sample to reduce its
temperature (Sec.~\ref{enhancingf}).  As indicated in Sec.~\ref{BTA}, other
one-photon processes can also have  deleterious effects. Optical filtering
can suppress residual pump light and out-of-band background radiation, but
cannot generally eliminate one-photon absorption or scattering at the
wavelengths of the downconverted photons.

The identification of ETPA has generally relied on measuring fluorescence or
transmission rates and distinguishing two-photon contributions from competing
one-photon processes. An alternative  approach is to seek an entanglement-dependent
spectroscopic signature rather than an absolute rate enhancement.
In a theoretical study, Jang and Yoo~\cite{jang2026PRB} predicted that
the Bell-state phase of polarization-entangled, frequency-nondegenerate photon pairs can
control the excitation of bright and exchange-dark biexciton eigenstates in
monolayer WSe$_2$. In their valley-symmetric model, the
antisymmetric Bell state selectively excites an exchange-dark biexciton state
at a rate twice the upper bound for separable polarization states having
the same pair flux. The proposed coincidence-transmission
and pump--probe measurements exploit this phase dependence to distinguish
ETPA from phase-independent one-photon backgrounds, although measurement noise
and the small absolute absorption rate remain limiting factors.
Similar approaches may be applicable  to other transition-metal dichalcogenide
monolayers, including MoSe$_2$, WS$_2$, and MoS$_2$~\cite{erkensten2021PRB}.

\subsubsection{Beamsplitter-Interferometer Configurations}\label{impabmic-02}

\begin{figure}[htb!]
\centering
\includegraphics[width=4.50in]{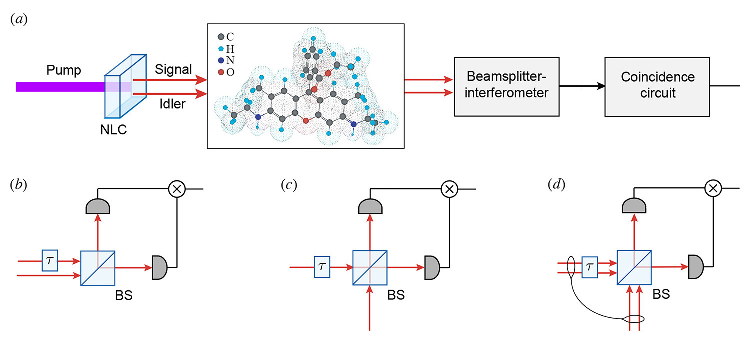}
\caption{Experimental arrangements for transmission-mode entangled-two-photon
absorption using different beamsplitter-interferometer (BI) configurations.
($a$)~The signal and idler photons generated via SPDC in a nonlinear optical crystal
(NLC) are transmitted through the sample, illustrated here as rhodamine 6G, and
thence to a particular BI.  The photons emerging from the output ports of
the beamsplitter (BS) are detected
within the BI, and the resulting electrical pulses are fed to a coincidence circuit. ($b$)~In this BI configuration, a
controllable time delay $\tau$ is imparted to one of the photons and both photons
impinge on a single input port of the 50/50 beamsplitter
(BS)~\cite{cushing2022JPCL,corona2022JPC}. ($c$)~Here, a controllable time
delay $\tau$ is imparted to one of the photons and the two photons enter different
input ports of the 50/50 beamsplitter. This is the Hong--Ou--Mandel (HOM)
interferometer~\cite{Hong87} used in quantum-optical coherence tomography
(QOCT)~\cite{teich2012,Nasr03}. ($d$)~The N00N-state interferometer
makes use of a coherent superposition
of both photons occupying one input port of a 50/50 beamsplitter,
and both occupying the other input port with a controllable time delay
$\tau$~\cite{2023-N00N-TRICK-URen-APLphotonics}.  Under suitable spectral
conditions, the normalized interferogram is relatively insensitive to
one-photon losses.}
\label{fig29}
\end{figure}

Different measurement configurations exhibit different sensitivities to
one-photon losses. For transmission-mode ETPA experiments, a
number of beamsplitter-interferometer (BI) configurations incorporating
controllable photon delays have been considered, as illustrated in
Fig.~\ref{fig29}. In 2023, Mart{\'i}nez-Tapia
\emph{et al.}~\cite{2023-N00N-TRICK-URen-APLphotonics} showed
theoretically that the normalized coincidence interferogram associated
with a N00N-state configuration can be relatively insensitive to
one-photon losses while remaining sensitive to ETPA. Investigations
based on Hong--Ou--Mandel (HOM) interferometry~\cite{Hong87}
\seccite{9.1}{saleh25} have elucidated the spectral conditions that
are conducive to the observation of ETPA along with the complications
arising from one-photon losses~\cite{2023-HOM-TRICK-Triana-JPCA,
2024-HOM-TRICK-Triana-JPCA,2025-HOM-TRICK-RamirezRamos-JPCA}.

An alternative approach to discriminating between ETPA and
one-photon losses involves modifying the transverse spatial
correlations of entangled-photon pairs while maintaining a
constant incident photon flux. Preliminary evidence consistent
with ETPA has been reported using this approach~\cite{pandyadefienne2024}.

\par

\subsubsection{Incident Photon-Flux Density}\label{subsubsec:phiregion}

Within the heuristic framework developed in Sec.~\ref{measidencrossPFD},
experiments can be
carried out using a photon-flux density for which the distinction
between the flux dependences of the absorption rates in the
presence and absence of ETPA is most clearly manifested,
namely in the range $\phi_{\scriptscriptstyle BC} < \phi  <
\phi_{\scriptscriptstyle EC}$, as specified in Eq.~(\ref{IFCECphi}).

\subsubsection{Nondegenerate Entangled-Photon Pairs}\label{subsubsec:nondeg}

Calculations carried out by Fei \emph{et al.}~\cite{Fei97} for the
entangled-two-photon absorption cross section $\sigma_{\!\scriptscriptstyle E}$
for the $1S\!\to\!2S$ transition in atomic hydrogen demonstrated that
$\sigma_{\!\scriptscriptstyle E}$ can be enhanced for strongly nondegenerate
photons. A computational framework for organic molecular systems, based on
time-dependent density functional theory, also suggested that the ETPA cross
section can be significantly enhanced when the energy of one of the photons
of a nondegenerate pair approaches resonance with a one-photon absorption
transition~\cite{graupnerMukamel2026}. However, this can also enhance
competing one-photon absorption.

This prospect was examined by carrying out fluorescence-mode ETPA
experiments with rhodamine 6G using strongly nondegenerate photon pairs
(545.3 and 1514.6~nm) generated via SPDC in PPKTP. The measured fluorescence
intensity exhibited a peak (of magnitude $\approx 3.5\%$ above the background)
as the delay between the signal and idler photons was scanned~\cite{graupner2025}.
This observation is consistent with, but does not unequivocally establish,
the presence of ETPA.

\subsubsection{Bright Squeezed Vacuum}\label{subsubsec:BSV}

 When operation in the low-gain domain is not feasible for observing ETPA
because of insufficient photon flux, it is natural to wonder whether bright
squeezed vacuum, discussed in Secs.~\ref{ETPAquantummodelBSV} and
\ref{ETPAquantummodelCrossRate}, might serve as an alternative source for
gaining a quantum advantage. A nonperturbative treatment of the
parametric-downconversion process, together with a perturbative treatment
of the light--matter interaction, predicts a transition from linear to
super-quadratic scaling of the two-photon absorption rate with increasing
SPDC pump intensity~\cite{2025-DUAL-SPDC-TRICK-Mukamel-JPCL}.
It has also been demonstrated theoretically that the transmission-mode
two-photon-absorption sensitivity for squeezed vacuum can become independent
of single-photon losses occurring subsequent to two-photon absorption at
high incident mean photon numbers, although coherent probes can still offer
superior sensitivity in photon-number measurements~\cite{schlawin2022PRA}.

In a related experimental investigation, Li \emph{et al.}~\cite{agarwal2020}
reported a 47-fold enhancement of two-photon
fluorescence from fluorescein relative to CW coherent-light excitation,
at the same incident optical power of 8~mW,
using bright two-mode squeezed light generated via seeded
four-wave mixing in rubidium vapor.

At the same time, it has been predicted that the range of pump intensities
over which linear scaling prevails increases as the pump bandwidth
grows~\cite{2025-DUAL-SPDC-TRICK-Mukamel-JPCL}. Moreover, it has been pointed
out that increasing the beam diameter relative to the entanglement diameter,
thereby increasing $A/A_{\scriptscriptstyle E}$, can under suitable
conditions extend the range of optical powers over which the linear ETPA
contribution dominates~\cite{klyshko82,jimenez2021,thew22,Schlawin2024}.
Indeed, in a recently reported experiment, a type-0 PPLN source was used
to produce high-gain (BSV-like) downconversion near 1064~nm, with an average
downconverted power reaching 10~mW and a beam diameter at the sample of
150--200~$\mu$m. Fluorescence measurements with rhodamine 6G exhibited
approximately linear and quadratic scaling with pre-SPDC and post-SPDC
attenuation, respectively~\cite{kasamatsu2026APL}. These observations
are consistent with ETPA in the high-gain domain, although the underlying
mechanism warrants further investigation.

An intriguing aspect of inducing ETPA with BSV is that the process can be
driven using broadband downconverted light whose photon-pair sum-frequency
spectrum is narrow. This permits selective excitation of narrow two-photon
resonances. Nevertheless, the high photon-flux densities required to obtain
measurable absorption can pose problems associated with optical damage.

Interestingly, the observability of ETPA in the high-gain domain can,
in principle, be enhanced by making use of a scheme in which the
uncorrelated contribution to two-photon absorption is determined using
the signal and idler beams from two independent, statistically matched
SPDC sources. Subtracting this contribution from the total signal
obtained using the signal and idler beams from a single SPDC source
should, in principle, permit the entanglement-dependent contribution
to be isolated~\cite{2025-DUAL-SPDC-TRICK-Mukamel-JPCL}.
A related subtraction methodology, based on different pulse orderings,
has been proposed for two-dimensional electronic spectroscopy using
intense entangled-photon beams~\cite{2025JadounMukamelPRL}.

\subsubsection{Coherent Control}\label{subsubsec:cohcon}

The correlations of entangled-photon pairs can be leveraged to optimize
the probability of two-photon absorption. Dayan \emph{et al.}
experimentally demonstrated coherent control of two-photon absorption
in atomic Rb by the spectral-phase shaping of broadband downconverted
light, exploiting the separation of temporal and spectral
resolutions~\cite{dayan2004PRL,dayan2007PRA}.

Schlawin and Buchleitner~\cite{schlawin2017NJP} subsequently developed
a variational framework, based on perturbation theory, in which the
optimal two-photon spectral wavefunction is proportional to the complex
conjugate of the material's two-photon response function. When this
response function is nonseparable, suitably shaped entangled-photon
pairs can induce resonant two-photon transitions more efficiently
than optimally shaped classical pulses. Carnio \emph{et al.}
extended this approach to accommodate experimental constraints,
demonstrating theoretically that substantial enhancements to the
two-photon transition probability can be achieved by spectral
shaping of downconverted photon pairs~\cite{schlawin2021arxiv}.

\section{CONCLUSION} \label{conclusion}

    Subthreshold absorption and photoemission under classical and
entangled-photon-pair illumination have been reviewed, with particular
emphasis on the isolated-photon-pair domain. While observations of
entangled-two-photon absorption in molecular solutions and entangled-two-photon
fluorescence microscopy have been reported, none has, to our knowledge,
been independently and reliably replicated. Entangled-two-photon cross
sections for typical organic molecules in solution appear to be so small
that absorption rates lie below the detection limits of many current
experimental systems~\cite{jimenez2021}. Entangled-two-photon photoemission
in the isolated-photon-pair domain, on the other hand, benefits from
the high efficiency with which photoelectrons can be collected,
multiplied, and detected.

After reviewing the properties, generation, and interactions of entangled-photon
pairs, we examined the current state of subthreshold photoemission from the
semiconductor cesium-potassium antimonide and the metal sodium. Three forms of
subthreshold photoemission were investigated: 1)~one-photon Fermi-tail
photoemission (FTP) from CsK$_2$Sb; 2)~two-photon photoemission (TPP) from Na
and CsK$_2$Sb; and 3)~entangled-two-photon photoemission (ETPP) from
CsK$_2$Sb. The associated quantum efficiencies, denoted $\eta_{\scriptscriptstyle
F}$, $\eta_{\scriptscriptstyle C}$, and $\eta_{\scriptscriptstyle E}$, respectively,
were found to have the following values:

\begin{enumerate}

\item \textbf{FTP from CsK$_2$Sb:}\,
In a photomultiplier tube, $\eta_{\scriptscriptstyle F}
\approx 4.8 \times 10^{-10}$ electrons/photon under coherent-light
illumination and $3.7 \times 10^{-10}$ electrons/photon under
entangled-photon-pair illumination, at a wavelength $\lambda=800$~nm
and a photocathode temperature $\mathsfit{T\/}=-20~^\circ$C.
At $\lambda=800$~nm and $\mathsfit{T\/}=27~^\circ$C,
$\eta_{\scriptscriptstyle F}\approx 2.3\times10^{-9}$
electrons/photon under coherent-light illumination.
In a channel photomultiplier, Fermi-tail photoemission can be
sufficiently small as to be undetectable.

\item \textbf{TPP from Na:}
A reexamination of mid-1960s two-photon photoemission
data from thick and thin Na photocathodes
in a specially constructed
photomultiplier tube~\cite{teich66PhD} supports a substantial volume
contribution for the thick
sample; the much weaker quadratic response of the thin sample is compatible
with a surface contribution, although a reduced-volume contribution cannot
be excluded. Revised CW-equivalent quantum efficiencies are
$\eta_{\scriptscriptstyle C}(\mathrm{thick})\approx
5.6\times10^{-20}\,I$ and
$\eta_{\scriptscriptstyle C}(\mathrm{thin})\approx
9.5\times10^{-23}\,I$ electrons/photon, respectively, where $I$ is the
intensity of the corresponding CW coherent illumination at the sample,
specified in W/m$^2$.
The measurements were carried out at a wavelength $\lambda=845$~nm
and a photocathode temperature $\mathsfit{T\/}=27~^\circ$C, under
illumination with pulsed multimode classical light.
The volume- and surface-theoretical quantum efficiencies lie within factors
of approximately two and four of the revised thick- and thin-film experimental
values, respectively~\cite{teich66PhD,Bloch64,Smith62,bowers64,
marinchuk66,marinchuk71,barashev1972I}.

  \item \textbf{TPP from CsK$_2$Sb:}\, In a channel photomultiplier,
      $\eta_{\scriptscriptstyle C} \approx 6.8 \times 10^{-16}\,I$\,\,
      electrons/photon, where \,$I$\, is specified in W/m$^2$, at a wavelength
      $\lambda = 1064$~nm and a photocathode temperature $\mathsfit{T\/}=
      27~^\circ$C, under illumination with either coherent light or entangled-photon
      pairs.   The common value was inferred from the overlap of the high-intensity
      data sets for the two forms of illumination.

      \item \textbf{ETPP from CsK$_2$Sb:}\,   In a channel photomultiplier,
$\eta_{\scriptscriptstyle E} \approx 2.3 \times 10^{-13}$ electrons/photon
at a wavelength $\lambda=1064$~nm and a photocathode temperature
$\mathsfit{T\/}=27~^\circ$C, under illumination with entangled-photon
pairs for which the entanglement area--time product is estimated
to be $A_{\scriptscriptstyle E}T_{\scriptscriptstyle E}
\approx 5.4\times10^{-22}$~m$^2$\,s within the heuristic crossover
model. The quantum theory formulated by Lissandrin \emph{et al.}
in 1999/2004~\cite{Lissandrin99,lissandrin2004}  yields an estimate for
$\eta_{\scriptscriptstyle E}$ that accords with the value inferred from
the data of Kobayashi \emph{et al.}~\cite{kobayashi2007} when the
response ratio  $\mathcal{Q} \approx 0.08$, as described in
Sec.~\ref{sssec:Kobalissandrin}.
Independent characterization of the entangled-photon field
at the photocathode, together with a calculation of the full
material response, would allow $\mathcal{Q}$ to be determined.
It would also be of interest to conduct analogous experiments using
entangled-two-photon-induced carrier generation in single-photon avalanche
diodes (SPADs)~\chapcite{19}{saleh2019}.
\end{enumerate}

Subthreshold absorption is manifested in three analogous forms: 1)~one-photon
Boltzmann-tail absorption (BTA); 2)~two-photon absorption (TPA); and
3)~entangled-two-photon absorption (ETPA). The widely used heuristic framework for
entangled-two-photon absorption developed by Fei \emph{et al.} in
1997~\cite{Fei97} has been extended to incorporate BTA and optical loss. In
appropriate limits, the formulas associated with the heuristic approach are recovered
from the quantum-theoretic absorption models developed in 2007 by
Dayan~\cite{dayan2007PRA} and in the early 2020s by Raymer, Sipe, Schlawin, and
colleagues~\cite{raymer2022PRA,sipe2022PRA,Schlawin2024}.

The high level of interest in entangled-two-photon absorption stems
from its linear scaling with photon-flux density in the
isolated-photon-pair domain and the possibility of substantially
enhancing two-photon absorption rates relative to those obtained
with coherent light at the same photon-flux density, at least over selected
parameter ranges. If sufficiently large absorption rates can be
achieved, entangled-two-photon spectroscopy and fluorescence
microscopy could be implemented at low photon-flux levels,
thereby mitigating photodamage to delicate specimens.

Recent theoretical studies of entangled-photon stimulated Raman
scattering (SRS) from polyatomic molecules indicate that this
modality can, under suitable conditions, yield spectral-line
intensities comparable to those predicted for
ETPA~\cite{Dorfman2014,Svidzinsky2021,Zhang2022LSA,schlawin2026JPCL}.
Quantum-enhanced broadband SRS bioimaging has also been
implemented using amplitude-squeezed light to improve the
signal-to-noise ratio~\cite{chekhova2026OQ}.
In a recently conducted theoretical investigation, it has been demonstrated
that the resonant two-photon ionization cross section for an atom such as sodium
is expected to be enhanced by more than two orders of magnitude,
relative to an uncorrelated reference state,  when driven by
momentum-entangled photon pairs emitted by ultrathin nonlinear
crystals~\cite{Kosheleva2026}. This enhancement is a consequence of
the broad distribution of source-photon transverse momentum and is
therefore revealed only in a nonparaxial theoretical treatment of the process.
This theoretical prediction highlights the potential benefits of using emerging nonclassical
light sources based on ultrathin crystals, metasurfaces, and
subwavelength nonlinear films, which naturally operate in
this regime (Sec.~\ref{sssec:EPPs-prop-novel}).

\begin{backmatter}

\bmsection{FUNDING} U.S. National Science Foundation (ECS-9800300); Boston
University (Photonics Center); U.S. National Science Foundation (CenSSIS ERC);
United States Army Research Office (MURI); David and Lucile Packard Foundation
(1999-8305).

\bmsection{ACKNOWLEDGMENT} We are grateful for the constructive
feedback provided by Roberto de J. León Montiel, Jan Pe{\v r}ina Jr., John Rarity,
Michael Raymer, Frank Schlawin, David Simon, and Dmitry Strekalov.
We are indebted to Professor Hirokazu Kobayashi at Kochi University of Technology
for graciously providing us with English translations of
Refs.~\cite{kobayashi2006,kobayashi2007}.

\bmsection{DISCLOSURES} The authors declare no conflicts of interest.

\bmsection{DATA AVAILABILITY} Data underlying the results presented in this paper
are not publicly available at this time but may be obtained from the authors upon
reasonable request.
\end{backmatter}

\bibliography{BIB-2026-AOP}

\end{document}